\documentclass[fleqn,usenatbib]{mnras}
\usepackage{natbib}
\usepackage{amsmath}
\usepackage{amssymb}
\usepackage{graphicx}
\usepackage[flushleft]{threeparttable}

\def\agridline#1{\vskip0pt\hbox to\hsize{#1}\vskip0pt}
\def\afig#1#2{\hfill\vbox{\parskip=0pt\hsize=#2
\includegraphics[width=#2]{#1}\vskip0pt{\centering
\footnotesize
\hsize=#2
\vskip0pt
}}\hfill}

\defcitealias{2018ApJ...858...33C}{Paper I}

\title[X-ray Signature of IMBHs in Globular Clusters]{A {\it Chandra} Survey of Milky Way Globular Clusters. III. Searching for X-ray Signature of Intermediate-mass Black Holes}
\author[Su et al.]{
Zhao Su$^{1,2}$\thanks{Email: suzhao@smail.nju.edu.cn},
Zhiyuan Li$^{1,2}$\thanks{Email: lizy@nju.edu.cn},
Meicun Hou$^{1,2,3}$,
Mengfei Zhang$^{1,2}$, and
Zhongqun Cheng$^{4,5}$\\
$^{1}$School of Astronomy and Space Science, Nanjing University, Nanjing 210046, China\\
$^{2}$Key Laboratory of Modern Astronomy and Astrophysics (Nanjing University), Ministry of Education, Nanjing 210046, China\\
$^{3}$Kavli Institute for Astronomy and Astrophysics, Peking University, Beijing 100871, China\\
$^{4}$School of Physics and Technology, Wuhan University, Wuhan 430072, China\\
$^{5}$WHU-NAOC Joint Center for Astronomy, Wuhan University, Wuhan 430072, China}
\date{Accepted XXX. Received YYY; in original form ZZZ}

\begin{document}
\label{firstpage}
\pagerange{\pageref{firstpage}--\pageref{lastpage}}
\maketitle

%
%
%
%
%
%

\begin{abstract}
Globular clusters (GCs) are thought to harbor the long-sought population of intermediate-mass black holes (IMBHs). We present a systematic search for a putative IMBH in 81 Milky Way GCs, based on archival {\it Chandra} X-ray observations. 
We find in only six GCs a significant X-ray source positionally coincident with the cluster center, which have 0.5--8 keV luminosities between $\sim1\times 10^{30}~{\rm erg~s^{-1}}$ to $\sim 4\times10^{33}~{\rm erg~s^{-1}}$.
However, the spectral and temporal properties of these six sources can also be explained in terms of binary stars.  
The remaining 75 GCs do not have a detectable central source, most with $3\sigma$ upper limits ranging between $10^{29-32}~{\rm erg~s^{-1}}$ over 0.5--8 keV, which are significantly lower than predicted for canonical Bondi accretion.
To help understand the feeble X-ray signature, we perform hydrodynamic simulations of stellar wind accretion onto a $1000~{\rm M_\odot}$ IMBH from the most-bound orbiting star, for stellar wind properties consistent with either a main-sequence (MS) star or an asymptotic giant branch (AGB) star. 
We find that the synthetic X-ray luminosity for the MS case ($\sim 10^{19}\rm~erg~s^{-1}$) is far below the current X-ray limits. 
The predicted X-ray luminosity for the AGB case ($\sim 10^{34}\rm~erg~s^{-1}$), on the other hand, is compatible with the detected central X-ray sources, in particular the ones in Terzan 5 and NGC 6652.
However, the probability of having an AGB star as the most-bound star around the putative IMBH is very low.
Our study strongly suggests that it is very challenging to detect the accretion-induced X-ray emission from IMBHs, even if they were prevalent in present-day GCs. 
\end{abstract}


\section{Introduction}\label{sec:introduciton}

Intermediate-mass black holes (IMBHs), generally defined as black holes having a mass of $10^2~{\rm M_\odot} \lesssim M \lesssim 10^5~{\rm M_\odot}$, are the long-sought population to fill a substantial gap in the black hole mass function, which is currently characterized by stellar-mass black holes ($M \sim10~\rm M_\odot$) and super-massive black holes (SMBHs; $10^5~{\rm M_\odot} \lesssim M \lesssim 10^{10}~{\rm M_\odot}$).
It is generally thought that present-day SMBHs have grown from seed BHs of much lower masses in the early universe, where physical conditions could favor the formation of IMBHs (see recent review by \citealp{2020ARA&A..58...27I}). 
One of the promising formation channels of IMBHs is via gravitational runaway processes in a dense, massive stellar cluster (see recent review by \citealp{2020ARA&A..58..257G}), in which heavy objects tend to segregate to the cluster center due to dynamic friction.
In particular, an IMBH with $10^2~{\rm M_\odot} \lesssim M \lesssim 10^4~{\rm M_\odot}$ may form via runaway mergers of massive stars and the subsequent collapse of the hyper-massive merger product \citep{2002ApJ...576..899P, 2004Natur.428..724P}, or form via successive mergers of 
a $\gtrsim50~{\rm M_\odot}$ seed black hole with lower-mass black holes that have mass-segregated to the cluster center \citep{2002MNRAS.330..232C}. 


If massive star clusters, which can be the progenitor of present-day globular clusters (GCs), were indeed an efficient IMBH factory, a fraction of the IMBHs thus formed may still reside in their birthplace.
This has motivated systematic searches for IMBHs in present-day GCs, especially those of our own Galaxy.
Two primary approaches are in line with this continued effort.
The {\it dynamical} approach probes the gravitational influence of a putative IMBH on the kinematic and photometric properties of the cluster stars \citep{2005ApJ...620..238B}, as well as on pulsar timing \citep{1993ASPC...50..141P}. 
This has led to a number of IMBH candidates in typically non-core-collapsed GCs (e.g., $\omega$ Cen, \citealp{2008ApJ...676.1008N, 2010ApJ...719L..60N}; NGC\,6388, \citealp{2013A&A...552A..49L}; M54, \citealp{2009ApJ...699L.169I}; NGC\,5286, \citealp{2013A&A...554A..63F}).
However, an IMBH is often not the only solution to the observed kinematics \citep[e.g.,][]{2019MNRAS.482.4713Z,2019MNRAS.488.5340B}.
Moreover, \citet{2021MNRAS.508.4385A} cautioned the potential effect of binary orbital motion causing an overestimation of the velocity dispersion within the cluster core, which may mimic the presence of an IMBH.

An IMBH, if existed in the cluster center, may also uncover itself when inevitably accreting from the ambient material. 
The {\it electromagnetic} approach thus attempts to capture the electromagnetic radiation from an accreting IMBH, which, for typical present-day cluster conditions, is expected to have a rather low accretion rate \citep[e.g.,][]{2003ApJ...587L..35H, 2018MNRAS.481..627A}.
Most efforts of this kind have focused on the radio or X-ray band, by analogy with weakly accreting SMBHs, of which the bolometric luminosity is only a small fraction of the Eddington luminosity ($L_{\rm bol}\lesssim 0.01L_{\rm Edd}$). 
Radio and X-ray emission are robust and direct tracers of such dormant black holes, with the former produced by a relativistic jet and the latter produced by the accretion flow and/or the jet  \citep{2003MNRAS.345.1057M, 2004A&A...414..895F,2009ApJ...703.1034Y}. 
In the radio perspective, 
\citet{2012ApJ...750L..27S} reported non-detection of radio emission from the center of three Galactic GCs (M15, M19 and M22) in deep Very Large Array (VLA) images.
More recently, \citet{2018ApJ...862...16T} conducted a deep radio survey for the signatures of IMBHs in 50 Galactic GCs, but no significant central radio source was detected. 
The GC systems of several nearby galaxies have also been searched for radio emission from a putative IMBH, but no significant signal was detected from either individual GCs or the stacked VLA image (NGC\,1023, \citealp{2015AJ....150..120W}; M81, \citealp{2016AJ....152...22W}; NGC\,3115, \citealp{2020ApJ...900..134W}).
An upper limit on the black hole mass, typically on the order of $\sim 1000\rm~M_\odot$, was yielded from these radio non-detections, invoking the so-called fundamental plane of black hole activity \citep{2003MNRAS.345.1057M,2004A&A...414..895F}, an empirical relation among black hole mass, radio luminosity and X-ray luminosity.
This methodology, however, heavily depends on {\it ad hoc} assumptions about the intra-cluster medium (ICM), the accretion rate and the accretion-induced X-ray luminosity \citep{2004MNRAS.351.1049M,2013ApJ...776..118S}, of which strong observational constraints are still absent for most Galactic GCs.

The {\it Chandra X-ray Observatory} is most useful for probing IMBHs in the X-ray band, thanks to its superb angular resolution and sensitivity that are crucial for probing faint X-ray sources in the crowding cluster center.
{\it Chandra} observations have ruled out a detectable X-ray counterpart in several Galactic GCs that are suggested to host an IMBH via dynamical measurements (e.g., 47 Tuc, \citealp{2001Sci...292.2290G}; M 15, \citealp{2003ApJ...587L..35H}; M 54, \citealp{2011AJ....142..113W}; $\omega$ Cen, \citealp{2013ApJ...773L..31H}).
\citet{2008A&A...478..763N} found X-ray emission coincident with the center of NGC\,6388, but no radio counterpart was detected \citep{2010MNRAS.406.1049C}.
IMBH-induced X-ray emission has also been searched for in extragalactic GCs.
Perhaps the best-known such case is related to G1, one of the most massive GCs in M31. 
{\it Chandra} observations revealed an X-ray source coincident with the core of G1 and having a 0.3--7 keV luminosity of $\sim2\times10^{36}{\rm~erg~s^{-1}}$, which may be due to a putative IMBH or an ordinary low-mass X-ray binary (LMXB; \citealp{2010MNRAS.407L..84K}).  
However, high-resolution VLA observations did not detect any significant radio emission from the core of G1, casting doubt on the IMBH interpretation \citep{2012ApJ...755L...1M}.

Besides an IMBH, whether present-day GCs can retain a significant number of stellar-mass BHs has also been debated for decades.
Early theoretical works \citep{1993Natur.364..421K, 1993Natur.364..423S} suggested that nearly all stellar-mass BHs would escape the GC in the first Gyr due to \citet{1969ApJ...158L.139S} instability.
However, recent numerical simulations showed that stellar-mass BHs can survive for a long time, forming a BH subsystem in the GC center due to mass segregation \citep[e.g.,][]{2013MNRAS.432.2779B, 2015ApJ...800....9M, 2018MNRAS.478.1844A}.
Recent observational studies also identified several stellar-mass BH candidates in Galactic (e.g., 47 Tuc, \citealp{2015MNRAS.453.3918M}; NGC\,3201, \citealp{2018MNRAS.475L..15G,2019A&A...632A...3G}; NGC\,6254 [M10], \citealp{2018ApJ...855...55S}; NGC\,6266 [M62], \citealp{2013ApJ...777...69C}; NGC\,6397, \citealp{2020MNRAS.493.6033Z}; NGC\,6656 [M22], \citealp{2012Natur.490...71S}) and extragalactic \citep[e.g.,][]{2007Natur.445..183M, 2019MNRAS.485.1694D} GCs.
This indicates that many GCs may host a large population of stellar-mass BHs at present, although the exact retention fraction is still unclear and might be strongly influenced by an IMBH, if existed \citep[e.g.,][]{2014MNRAS.444...29L,2022MNRAS.514.5879M}.

A systematic search for the putative X-ray signals from accreting IMBHs in Galactic GCs is still missing. 
We are thus motivated to take up such a task in this work, utilizing archival {\it Chandra} observations. This is also the third paper in a series of an archival {\it Chandra} survey of X-ray emission from Galactic GCs. In our previous work \citep{2018ApJ...858...33C,2018ApJ...869...52C},
we studied the emissivity and abundance of faint X-ray sources, mostly cataclysmic variables (CVs) and coronally active binaries (ABs), in a sample of 69 Galactic GCs.

The remainder of this work is organized as follows. 
Section \ref{sec:data} describes our GC sample selection and reduction of the relevant {\it Chandra} data. 
Section \ref{sec:analysis} presents an analysis of the X-ray emission from the cluster center, deriving photometric, timing and spectral properties. It is found that the majority of our sample GCs exhibit no significant X-ray emission from the position of a putative IMBH. 
Therefore, in Section \ref{sec:simulation}, we perform hydrodynamic simulations of IMBH accretion in a physically motivated GC environment, to help understand the feeble X-ray emission.
Implications of the results are addressed in Section \ref{sec:discussion}, followed by a summary in Section \ref{sec:summary}.
\\

\section{Sample selection and data preparation}\label{sec:data}

Our sample selection started from the catalog of \citet[2010 edition]{1996AJ....112.1487H}, which contains 157 GCs.
We cross-correlated this catalog with the {\it Chandra} public archive, finding 87 GCs with at least one observation taken with the Advanced CCD Imaging Spectrometer (ACIS) as of June 2021.
After visual examination, we removed several observations, which include all exposures for NGC 1851, Terzan 2, NGC 6441, Terzan 6, NGC 6712, and NGC 7078, and partial exposures for NGC 6388, NGC 6440, Terzan 1, and Terzan 5, in which a bright source, most likely a LMXB \citep{2007A&A...469..807L}, is present near the cluster center and outshines other sources.
Our final sample contains 81 GCs with a total of 169 observations. 
Compared to \citet[][hereafter Paper I]{2018ApJ...858...33C}, the current sample includes 12 more GCs.
Basic information of the sample GCs, primarily obtained from \citet[2010 edition]{1996AJ....112.1487H} and \citet{2019MNRAS.482.5138B}, as well as the {\it Chandra} observations, are given in Table \ref{tab:observation}.

We downloaded and reprocessed the {\it Chandra} data with CIAO v4.12 and calibration files CALDB v4.9.5, following the standard procedure\footnote{\url{https://cxc.harvard.edu/ciao/}}. 
After generating the level-2 events files, we examined the light curve of each observation and filtered time intervals that suffer from significant particle flares, if any.
The cleaned exposure time of each ObsID is given in Table~\ref{tab:observation}.
For each observation, we then produced counts maps and exposure maps, at the natal pixel scale
of 0\farcs492, in three energy bands: 0.5--2 ($S$-band), 2--8 ($H$-band) and 0.5--8 ($F$-band) keV. The exposure maps were weighted by an assumed incident spectrum of an absorbed power-law, with photon-index of 1.7 and absorption column density $N_{\rm H}=10^{21}\rm~cm^{-2}$. 
We also generated for each band and each ObsID maps of the point-spread function (PSF), using the same spectral weighting as for the exposure map. 
For 27 GCs with more than one exposures, 
we calibrated the relative astrometry by matching the centroid of commonly detected point sources, using the CIAO tool {\it reproject\_aspect}. 
We then reprojected the individual counts maps or exposure maps to a common tangential point, i.e., the cluster center, to produce a combined image. The PSF maps were similarly combined, with weights according to the local effective exposure.


Following the methods of \citet{2019ApJ...876...59C}, we performed source detection over the ACIS field-of-view (FoV) of each GC. 
For the 27 GCs with more than one exposures, source detection was performed over the combined image of enhanced signal-to-noise ratio (S/N).
Specifically, we run the CIAO tool {\it wavdetect} to detect point sources in each of the three bands. 
A 50\% enclosed count radius (ECR) was adopted for the PSF map in this step, along with a false detection probability of $10^{-6}$, to give tolerance to source crowing in the cluster core, which is the focus of this work.
We defer a detailed presentation of the detected X-ray sources to a future work (Z. Cheng et al. in preparation).
We note in passing that \citet{2020ApJ...901...57B} have recently presented a catalog of X-ray point sources in 38 GCs based on {\it Chandra}/ACIS observations, but they did not specifically examine  
potential X-ray counterparts of the putative IMBHs.
We contrast our detection of cluster central sources with the catalog of \citet{2020ApJ...901...57B} in Section~\ref{subsec:detection}.
\\

\section{X-ray Emission from the Cluster Center}
\label{sec:analysis}
\subsection{Detection of central X-ray sources} \label{subsec:detection}
The identification of an X-ray source coincident with the cluster center can be subject to three terms of uncertainty: (i) the exact position of the cluster center, (ii) the relative astrometry of the {\it Chandra} image, and (iii) a random drift of the putative IMBH due to Brownian motion induced by the surrounding stars \citep{1980ApJ...242..789L}, which is a physical effect. 

For (i), we follow the common practice of adopting the cluster center coordinates given in the catalog of \citet[2010 edition]{1996AJ....112.1487H}.
Among our sample GCs, $\sim 50\%$ have a centroid position determined from high-resolution image of the {\it Hubble Space Telescope} (HST), which has a typical uncertainty below $1\arcsec$ \citep{2010AJ....140.1830G}.
For (ii), the accuracy of {\it Chandra}/ACIS pointing is empirically determined to be $\lesssim0\farcs6$ (68\% limit)\footnote{\url{https://cxc.cfa.harvard.edu/proposer/POG/}}. 
In principle, the astrometry of the {\it Chandra} images can be improved given a sufficient number of X-ray-optical counterparts within the FoV, but this is not the case for a large fraction of our sample GCs, hence we adopt the natal astrometry of the {\it Chandra} images.
As for (iii), we estimate the root-mean-square offset of the putative IMBH following \citet{1980ApJ...242..789L},
\begin{equation}
R_{\rm BH,rms} = (\frac{{\pi}m}{6M})^{\frac{1}{2}}R_c \approx 0.02 R_{\rm c},
\label{eqn:Roff}
\end{equation}
where we have assumed a fiducial black hole mass of $M = 1000\rm~M_\odot$ and a stellar mass of $m = 1\rm~M_\odot$.
Given the core radius $R_{\rm c}$ of our sample GCs (Table~\ref{tab:observation}), $R_{\rm BH,rms}$ is found to range from $0\farcs01$ to $3\farcs7$, with a median value of $0\farcs32$.
For 16 GCs, $R_{\rm BH,rms}$ is greater than $1\arcsec$ due primarily to their proximity.

\begin{figure*}
     \agridline{
        \afig{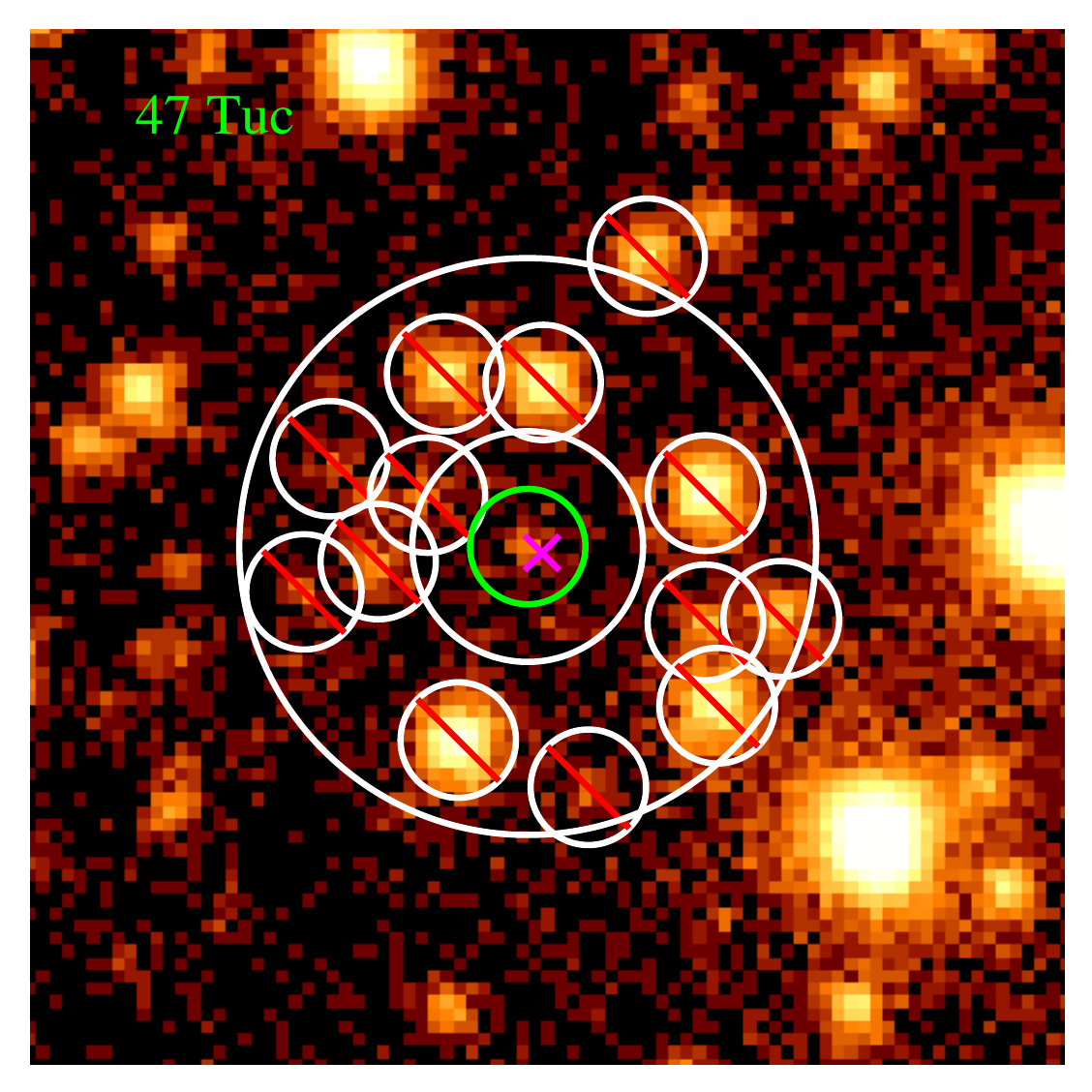}{0.33\textwidth}
        \afig{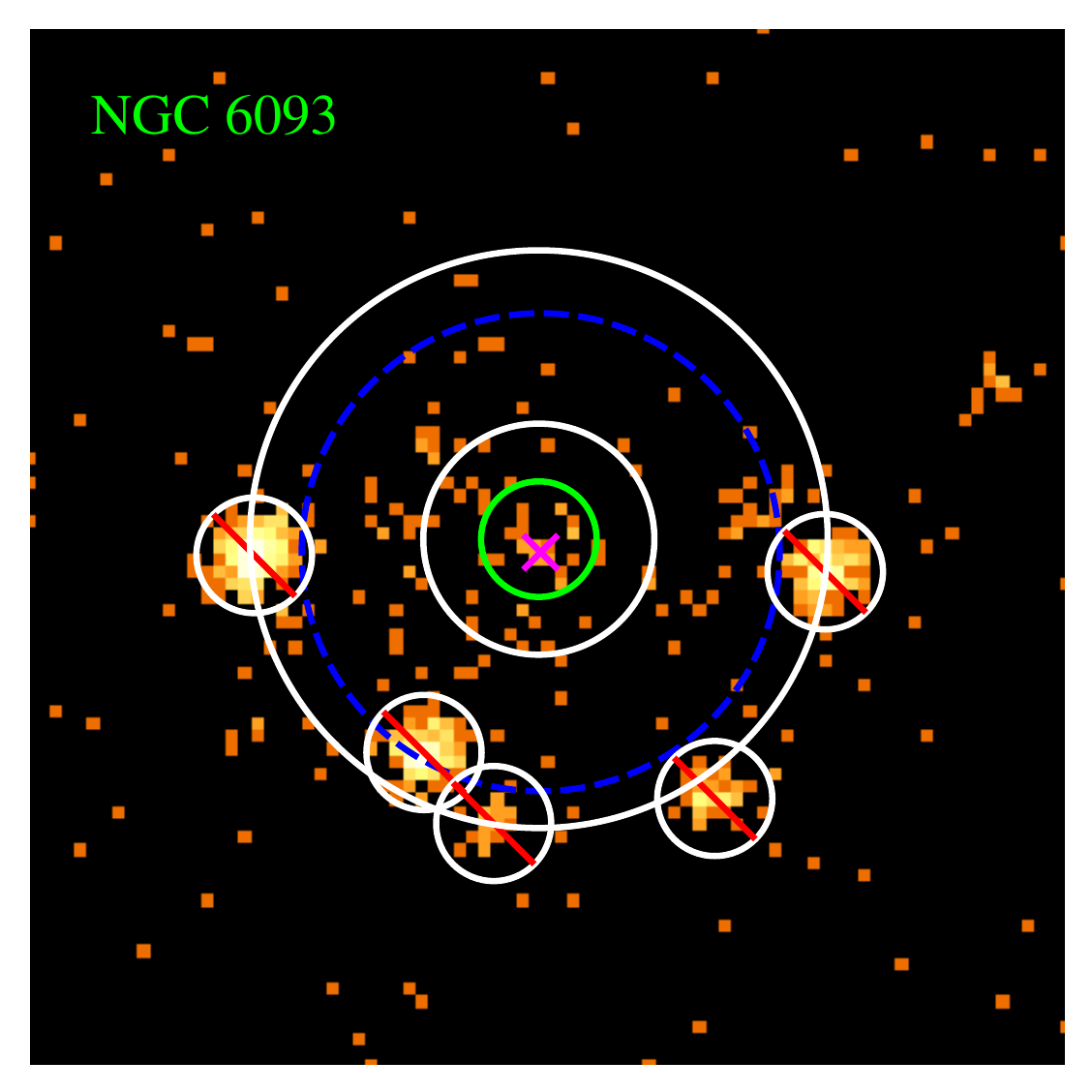}{0.33\textwidth}
        \afig{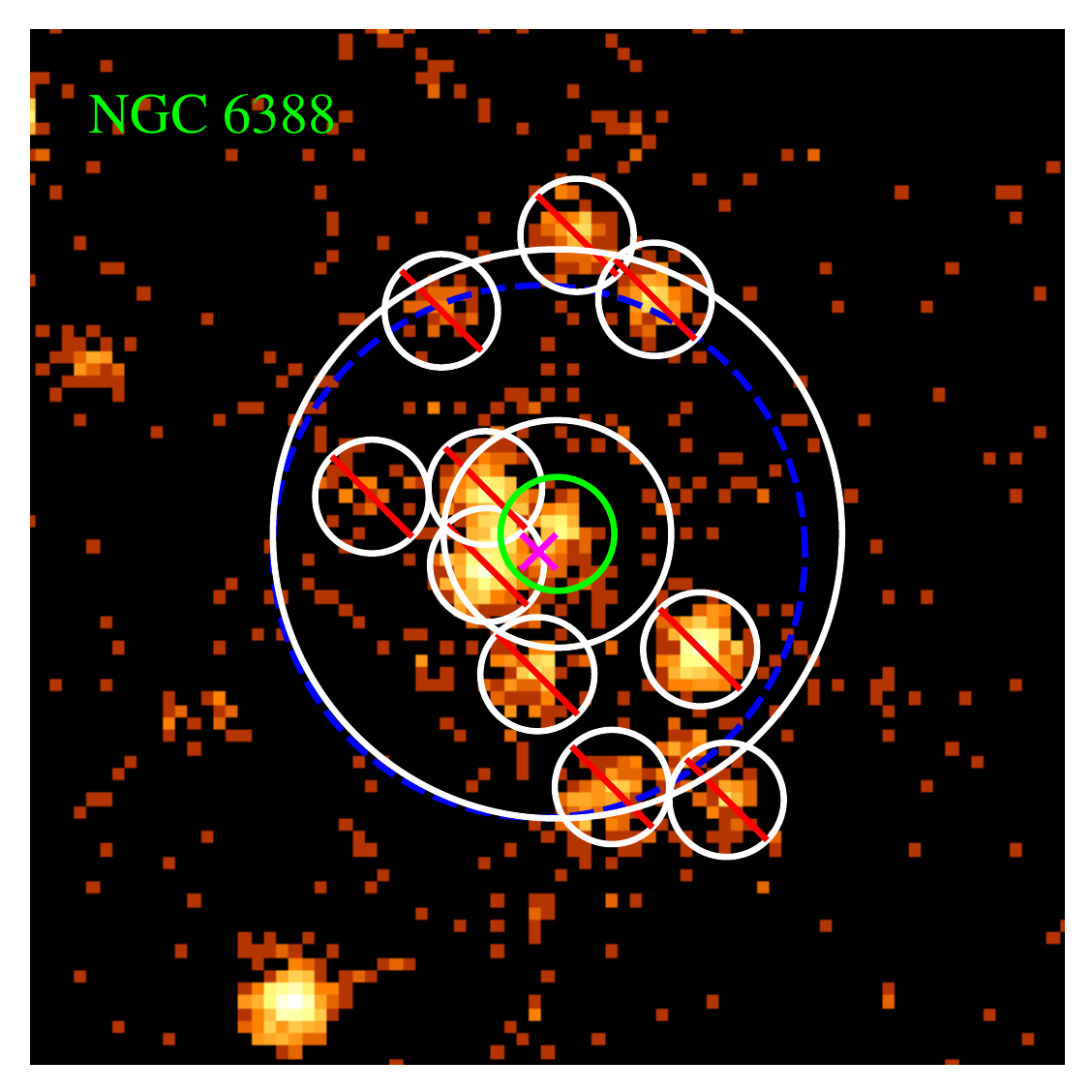}{0.33\textwidth}
        }
        \agridline{
        \afig{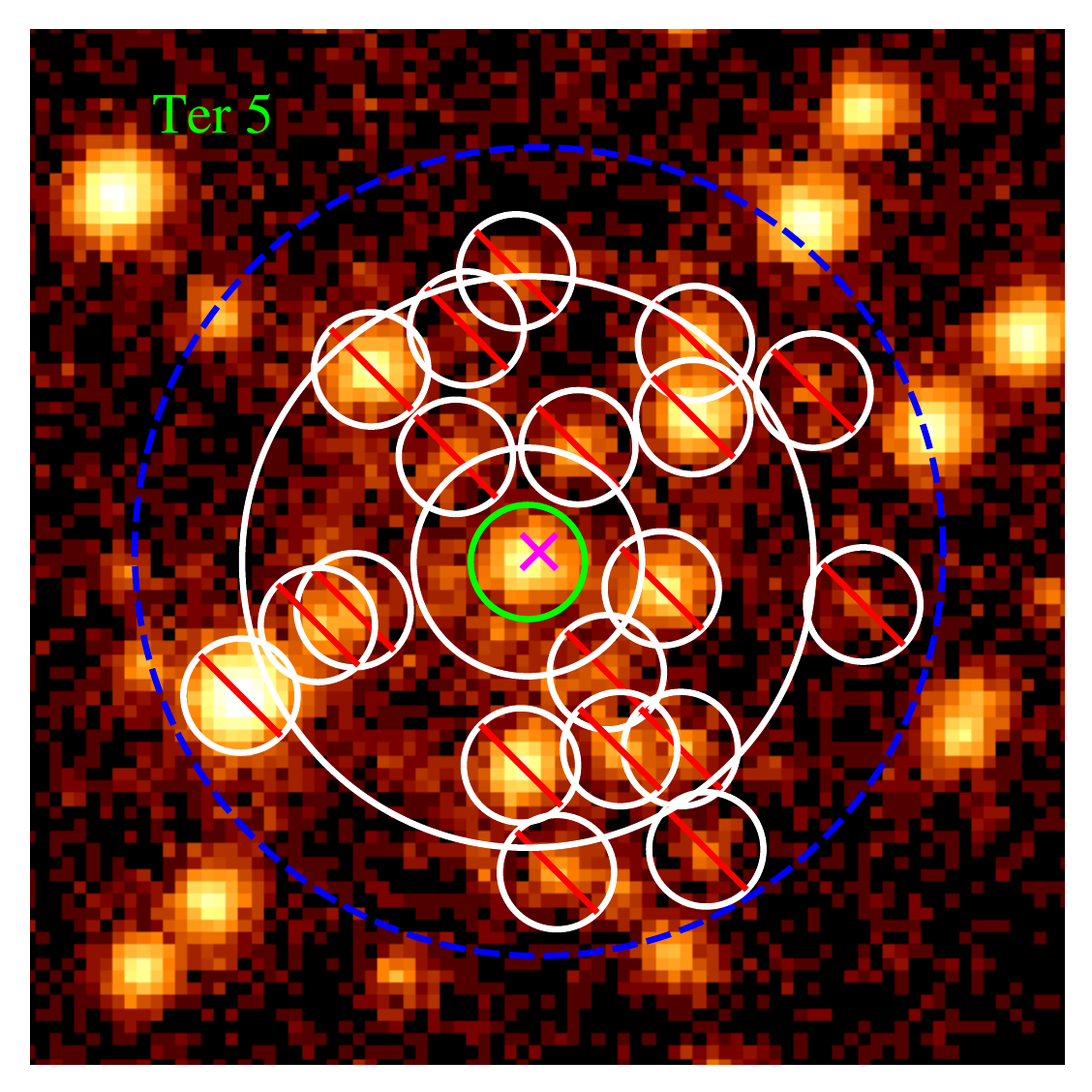}{0.33\textwidth}
        \afig{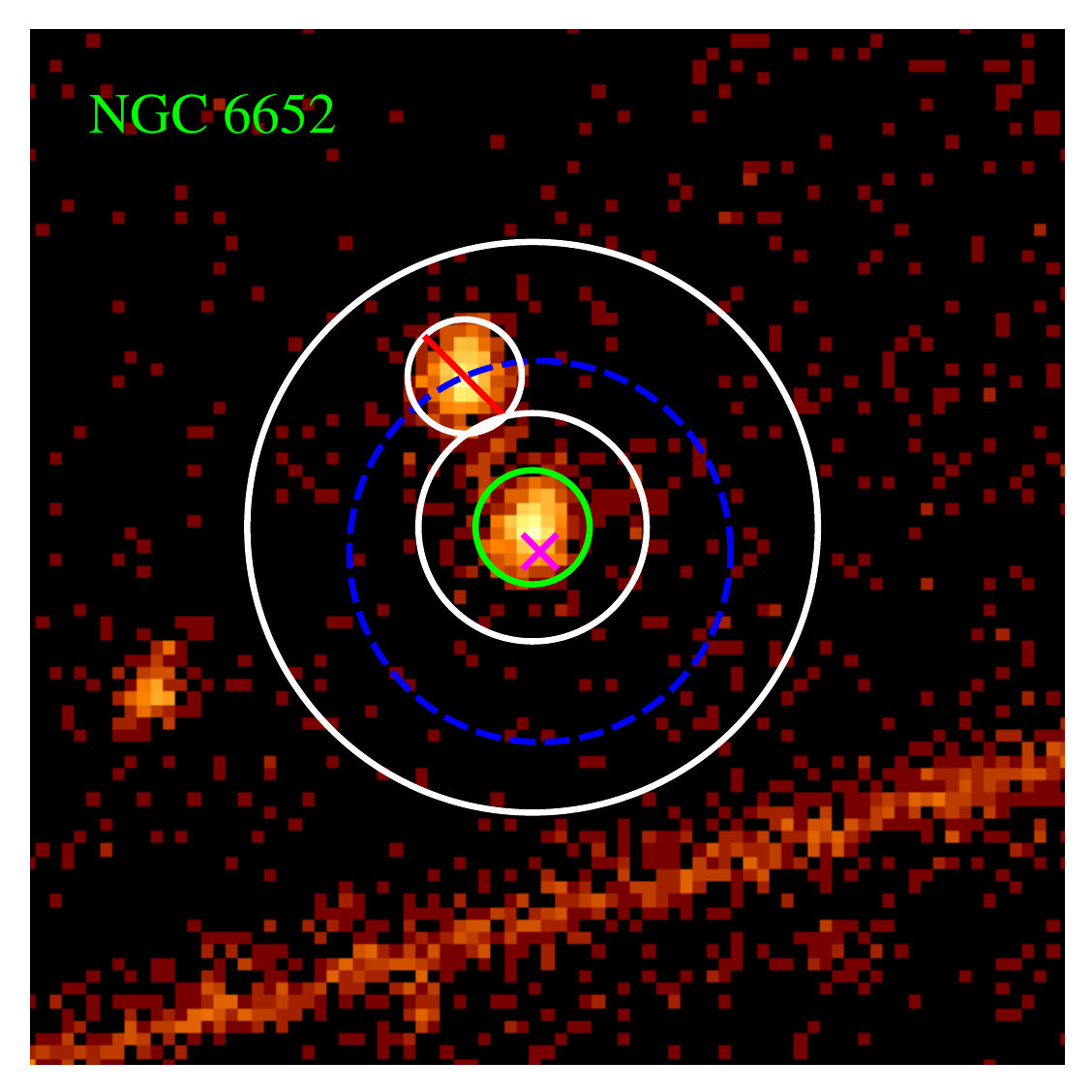}{0.33\textwidth}       
        \afig{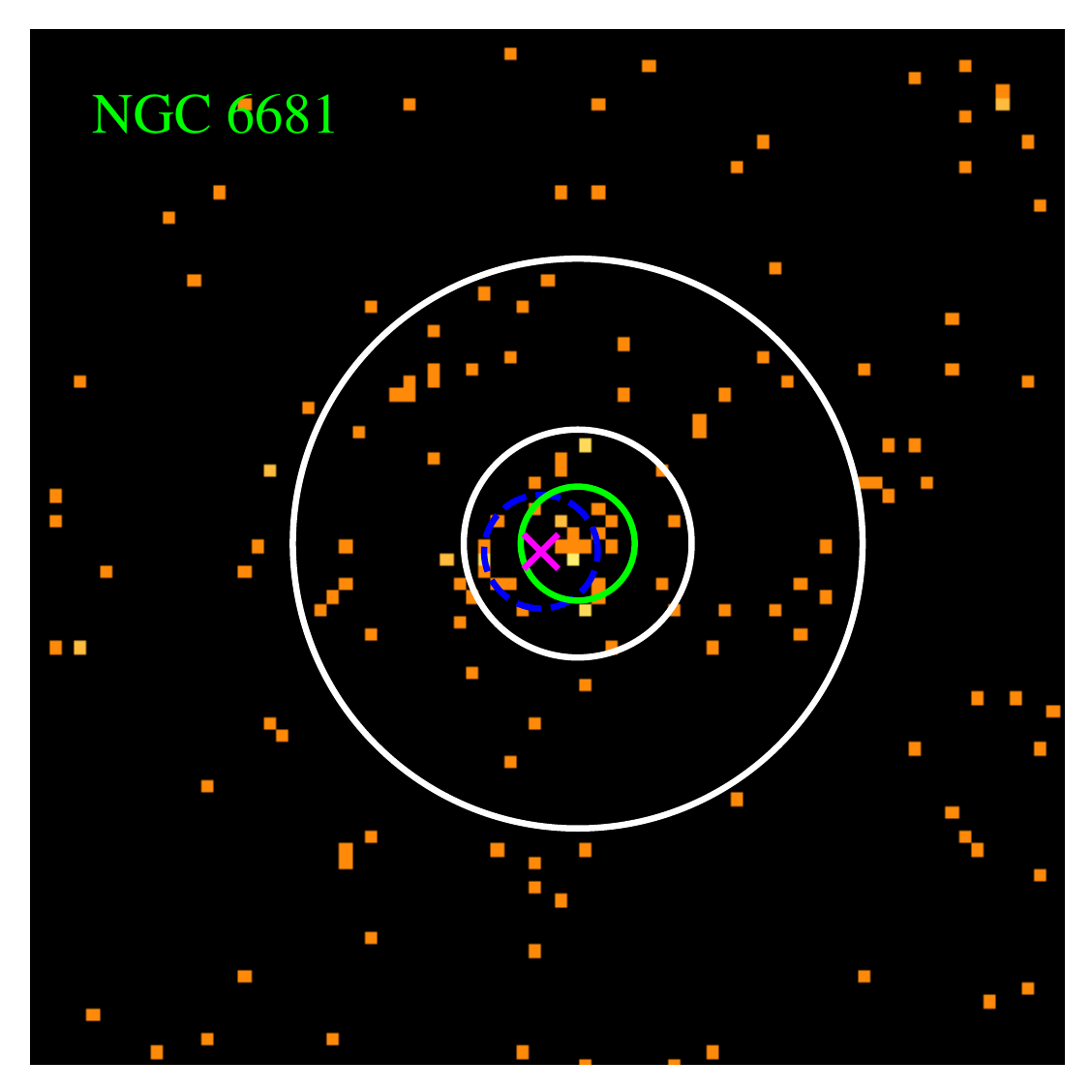}{0.33\textwidth}
        }
    \caption{\textit{Chandra}/ACIS 0.5--8 keV counts images of 6 GCs with a detected source coincident with the cluster center. The images have a size of $20\arcsec\times 20\arcsec$ and a binning of 1/2 natal ACIS pixel. 
    In each panel, the magenta cross marks the putative cluster center and the blue dashed circle marks the core radius (not shown for 47 Tuc, whose core radius is located beyond the image). 
    The green circle outlines the 90\% ECR of the central source for spectral extraction, while the pairs of large white circles enclose the background region.  The small white circles mark the neighboring point sources, which are masked from spectral extraction.
    \label{fig:GCs}}
\end{figure*}

Hence we adopt a matching radius of $max\{1\arcsec, R_{\rm BH,rms}\}$ to search for X-ray point sources positionally coincident with the cluster center. 
Six GCs, including 47 Tuc (=NGC\,104), NGC\,6093, NGC\,6388, Terzan 5, NGC\,6652, and NGC\,6681, are thus found to have one X-ray counterpart detected in at least one of the three energy bands. 
This is illustrated in Figure~\ref{fig:GCs}, which shows the 0.5--8 keV counts image of the core region of the six GCs, with the detected X-ray sources highlighted.
We note that all these six GCs have $R_{\rm BH,rms} < 1\arcsec$.

\begin{table*}
    \centering
    \renewcommand{\thetable}{1}
    \caption{Key information of the GC sample\label{tab:observation}}
   	{\scriptsize
        \begin{tabular}{lllcccclccccc}
       
        \hline
        GC    &          R.A.  & Dec &$D$   &  $M$  &
        $R_{\rm c}$   & $R_{\rm BH,rms}$ &ObsID & Exp.  & $C_{\rm S}$  &$C_{\rm B}$  &
        $L_{\rm X}$ & $L_{\rm R,lim}$ \\
        (1)   &   (2) &  (3) &    (4)  &   (5)  &  (6)   & (7)   & (8)   & (9) &    (10) &  (11) &  (12)& (13) \\
        \hline
NGC 104 &$6.02363$ &$-72.08128$ &$4.4$ &$7.61\pm0.05$ &$23.7$ &$0.54$ &78,953,954 &$522.9$ &$84$ &$47.0$ &$1.27^{+0.37}_{-0.34}$ &$1.4$\\
  &  &  &  &  &  &  &955, 956, 2735  &  &  &  &  &  \\
  &  &  &  &  &  &  &2736, 2737, 2738  &  &  &  &  &  \\
  &  &  &  &  &  &  &3384, 3386, 3387  &  &  &  &  &  \\
  &  &  &  &  &  &  &15747, 15748, 16527  &  &  &  &  &  \\
  &  &  &  &  &  &  &16528, 16529, 17420  &  &  &  &  &  \\
NGC 288 &$13.18850$ &$-26.58261$ &$10.0$ &$1.21\pm0.03$ &$101.7$ &$2.33$ &3777 &$54.3$ &$0$ &$0.3$ &$<10.9$ & \\
NGC 362 &$15.80942$ &$-70.84878$ &$9.2$ &$3.33\pm0.05$ &$8.1$ &$0.19$ &4529,5299 &$76.3$ &$6$ &$5.7$ &$<13.7$ & \\
Pal 2 &$71.52463$ &$+31.38150$ &$27.2$ &$2.32\pm0.42$ &$16.5$ &$0.38$ &9028 &$10.0$ &$0$ &$0.1$ &$<448$ & \\
NGC 1904 &$81.04621$ &$-24.52472$ &$13.3$ &$1.70\pm0.11$ &$6.5$ &$0.15$ &9027 &$10.0$ &$0$ &$0.3$ &$<110$ & \\
NGC 2419 &$114.53529$ &$+38.88244$ &$83.2$ &$12.50\pm1.82$ &$22.2$ &$0.51$ &19490 &$27.0$ &$0$ &$0.3$ &$<2300$ & \\
NGC 2808 &$138.01292$ &$-64.86350$ &$10.2$ &$8.19\pm0.06$ &$14.5$ &$0.33$ &7453,8560 &$56.2$ &$0$ &$0.9$ &$<16.8$ &$1.0$\\
E 3 &$140.23779$ &$-77.28189$ &$8.1$ &$0.03\pm0.01$ &$43.9$ &$1.00$ &7459 &$19.9$ &$0$ &$0.1$ &$<20.1$ & \\
NGC 3201 &$154.40342$ &$-46.41247$ &$4.6$ &$1.34\pm0.05$ &$78.0$ &$1.79$ &11031 &$83.5$ &$0$ &$0.8$ &$<1.91$ &$2.7$\\
NGC 4372 &$186.43917$ &$-72.65900$ &$5.8$ &$2.19\pm0.09$ &$150.0$ &$3.43$ &17843 &$10.3$ &$0$ &$0.1$ &$<27.0$ &$4.9$\\
NGC 4833 &$194.89133$ &$-70.87650$ &$6.2$ &$1.76\pm0.11$ &$55.5$ &$1.27$ &17846 &$11.8$ &$0$ &$0.0$ &$<28.6$ &$2.6$\\
NGC 5024 &$198.23021$ &$+18.16817$ &$17.9$ &$3.44\pm0.31$ &$25.0$ &$0.57$ &6560 &$24.6$ &$0$ &$0.1$ &$<78.5$ & \\
NGC 5139 &$201.69683$ &$-47.47958$ &$5.2$ &$34.00\pm0.25$ &$163.0$ &$3.73$ &653,1519,13726 &$289.3$ &$0$ &$0.8$ &$<0.86$ &$1.3$\\
  &  &  &  &  &  &  &13727  &  &  &  &  &  \\
NGC 5272 &$205.54842$ &$+28.37728$ &$9.6$ &$3.79\pm0.17$ &$20.6$ &$0.47$ &4542,4543,4544 &$29.2$ &$3$ &$0.2$ &$<36.9$ &$3.5$\\
NGC 5286 &$206.61171$ &$-51.37425$ &$11.4$ &$3.85\pm0.18$ &$12.2$ &$0.28$ &8964,9852 &$13.1$ &$1$ &$0.4$ &$<80.4$ & \\
NGC 5824 &$225.99429$ &$-33.06822$ &$31.8$ &$7.43\pm0.41$ &$0.5$ &$0.01$ &9026 &$10.7$ &$3$ &$0.3$ &$<1100$ & \\
NGC 5904 &$229.63842$ &$+2.08103$ &$7.6$ &$3.66\pm0.06$ &$32.4$ &$0.74$ &2676 &$44.3$ &$0$ &$0.6$ &$<8.73$ &$1.5$\\
NGC 5927 &$232.00287$ &$-50.67303$ &$9.1$ &$2.54\pm0.02$ &$32.0$ &$0.73$ &8953,13673 &$55.2$ &$0$ &$0.6$ &$<9.67$ &$4.0$\\
NGC 5946 &$233.86883$ &$-50.65967$ &$10.6$ &$1.15\pm0.22$ &$6.4$ &$0.15$ &9956 &$24.7$ &$2$ &$0.7$ &$<44.5$ & \\
NGC 6093 &$244.26004$ &$-22.97608$ &$8.9$ &$2.80\pm0.08$ &$4.7$ &$0.11$ &1007 &$48.6$ &$27$ &$4.9$ &$33.5^{+8.3}_{-7.5}$ & \\
NGC 6121 &$245.89675$ &$-26.52575$ &$1.9$ &$0.91\pm0.02$ &$56.6$ &$1.30$ &946,7446,7447 &$119.2$ &$2$ &$1.7$ &$<0.27$ &$0.1$\\
NGC 6144 &$246.80775$ &$-26.02350$ &$8.9$ &$0.53\pm0.19$ &$44.0$ &$1.01$ &7458 &$54.1$ &$0$ &$0.4$ &$<8.99$ & \\
NGC 6139 &$246.91821$ &$-38.84875$ &$9.8$ &$3.41\pm0.53$ &$6.3$ &$0.14$ &8965 &$17.7$ &$2$ &$1.8$ &$<48.5$ &$7.0$\\
Ter 3 &$247.16700$ &$-35.35347$ &$8.1$ &$0.50\pm0.19$ &$60.6$ &$1.39$ &11026 &$65.2$ &$0$ &$0.5$ &$<6.43$ & \\
NGC 6171 &$248.13275$ &$-13.05378$ &$6.0$ &$0.88\pm0.05$ &$29.4$ &$0.67$ &17845 &$11.8$ &$0$ &$0.1$ &$<25.8$ &$1.3$\\
NGC 6205 &$250.42183$ &$+36.45986$ &$6.8$ &$4.61\pm0.20$ &$49.1$ &$1.12$ &5436,7290 &$54.7$ &$0$ &$0.4$ &$<5.01$ &$1.7$\\
NGC 6218 &$251.80908$ &$-1.94853$ &$4.7$ &$0.81\pm0.04$ &$8.0$ &$0.18$ &4530 &$26.6$ &$0$ &$0.2$ &$<5.05$ &$0.7$\\
NGC 6254 &$254.28771$ &$-4.10031$ &$5.0$ &$1.88\pm0.04$ &$37.4$ &$0.86$ &16714 &$32.3$ &$0$ &$0.4$ &$<5.86$ &$0.6$\\
NGC 6256 &$254.88592$ &$-37.12139$ &$6.4$ &$1.02\pm0.32$ &$2.6$ &$0.06$ &8951 &$9.4$ &$4$ &$0.8$ &$<56.6$ & \\
NGC 6266 &$255.30333$ &$-30.11372$ &$6.4$ &$6.74\pm0.05$ &$11.6$ &$0.27$ &2677,15761 &$144.0$ &$30$ &$23.2$ &$<9.44$ &$1.8$\\
NGC 6273 &$255.65750$ &$-26.26797$ &$8.3$ &$6.57\pm0.34$ &$24.9$ &$0.57$ &17848 &$22.7$ &$0$ &$0.1$ &$<25.2$ &$1.9$\\
NGC 6287 &$256.28804$ &$-22.70836$ &$9.4$ &$1.33\pm0.28$ &$9.4$ &$0.22$ &13734 &$40.0$ &$0$ &$0.7$ &$<15.0$ & \\
NGC 6293 &$257.54250$ &$-26.58208$ &$9.8$ &$2.65\pm0.21$ &$6.1$ &$0.14$ &8962 &$9.9$ &$1$ &$0.5$ &$<77.4$ & \\
NGC 6304 &$258.63437$ &$-29.46203$ &$5.8$ &$1.46\pm0.13$ &$15.7$ &$0.36$ &8952,11073 &$102.4$ &$2$ &$1.3$ &$<4.48$ &$1.2$\\
NGC 6341 &$259.28079$ &$+43.13594$ &$8.4$ &$3.11\pm0.04$ &$15.4$ &$0.35$ &3778,5241 &$51.6$ &$3$ &$3.2$ &$<13.8$ &$1.7$\\
NGC 6325 &$259.49671$ &$-23.76600$ &$7.8$ &$0.75\pm0.15$ &$8.7$ &$0.20$ &8959 &$16.7$ &$0$ &$0.2$ &$<22.5$ &$1.1$\\
NGC 6333 &$259.79692$ &$-18.51594$ &$8.4$ &$3.06\pm0.24$ &$22.6$ &$0.52$ &8954 &$8.4$ &$0$ &$0.3$ &$<51.7$ &$1.4$\\
NGC 6342 &$260.29200$ &$-19.58742$ &$8.4$ &$0.61\pm0.14$ &$6.1$ &$0.14$ &9957 &$15.8$ &$1$ &$0.5$ &$<36.5$ & \\
NGC 6355 &$260.99412$ &$-26.35342$ &$8.7$ &$1.26\pm0.29$ &$6.4$ &$0.15$ &9958 &$22.8$ &$0$ &$0.3$ &$<20.4$ & \\
NGC 6352 &$261.37129$ &$-48.42217$ &$5.9$ &$0.61\pm0.06$ &$23.5$ &$0.54$ &13674 &$19.8$ &$0$ &$0.2$ &$<11.7$ &$1.7$\\
NGC 6366 &$261.93433$ &$-5.07986$ &$3.7$ &$0.50\pm0.03$ &$120.0$ &$2.75$ &2678 &$22.0$ &$0$ &$0.5$ &$<4.04$ & \\
NGC 6362 &$262.97913$ &$-67.04833$ &$7.4$ &$1.13\pm0.03$ &$77.1$ &$1.76$ &11024,12038 &$39.5$ &$0$ &$0.6$ &$<8.60$ &$2.7$\\
Liller 1 &$263.35208$ &$-33.38900$ &$8.1$ &$6.56\pm1.20$ &$3.1$ &$0.07$ &22392,23242 &$78.1$ &$3$ &$3.5$ &$<14.8$ &$2.7$\\
Ter 1 &$263.94917$ &$-30.46972$ &$6.7$ &$3.00\pm0.68$ &$43.4$ &$0.99$ &5464,17847,20075 &$47.2$ &$0$ &$0.5$ &$<7.09$ &$1.9$\\
NGC 6388 &$264.07179$ &$-44.73550$ &$10.7$ &$10.50\pm0.09$ &$5.2$ &$0.12$ &5505 &$44.6$ &$115$ &$6.0$ &$560^{+78}_{-74}$ &$6.0$\\
NGC 6402 &$264.40042$ &$-3.24592$ &$9.3$ &$7.49\pm0.39$ &$50.5$ &$1.16$ &8947 &$12.1$ &$0$ &$0.1$ &$<44.0$ &$2.0$\\
NGC 6401 &$264.65250$ &$-23.90950$ &$7.7$ &$2.83\pm0.53$ &$14.7$ &$0.34$ &8948 &$11.1$ &$0$ &$0.3$ &$<33.0$ & \\
NGC 6397 &$265.17537$ &$-53.67433$ &$2.4$ &$0.89\pm0.01$ &$2.5$ &$0.06$ &79,2668,2669 &$338.7$ &$48$ &$45.8$ &$<0.55$ &$0.3$\\
  &  &  &  &  &  &  &7460, 7461  &  &  &  &  &  \\
Ter 5 &$267.02000$ &$-24.77917$ &$5.5$ &$3.89\pm0.42$ &$7.9$ &$0.18$ &3798,10059,13225 &$733.6$ &$1782$ &$125.5$ &$1900^{+88}_{-86}$ &$2.0$\\
  &  &  &  &  &  &  &13252, 13705, 13706  &  &  &  &  &  \\
  &  &  &  &  &  &  &14339, 14475, 14476  &  &  &  &  &  \\
  &  &  &  &  &  &  &14477, 14478, 14479  &  &  &  &  &  \\
  &  &  &  &  &  &  &14625, 15615, 15750  &  &  &  &  &  \\
  &  &  &  &  &  &  &16638, 17779, 18881  &  &  &  &  &  \\
NGC 6440 &$267.21958$ &$-20.36025$ &$8.2$ &$3.83\pm0.51$ &$7.3$ &$0.17$ &947,3799,11802 &$53.2$ &$28$ &$14.5$ &$<39.1$ &$2.8$\\
  &  &  &  &  &  &  &18960  &  &  &  &  &  \\
NGC 6453 &$267.71542$ &$-34.59917$ &$11.6$ &$3.21\pm1.01$ &$12.8$ &$0.29$ &9959 &$20.9$ &$0$ &$1.5$ &$<39.8$ & \\
Ter 9 &$270.41167$ &$-26.83972$ &$7.1$ &$0.04$ &$4.4$ &$0.10$ &9960 &$16.8$ &$1$ &$0.6$ &$<23.9$ & \\
Djorg 2 &$270.45458$ &$-27.82583$ &$6.3$ &$0.63\pm0.34$ &$10.1$ &$0.23$ &17844 &$22.3$ &$0$ &$0.4$ &$<15.9$ &$3.1$\\
NGC 6517 &$270.46050$ &$-8.95878$ &$10.6$ &$3.64\pm0.74$ &$3.1$ &$0.07$ &9597 &$23.6$ &$2$ &$0.6$ &$<47.0$ & \\
NGC 6522 &$270.89175$ &$-30.03397$ &$5.8$ &$3.64\pm0.39$ &$4.3$ &$0.10$ &8963 &$8.0$ &$0$ &$0.0$ &$<25.8$ &$3.4$\\
NGC 6535 &$270.96046$ &$-0.29764$ &$6.5$ &$0.09\pm0.01$ &$2.5$ &$0.06$ &11025 &$52.1$ &$0$ &$0.6$ &$<5.18$ & \\
NGC 6528 &$271.20683$ &$-30.05628$ &$7.5$ &$0.94\pm0.18$ &$4.2$ &$0.10$ &8961,12400 &$66.5$ &$4$ &$3.4$ &$<9.45$ & \\
NGC 6539 &$271.20700$ &$-7.58586$ &$7.8$ &$2.49\pm0.32$ &$22.6$ &$0.52$ &8949 &$15.0$ &$0$ &$0.1$ &$<25.1$ &$1.6$\\
NGC 6540 &$271.53583$ &$-27.76528$ &$5.2$ &$0.49\pm0.14$ &$17.1$ &$0.39$ &8956 &$5.1$ &$1$ &$0.2$ &$<43.6$ & \\
NGC 6544 &$271.83575$ &$-24.99733$ &$2.6$ &$1.16\pm0.11$ &$12.7$ &$0.29$ &5435 &$16.3$ &$0$ &$1.2$ &$<2.46$ &$0.3$\\
NGC 6541 &$272.00983$ &$-43.71489$ &$8.0$ &$2.50\pm0.08$ &$3.6$ &$0.08$ &3779 &$44.5$ &$15$ &$7.1$ &$<30.9$ &$3.9$\\
NGC 6553 &$272.32333$ &$-25.90869$ &$6.8$ &$3.31\pm0.20$ &$26.3$ &$0.60$ &8957,13671 &$36.4$ &$0$ &$0.6$ &$<8.42$ &$2.2$\\
NGC 6558 &$272.57333$ &$-31.76389$ &$7.2$ &$0.37\pm0.08$ &$13.2$ &$0.30$ &9961 &$11.0$ &$2$ &$0.3$ &$<47.7$ & \\
NGC 6569 &$273.41167$ &$-31.82689$ &$10.6$ &$2.33\pm0.24$ &$18.7$ &$0.43$ &8974 &$11.1$ &$1$ &$0.3$ &$<82.5$ & \\
    \hline
    \end{tabular}
    }
\end{table*}

\begin{table*}
    \centering
    \contcaption{}
   	{\scriptsize
        \begin{tabular}{lllcccclccccc}
       
        \hline
        GC    &          R.A.  & Dec &$D$   &  $M$  &
        $R_{\rm c}$   & $R_{\rm BH,rms}$ &ObsID & Exp.  & $C_{\rm S}$  &$C_{\rm B}$  &
        $L_{\rm X}$ & $L_{\rm R,lim}$ \\
        (1)   &   (2) &  (3) &    (4)  &   (5)  &  (6)   & (7)   & (8)   & (9) &    (10) &  (11) &  (12)& (13) \\
        \hline
NGC 6626 &$276.13671$ &$-24.86978$ &$5.4$ &$2.98\pm0.16$ &$6.5$ &$0.15$ &2683,2684,2685 &$321.7$ &$91$ &$88.4$ &$<3.52$ &$0.9$\\
  &  &  &  &  &  &  &9132, 9133, 16748  &  &  &  &  &  \\
  &  &  &  &  &  &  &16749, 16750  &  &  &  &  &  \\
NGC 6638 &$277.73375$ &$-25.49747$ &$10.3$ &$1.82\pm0.29$ &$9.4$ &$0.21$ &8950 &$8.9$ &$0$ &$0.3$ &$<73.1$ & \\
NGC 6637 &$277.84625$ &$-32.34808$ &$8.8$ &$1.46\pm0.18$ &$16.2$ &$0.37$ &8946 &$7.4$ &$1$ &$0.1$ &$<87.4$ & \\
NGC 6642 &$277.97542$ &$-23.47519$ &$8.1$ &$0.38\pm0.11$ &$3.3$ &$0.08$ &8955 &$7.6$ &$0$ &$0.5$ &$<52.1$ & \\
NGC 6652 &$278.94013$ &$-32.99072$ &$10.0$ &$0.63\pm0.14$ &$3.7$ &$0.08$ &9025,12461,18987 &$59.4$ &$700$ &$7.2$ &$3670^{+270}_{-260}$ &$6.4$\\
NGC 6656 &$279.09975$ &$-23.90475$ &$3.2$ &$4.05\pm0.04$ &$58.8$ &$1.34$ &5437,14609 &$100.0$ &$0$ &$0.8$ &$<0.73$ &$0.3$\\
NGC 6681 &$280.80317$ &$-32.29211$ &$9.3$ &$1.15\pm0.02$ &$1.1$ &$0.03$ &8958,9955 &$76.0$ &$16$ &$2.1$ &$16.5^{+5.2}_{-4.5}$ & \\
Glim 01 &$282.20708$ &$-1.49722$ &$4.2$ &$\cdots$ &$35.4$ &$0.81$ &6587,21641 &$75.6$ &$10$ &$3.5$ &$<5.84$ & \\
NGC 6715 &$283.76387$ &$-30.47986$ &$24.1$ &$15.80\pm0.19$ &$5.2$ &$0.12$ &4448 &$29.4$ &$7$ &$1.8$ &$<465$ &$1.0$\\
NGC 6717 &$283.77517$ &$-22.70147$ &$7.1$ &$0.18\pm0.03$ &$5.8$ &$0.13$ &13733 &$18.2$ &$0$ &$0.9$ &$<18.7$ & \\
NGC 6752 &$287.71713$ &$-59.98456$ &$4.2$ &$2.29\pm0.03$ &$7.3$ &$0.17$ &948,6612,19013 &$342.8$ &$25$ &$17.6$ &$<1.95$ &$1.0$\\
  &  &  &  &  &  &  &19014, 20121, 20122  &  &  &  &  &  \\
  &  &  &  &  &  &  &20123  &  &  &  &  &  \\
NGC 6760 &$287.80004$ &$+1.03047$ &$8.0$ &$2.71\pm0.38$ &$20.8$ &$0.47$ &13672 &$51.4$ &$1$ &$0.4$ &$<10.8$ &$1.5$\\
Pal 10 &$289.50875$ &$+18.57167$ &$5.9$ &$0.55\pm0.30$ &$40.6$ &$0.93$ &8945 &$11.1$ &$0$ &$0.1$ &$<19.1$ & \\
NGC 6809 &$294.99879$ &$-30.96475$ &$5.3$ &$1.87\pm0.07$ &$100.4$ &$2.30$ &4531 &$33.7$ &$0$ &$0.3$ &$<4.89$ &$1.0$\\
NGC 6838 &$298.44371$ &$+18.77919$ &$4.0$ &$0.53\pm0.03$ &$49.6$ &$1.14$ &5434 &$52.1$ &$0$ &$0.5$ &$<1.92$ & \\
NGC 7089 &$323.36258$ &$-0.82325$ &$10.5$ &$5.09\pm0.10$ &$12.8$ &$0.29$ &8960 &$11.5$ &$0$ &$0.3$ &$<58.8$ &$3.1$\\
NGC 7099 &$325.09217$ &$-23.17986$ &$8.0$ &$1.31\pm0.07$ &$1.5$ &$0.04$ &2679,18997,20725 &$321.7$ &$55$ &$9.1$ &$<19.1$ &$1.7$\\
  &  &  &  &  &  &  &20726, 20731, 20732  &  &  &  &  &  \\
  &  &  &  &  &  &  &20792, 20795, 20796  &  &  &  &  &  \\
    \hline
    \end{tabular}
    }
    {\footnotesize 
    {\bf Notes.} (1) GC name. (2)-(3) Right ascension and declination of the cluster center, from \citet[2010 edition]{1996AJ....112.1487H}. (4)-(6) Distance from the Sun in units of kpc, total mass in units of $10^{5}~{\rm M_\odot}$, and core radius in units of arcsecond, from \citet{2019MNRAS.482.5138B} except for Glim 01, which is adopted from \citet[2010 edition]{1996AJ....112.1487H}. (7) Root-mean-squre angular offset of the putative IMBH due to Brownian motion, in units of arcsecond. (8)-(9): {\it Chandra} Observation ID and total cleaned exposure in ks. (10) 0.5--8 keV total counts in the source region. (11) 0.5-8 keV scaled background counts. (12) 0.5-8 keV X-ray luminosity  in units of $10^{30}\ {\rm erg\ s^{-1}}$. Errors are of 1$\sigma$, while upper limits are of 3$\sigma$. (13) Upper limts of radio luminosity in units of $10^{27}\ {\rm erg\ s^{-1}}$, from \citet{2018ApJ...862...16T}.}
\end{table*}

\begin{figure*}
    \agridline{      
        \afig{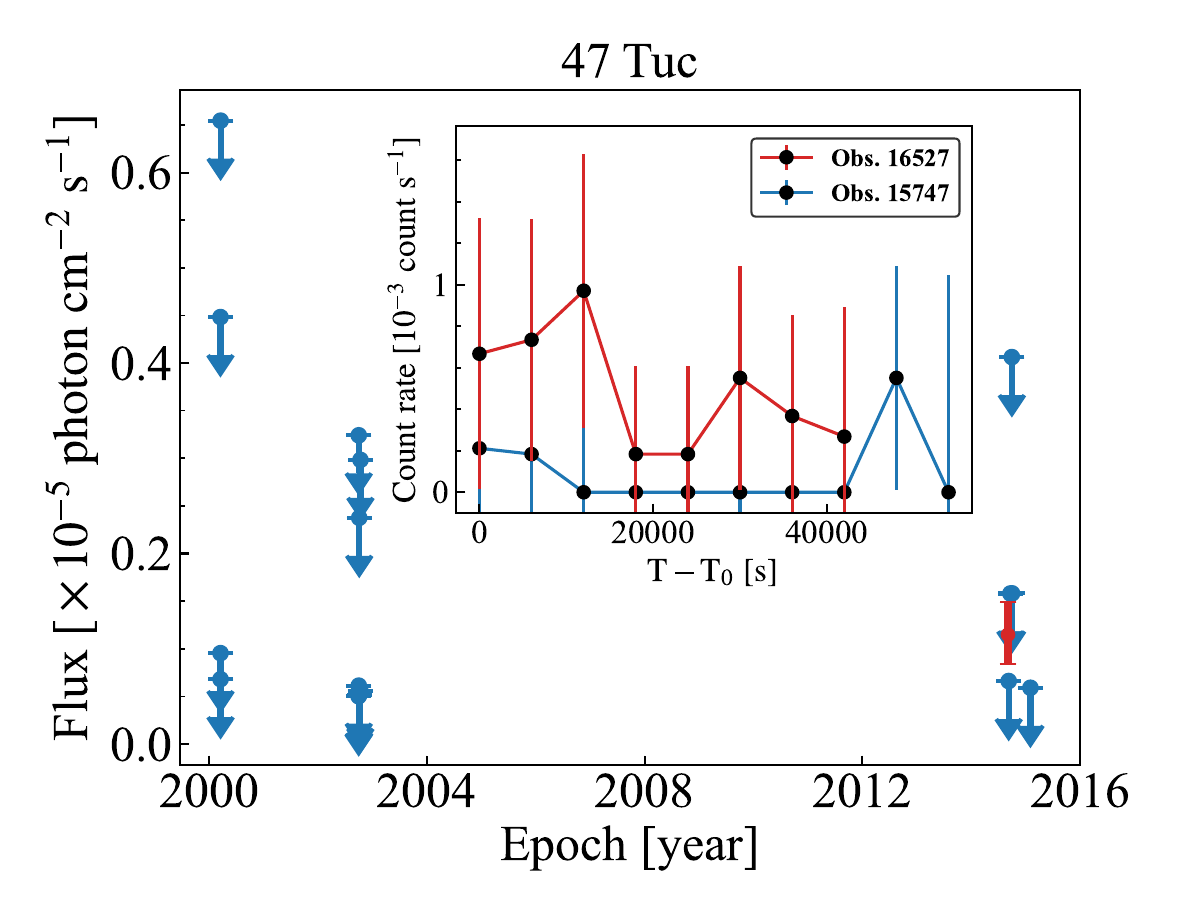}{0.5\textwidth}
        \afig{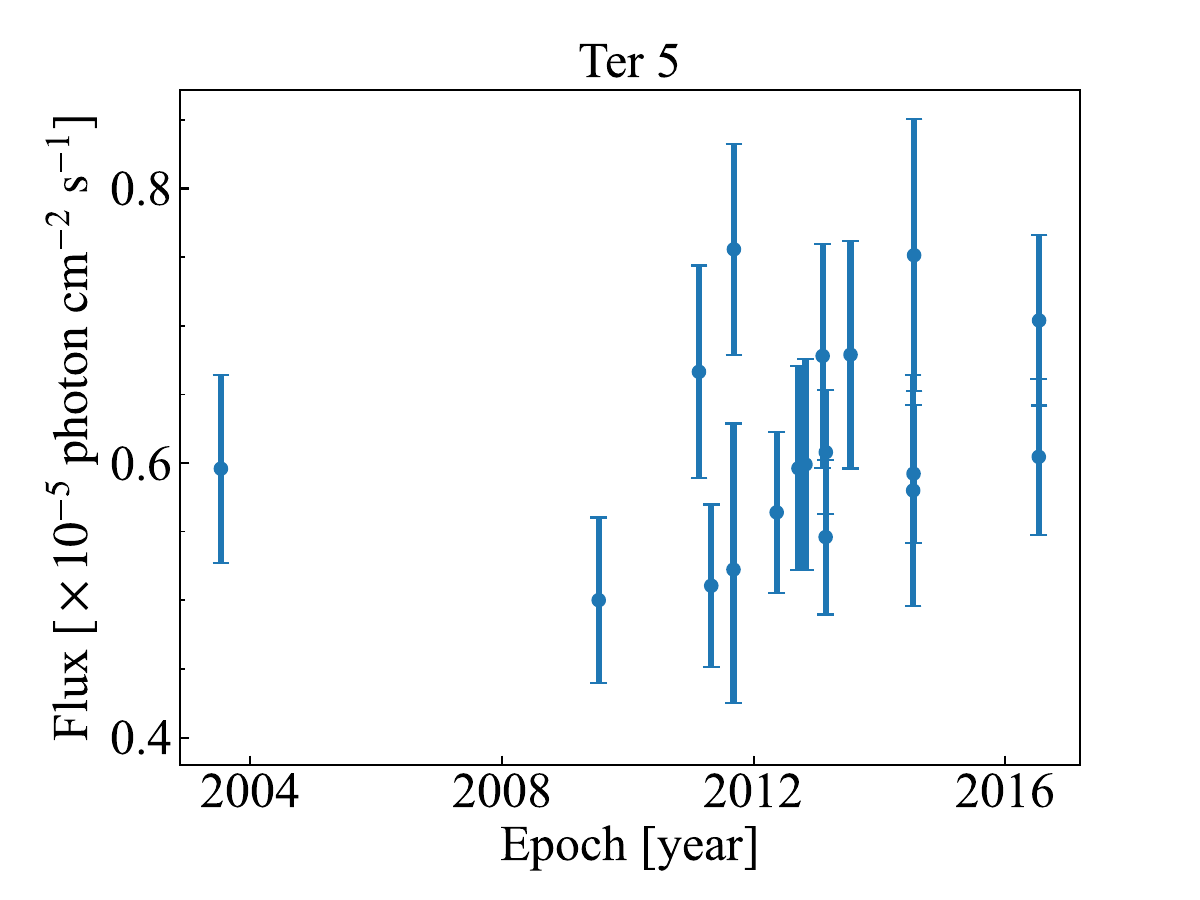}{0.5\textwidth}
        }
     \agridline{
     	\afig{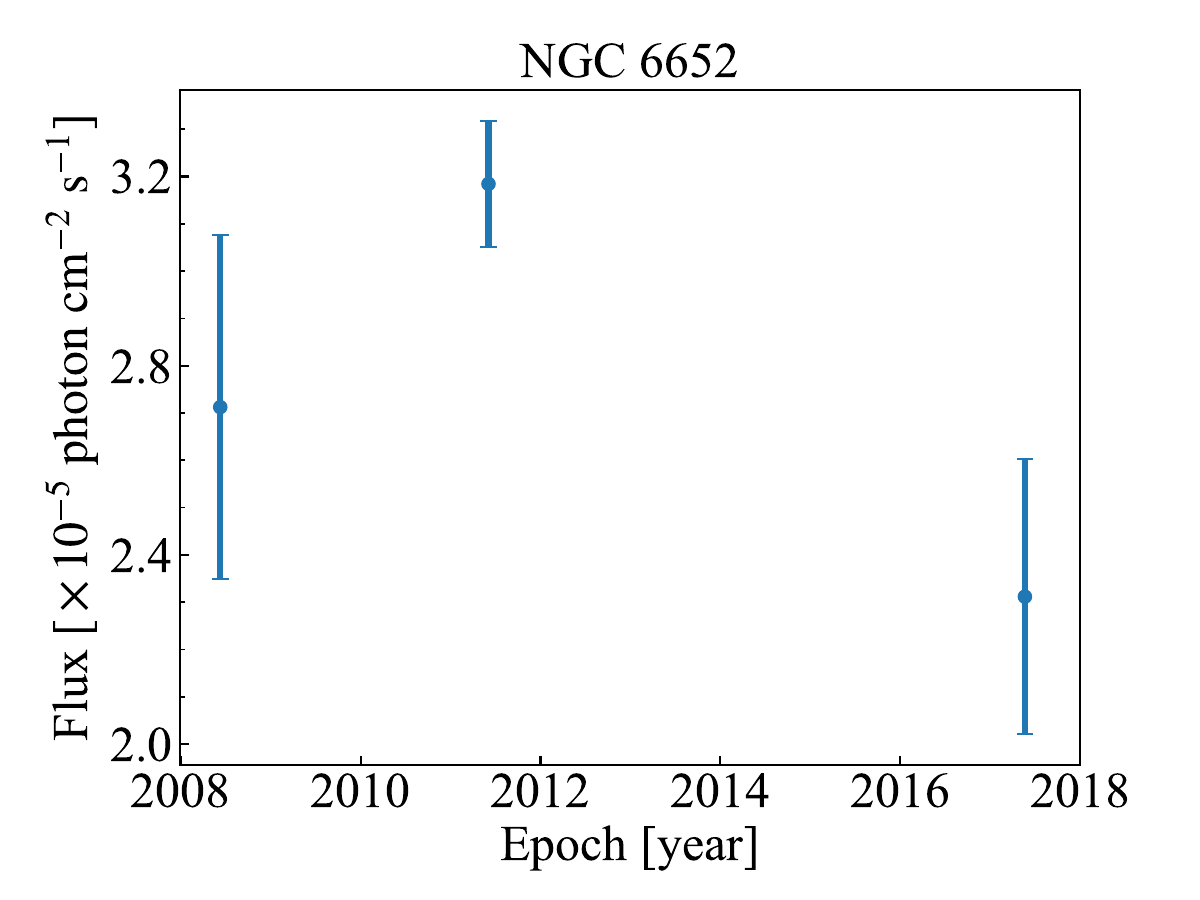}{0.5\textwidth}
     	\afig{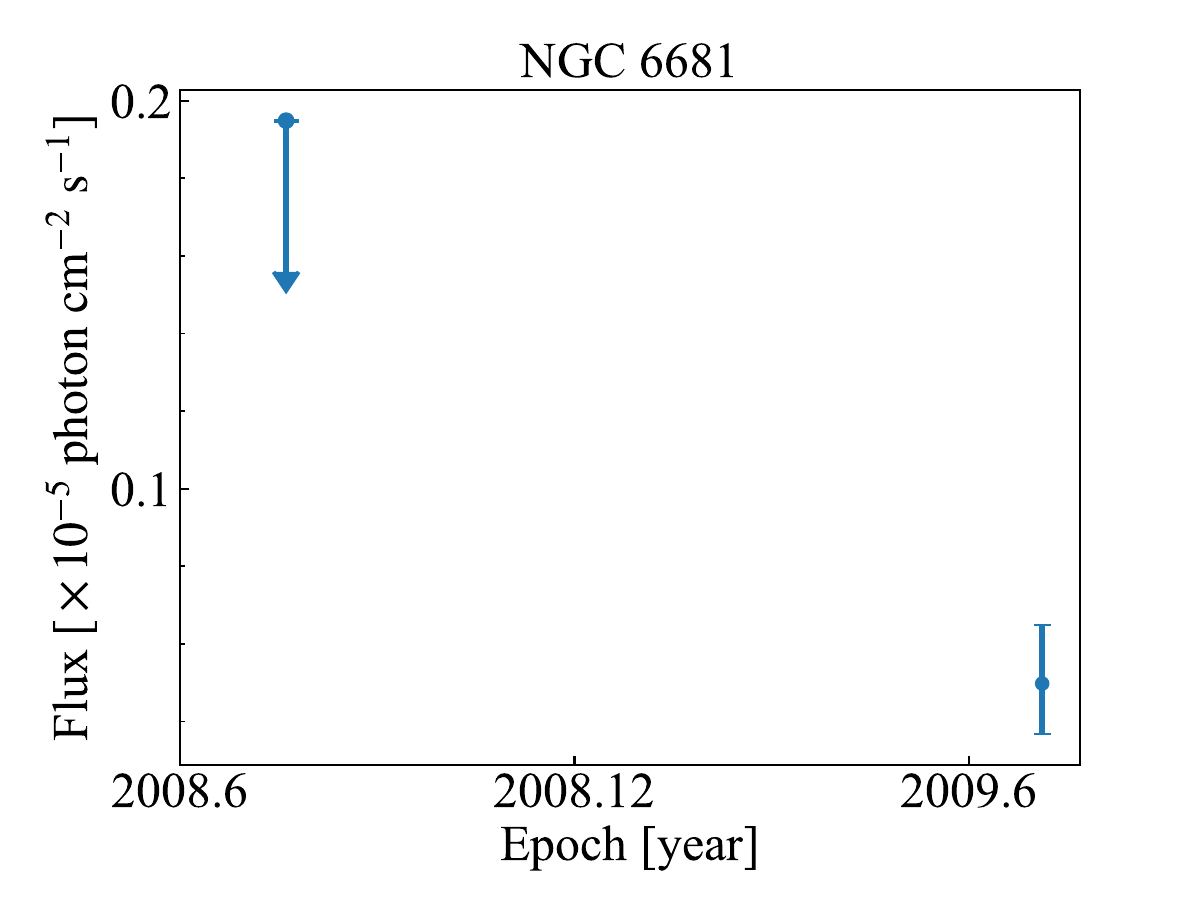}{0.5\textwidth}
     }
    \caption{Long-term light curves (0.5--8 keV photon flux versus observation date) of four central sources (47 Tuc, Terzan 5, NGC 6652, and NGC 6681) with multiple observations. 
    Error bars are of 1-$\sigma$. Arrows represent 3-$\sigma$ upper limits.
    For 47 Tuc, the only observation (ObsID 16527) having the central source detected is represented by a red symbol.
    The inset compares the 0.5--8 keV light curves of the central source in ObsID 16527 and ObsID 15747, which were separated by 5 days.
    \label{fig:lc}}
\end{figure*}


Among the six GCs, four (47 Tuc, Terzan 5, NGC\,6652, and NGC\,6681) have more than one observations. 
For their central X-ray source, we examine possible flux variability by quantifying the 0.5--8 keV net photon flux in the individual observations. To do so, we define the source region as a circle with a radius of the 90\% ECR, and the background region as a concentric annulus with inner-to-outer radii of 2--5 times the 90\% ECR. We exclude pixels falling within the 90\% ECR of neighbouring sources, if any, for both the source and background regions.  
The total source counts, background counts, source area, background area and exposure are then fed to the CIAO tool {\it aprates} to calculate the net photon flux and associated error, which accounts for the Poisson statistics in the low-count regime. 
If the 3-$\sigma$ lower limit of the photon flux hits zero in a certain observation, the source is considered non-detected and we provide a 3-$\sigma$ upper limit for this observation. 
The resultant long-term light curves of the four central X-ray sources are shown in Figure~\ref{fig:lc}.
Moderately significant ($\lesssim40\%$) inter-observation variability is seen in the central source of Terzan 5 and NGC\,6652. 
In the case of 47 Tuc, the central source is too faint to have a solid detection (i.e., 3-$\sigma$ lower limit of net photon flux greater than zero) in any of the individual observations except for ObsID 16527, in which the net photon flux is at least a factor of 1.7 higher than that in ObsID 15747 (the two observations were separated by 5 days).  
The central source in NGG\,6681 is undetected in one of the two observations, but the flux upper limit is consistent with little or no inter-observation variability.

Five of the above six central sources had been reported by previous work. However, except in one case \citep[NGC\,6388,][]{2008A&A...478..763N}, their possible relation with a putative IMBH was not explicitly discussed.
The central source in 47 Tuc was detected by
\citet{2005ApJ...625..796H} (named W286 therein), based on a subset of {\it Chandra} observations used here. 
\citet{2005ApJ...625..796H} associated the source with a BY Dra variable detected from HST images \citep{2001ApJ...559.1060A}.
The central source in Terzan 5 was detected by 
\citet{2006ApJ...651.1098H} (named CX9 therein), based on the first of the 18 {\it Chandra} observations used here as well as two observations taken in 2000 which we have discarded. \citet{2006ApJ...651.1098H} suggested this source to be a quiescent LMXB for its soft X-ray spectrum.
The central source in NGC\,6388 was detected by \citet{2012ApJ...756..147M} (named CX7 therein), based on the same {\it Chandra} observation used here, and was suggested to be a CV based on a tentative optical counterpart with a blue color.
Two neighboring sources are present to the immediate east of this central source (Figure~\ref{fig:GCs}), and the three sources were collectively identified by \citet{2008A&A...478..763N} (named 14* therein). 
The central source in NGC\,6652 was detected by \citet{2012ApJ...751...62S} (named source D therein) based on the second of the three {\it Chandra} observations used here, and was suggested to be a quiescent LMXB due to a rather soft X-ray spectrum.
The central source in NGC\,6093 was detected by \citet{2003ApJ...598..516H} (named CX9 therein) based on the same {\it Chandra} observation used here, but no comment was provided about its possible nature.
The central source in NGC\,6681 has not been reported elsewhere, to our best knowledge.

We note that the central X-ray source found in two of the six GCs (NGC\,6388 and Terzan 5) were also detected and catalogued by \citet{2020ApJ...901...57B}. The other four GCs were not included in \citet{2020ApJ...901...57B}.

\subsection{Spectral analysis} 
\label{subsec:spec}
Among the six central sources, four (i.e., those residing in 47 Tuc, NGC\,6388, Terzan 5, and NGC\,6652) have sufficient counts for a meaningful spectral analysis. 
We extract the source and background spectra from the same regions as defined in Figure~\ref{fig:GCs}, using the CIAO tool {\it specextract}.
For sources with multiple observations, the spectra are combined using the CIAO tool {\it combine\_spectra}, neglecting the possible moderate inter-observation variability.
The spectra are binned to have at least 10 counts and a S/N greater than 3 per bin over the energy range of 0.5--8 keV, except for 47 Tuc, for which the minimal S/N per bin is set as 2 due to its relatively small number of counts.

\begin{figure*}
    \agridline{
    	\afig{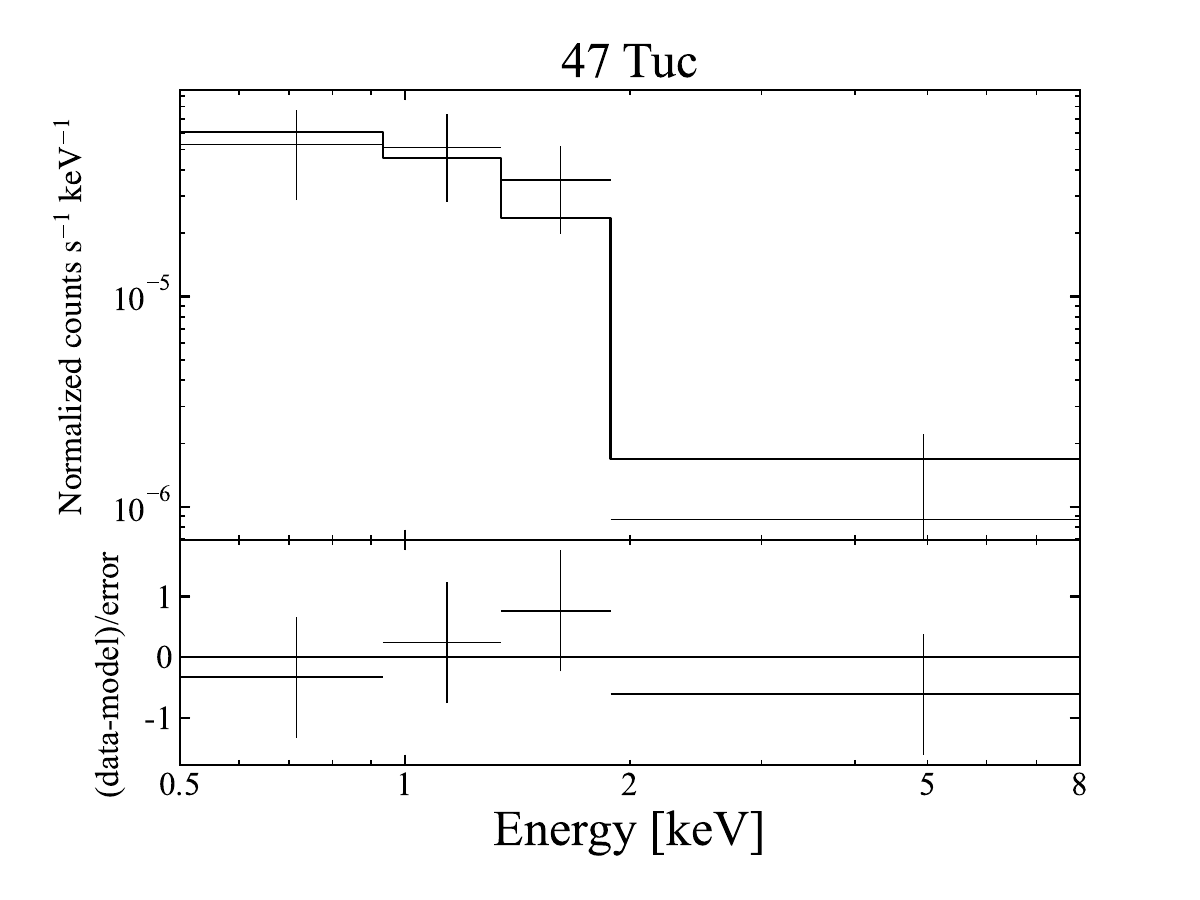}{0.5\textwidth}
        \afig{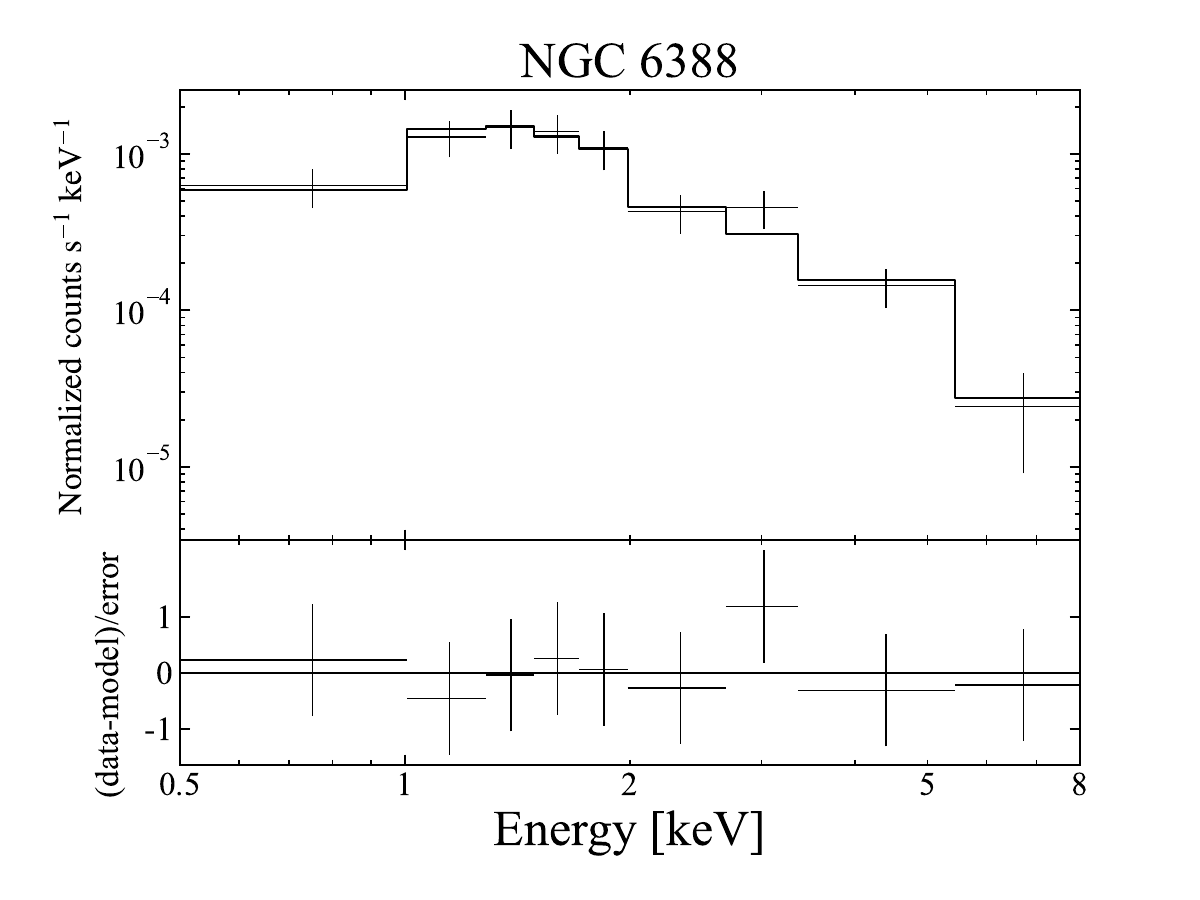}{0.5\textwidth}
        }
    \agridline{
    	\afig{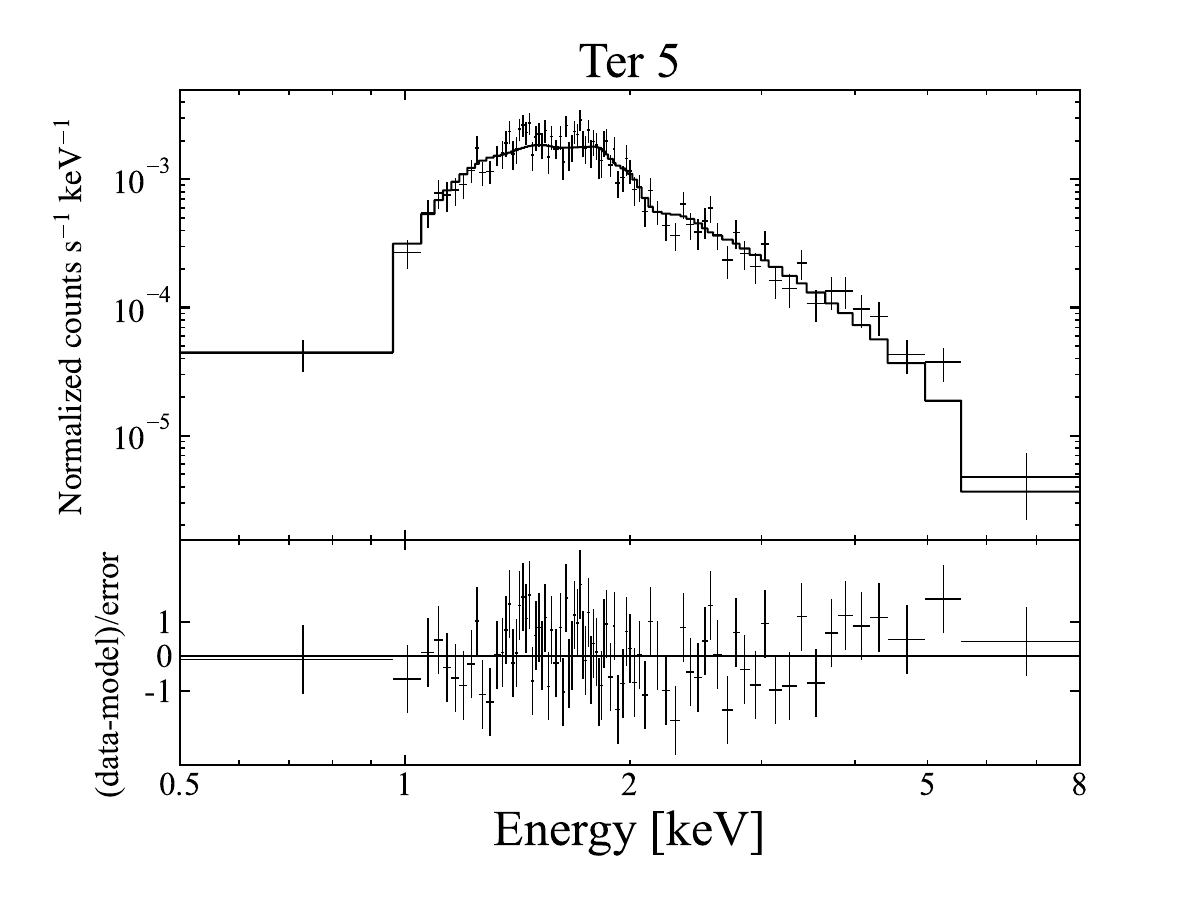}{0.5\textwidth}
        \afig{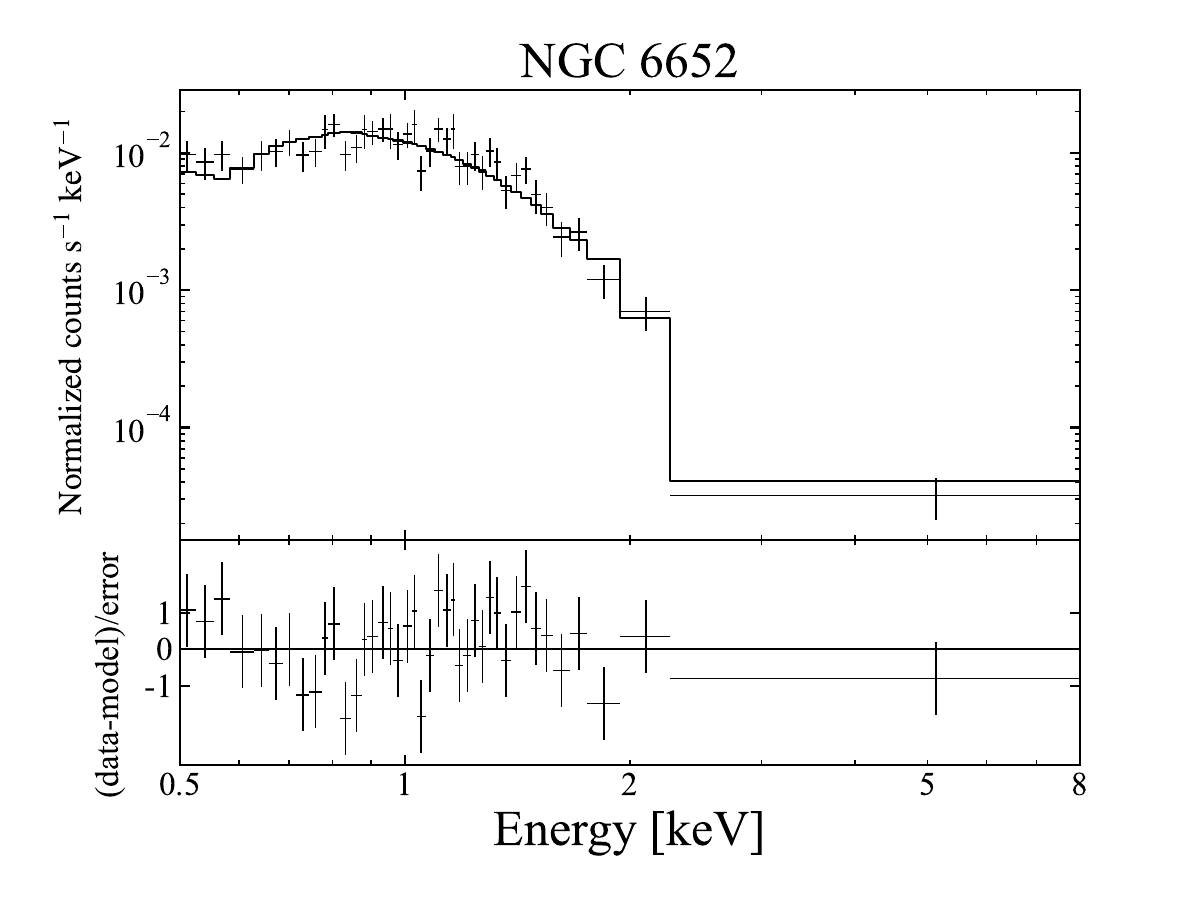}{0.5\textwidth}
        }
    \caption{Spectra and best-fit absorbed power-law models of the central sources in 47 Tuc, NGC\,6388, Terzan 5, and  NGC\,6652. The spectra are binned to have at least ten counts and a S/N greater than three (two for 47 Tuc) per bin. \label{fig:spectra}}
\end{figure*}

Spectral analysis is performed with XSPEC v12.10.1 \citep{1996ASPC..101...17A},
employing \citet{2000ApJ...542..914W} elemental abundances, \citet{1996ApJ...465..487V} photoionization cross sections and $\chi^2$ statistics.
In view that all four spectra have a moderate S/N and show no obvious line-like features, we adopt a simple absorbed power-law model (XSPEC model {\tt phabs*powerlaw}).
The fit for all four spectra is acceptable with null hypothesis probability $>$50\%.
For 47 Tuc, which has the lowest S/N, the absorption column density ($N_{\rm H}$) is fixed at $3.56\times10^{20}~{\rm cm^{-2}}$ which is in accordance with its foreground reddening $E(\mbox{$B\!-\!V$})$ \citep{2016ApJ...826...66F}.
For the other three GCs, $N_{\rm H}$ is kept as a free parameter, and the best-fit values are consistent with foreground reddening of NGC\,6388 and NGC\,6652, but suggest additional absorption intrinsic to Terzan\,5.
The spectra and best-fit models are shown in Figure~\ref{fig:spectra}. 
Both Terzan 5 and NGC\,6652 exhibit a steep power-law, with a best-fit photon-index of $\sim$4.5 and $\sim$4.9, respectively. 
47 Tuc and NGC\,6388, on the other hand, have a photon-index of $\sim$2.
The unabsorbed 0.5--8 keV luminosity ($L_{\rm 0.5-8}$) is estimated to be $1.3\times 10^{30}~{\rm erg~s^{-1}}$, $5.6\times10^{32}~{\rm erg~s^{-1}}$, $1.9\times10^{33}~{\rm erg~s^{-1}}$, and $3.7\times 10^{33}~{\rm erg~s^{-1}}$,
for 47 Tuc, NGC\,6388, Terzan 5 and NGC\,6652, respectively, spanning more than three orders of magnitude.  
The fit results are summarized in Table~\ref{tab:fit}.

\begin{table}
    \centering
    \renewcommand{\thetable}{2}
    \caption{Spectral fit results \label{tab:fit}}
    \renewcommand{\arraystretch}{1.4}
    \setlength{\tabcolsep}{2pt}
    {\footnotesize
    \begin{tabular}{lccccc}
	\hline
        GC    & $C_{\rm net}$ & $N_{\rm H}$ & $\Gamma$ & $L_{\rm 0.5-8}$  & $\chi_\nu^2$/d.o.f. \\
        (1)   & (2)        &  (3)     & (4)                  & (5) & (6)\\
    \hline 
    47 Tuc  & 33.5 & $0.036^*$ & $2.43^{+1.15}_{-0.78}$ & $0.013^{+0.006}_{-0.006}$ & 0.506/2\\
    NGC\,6388 & 113.6 & $0.64^{+0.51}_{-0.35}$ & $1.85^{+0.62}_{-0.51}$ & $5.6^{+1.3}_{-1.2}$ & 0.279/7\\
    Terzan 5 & 1588.0 & $3.14^{+0.30}_{-0.27}$ & $4.51^{+0.34}_{-0.31}$ & $19.0^{+1.5}_{-1.4}$ & 0.906/79 \\
    NGC\,6652 & 695.0 & $0.50^{+0.13}_{-0.12}$ & $4.88^{+0.50}_{-0.43}$ & $36.7^{+5.6}_{-4.2}$ & 0.947/38\\
    \hline
	\end{tabular}
	}
    \footnotesize 
    {\bf Notes.}
    Results of spectral fit with Xspec model {\tt phabs*powerlaw}. (1) GC name. (2) Net counts of the spectrum. (3) Line-of-sight absorption column density in units of $10^{22}~{\rm cm^{-2}}$. 
$^*$ indicates that the parameter is fixed during the fit. (4) Photon index of the power-law. (5) The unabsorbed 0.5--8 keV luminosity in units of $10^{32}~{\rm erg~s^{-1}}$. (6) Reduced $\chi^2$ and degree-of-freedom.
All errors are at 90\% confidence level.
\end{table}

\subsection{Upper limit of non-detected GCs} \label{subsec:non}
None of the remaining 75 GCs has a significant central X-ray source reported by {\it wavdetect}. This is confirmed by a visualization of the {\it Chandra} image of individual clusters presented in the Appendix.
For these 75 GCs,
we employ the CIAO tool {\it aprates} to estimate a 3-$\sigma$ upper/lower limit in the 0.5--8 keV net count rate for the putative central source.  
To do so, we again define a source region as a circle with a radius of the 90\% ECR, and a background region as a concentric annulus with inner-to-outer radii of 2--5 times the 90\% ECR, excluding pixels falling within the 90\% ECR of neighbouring sources, if any.  
It is found that 73 of the 75 GCs have a 3-$\sigma$ lower limit consistent with zero net count. 
In the remaining two cases, NGC\,6440 and NGC\,7099, the cluster center is heavily contaminated by the PSF wing of bright neighboring sources. Therefore, even though the 3-$\sigma$ lower limit of these two GCs reported by {\it aprates} is marginally above zero, we still classify them as exhibiting no significant central emission. 

\begin{figure*}
    \agridline{
    \afig{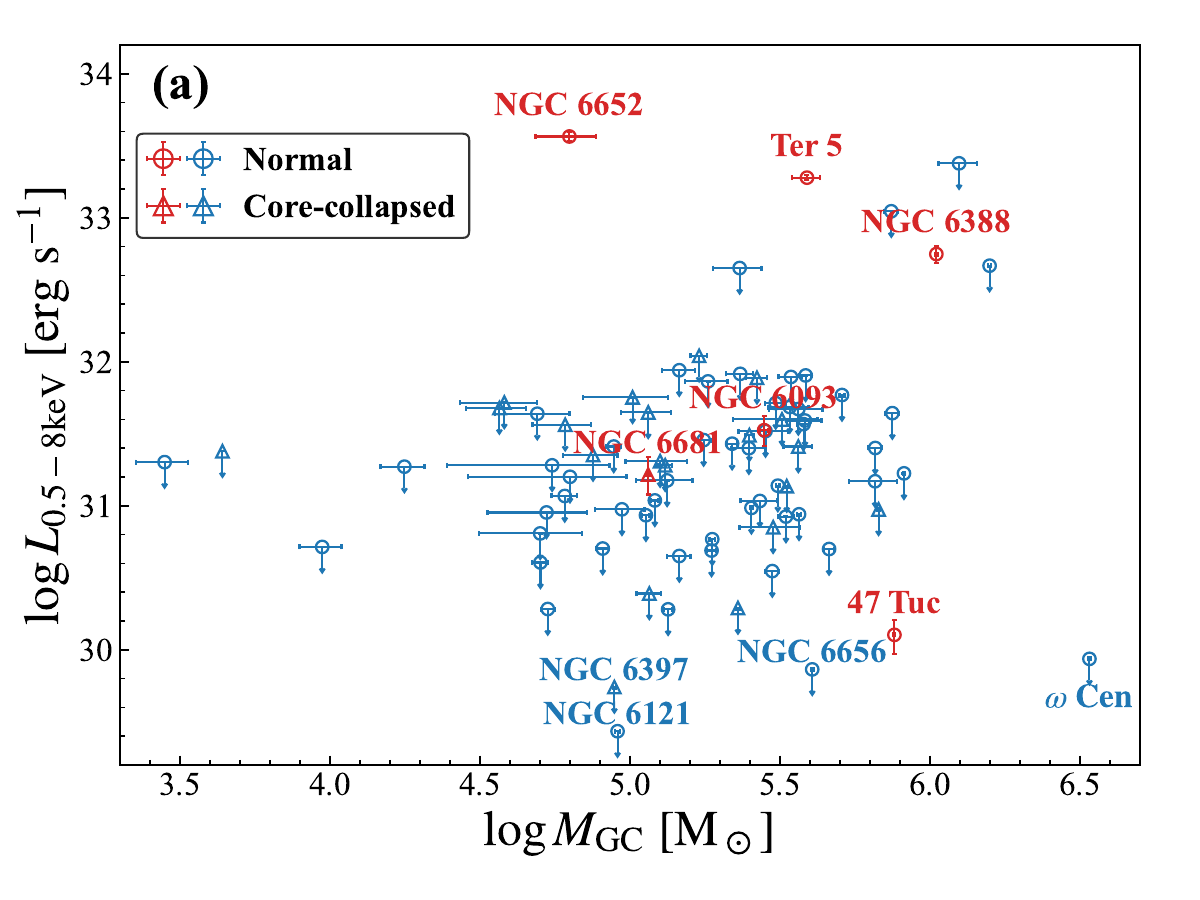}{0.5\textwidth}
    \afig{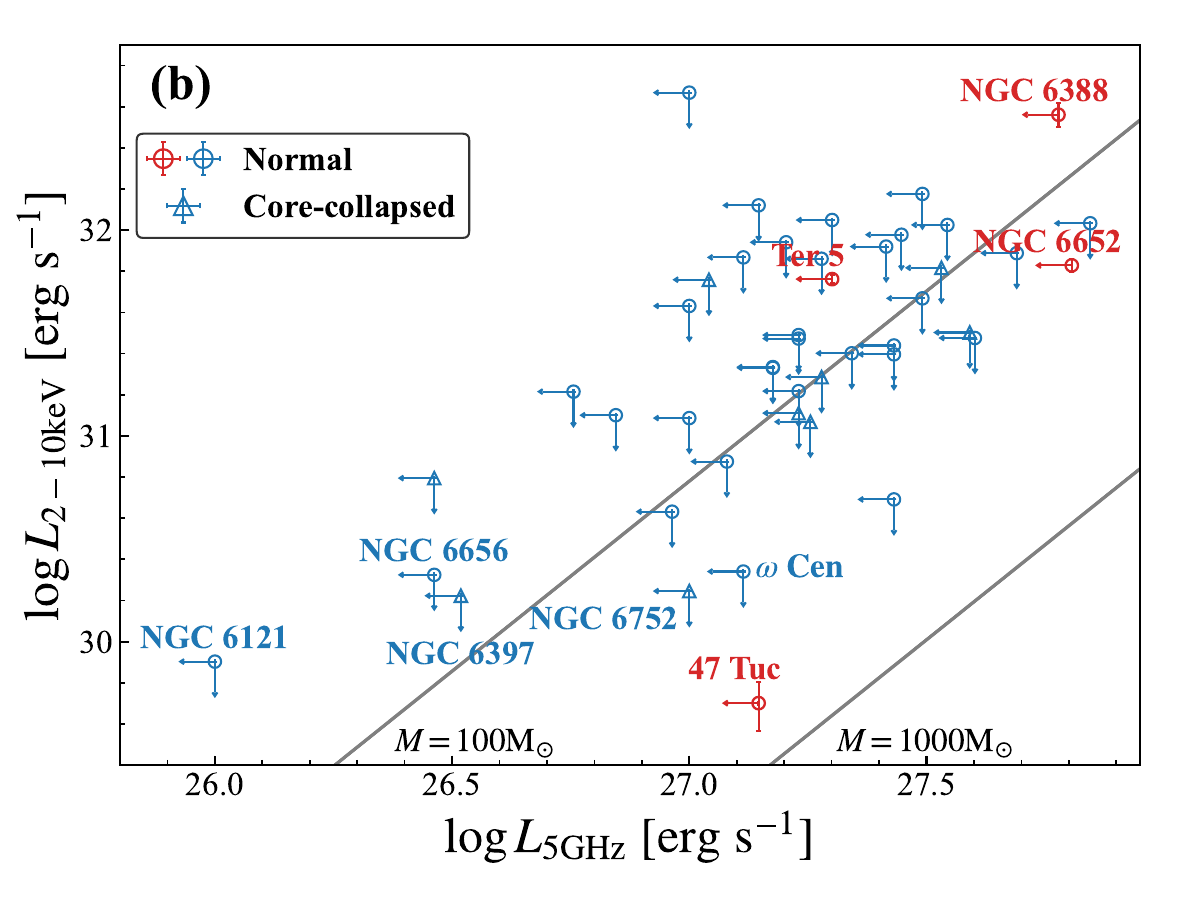}{0.5\textwidth}
    }
    \caption{(a) 0.5--8 keV X-ray luminosity versus GC mass adopted from \citet{2019MNRAS.482.5138B}. All 81 GCs except Glim 01 are included. (b) 2--10 keV X-ray luminosity versus radio luminosity at 5 GHz, for a subset of 45 GCs that have an available upper limit of the radio luminosity from \citet{2018ApJ...862...16T}. Grey solid lines indicate the fundamental plane of black hole activity adopted from \citet{2019ApJ...871...80G} (see Equation~\ref{equ:funda1}) for black hole masses of $100~{\rm M_\odot}$ and $1000~{\rm M_\odot}$. In both panels, the red and blue symbols represent GCs with and without a detected central X-ray source, respectively. The circles and triangles represent normal and core-collapsed GCs from \citet[2010 edition]{1996AJ....112.1487H}, respectively. Error bars are of 1-$\sigma$ and arrows represent 3-$\sigma$ upper limits.
    \label{fig:lx_relation}}
\end{figure*}

The 3-$\sigma$ upper limit of each GC is further converted into an unabsorbed luminosity, by adopting the same fiducial power-law spectrum as for creating the exposure map (Section~\ref{sec:data}) and the distance from the Sun \citep{2019MNRAS.482.5138B}. 
The resultant 0.5--8 keV luminosity limits of the 75 GCs, 
along with the measured luminosities of the six detected central sources, are listed in Table \ref{tab:observation}.
For the six detected central sources, the luminosity is preferentially calculated from the best-fit spectral model (Section~\ref{subsec:spec}), or when a spectral fit is absent (for NGC\,6093 and NGC\,6681), the same fiducial power-law spectrum is adopted. 
Figure~\ref{fig:lx_relation}a shows the 0.5--8 keV luminosity or luminosity limit versus the GC mass taken from \citet{2019MNRAS.482.5138B}. 
We note that Glim 01 is omitted in this plot, since it does not have a well-constrained mass. 
It can be seen that a sensitivity of $10^{30}-10^{32}\rm~erg~s^{-1}$ is achieved for the vast majority of our sample GCs, and for four GCs ($\omega$ Cen [=NGC 5139], NGC 6656, NGC 6397, NGC 6121) the sensitivity reaches below $10^{30}\rm~erg~s^{-1}$, thanks to their proximity and the depth of the {\it Chandra} data.
The sensitivity of $\omega$ Cen is consistent with that reported by \citet{2013ApJ...773L..31H}, who used the same set of {\it Chandra} observations.
There is no significant correlation between the X-ray luminosity (or upper limit) and GC mass.

\subsection{Nature of the central X-ray sources}
\label{subsec:nature}
Despite the excellent sensitivity afforded by the {\it Chandra} observations, an X-ray source positionally coincident with the cluster center is found in only
six of the 81 Galactic GCs. 
The X-ray spectra of these sources appear featureless and can be well-fitted by a power-law model, leading to unabsorbed 0.5--8 keV luminosities ranging from $\sim10^{30}\rm~erg~s^{-1}$ to a few $10^{33}\rm~erg~s^{-1}$.
Significant inter-observation variability is seen in three sources, which belong to 47 Tuc, Terzan 5 and NGC\,6652. 
These spectral and temporal properties, however, do not immediately reveal the nature of the X-ray sources.
Indeed, the most abundant X-ray sources in GCs, namely, quiescent LMXBs, CVs and ABs, have quite comparable X-ray luminosities (\citealp{2010AIPC.1314..135H}, \citetalias{2018ApJ...858...33C}). 
We note that the X-ray spectra of CVs and ABs are dominated by a collisionally ionized plasma and thus expected to exhibit metal emission lines \citep{2016ApJ...818..136X}. 
But this is not to be considered a serious discrepancy with the featureless observed spectra, given the moderate spectral S/N and the fact that 
GC sources generally have a very low metallicity \citep{1996AJ....112.1487H}.

As mentioned in Section~\ref{subsec:detection}, the central sources in 47 Tuc and NGC\,6388 were associated with a BY Dra variable \citep{2005ApJ...625..796H} and a CV \citep{2012ApJ...756..147M}, respectively, based on tentative identification of an optical counterpart. 
While caution must be taken with any optical counterpart drawn from the crowded cluster core, at least the short-term X-ray variability found in the central source of 47 Tuc (Figure~\ref{fig:lc}) appears to be consistent with the behavior of a BY Dra variable, and the X-ray spectrum of the central source in NGC\,6388 is not atypical of a CV.
The two sources in Terzan\,5 and NGC\,6652 have the highest luminosities (a few $10^{33}\rm~erg~s^{-1}$), which are empirically too high to be compatible with ABs.
The central source in Terzan\,5 was suggested to be a quiescent LMXB for its soft spectrum \citep{2006ApJ...651.1098H}.
This possibility holds with the much deeper {\it Chandra} data here confirming a steep spectrum and revealing mild variability (Figure~\ref{fig:lc}).
The same tentative classification can be argued for the central source in NGC\,6652, as done by \citet{2012ApJ...751...62S}.
Empirically, intermediate polars, a sub-class of magnetic CVs, can also reach an X-ray luminosity as high as a few $10^{33}\rm~erg~s^{-1}$ \citep{2016ApJ...818..136X}, but their spectra are typically much harder than found in the central sources of Terzan\,5 and NGC\,6652 \citep{2009MNRAS.392..630L}.
The two remaining sources (NGC\,6093 and NGC\,6681) have too limited X-ray information to provide a strong diagnostic of their nature.

While the above arguments favor a stellar origin for at least four of the six central X-ray sources, we are still left with the possibility that some, if not all, of the six sources are related to an IMBH.
Indeed, by analogy to low-luminosity active galactic nuclei \citep[LLAGNs;][]{2008ARA&A..46..475H}, a weakly accreting IMBH can in principle produce a power-law X-ray spectrum with a mild variability within the observed range of luminosities ($\lesssim10^{34}\rm~erg~s^{-1}$). 
In this regard, we confront the detected central X-ray emission (or upper limits in the case of non-detection) with the black hole fundamental plane (FP) relation \citep{2003MNRAS.345.1057M}, which spans the range from stellar-mass black holes in X-ray binaries to SMBHs in LLAGNs. 
Following Equation (8) of \citet{2019ApJ...871...80G}, the FP takes the form of
\begin{equation}
    \mu=0.55 \pm 0.22 + (1.09 \pm 0.10)R + (-0.59^{+0.16}_{-0.15})X,
    \label{equ:funda1}
\end{equation}
where $\mu=\log(M/10^8{\rm~M_{\odot}})$, $R=\log(L_{\rm R}/10^{38}{\rm~erg~s^{-1}})$ and $X=\log(L_{\rm X}/10^{40}{\rm~erg~s^{-1}})$.
The X-ray luminosity $L_{\rm X}$ is conventionally evaluated over 2--10 keV, which is converted for each GC from the 0.5--8 keV luminosity or upper limit derived in Section~\ref{subsec:non}. 
The radio luminosity $L_{\rm R}$ is conventionally evaluated at 5 GHz.
Currently no firm detection of central radio source is known for any GC. Hence we focus on a subset of 45 GCs, which have available  upper limits in $L_{\rm R}$ from \citet{2018ApJ...862...16T}.
These include four GCs with a firmly detected central X-ray source, namely, 47 Tuc, Terzan\,5, NGC\,6388 and NGC\,6652.
It is noteworthy that an X-ray luminosity was also estimated by \citet{2018ApJ...862...16T} in order to construct a FP for their sample GCs.
However, we find that our X-ray luminosities or upper limits measured from the {\it Chandra} observations are generally higher than the predicted values in \citet{2018ApJ...862...16T} by a factor of a few to $\sim$100 for the 45 GCs in common. 
This invokes caution when using {\it ad hoc} assumptions, e.g., Bondi-like accretion and radiative efficiency typical of LLAGNs (also see \citealp{2012ApJ...750L..27S}, \citealp{2018ApJ...862...16T} for discussions on the uncertainty of these assumptions), to predict the accretion-induced X-ray luminosity.

Figure~\ref{fig:lx_relation}b displays the $L_{\rm X}-L_{\rm R}$ relation for the 45 GCs. Also plotted are two lines following Equation~(\ref{equ:funda1}) for fiducial masses of $M = 100~{\rm M_\odot}$ and $1000~{\rm M_\odot}$.
It can be seen that the four GCs with a detected central source already provide an interesting constraint on the putative IMBH, in the sense that $M \sim 1000\rm~M_\odot$ is ruled out by all four GCs, and $M \sim 100\rm~M_\odot$ is ruled out by Terzan\,5 and NGC\,6388.
Such constraints, of course, rely on the assumption that the detected X-ray emission arises from an accreting IMBH, which is not necessarily true for any of the four GCs. 
The other GCs do not lead to a formal constraint on the black hole mass, since they are non-detected in both X-rays and radio. 
Nevertheless, we emphasize the complementary role of X-ray and radio measurements: further enhancing the X-ray sensitivity helps to rule out lower masses ($\sim 100\rm~M_\odot$), whereas enhancing the radio sensitivity tends to rule out higher masses ($\sim 1000\rm~M_\odot$), an interesting outcome of the FP (Equation~\ref{equ:funda1}).

\section{Hydrodynamic Simulation of IMBH accretion}\label{sec:simulation}
The previous section shows that the vast majority of the Galactic GCs surveyed by {\it Chandra} exhibit no significant X-ray emission from the cluster center.
A simple, albeit somewhat disappointing, explanation is a low occupation fraction of IMBH in GCs.
An alternative explanation, regardless of the intrinsic IMBH occupation fraction, is that current X-ray observations lack the sensitivity to reveal the otherwise existent IMBH due to a generally low accretion rate (hence a feeble X-ray flux) in a typical GC environment.
In this section we investigate this latter possibility. 

\subsection{The physical picture} \label{subsec:physics}
The accretion of a putative IMBH in a typical GC environment can take place via several channels. 
Here we shall not consider IMBH accretion by ingesting individual stars, a fraction of which could be associated with tidal disruption events \citep{2009ApJ...697L..77R}, or through Roche lobe overflow from a tightly-bound companion star \citep{2004ApJ...604L.101H}, because both of these two channels are rare events and are characterized by a high luminosity inconsistent with the observational result presented in Section~\ref{sec:analysis}.
Instead, we focus on the more gentle and continuous accretion from the ambient gas.
Even in this case, quantification of the (time-dependent) accretion rate is often hampered by the lack of precise knowledge about the density and temperature of the ICM, which, along with the black hole mass, determines the rate of the classic {\it Bondi accretion}.
\citet{2011ApJ...730..145V} modeled the Bondi accretion from gas supplied by stellar winds, for various combinations of central black holes and stellar spheroids, including GCs. 
Approximating the stellar wind injection and accretion with a one-dimensional smooth profile within the gravitational influence radius of the black hole, \citet{2011ApJ...730..145V} found in general very low accretion rates, $\sim 10^{-9}-10^{-6}$ of the Eddington limit\footnote{The Eddington limit of accretion rate scales with black hole mass, $\dot{M}_{\rm Edd} \approx 2.3\times10^{-5}(M/1000 \rm~M_\odot)\rm~M_\odot~yr^{-1}$.}, for black hole masses $M = 10^{2}-10^{4}\rm~M_\odot$ suitable for the case of GCs.
The corresponding X-ray luminosity was estimated to be $\lesssim10^{29}\rm~erg~s^{-1}$ in the case of a radiatively inefficient accretion flow, which is believed to operate at very sub-Eddington rates \citep{2014ARA&A..52..529Y}.
This dormant accretion is apparently consistent with the deficiency of central X-ray emission in our sample GCs. 
\citet{2013MNRAS.430.2789P}, on the other hand, modeled the IMBH accretion by taking into account the global dynamics of the ICM. For presumed cluster conditions, they found accretion rates of $10^{-8}-10^{-5}\rm~M_\odot~yr^{-1}$, which are several orders of magnitude higher than the canonical Bondi accretion rate. 
Such high accretion rates, however, would lead to high X-ray luminosities incompatible with the upper limits shown in Figure~\ref{fig:lx_relation}.

In reality, accretion onto the IMBH might be dominated by the innermost orbiting star(s), which would break the assumption of a smooth radial inflow by \citet{2011ApJ...730..145V} and \citet{2013MNRAS.430.2789P}. 
This is supported by the work of \citet{2016ApJ...819...70M}, which used direct $N$-body simulations
to study the dynamical hierarchy of cluster stars around a central IMBH.
These simulations reveal that the IMBH would have a stellar companion (which can be a main-sequence star, a giant star, a white dwarf, a neutron star or a stellar-mass black hole) with an orbital semi-major axis at least three times tighter than the second-most-bound stellar object over 90\% of the simulation time.
Specifically, for an adopted black hole mass $\sim 100\rm~M_\odot$, the most-bound star is predicted to have a characteristic semi-major axis of $\sim$100 AU and an orbital period of $\sim 10$ yr. These correspond to an orbital velocity of $\sim 100\rm~km~s^{-1}$, which is much higher than the typical velocity dispersion of stars in the cluster core ($\sim 10\rm~km~s^{-1}$).
Winds from this most-bound star thus carry a substantial orbital energy and angular momentum.
Moreover, the outward propagating wind may impede the inflow from larger radii.
One expects that both effects work to significantly reduce the accretion rate as in a smooth radial inflow \citep{2013MNRAS.430.2789P}. 

Motivated by the predicted stellar hierarchy in \citet{2016ApJ...819...70M}, 
here we conduct hydrodynamic simulations to examine how the stellar wind of the innermost orbiting star may control the IMBH accretion, and whether the resultant accretion rate is compatible with the observed X-ray luminosities (or upper limits) in our GC sample.

\subsection{Simulation setup}
\label{subsec:setup}

We shall remark that the main purpose of our simulations is to capture the basic picture of stellar wind accretion outlined in Section~\ref{subsec:physics} and to make order-of-magnitude predictions about detectability of the accretion-induced X-ray emission. Therefore, we perform simulations with a restricted but representative set of physical parameters, as described below, and defer a more thorough exploration of the parameter space and relevant physics to future work.   

The simulations are performed using the publicly available hydrodynamics (HD) code, PLUTO\footnote{\url{http://plutocode.ph.unito.it/}} \citep{2007ApJS..170..228M}.
This grid-based HD code, with a second-order Runge–Kutta time integrator and
a Harten-Lax-van Leer Riemann solver for middle contact discontinuities, is well suited for simulating the stellar orbital motion and the self-interaction of stellar winds.


We first examine the case of a single star (i.e., the most-bound star) orbiting the IMBH, using three-dimensional (3D) simulations.
We assume that the gravitational potential is static and completely determined by a central IMBH, for which a fiducial mass of $10^3\rm~M_\odot$ is adopted.
The star is placed at a circular orbit with a radius of 30 AU. 
We note that the simulations of \citet{2016ApJ...819...70M} predict a nearly isothermal probability distribution of the orbital eccentricity and a wide range of semi-major axis (5--1000 AU) peaking at $\sim100$ AU, for their fiducial black hole mass of $150\rm~M_\odot$.
Our chosen black hole mass and circular orbit should slightly favor the wind accretion while still being consistent with the orbital distribution predicted by \citet{2016ApJ...819...70M}.

The dynamical hierarchy of \citet{2016ApJ...819...70M} consists of additional bound stars, which may also feed the IMBH with their own stellar winds.
Therefore, we further examine the case of two stars orbiting the IMBH to understand whether the inner star can effectively shield the winds from the outer star. 
This should be representative of the more general case of the most-bound star shielding a gas inflow from large radii. 
For the two-star case we employ two-dimensional (2D) simulations, as a compromise between computational cost and accuracy.
A circular orbit is adopted for both stars.
The orbital radius of the inner star is again set as 30 AU, while the outer star has an orbital radius of 150 AU.
The semi-major axis ratio between the second-most-bound star and the innermost star is consistent with the prediction of \citet{2016ApJ...819...70M}, in which this ratio is greater than three for about 90\% of the simulation time. 
We consider only the prograde situation, i.e., the two stars orbit in the same direction (counterclockwise).

To determine the stellar wind properties, we consider two cases for the orbiting star: (i) a low-mass main-sequence (MS) star, and (ii) a giant star. 
The simulations of \citet{2016ApJ...819...70M} predict that $\sim$40\% of the most-bound star is an MS star, while $\sim$10\% is a giant star.  
For the MS star case, a wind mass loss rate of $10^{-12}{\rm~M_\odot~yr^{-1}}$ is adopted, 
along with a wind terminal velocity of $500 \rm~km~s^{-1}$ and a wind temperature of $10^5$ K. 
Such a wind is analogous to the solar wind.
For the case of the giant star, we adopt a mass loss rate of $10^{-6}{\rm~M_\odot~yr^{-1}}$, along with a wind terminal velocity of $10\rm~km~s^{-1}$ and a wind temperature of 3000 K, which are analogous to strong winds from asymptotic giant branch (AGB) stars \citep{2005ARA&A..43..435H}. 
The latter case is expected to result in a maximally possible accretion rate (hence a high X-ray luminosity), although the AGB phase with a strong wind is generally short ($\lesssim$ a few Myr).
The orbital motion is realized by placing the star at the expected position at each time step of simulation, which is adaptively determined.
In the meantime, stellar winds are injected at a radius of 2 AU from the star, satisfying the above wind conditions in the star rest frame and taking into account the instantaneous orbital motion. 

The 3D simulation for the MS star case (3D1MS) is run on a static Cartesian grid of $512^3$ corresponding to a physical box of $200^3~{\rm AU^3}$ for three orbital periods; the AGB star case (3D1AGB) is run on a grid of $512\times512\times256$ corresponding to a box of $200\times200\times80~{\rm AU^3}$ for 20 orbital periods. 
The orbital period ($\sim 5.2$ yr) is determined by Kepler's third law given the mass of the IMBH and the orbital radius.  
The run time is sufficiently long to establish a quasi-equilibrium state for both cases, and is also sufficiently short such that the most-bound star is not perturbed or replaced by other stars \citep{2016ApJ...819...70M}.
The 2D simulations for the two-star case assume either a MS-MS or a AGB-AGB combination (2D2MS and 2D2AGB), which are run on a static Cartesian grid of $1280^2$ corresponding to a physical box of $400^2~{\rm AU^2}$ and for 2 orbital periods of the outer star.
To better resolve the vicinity of the black hole, we employ static grid refinement,
using $128^3$ ($256^2$) cells for a zoom-in domain of $10^3~{\rm AU^3}$ ($10^2~{\rm AU^2}$) in the 3D (2D) simulations, corresponding to a resolution of $0.08^3~{\rm AU^3}$ ($0.04^2~{\rm AU^2}$).
The key parameters of the simulations are summarized in Table~\ref{tab:param}.

\begin{table}
	\centering
	\renewcommand{\thetable}{3}
    \caption{Simulation parameters\label{tab:param}}
    \centering
    {\scriptsize
    \begin{tabular}{lcc}
     \hline
     Parameters & MS case & AGB case \\
     \hline
     IMBH mass & $1000~{\rm M_\odot}$ & $1000~{\rm M_\odot}$\\
     \hline
     Wind mass loss rate &  $10^{-12}~{\rm M_\odot}~{\rm yr^{-1}}$	&  $10^{-6}~{\rm M_\odot}~{\rm yr^{-1}}$\\
     Wind terminal velocity & $500~{\rm km~s^{-1}}$ & $10~{\rm km~s^{-1}}$\\
     Wind temperature & $10^5~{\rm K}$ & $3\times10^3~{\rm K}$\\
     Wind injection radius & $2~{\rm AU}$ & $2~{\rm AU}$\\
     Orbital radius (3D) & $30~{\rm AU}$ & $30~{\rm AU}$ \\
     Orbital radius (2D) & $30~{\rm AU}/150~{\rm AU}$ & $30~{\rm AU}/150~{\rm AU}$\\
     \hline
     ICM density & $0.1~{\rm cm^{-3}}$ & $0.1~{\rm cm^{-3}}$\\
     ICM temperature & $10^4~{\rm K}$ & $10^4~{\rm K}$ \\
     \hline \hline
     Runs & Simulation Box & Cells \\
          & $x\times y\times z$ or $x\times y$  & \\
     \hline
     3D1MS & $200\times200\times200~{\rm AU^3}$ & $512\times512\times512$\\
     3D1AGB & $200\times200\times80~{\rm AU^3}$ & $512\times512\times256$ \\
     2D2MS & $400\times400~{\rm AU^2}$ & $1280\times1280$\\
     2D2AGB & $400\times400~{\rm AU^2}$ & $1280\times1280$\\
     \hline
    \end{tabular}
    }
\end{table}

The simulations apply an ideal equation of state, neglect magnetic field, viscosity and thermal conduction, but take into account radiative cooling. 
We adopt the \texttt{TABULATED} radiative cooling module in PLUTO, of which the cooling function $\Lambda_{\rm H}(T,Z)$ is generated with \textsc{Cloudy} \citep[version C17.02;][]{2017RMxAA..53..385F} for an optically thin plasma with metal abundances of $0.1~Z_{\rm \odot}$ and covering temperatures between $10^3 - 10^9$ K.
However, with this \textsc{tabulated} cooling module, the run time of the 3D1AGB simulation becomes prohibitively long, as the time step is strongly constrained by the short cooling timescale.
Therefore, for this particular case we adopt the exact integration cooling scheme proposed by \citet{2009ApJS..181..391T},
which analytically integrates energy loss with a piecewise power law approximation of the cooling function at each time step, regardless of the cooling timescale.
For the initial conditions, we adopt an ICM temperature of $10^4$ K and an ICM density of $0.1\rm~cm^{-3}$, which are typical of GC models \citep[e.g.,][]{2013MNRAS.430.2789P} and supported by observations of 47 Tuc \citep{2018MNRAS.481..627A}.
We note that our simulation results are insensitive to these initial conditions, because the stellar wind carries a sufficiently strong momentum to expel the ICM from the simulation box in a small fraction of the simulation run time.
To mimic accretion onto the IMBH and to avoid unphysical gas pileup, after each time step we reset the density and temperature of the central $2^3$ ($2^2$) cells in the 3D (2D) simulations to $10^{-4}\rm~cm^{-3}$ and $10^3~{\rm K}$, to ensure that the resultant pressure in these cells is negligibly small compared to that of the infalling gas \citep[see also][]{1992MNRAS.255..183M,2018MNRAS.478.3544R}.
This {\it effective accretion radius} corresponds to about $1.6\times10^4~r_{g}$ ($8\times10^3~r_{g}$) in the 3D (2D) simulations, where $r_g = GM/c^2\approx10^{-5}~\rm AU$ is the gravitational radius of the IMBH. 
The effective accretion rate at each time step is calculated by summing up the mass thus removed from these central cells.
The accretion-induced radiation is low (see below) and its effect on the hydrodynamics is therefore neglected.

\subsection{Results}

\subsubsection{3D simulations: the most-bound star}
\label{subsubsec:3D}
\begin{figure*}
	\agridline{
	\afig{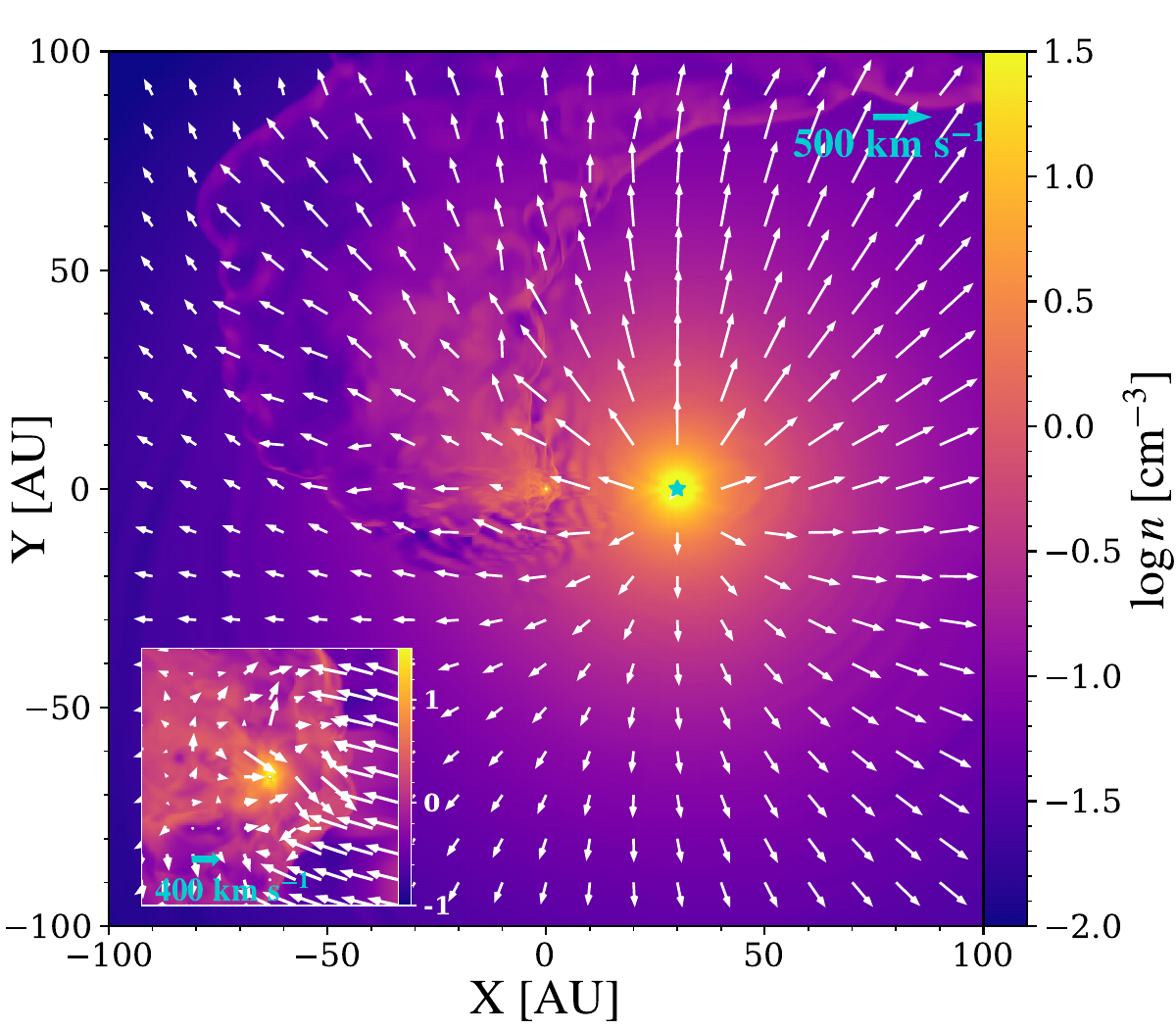}{0.5\textwidth}
    \afig{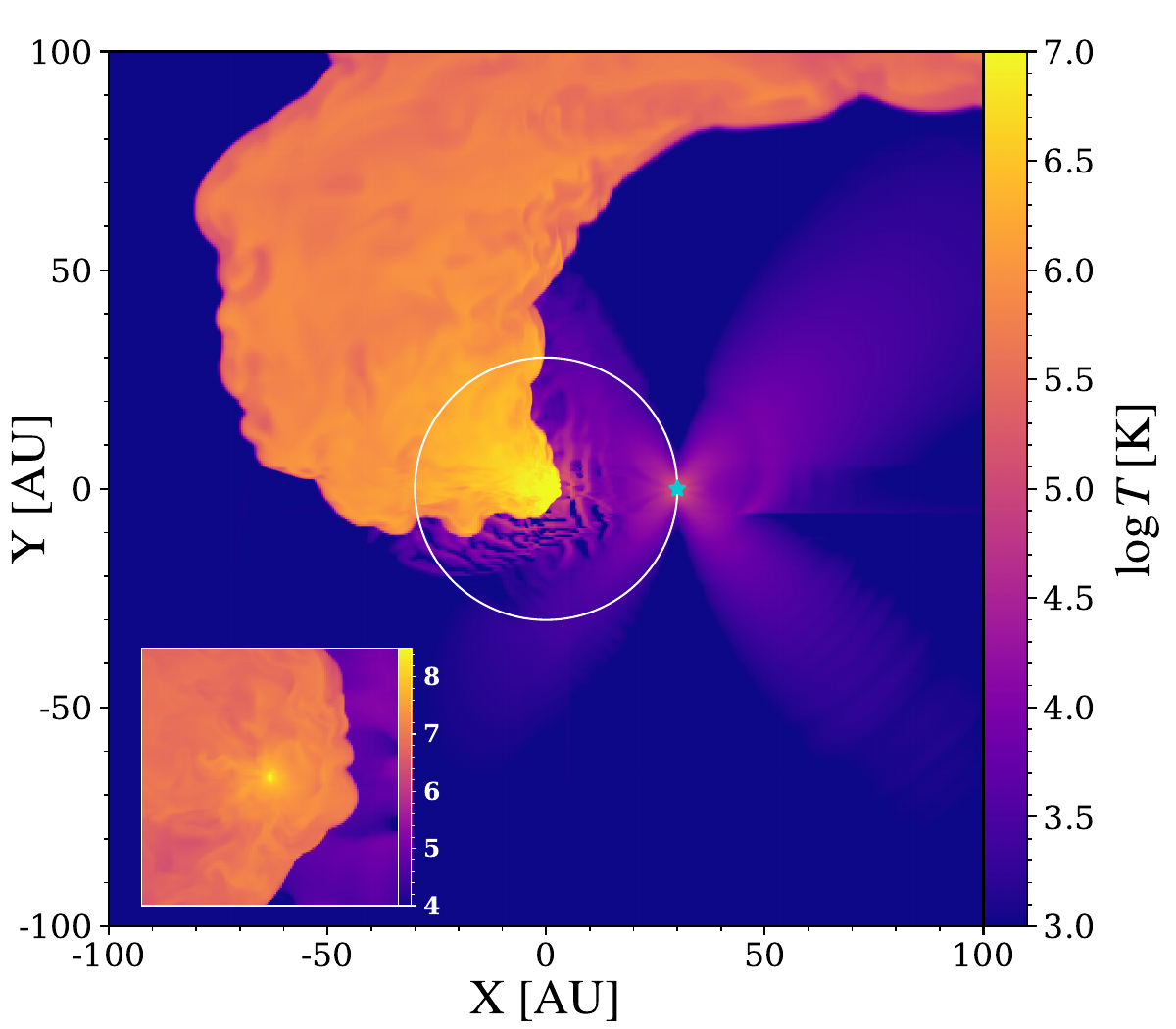}{0.5\textwidth}
		}
	\agridline{
	\afig{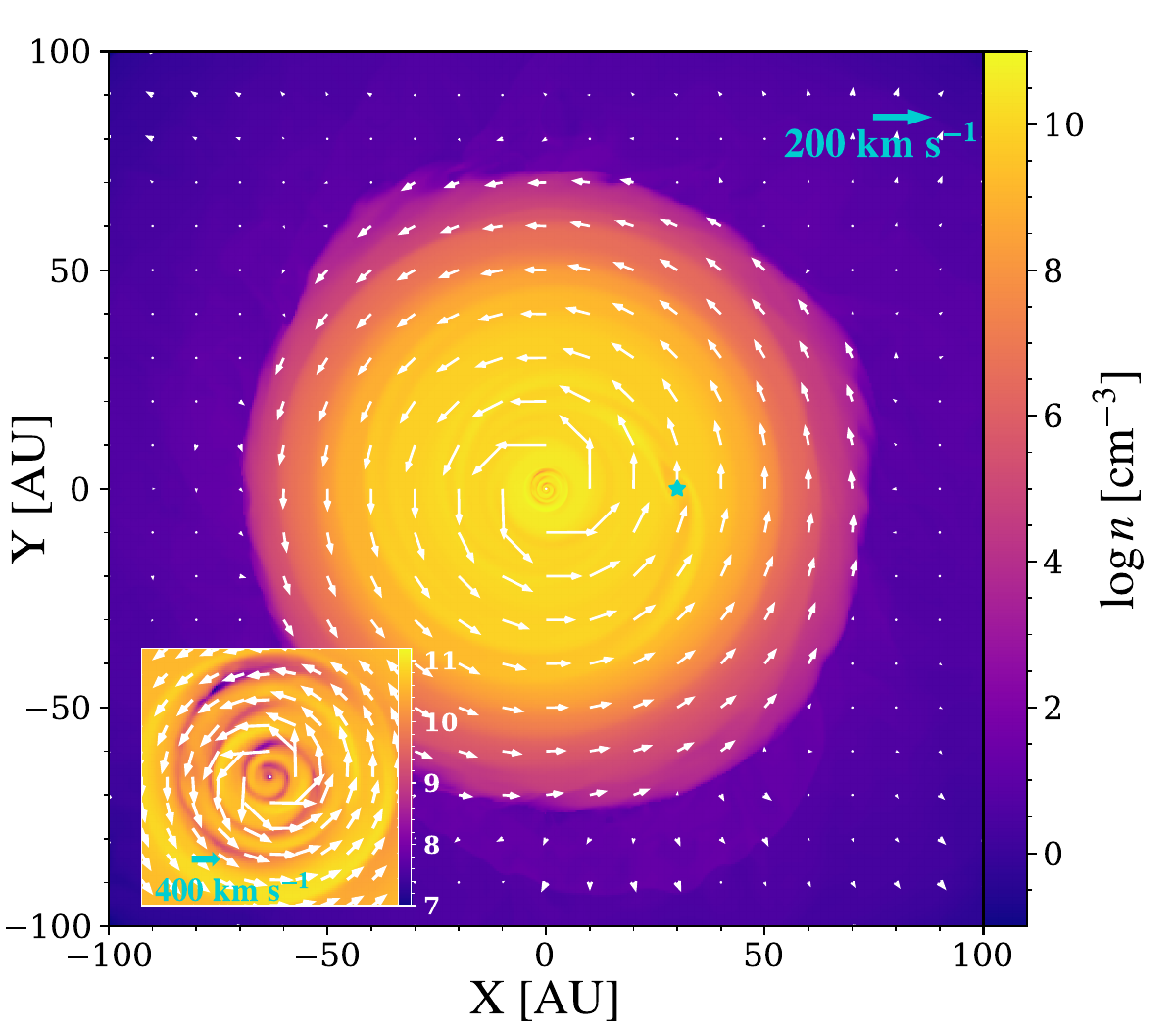}{0.5\textwidth}
    \afig{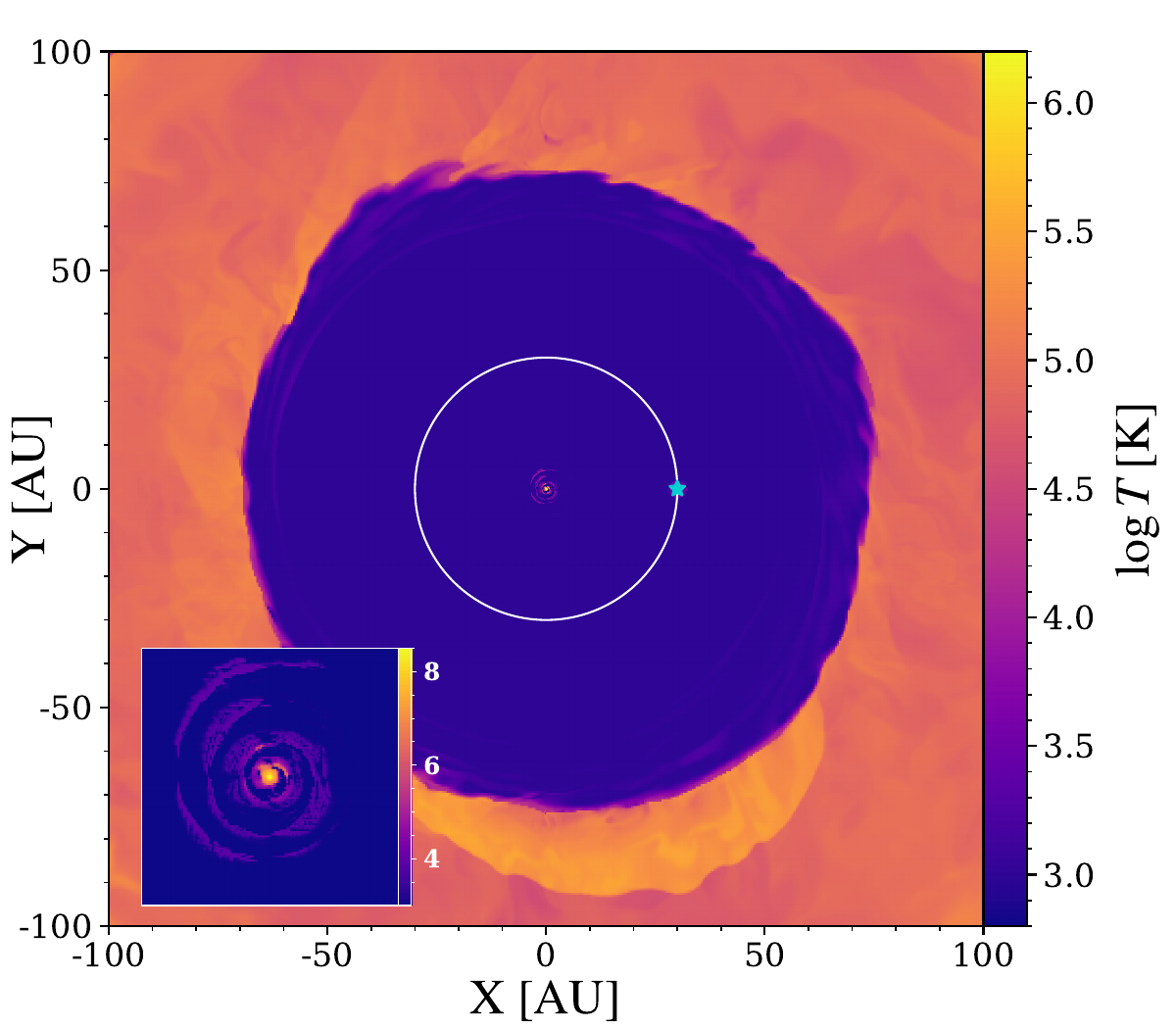}{0.5\textwidth}
		}
	\caption{Snapshots of the orbital plane in the 3D simulations of the most-bound, windy star orbiting the IMBH. {\it Left}: Number density distribution in units of ${\rm cm^{-3}}$. {\it Right}: Temperature distribution in units of K. {\it Top}: the MS case. {\it Bottom}: the AGB case. The inserts show a zoom-in view of the central 10 AU$\times$10 AU region. The star (marked by the pentagram) orbits the central IMBH counterclockwise at a radius of $30~{\rm AU}$, which is represented by the white circle in the right panels. 
	Arrows in the left panels indicate the gas velocity.
	In all panels, the color bar is in logarithm of 10. \label{fig:3d}}	
\end{figure*}

The density, temperature and velocity distributions in the orbital plane (i.e., the $X-Y$ plane) are shown in Figure~\ref{fig:3d} for the last snapshot of the 3D simulations, i.e., after the MS star and the AGB star has completed 3 and 20 orbits, respectively.
In the MS case (top panels), 
the wind speed ($V_{\rm wind} \sim 500\rm~km~s^{-1}$) is much higher than the orbital velocity ($V_{\rm orb}$ $\sim 170\rm~km~s^{-1}$),
hence the bulk of the wind escapes the simulation domain almost radially in a timescale much shorter than the orbital period. 
However, a small fraction of the wind, which shoots at the IMBH with a small impact parameter, is pulled by the latter's strong gravity.
A bow shock forms in front of the IMBH due to the supersonic motion,
as highlighted by the inserts in the upper panels of Figure~\ref{fig:3d}.
Downstream the bow shock the wind material forms a high-temperature ($T \gtrsim 10^6$ K) wake. Such a structure is characteristic of Bondi accretion \citep[e.g.,][]{2004NewAR..48..843E}.
The wake further develops into a spiral-like structure due to the orbital motion.
Much of the shocked-heated winds still eventually escape the simulation domain, but those with the smallest impact parameters ($\lesssim GM/V_{\rm wind}^2 \sim 3\rm~AU$) are gravitationally trapped by the IMBH. 

\begin{figure*}
	\agridline{
		\afig{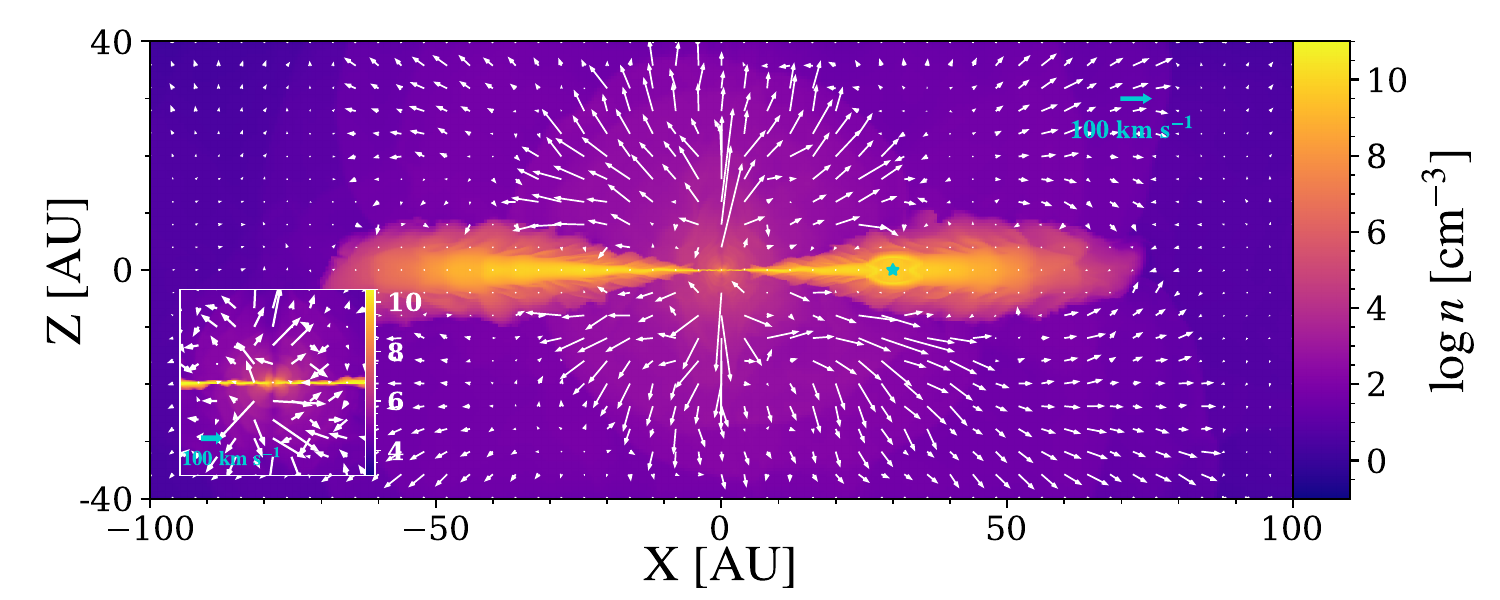}{0.8\textwidth}
		}
    \agridline{
    	\afig{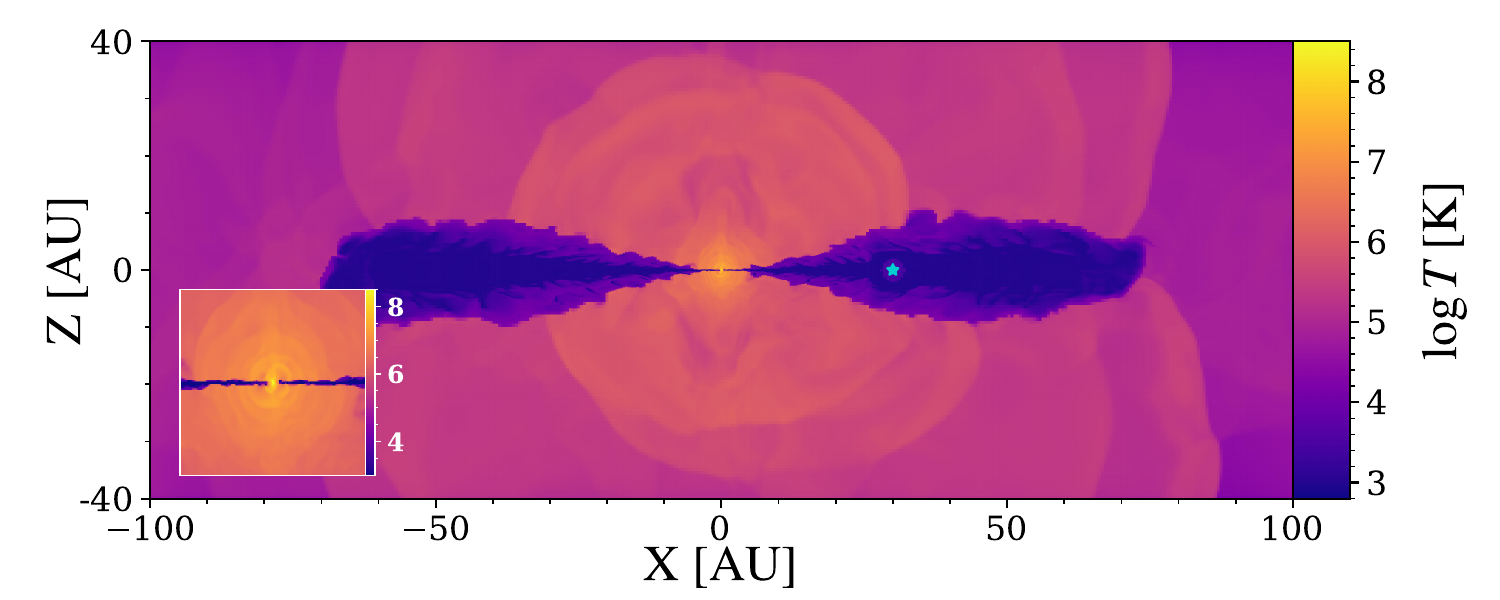}{0.8\textwidth}
    	}
	\caption{Snapshots of the $X-Z$ plane in the 3D simulation of AGB star orbiting the IMBH. {\it Upper panel}: density distribution overlaid by velocity vectors. {\it Bottom panel}: temperature distribution. The pentagram marks the position of the star. The inserts show a zoom-in view of the central 10 AU$\times$10 AU region. \label{fig:1agb3dxz}}	
\end{figure*}

In the AGB case (bottom panels), the wind speed ($V_{\rm wind} \sim 10\rm~km~s^{-1}$) is small compared to the orbital velocity. Instead of a high-temperature wake, the dominant structure formed in the orbital plane is a dense, cool ($T \sim 10^{3-4}~{\rm K}$) disk with a horizontal extent of approximately 140 AU, as shown in the lower panels of Figure~\ref{fig:3d}. 
This disk structure is an anticipated result of the predominant orbital motion.
The bulk of the disk maintains a low temperature because of efficient radiative cooling.
That the disk extends substantially beyond the stellar orbit indicates an effective angular momentum transport, which is probably due to the combined effect of stellar wind momentum and turbulence.
However, only a small fraction of the injected wind material flows into the central few AUs, where the gas temperature raises to $\gtrsim 10^5$ K, as shown by the inserts in the bottom panels of Figure~\ref{fig:3d}.
Interestingly, a tenuous, hot corona exists above and below the disk, which is best seen in Figure~\ref{fig:1agb3dxz}, which depicts the density, temperature and velocity distributions in the $X-Z$ plane.
The temperature of the corona is highest ($T \gtrsim 10^7$ K) within the central few AUs, and gradually declines to $\sim 10^6$ K at a radius of $\sim 50$ AU. 
The corona is turbulent, but exhibits an overall radial expansion, with a mean velocity of $\sim 100\rm~km~s^{-1}$.
Moreover, we observe an episodic, high-velocity (a few $100\rm~km~s^{-1}$) jet near the vertical axis (i.e. along the $Z$-axis), which typically lasts for a duration of $\sim1.5~{\rm yr}$.
In the absence of magnetic fields, thermal pressure gradient and centrifugal force are probably the main driver of the radial outflow \citep[e.g.,][]{2015ApJ...804..101Y}.
We note that the outflow rate, estimated at a radius of 30 AU and over latitudinal angles $> 15^{\circ} $, is only $\sim 10^{-11}-10^{-10}\rm~M_\odot~yr^{-1}$, 
but by face value this outflow carries a radial momentum flux comparable to that in the MS case. 
The small outflow rate means that the vast majority of the stellar winds are still trapped in the cool disk, indicating that the disk continues to grow. Nevertheless, we observe that a quasi-equilibrium is established after $\sim10$ orbital periods.

\begin{figure*}
	\afig{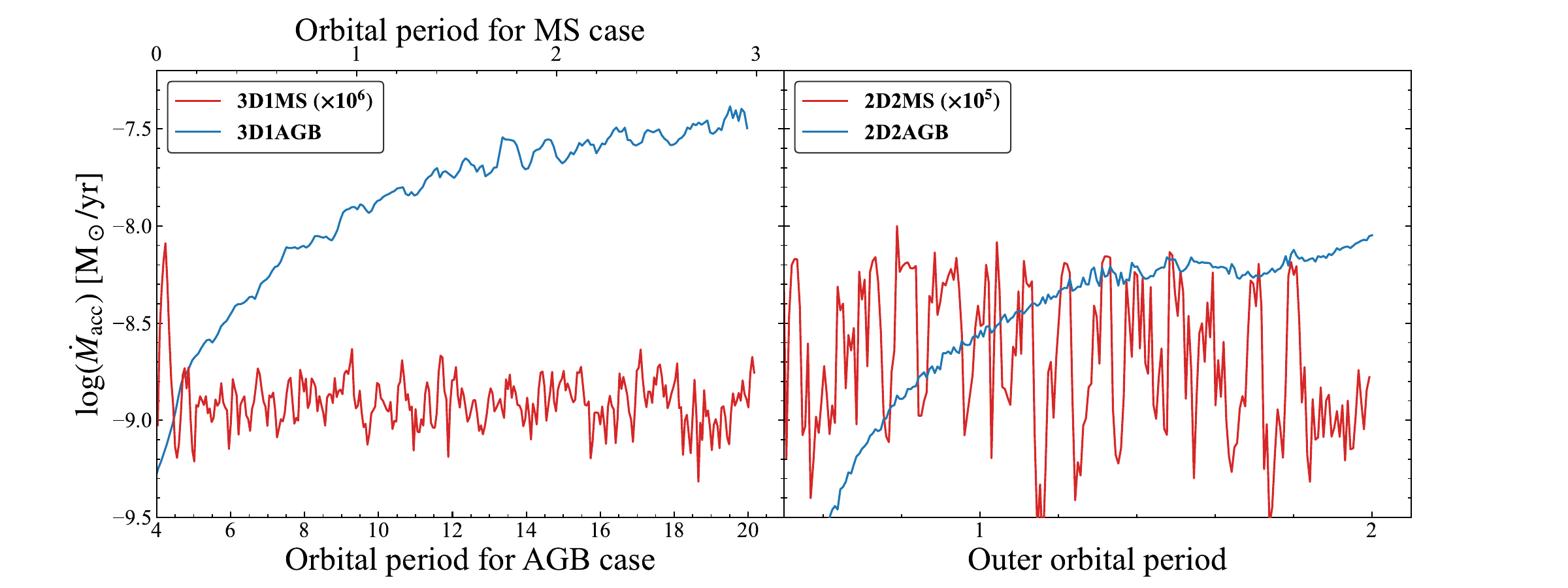}{1.0\textwidth}
    \caption{The time-dependent accretion rate. {\it Left}: the 3D one-star simulations. The red curve represents the MS case for three orbital periods, which is scaled up by a factor of $10^6$ for clarity. The blue curve represents the AGB case for 20 orbital periods. {\it Right}: the 2D two-star simulations for two orbital periods of the outer star. The red curve represents the MS case, which is scaled up by a factor of $10^5$ for clarity. The blue curve represents the AGB case.  
    \label{fig:accrete_3d}}
\end{figure*}

The left panel of Figure~\ref{fig:accrete_3d} displays the time-dependent  IMBH accretion rates in the MS (red curve) and AGB (blue curve) cases,
which are calculated as described in Section~\ref{subsec:setup}.
It can be seen that a quasi-steady state is reached in the MS case, which shows a mean accretion rate of $1.3 \times 10^{-15}\rm~M_\odot~yr^{-1}$, corresponding to 0.13\% of the stellar wind mass loss rate. 
In the AGB case, the accretion grows steadily as the cool disk itself accumulates mass with time. 
Nevertheless, near the end of the simulation the accretion rate increases only mildly, reaching a value of 
$3.5 \times 10^{-8}\rm~M_\odot~yr^{-1}$ or 3.5\% of the stellar wind mass loss rate.
Within one orbital period, the accretion rate fluctuates with an amplitude of $\sim0.2$ dex in both the MS and AGB cases.



\subsubsection{2D simulations: the second-most-bound star}
\label{subsubsec:3D}
\begin{figure*}
	\agridline{
	\afig{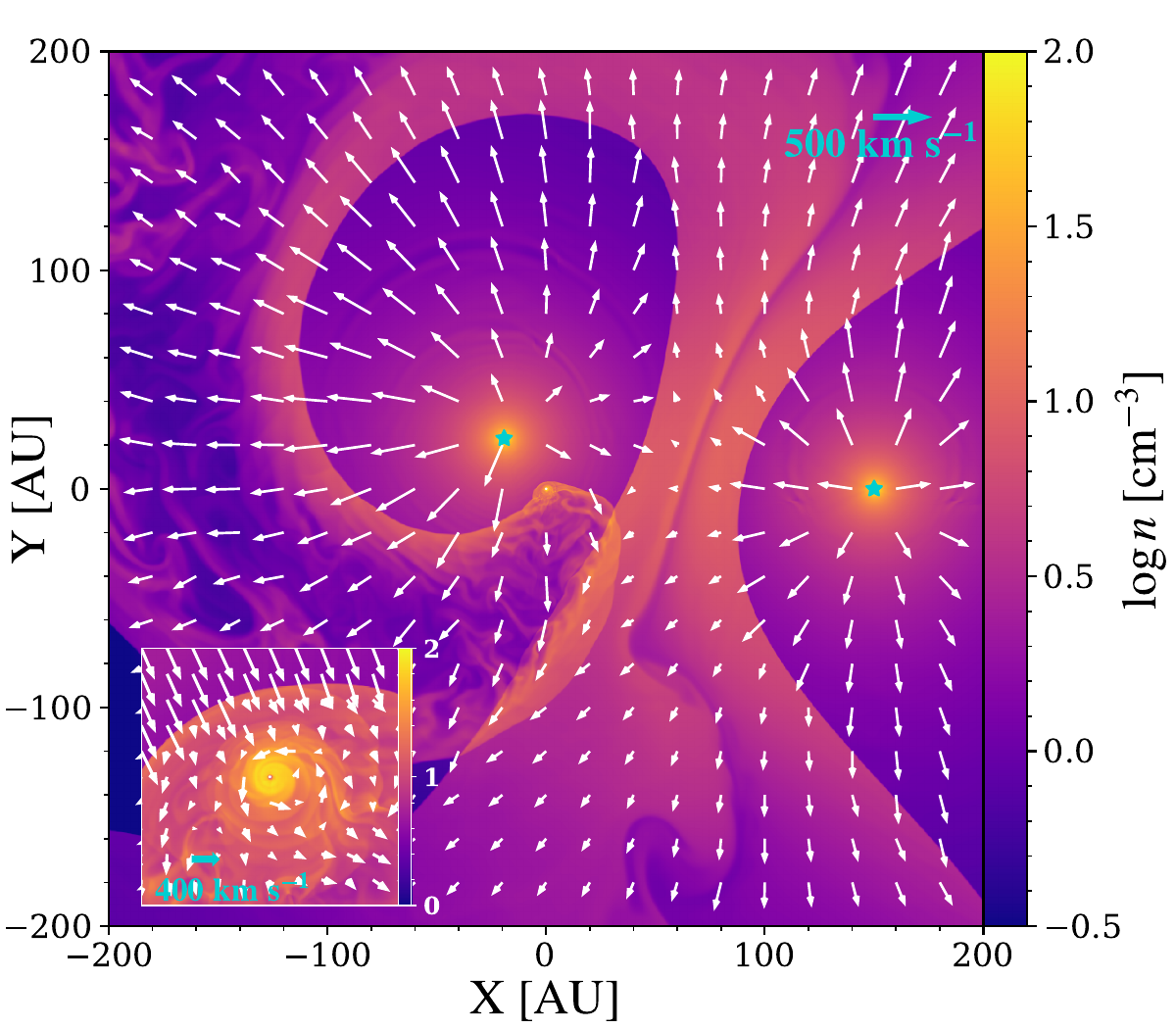}{0.5\textwidth}
    \afig{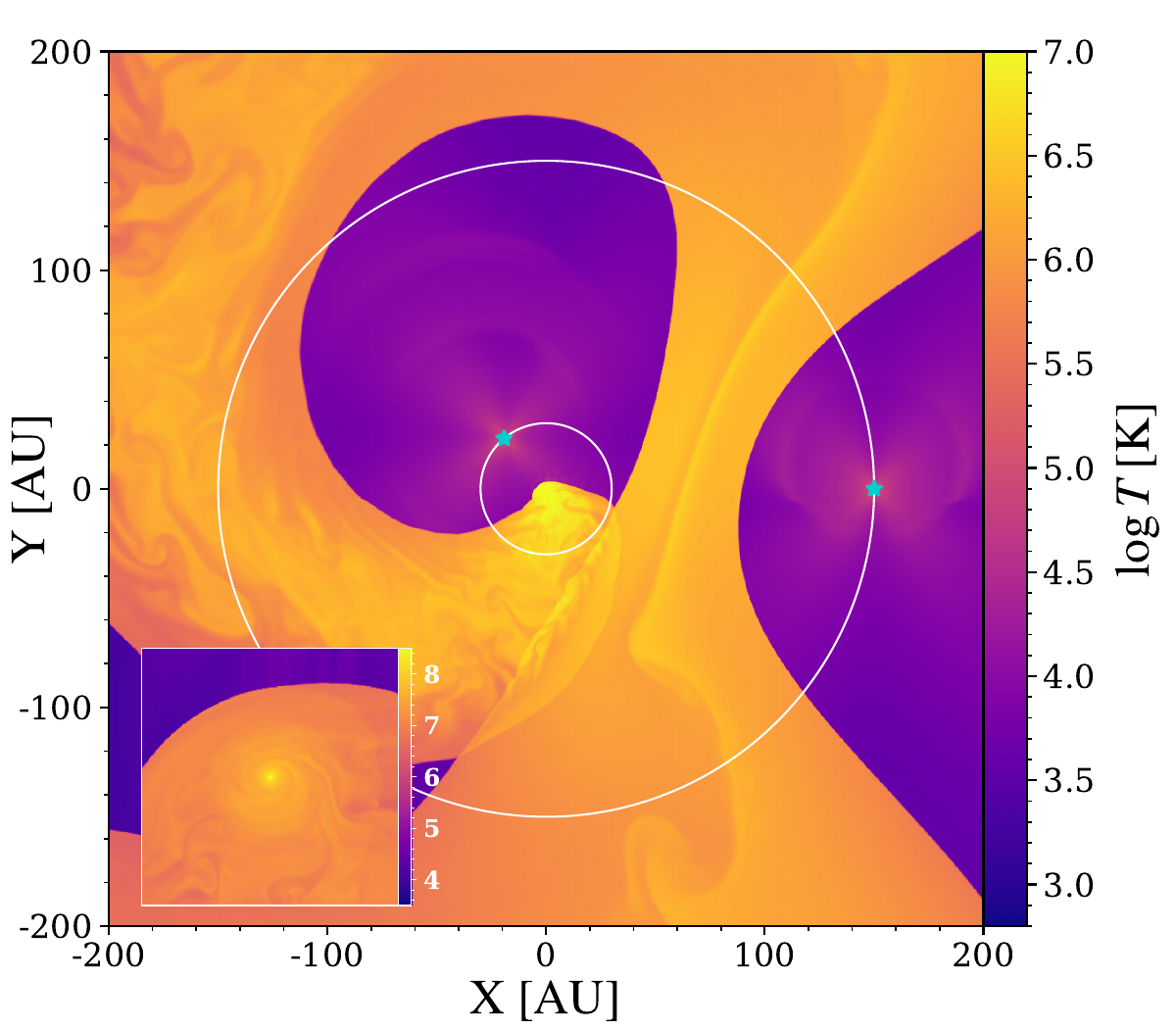}{0.5\textwidth}
		}
	\agridline{
	\afig{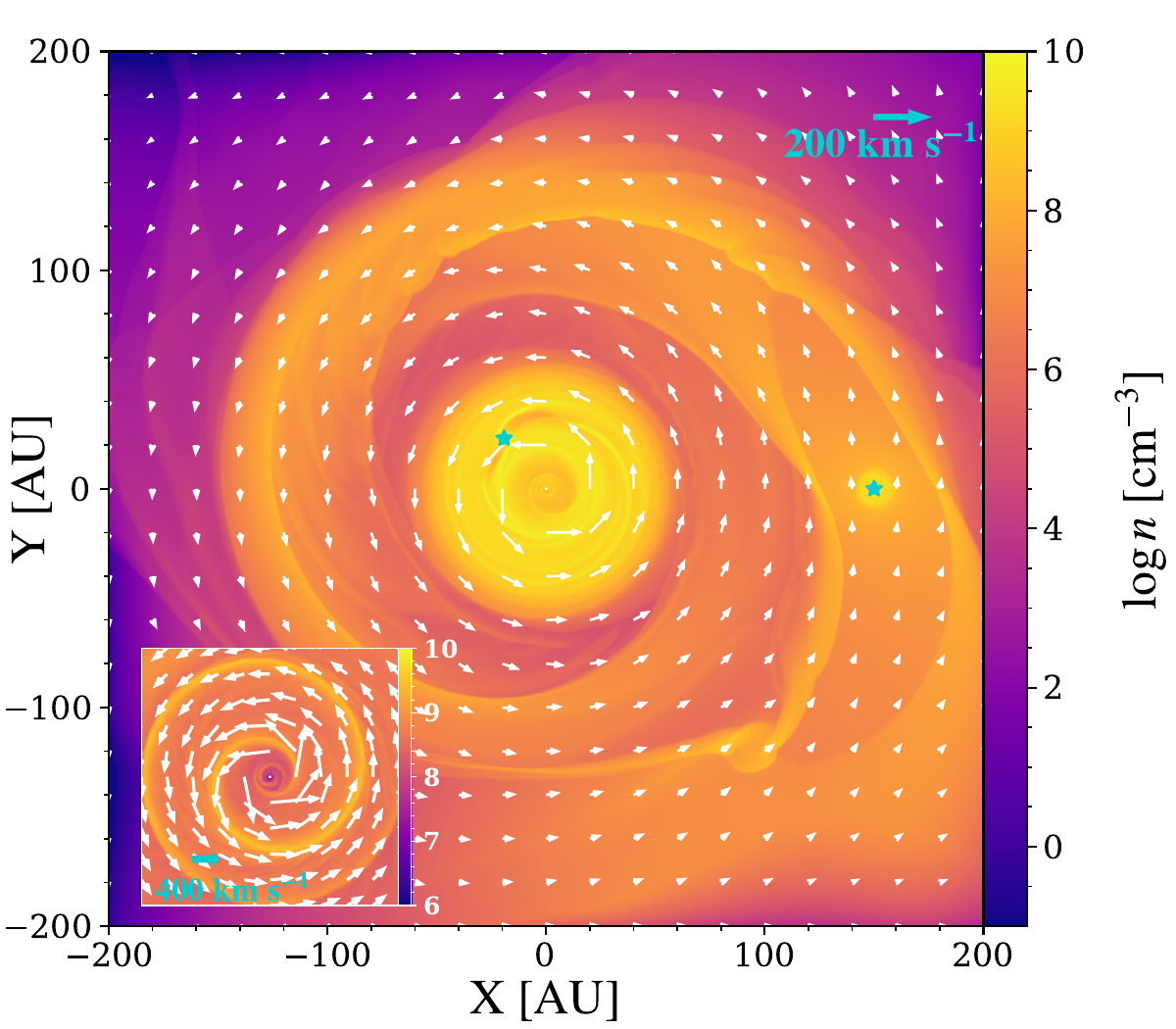}{0.5\textwidth}
    \afig{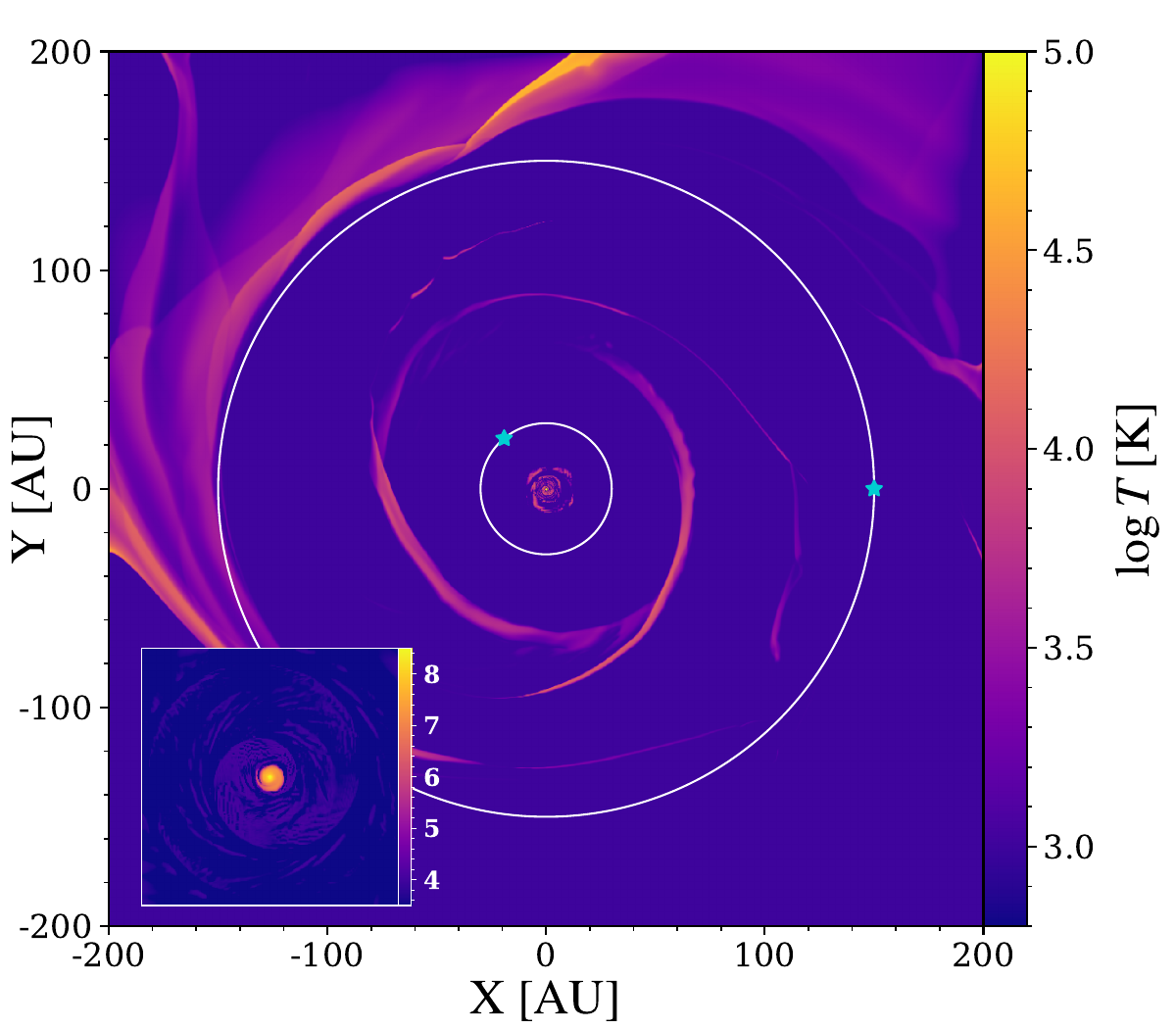}{0.5\textwidth}
		}
	\caption{Snapshots of the 2D simulations for two stars orbiting the IMBH. {\it Left}: Number density distribution in units of ${\rm cm^{-3}}$. {\it Right}: Temperature distribution in units of K. {\it Top}: the 2MS case. {\it Bottom}: the 2AGB case. The inserts show a zoom-in view of the  central 10 AU$\times$10 AU region. The two stars, marked by the pentagrams, have a circular orbit with a radius of $30~{\rm AU}$ and $150~{\rm AU}$, which are represented by the white circles. In all panels, the color bar is in logarithm of 10. \label{fig:2d}}	
\end{figure*}

Figure~\ref{fig:2d} displays the last snapshot of the 2D simulations, i.e., after the outer star has completed 2 orbits (the inner star has completed more than 22 orbits in the meantime) in both the 2MS and 2AGB cases.
In the 2MS case (top panels), both stars launch fast winds. The outward pointing winds from the inner star collide with the inward pointing winds from the outer star, creating a strong shock.
In the meantime, the inward pointing winds from the inner star have almost the same behavior as in the single MS star case, i.e., producing a bow shock in front of the IMBH and a high-temperature, trailing wake. 
The wind-wind collision effectively screens the entire winds from the outer star, and the IMBH accretes only from the inner star, with a mean accretion rate of $3.0 \times 10^{-14}\rm~M_\odot~yr^{-1}$(right panel of Figure~\ref{fig:accrete_3d}). This is about an order of magnitude higher than that found in the 3D1MS case. 
Such a difference is due to two factors. The primary factor is the fact that the winds can escape in the vertical direction in the 3D case. 
Moreover, the colliding winds 
between the two stars effectively boost the turbulent motions within the wake, resulting in an enhanced angular momentum transport. This latter effect also explains the larger fluctuation ($\sim$0.5 dex) around the mean accretion rate in the 2D case.


In the 2AGB case (bottom panels), interaction between the two stars also takes place, which is best seen as arc-like features of enhanced density and temperature as the result of shock compression and heating. The main structure in this case is still a cool disk, mainly supplied by the winds from the inner star. The outer star also produces a high-density shell structure near its orbit, which is separated from the inner disk by a low-density belt.
Also in this case the IMBH accretion is entirely fed by winds from the inner star, at least throughout the simulation time.  
The mean accretion rate 
averaged over the inner star's last orbital period is $\sim7 \times 10^{-9}\rm~M_\odot~yr^{-1}$ (right panel of Figure~\ref{fig:accrete_3d}),
about 5 times lower than in the 3D1AGB case.
Notably, such a trend is opposite to 
that between the 3D1MS and 2D2MS cases.
This can be understood as turbulence being suppressed in 2D simulations, which results in less efficient angular momentum transport in the cool disk.

\subsubsection{X-ray detectability}


To facilitate comparison with the X-ray observations, we calculate synthetic X-ray spectrum and luminosity at the last snapshot of the 3D, one-star simulations. 
To do so, we employ the emissivity, $\Lambda(T, Z, E)$, of an optically-thin thermal plasma in collisional ionization equilibrium extracted from ATOMDB\footnote{\url{http://www.atomdb.org}} version 3.0.9 and assuming a metallicity of $0.1~Z_\odot$.
For a single cell of volume $dV$, the X-ray luminosity follows $dL_{\rm X}=n_e n_{\rm H}\Lambda_{\rm X}dV$, where $\Lambda_{\rm X}=\int \Lambda(T,Z,E)dE$ is evaluated over 0.5--8 keV,
$n_{\rm e}$ is the electron density, $n_{\rm H}$ the hydrogen density, $T$ the temperature and $E$ the photon energy.
The total 0.5--8 keV luminosity is derived by summing up all cells in the simulation box, which is found to be $1.2\times 10^{19}~{\rm erg~s^{-1}}$ and $1.5\times 10^{34}~{\rm erg~s^{-1}}$, respectively, in the 3D1MS and 3D1AGB simulations.
The huge difference between the two cases stems from the $\sim 7$ orders of magnitude difference in the accretion rate and the density square dependency on the emission measure.
The predicted X-ray luminosity in the MS case is so small that it is impossible to detect this feeble accretion-induced signature with {\it Chandra}.
On the other hand,
the predicted X-ray luminosity in the AGB case is substantial, and interestingly, quite comparable to the X-ray luminosities of the detected central sources, in particular the ones in NGC\,6652 and Terzan\,5.
It is noteworthy that in both the MS and AGB cases the synthetic X-ray emission arises predominantly from the $\sim 1~{\rm AU}$ vicinity of the IMBH, where the gas temperature have values $\gtrsim 10^6$ K. 
Hence a remote observer should detect a point-like X-ray source -- indeed, at a distance of 1 kpc, the entire simulation box of 200 AU would span an angular size of only 0\farcs2, significantly smaller than the PSF of {\it Chandra}.

\begin{figure}
	\centering
	\includegraphics[width=0.5\textwidth]{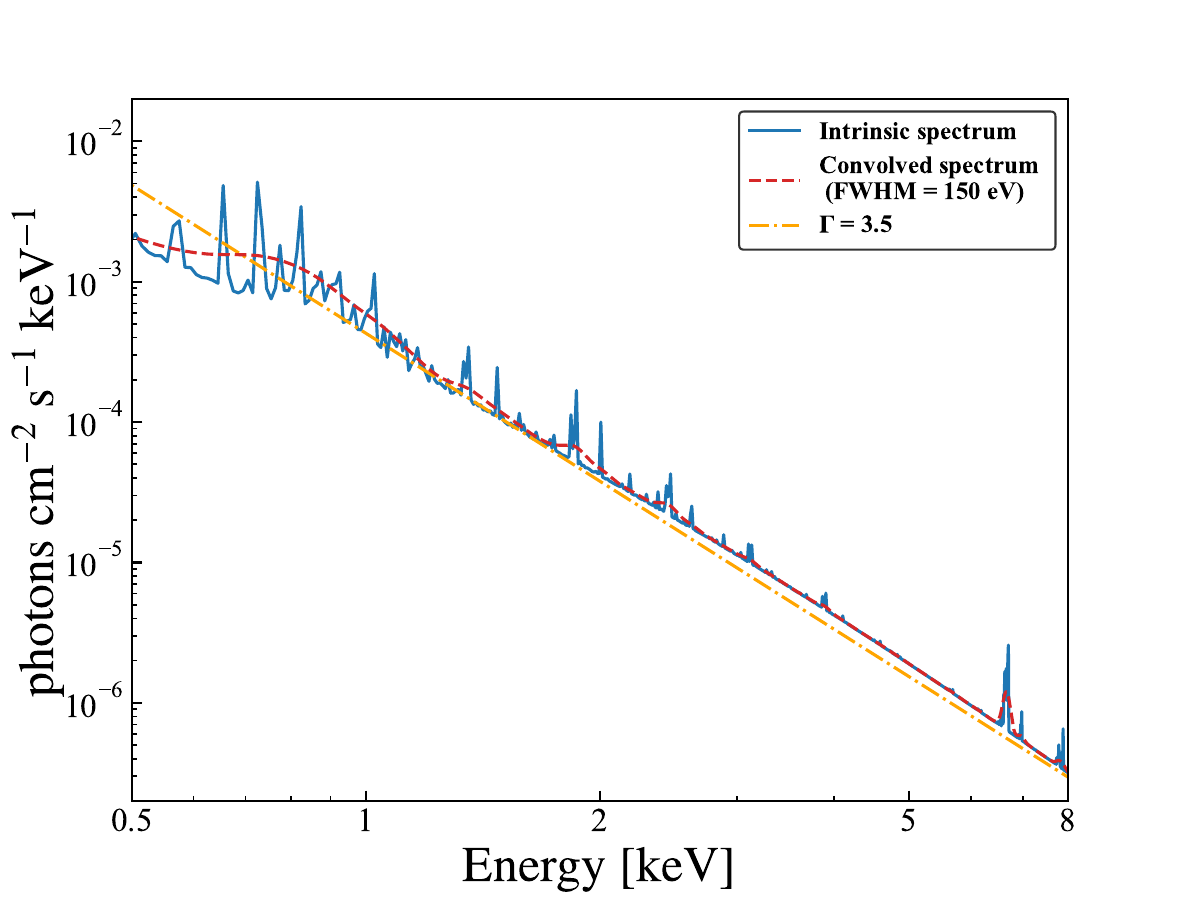}
	\caption{Synthetic X-ray spectrum of the 3D1AGB simulation. The blue solid curve represents the intrinsic spectrum, assuming a distance of 10 kpc. To mimic the energy resolution of {\it Chandra}/ACIS CCDs, the intrinsic spectrum is folded with a Gaussian function with FWHM of 150 eV, which is represented by the red dashed curve. For comparison, the orange dash-dotted line indicates a power-law spectrum with a photon-index of 3.5.
	\label{fig:synspec}}	
\end{figure}

We further show the synthetic X-ray spectrum of the 3D1AGB case in Figure~\ref{fig:synspec}.
Interestingly, under the typical energy resolution of {\it Chandra}, the synthetic spectrum appears smooth and can mimic a rather steep power-law, similar to those seen in NGC\,6652 and Terzan\,5 (Figure~\ref{fig:spectra}).
We note that the cool disk can cast significant line-of-sight absorption to the centrally-peaked X-rays, which is neglected in Figure~\ref{fig:synspec}. 
From the 3D1AGB simulation we estimate that the neutral hydrogen column density at an inclination angle of $0^\circ$ (i.e., along the orbital plane) and $15^\circ$ is $\sim 8\times10^{24}~\rm cm^{-2}$ and $\sim 7\times10^{20}~\rm cm^{-2}$, respectively. 
The column density drops rapidly above $15^\circ$, since the increased gas temperature in the corona ($T > 10^4$ K) leaves little room for neutral gas.  
Hence the probability for the disk obscuring the central X-ray emission is rather small.

\section{Discussion}\label{sec:discussion}

Thanks to the extensive archival {\it Chandra} data, we have conducted a survey of central X-ray emission for more than half of the currently known population of Galactic GCs. 
From this systematic search (Section~\ref{sec:analysis}), we reveal a general deficiency of X-ray emission from the central $\sim1
\arcsec$ region in these GCs,
obtaining an upper limit of $L_{\rm X,lim} < 10^{32}\rm~erg~s^{-1}$ for
the vast majority of them.
A central point source is firmly detected in only six clusters, which have $L_{\rm X} \lesssim 10^{34}\rm~erg~s^{-1}$. These values are many orders of magnitude lower than the Eddington luminosity of an IMBH, $\sim 10^{41}(M/10^3\rm~M_\odot)~erg~s^{-1}$.
An immediate implication of this result is a low occupation fraction of IMBH in GCs, which, however, appears at odds with efficient IMBH formation in GCs as hypothesized by early work \citep{2002ApJ...576..899P,2002MNRAS.330..232C}.
If instead IMBHs were prevalent in GCs, then the deficiency of central X-ray emission requires that these IMBHs have current accretion rates much lower than the Eddington limit. 
This is supported by the result of our numerical simulations (Section~\ref{sec:simulation}).

Indeed, highly sub-Eddington accretion rates were generally inferred by previous work, which were often based on the assumption of Bondi accretion from a uniform ambient \citep{2004MNRAS.351.1049M,2018ApJ...862...16T}, or from a more sophisticated modelling \citep{2011ApJ...730..145V,2013MNRAS.430.2789P}.
In these estimates of the accretion rate, often the entire volume under the IMBH's gravitational influence is involved, which is characterized by a canonical Bondi radius of $R_{\rm B} \sim (M/10^3{\rm~M_\odot})/(T/10^4\rm~K) \sim 10^4\rm~AU$.
Moreover, the above studies needed to assume a radiation efficiency to predict the X-ray luminosity from the accretion rate.  
Unfortunately, this had led to too pessimistic X-ray luminosities than actually constrained by the X-ray observations (Section~\ref{subsec:nature}).
In contrast, our modelling in Section~\ref{sec:simulation} builds on the recognition that a dynamical hierarchy exists among the stars gravitationally bound to the putative IMBH \citep{2016ApJ...819...70M}. 
Our modelling also has the advantage that the accretion-induced X-ray luminosity can be derived directly from the HD simulation and requires no empirical assumption on the radiation efficiency.
Our simulations demonstrate that the IMBH is essentially fed by the most-bound star, the orbital size of which is a factor of $\sim$100 smaller than the canonical Bondi radius.
Consequently, the predicted accretion rates for the most-bound star being a MS star are much lower than inferred by previous work, and the extremely low X-ray luminosities predicted provides a natural understanding to the non-detection of central X-ray source in most GCs. 
A potential caveat in drawing this conclusion is that our simulations have only considered limited physical configurations, i.e., involving the two most-bound stars and evolving in 2D. Nevertheless, given the substantial radial outflow in both the 3D1MS and 3D1AGB cases (Section~\ref{subsubsec:3D}), it seems safe to predict that the most-bound star with its winds can effectively shield an inflow from outer regions, provided the validity of the presumed stellar hierarchy around the IMBH.

Perhaps the most interesting finding of our simulations is that the predicted X-ray luminosity and spectrum for the most-bound star being an AGB star are in reasonable agreement with the detected central X-ray sources, in particular those found in Terzan\,5 and NGC\,6552 (Section~\ref{subsec:spec}). However, we consider it premature to claim that any of these two sources are tracing an accreting IMBH, for the following considerations. First, in reality the chance of having an AGB star with strong winds as the most-bound star is rather small. \citet{2016ApJ...819...70M} predicts a $\sim$10\% probability for the most-bound star being a giant star, but this should include both red giants and AGB stars, with the former being far more likely.
Moreover, only AGB stars during the thermally pulsing phase can hold this high mass loss rate ($10^{-6}~\rm M_\odot~yr^{-1}$), which lasts for $\rm \sim 1~Myr$, about one hundredth of the lifetime of red giants \citep{1993ApJ...413..641V}.
Second, while AGB stars are known to be present in GCs \citep[e.g.,][]{2010MmSAI..81.1004C}, from a literature search we find no known AGB star coincident with the center of Terzan\,5 or NGC\,6552.
We note, however, such an orbiting AGB star might be obscured by the cool disk (Section~\ref{subsubsec:3D}) depending on the viewing angle and the wavelength of observation. 
Third, as addressed in Section~\ref{subsec:nature}, the observed properties of all six central X-ray sources can be equally, if not better, explained in terms of a stellar object.
Nevertheless, our simulations clearly demonstrate how an IMBH might be illuminated by a stellar companion and becomes detectable in the X-ray band.

The non-detection of central X-ray sources in the vast majority of our sample GCs suggests that we must consider alternatives for the IMBH scenario.
A leading alternative, especially in non-core-collapsed GCs, is a black hole subsystem (BHS), namely, a centrally concentrated population of stellar-mass BHs, which have segregated to the cluster core and can survive for a sufficiently long time without further coalescing into an IMBH or being individually ejected   \citep{2013MNRAS.432.2779B}.
Recent dynamical models have shown that a BHS with a total mass falling within the IMBH mass range can mimic the dynamical behavior of an IMBH in the case of $\omega$ Cen \citep{2019MNRAS.488.5340B, 2019MNRAS.482.4713Z}.
Unfortunately, this also means that it is difficult to distinguish a BHS from an IMBH based on surface brightness and/or velocity dispersion distributions.
The radiative signature of a BHS is still an open question. 
Stellar wind accretion by individual members of the BHS is not expected to produce detectable X-ray emission, individually or collectively, because the smaller BH mass would result in a lower accretion rate than inferred for an IMBH in our simulations, other conditions being equal. 
Significantly stronger X-ray emission from the BHS might be realized if some of the BHs could capture a normal star to form a LMXB. Indeed, stellar-mass BH candidates have been proposed in a number of GCs, which are often associated with a LMXB (Section \ref{sec:introduciton}).
Most of these BH candidates are located within the cores of their host GCs, which is consistent with the strong mass-segregation effect of a stellar-mass BH. 
On the other hand, at least in the case of $\omega$ Cen, deep {\it Chandra} observations found no evidence for a central concentration of X-ray sources \citep{2020ApJ...904..198C}, indicating that at any given time only a minor fraction of the member BHs of a BHS, if existed, form LMXBs. 

Another alternative, which can be reconciled with the IMBH formation scenarios mentioned in Section \ref{sec:introduciton}, is that the IMBH can be subsequently ejected from the host GC due to dynamical encounters with stellar-mass BHs.
In the {\it N}-body simulation by \citet{2021MNRAS.507.5132D}, for young star clusters that may resemble GCs in their youth, about half of the IMBHs formed were ejected and the ejection was more likely for more massive clusters and less massive IMBHs. 
\citet{2022MNRAS.514.5879M} found a similar trend of ejection probability with IMBH mass: most escaping IMBHs in their simulations have masses less than $300~{\rm M_\odot}$, whereas a more massive IMBH will rapidly deplete stellar-mass BHs in the cluster core so that few or even no black holes can trigger the ejection. 
It is noteworthy that simulations focusing on gravitational wave recoil by which the IMBH is ejected predicted heavier ejected IMBHs with masses up to $3000~{\rm M_\odot}$ and a higher ejection fraction \citep{2008ApJ...686..829H}.
After ejection, a wondering IMBH is expected to be enclosed by a compact cluster of stars that were bound to it during the ejection.
\citet{2012MNRAS.421.2737O} used {\it N}-body simulations to show that a $10^4~{\rm M_\odot}$ wandering IMBH will host $\sim20$ stars today, with 80\% of the bound stars undergoing ejection or tidal disruption within 10 Gyr.
For a wandering IMBH with mass $\lesssim 1000~{\rm M_\odot}$, it is plausible to expect $\lesssim10$ bound stars today.
If this were the case, it would be very challenging to detect such compact clusters.
There is also chance that the wandering IMBH hosts a close stellar companion so that stellar wind accretion leads to some detectable signature.
However, \citet{2012MNRAS.421.2737O} showed no indication of companion hierarchy and predicted that the cluster would undergo persistent expansion. 

\section{Summary}\label{sec:summary}
In this work, we have searched for the accretion signature of putative IMBHs in 81 Galactic GCs using archival {\it Chandra} observations.
We have also performed HD simulations of IMBH accretion from winds of the tightly bound stars, to help understand the cause of low accretion rates and feeble accretion-induced radiation implied by the X-ray observations. 
Our main results are as follows:
\begin{itemize}
\item A central X-ray source is firmly detected in only six (47 Tuc, NGC\,6093, NGC\,6388, NGC\,6652, Ter 5, and NGC\,6681) of the 81 GCs, with 0.5--8 keV unabsorbed luminosities ranging between $1.3\times10^{30}~{\rm erg~s^{-1}}$ (47 Tuc) to $3.7\times10^{33}~{\rm erg~s^{-1}}$ (NGC\,6652).
Among them, four sources provide a statistically meaningful X-ray spectrum, all of which can be well-fitted by a power-law model;
three sources exhibit significant inter-observation variability.
The remaining 75 GCs show no significant central X-ray emission, but thanks to the superb sensitivity of {\it Chandra},  
a $3\sigma$ upper limit as low as a few $10^{29}~{\rm erg~s^{-1}}$ is achieved for these GCs. 

\item The nature of the detected central X-ray sources is uncertain. Their observed X-ray properties are consistent with an origin of close binaries, such as quiescent LMXBs, CVs and ABs.
But the possibility that some, if not all, of the six sources are related to an accreting IMBH cannot be ruled out by the observed X-ray properties alone.

\item The 3D simulations are run for two representative cases of the most-bound star, one having a MS star with a fast, low-density wind and the other having an AGB star with a slow, high-density wind. In both cases, the star orbits a $1000~{\rm M_\odot}$ IMBH on a circular orbit with a radius of 30 AU. 
In the MS case, most stellar winds escape the simulation domain as a radial outflow and only $\sim$0.13\% of the stellar mass loss ($10^{-12}~{\rm M_\odot~yr^{-1}}$) is ultimately accreted by the IMBH, producing a tiny X-ray luminosity of $\sim 10^{19}~{\rm erg~s^{-1}}$.
In the AGB case, the bulk of the stellar winds are trapped in a cool, thin disk and $\sim$3.5\% of the stellar mass loss ($10^{-6}~{\rm M_\odot~yr^{-1}}$) is accreted by the IMBH, producing an X-ray luminosity of $\sim10^{34}~{\rm erg~s^{-1}}$, which can be comfortably detected by {\it Chandra}. However, this detectability is strongly comprised by the low probability of having an AGB star as the most-bound star around the putative IMBH.
\item The 2D simulations are run for two representative cases of two most-bound stars, one having two MS stars and the other having two ABG stars. In both cases, the inner and outer stars orbit at a radius of 30 AU and 150 AU, respectively, around a $1000~{\rm M_\odot}$ IMBH. 
While the accretion rate differs quantitatively compared to that found in the corresponding 3D simulation, in both cases winds from the outer star are effectively shielded by winds from the inner star, and the IMBH accretes only from the latter.
\end{itemize}

The present study suggests that it is very difficult to detect IMBHs in GCs from the X-ray window, even if they were truly prevalent in GCs today.
Alternative and promising methods such as utilizing gravitational microlensing \citep[e.g.,][]{2016MNRAS.460.2025K} and gravitational waves \citep[e.g.,][]{2019MNRAS.486.5008A} deserve more future effort. 

\section*{Acknowledgements}
This work is supported by the National Key Research and Development Program of China (grant 2017YFA0402703) and the National Natural Science Foundation of China (grants 11873028 and 12003017).
M.H. acknowledges support by the fellowship of China National Postdoctoral Program for Innovation Talents (grant BX2021016).
The authors wish to thank an anonymous referee for helpful suggestions, and Drs. Feng Yuan and Zhengwei Liu for useful comments.

\section*{Data Availability}
The data underlying this article will be shared on reasonable request to the corresponding author.

\bibliography{draft}{}
\bibliographystyle{mnras}


\appendix
\section{Additional Figures of Non-detections}
In this appendix, we present {\it Chandra}/ACIS counts images of the remaining 75 GCs (Figure~\ref{fig:nondetection}), which do not exhibit a significant source coincident with the cluster center.

    \begin{figure*}
    \centering
\agridline{
        \afig{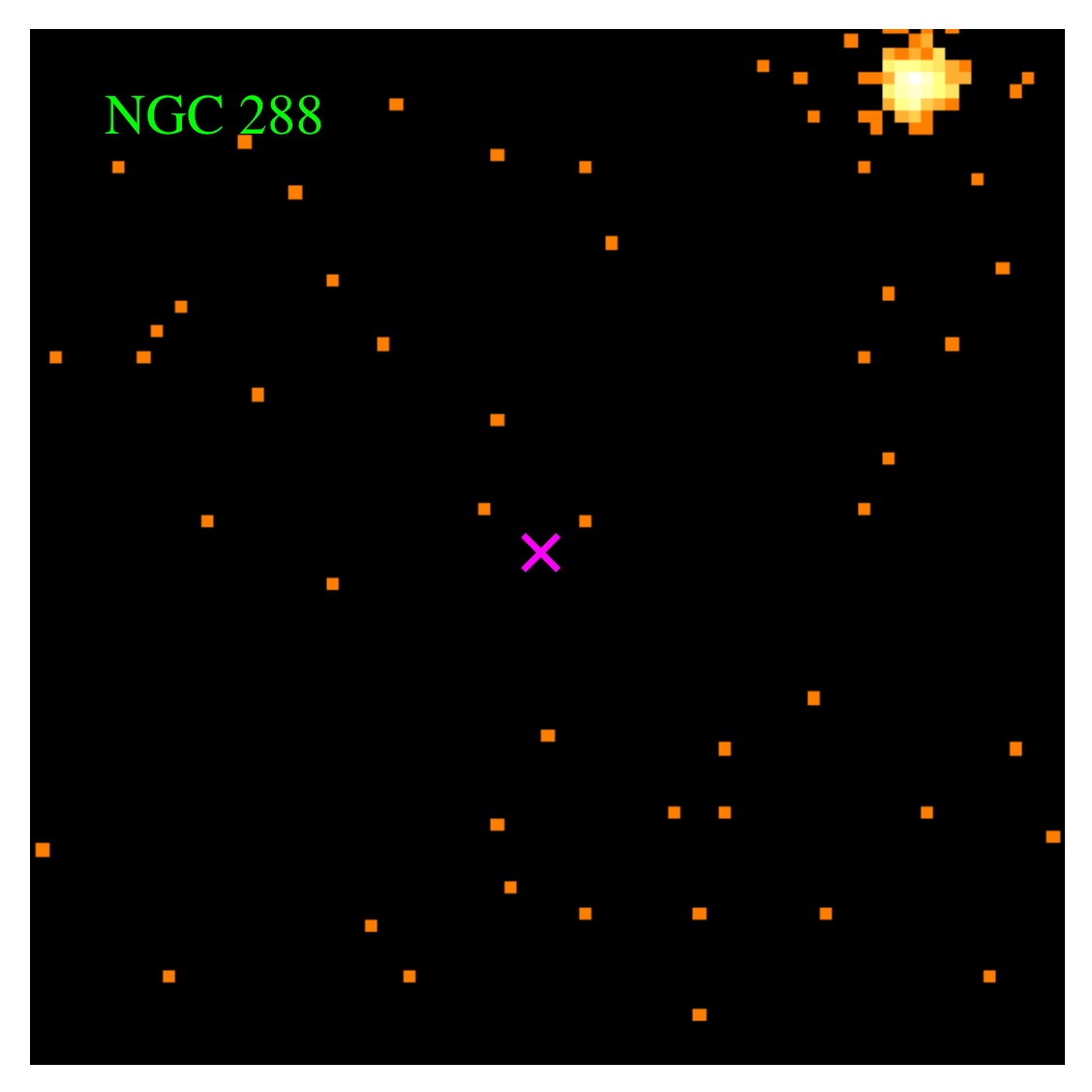}{0.2\textwidth}
        \afig{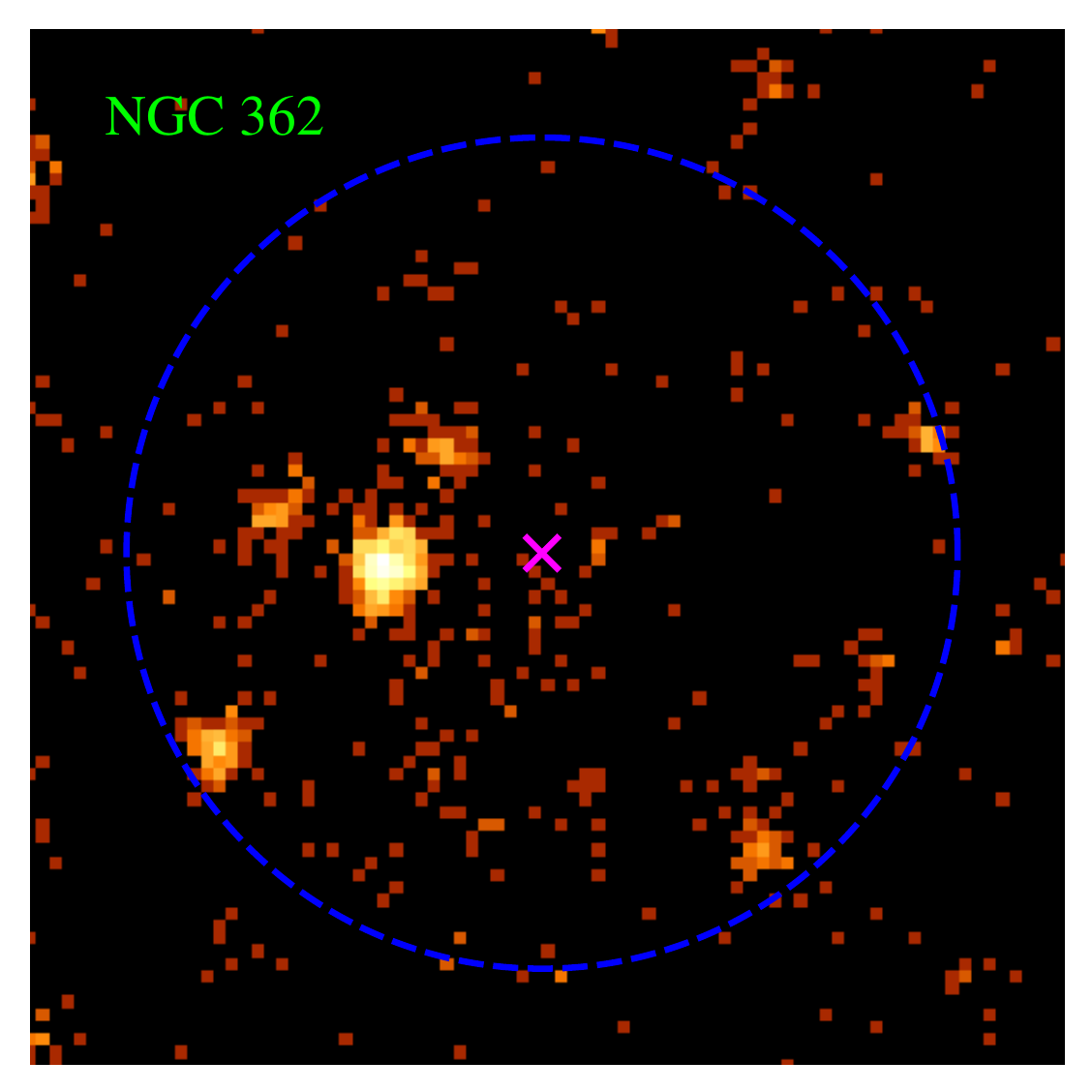}{0.2\textwidth}
        \afig{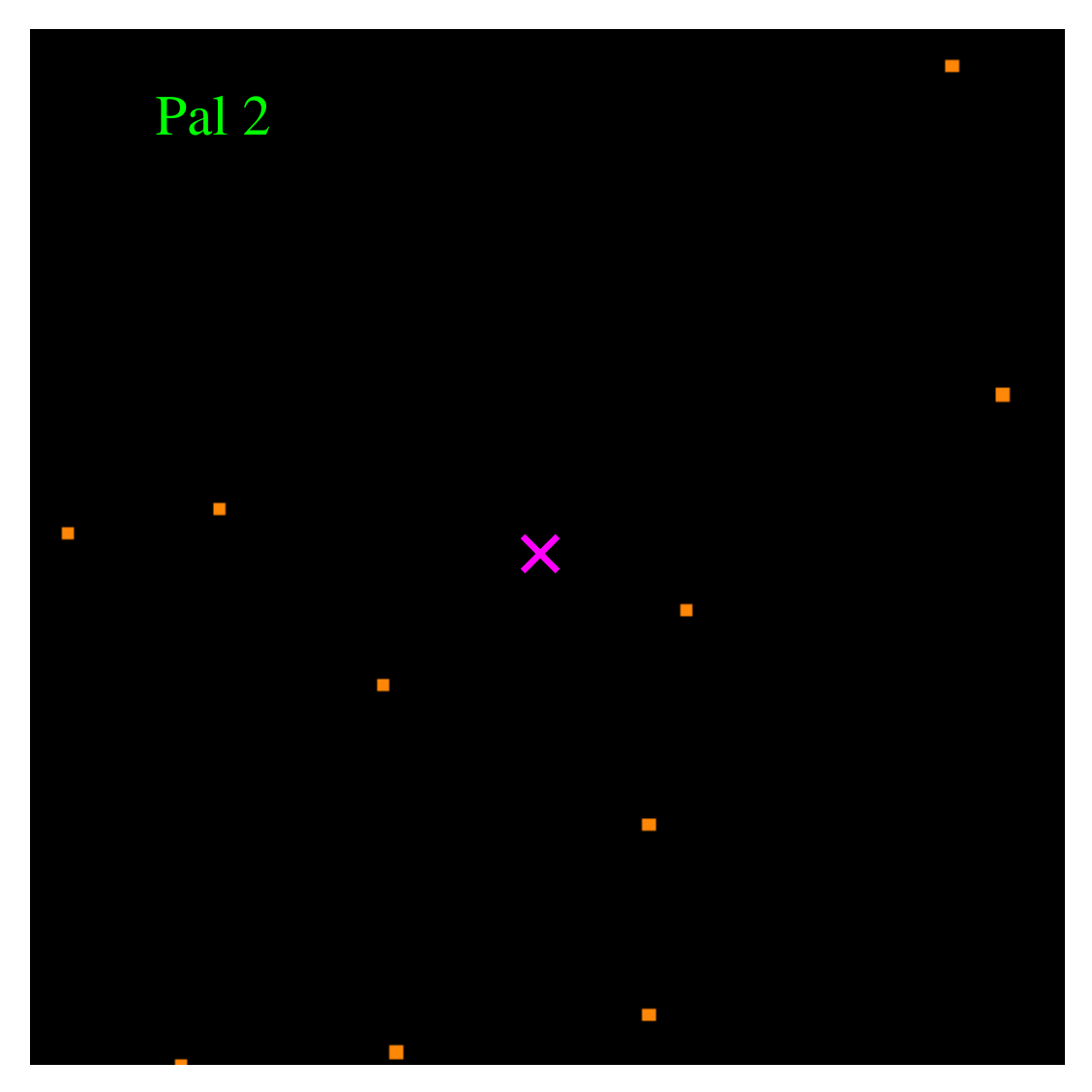}{0.2\textwidth}
        \afig{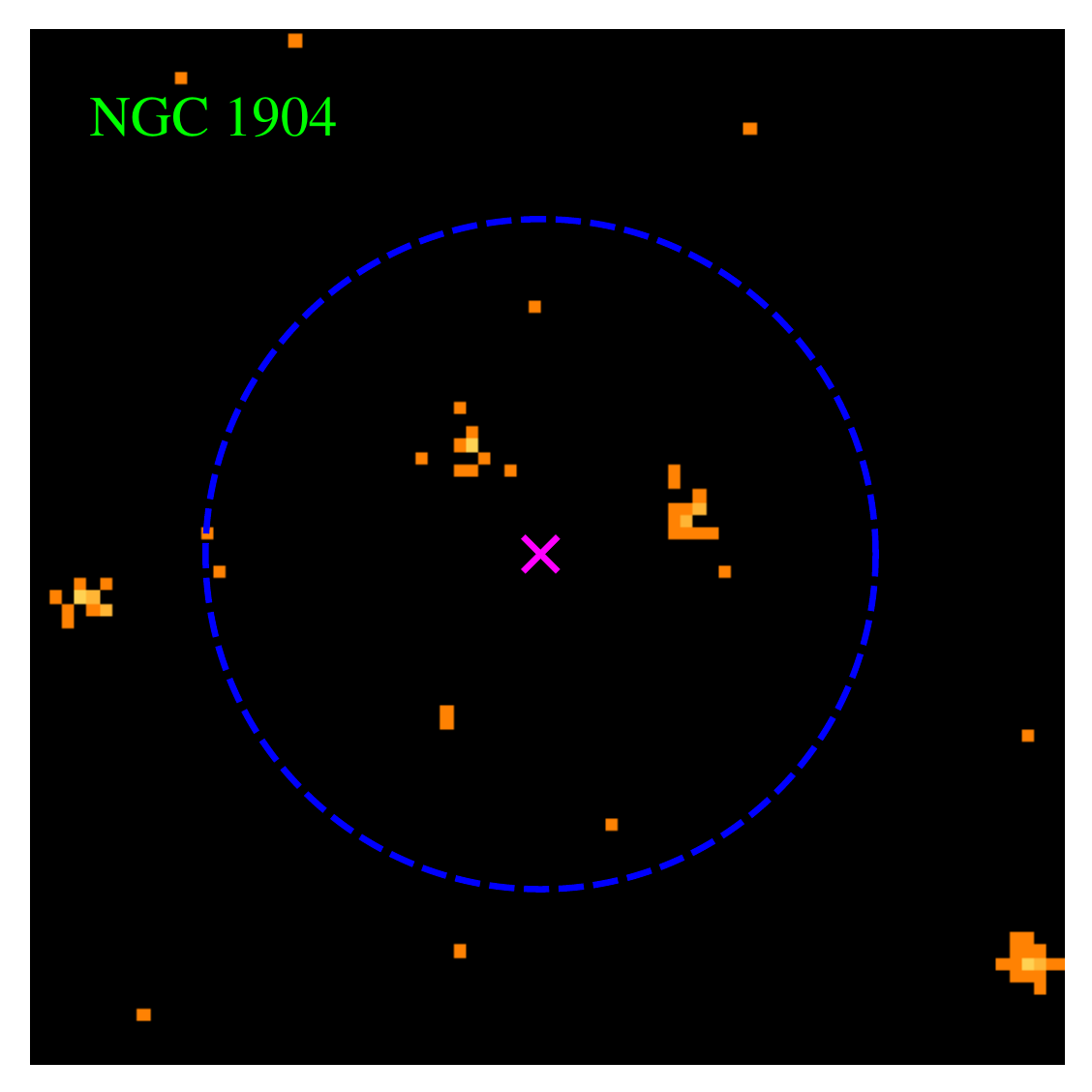}{0.2\textwidth}
        \afig{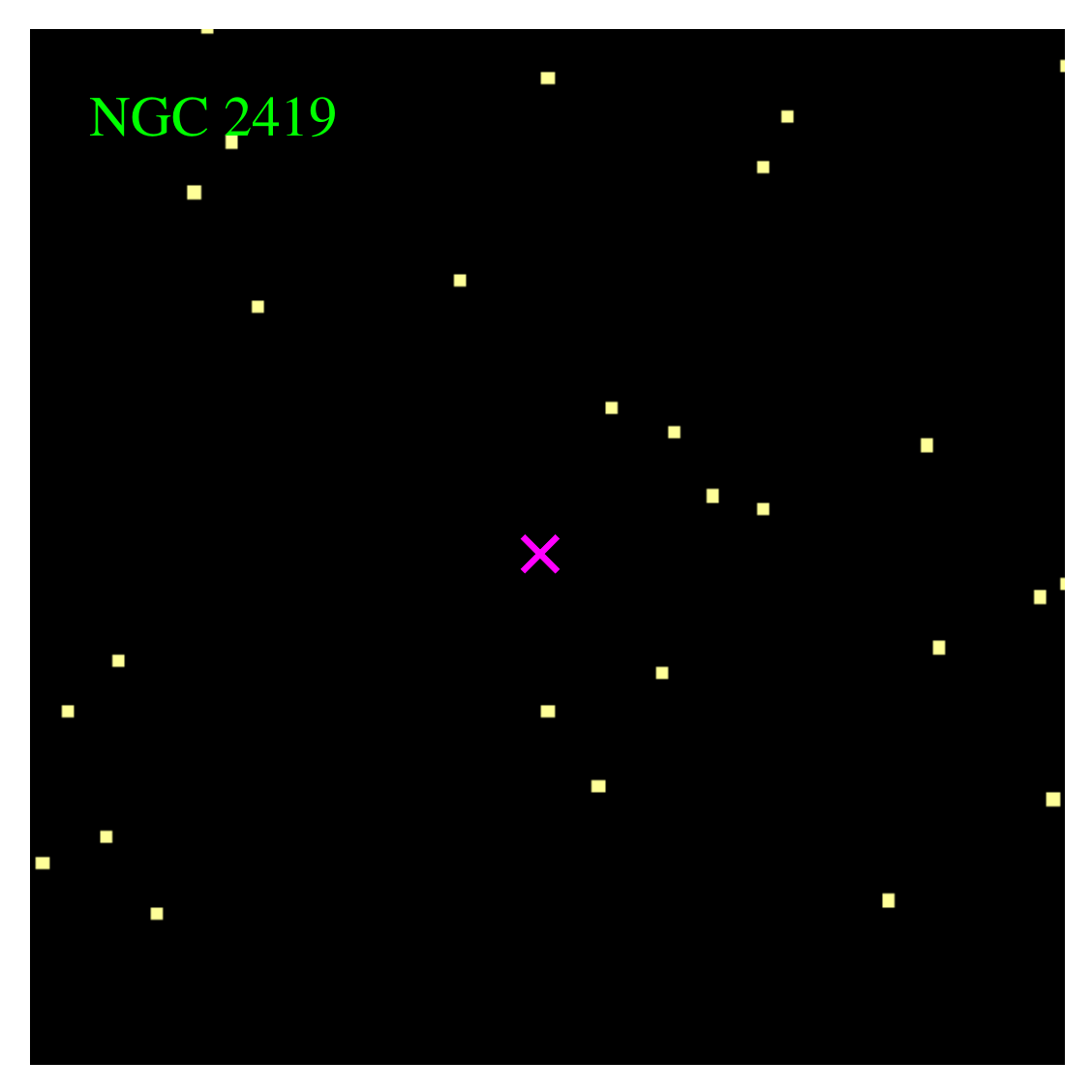}{0.2\textwidth}
    }
\agridline{
        \afig{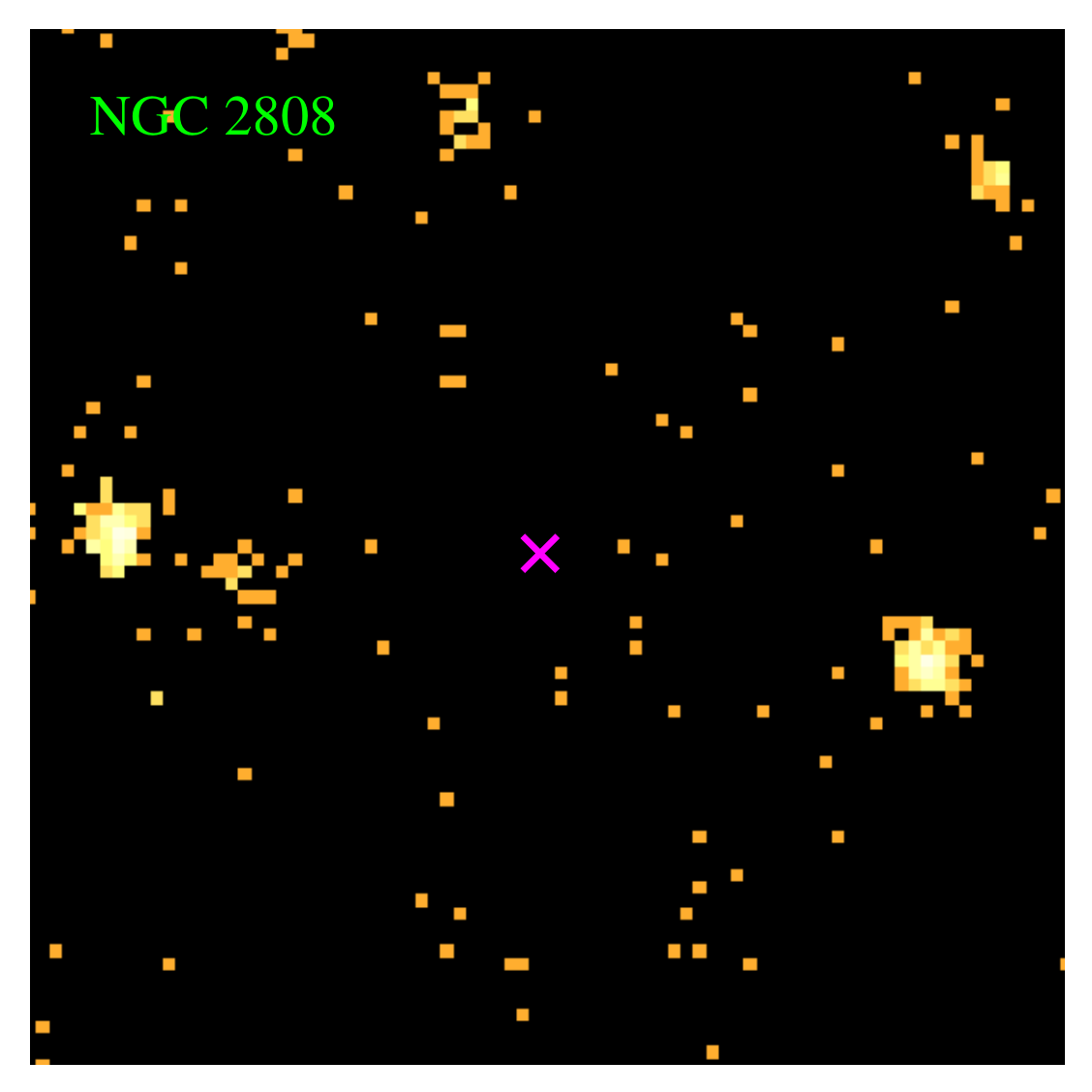}{0.2\textwidth}
        \afig{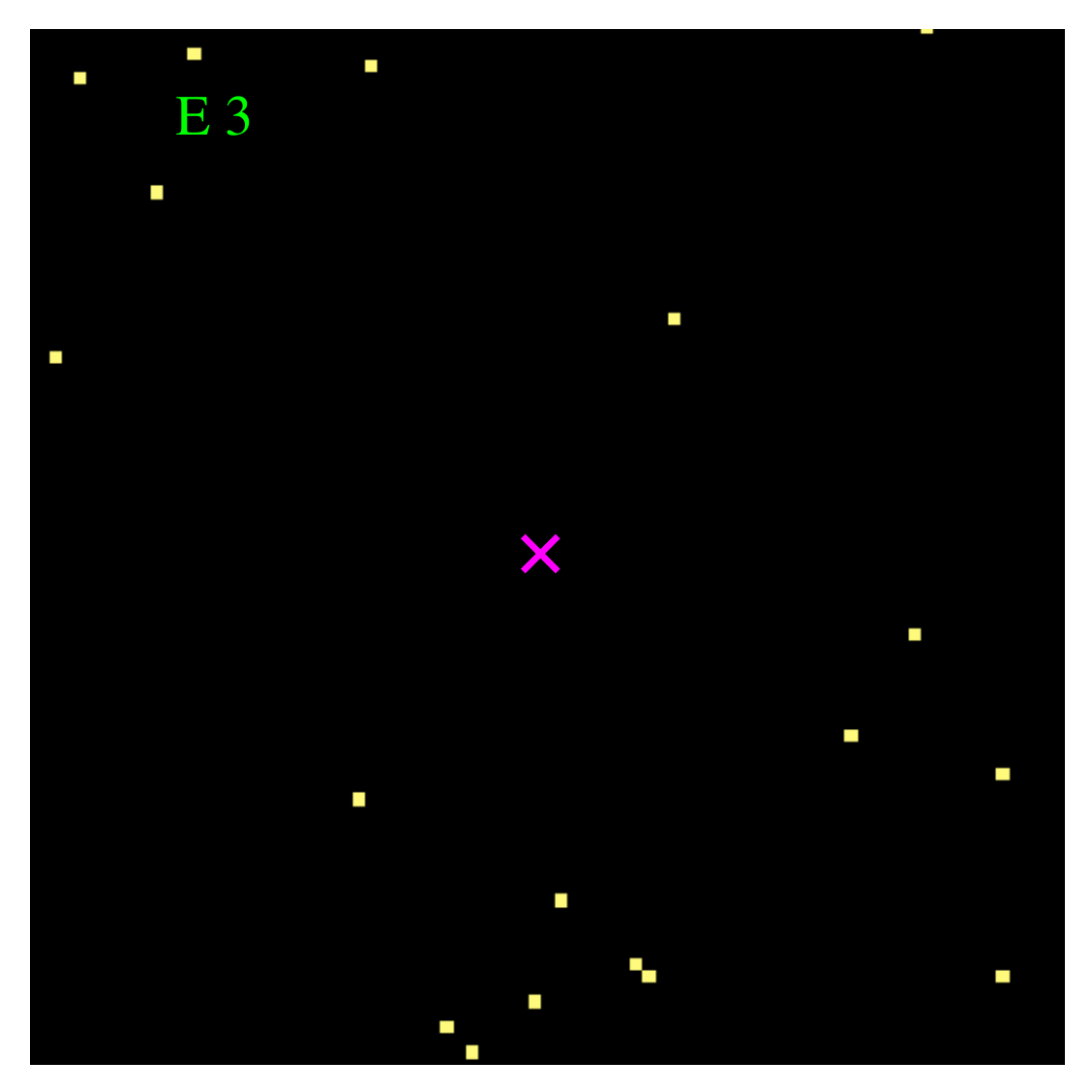}{0.2\textwidth}
        \afig{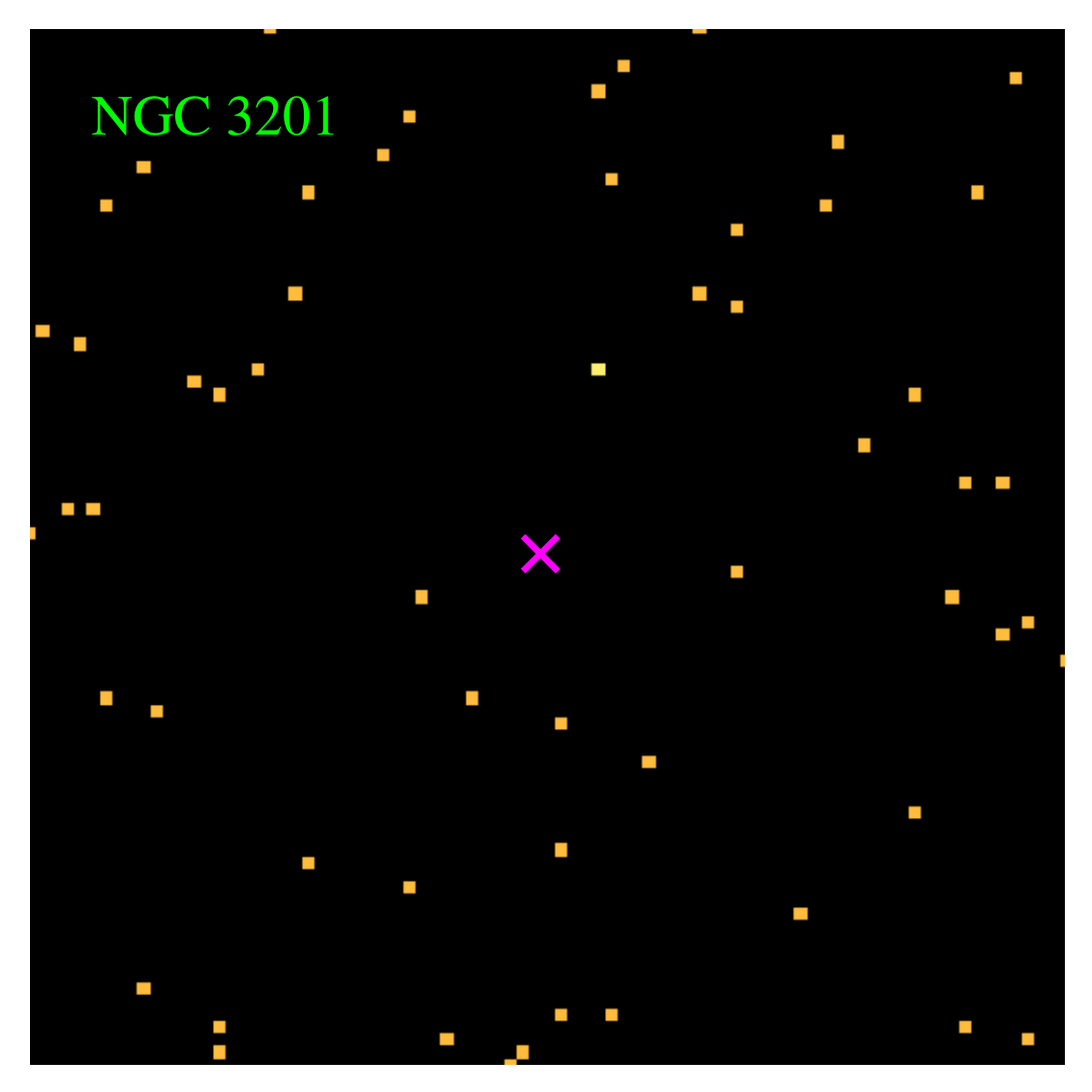}{0.2\textwidth}
        \afig{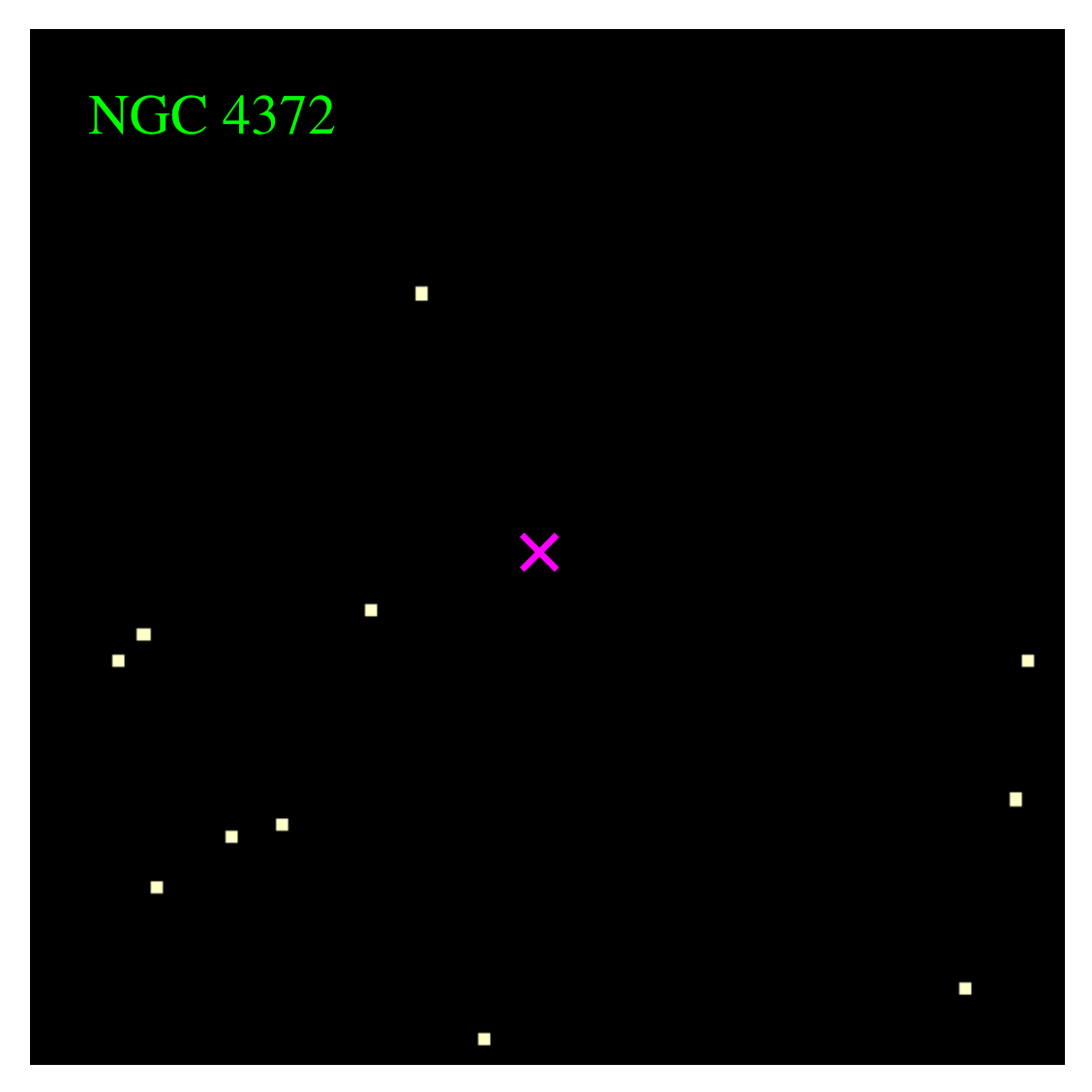}{0.2\textwidth}
        \afig{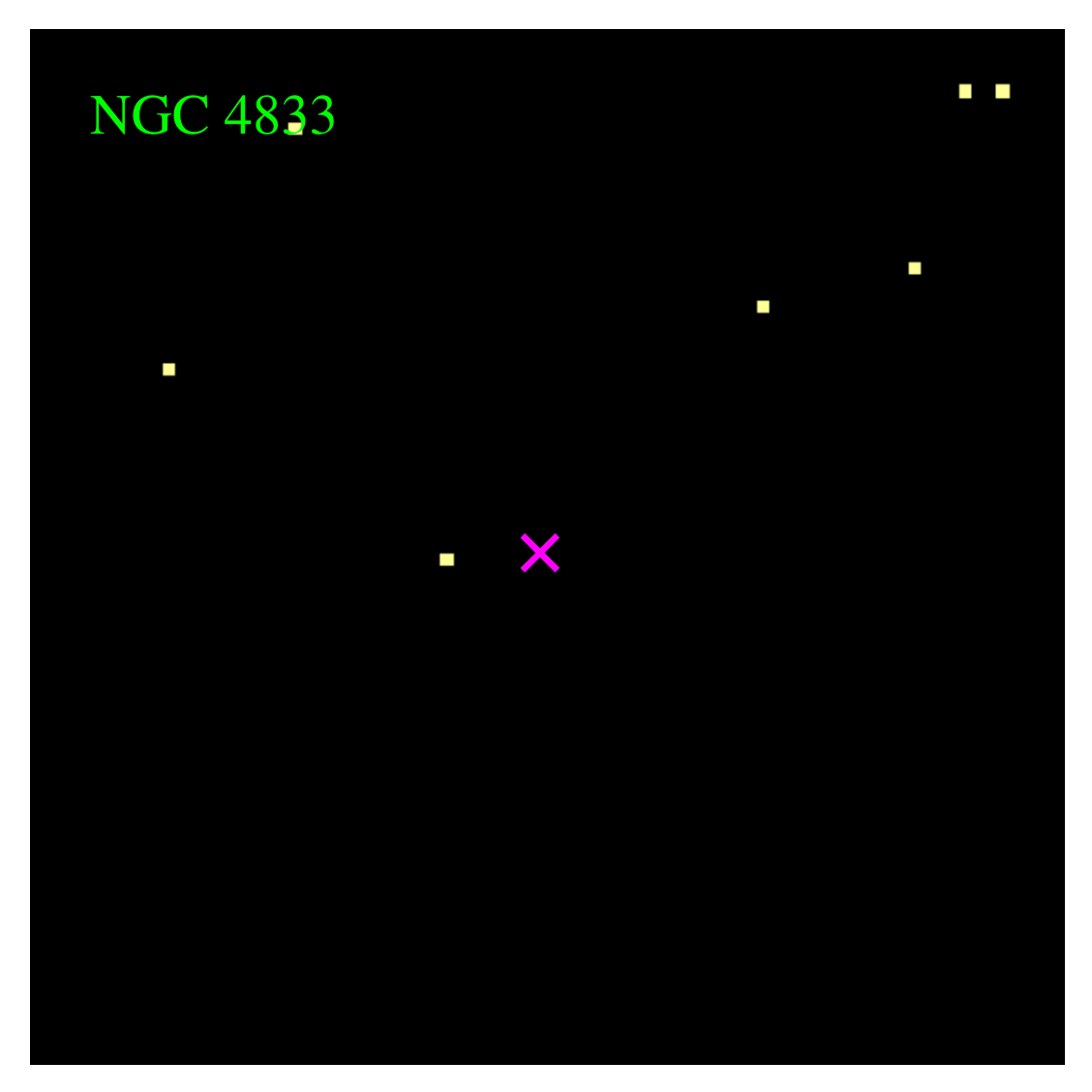}{0.2\textwidth}
    }
\agridline{
        \afig{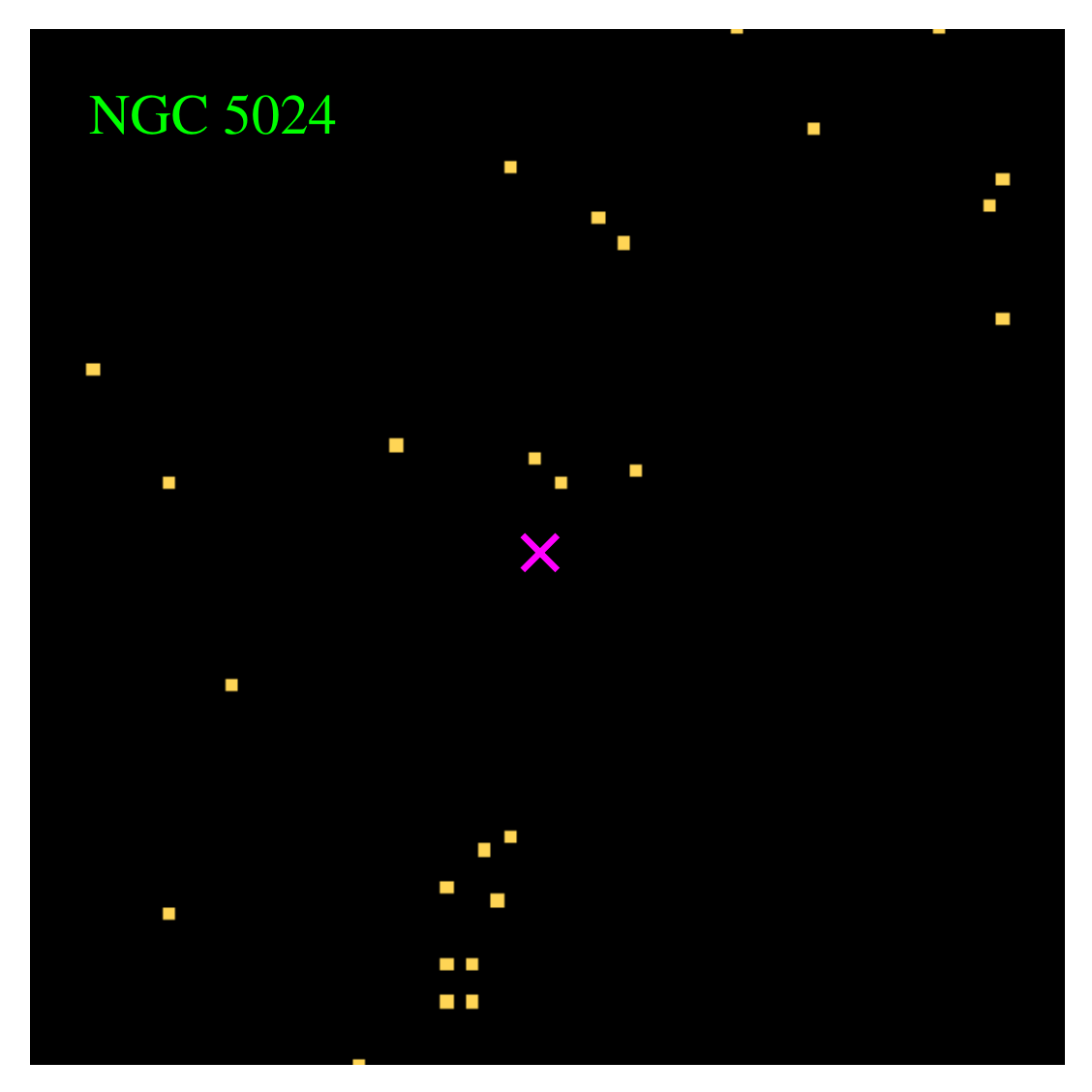}{0.2\textwidth}
        \afig{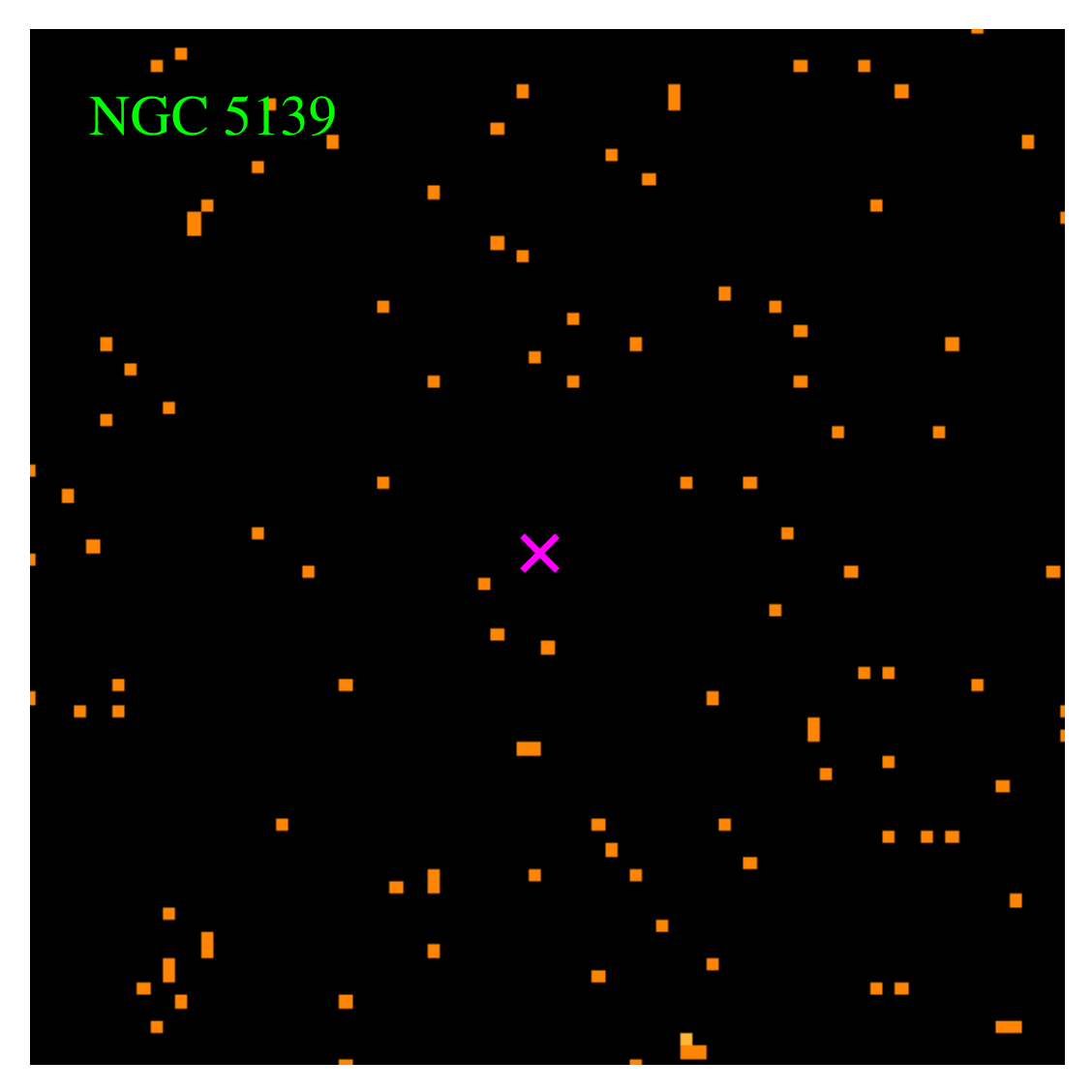}{0.2\textwidth}
        \afig{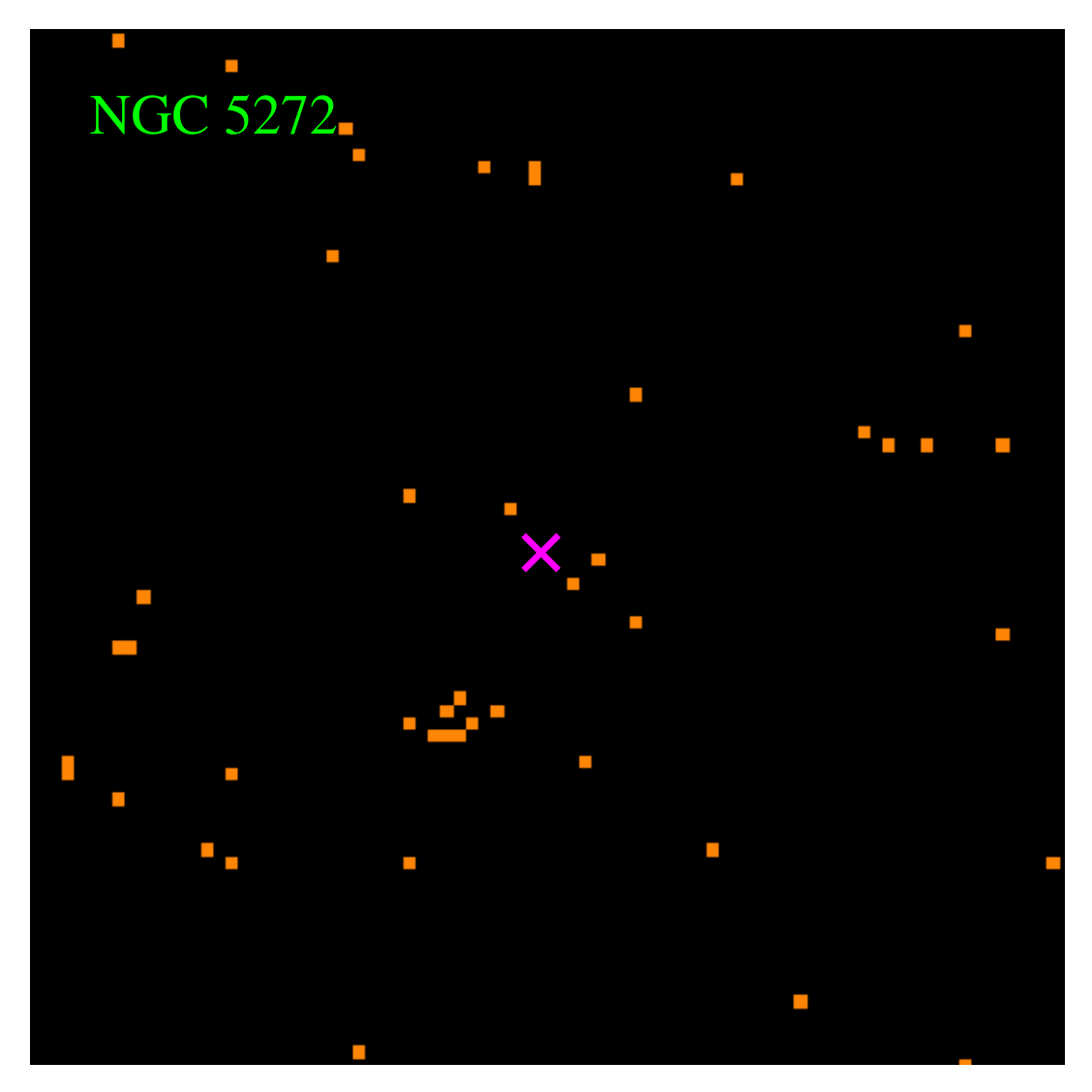}{0.2\textwidth}
        \afig{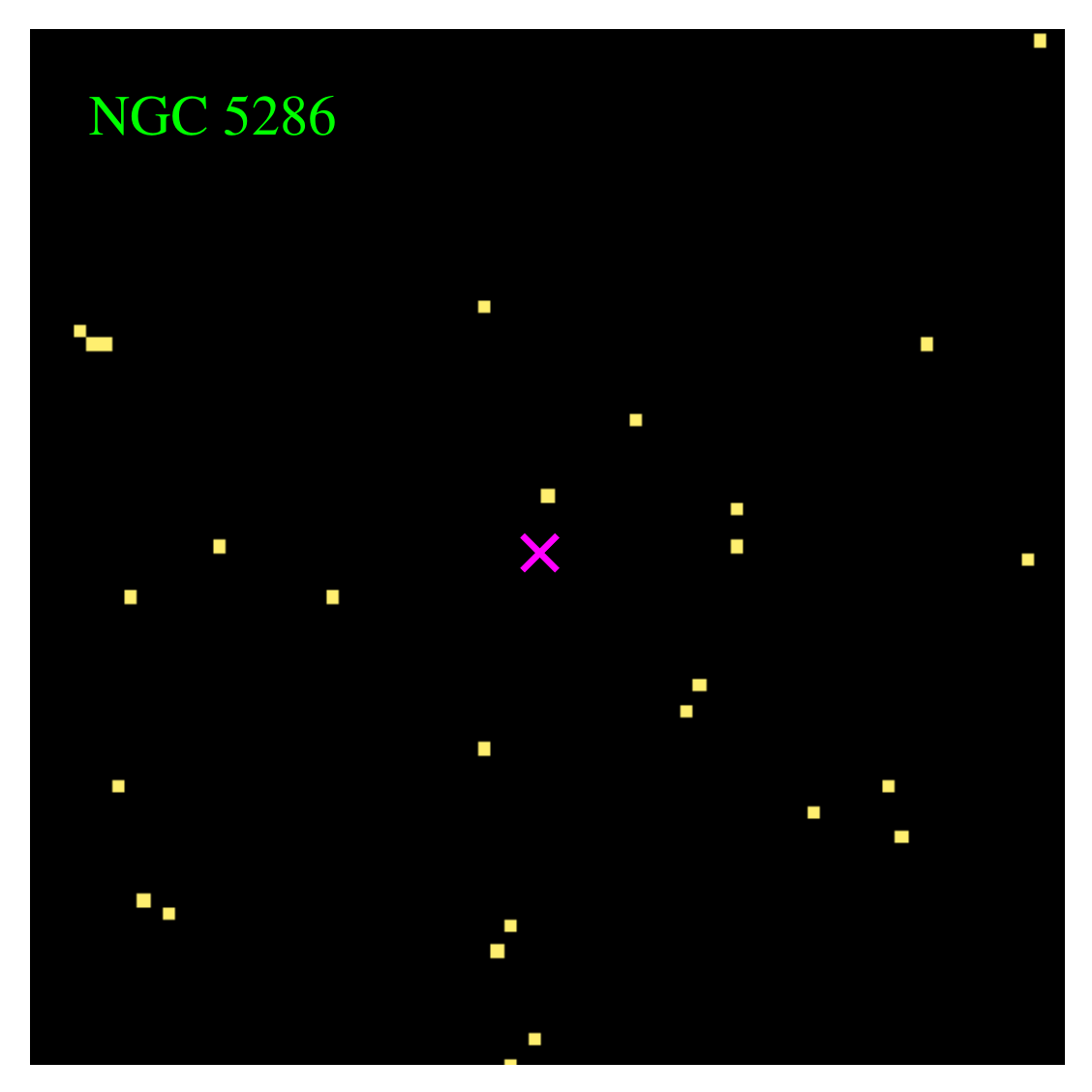}{0.2\textwidth}
        \afig{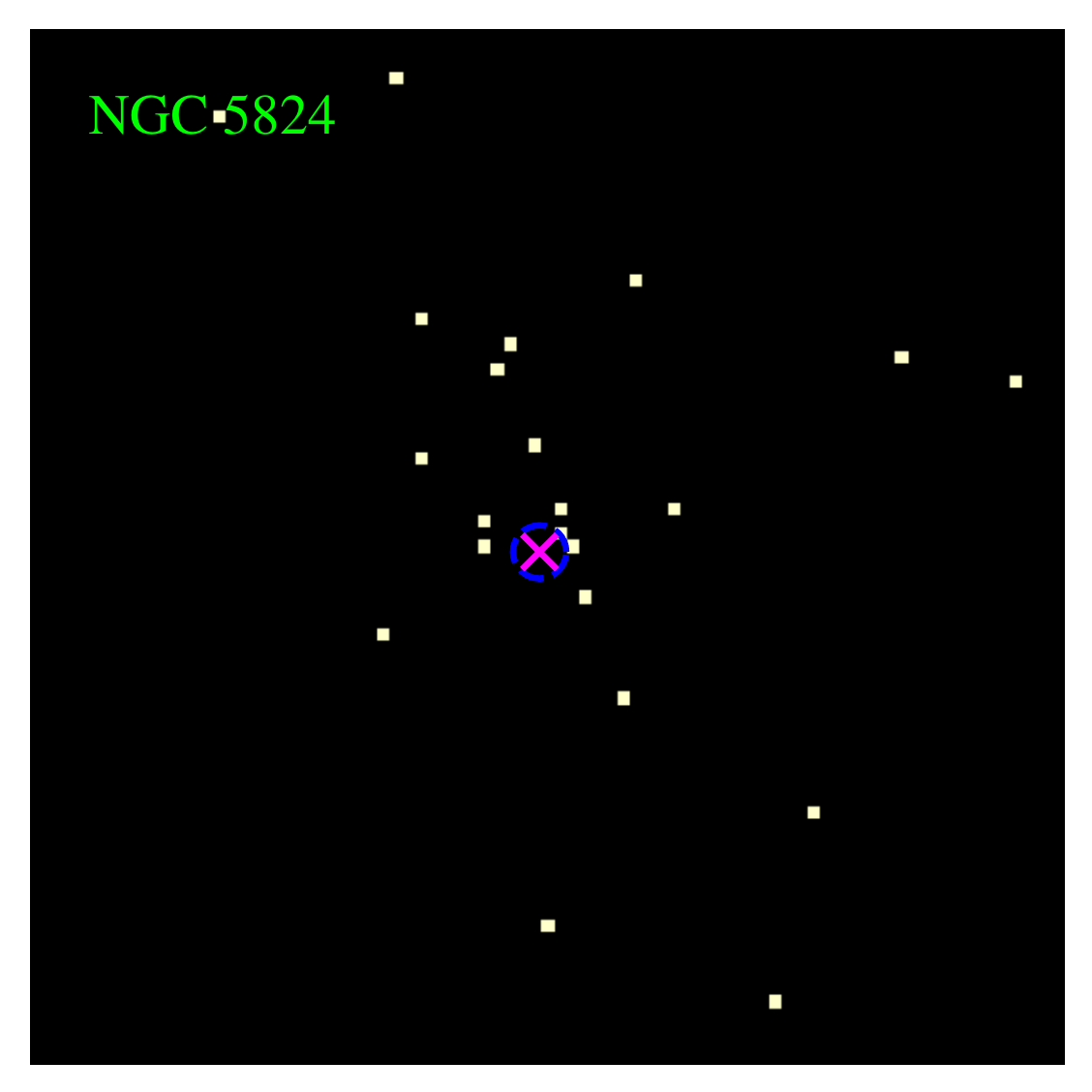}{0.2\textwidth}
    }
\agridline{
        \afig{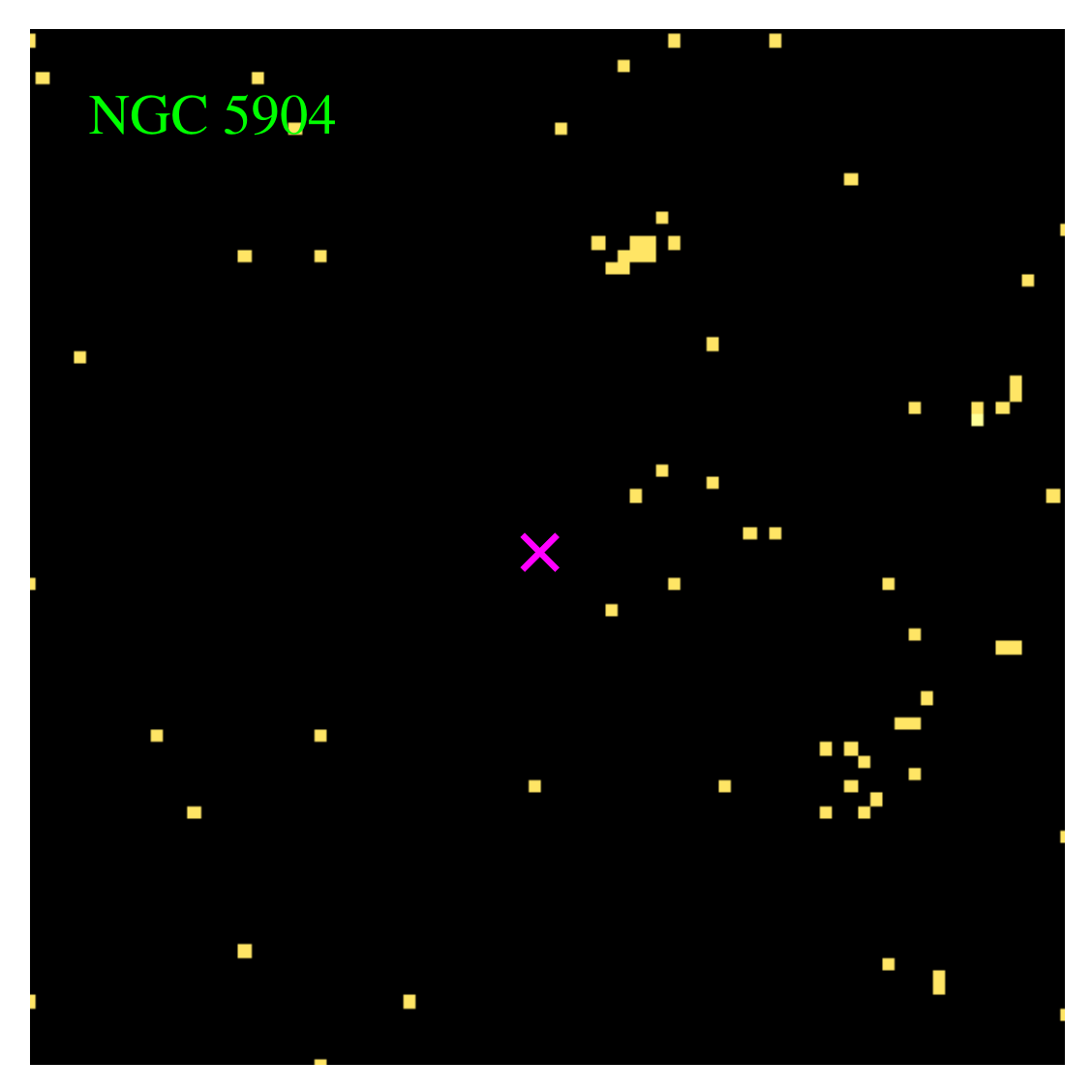}{0.2\textwidth}
        \afig{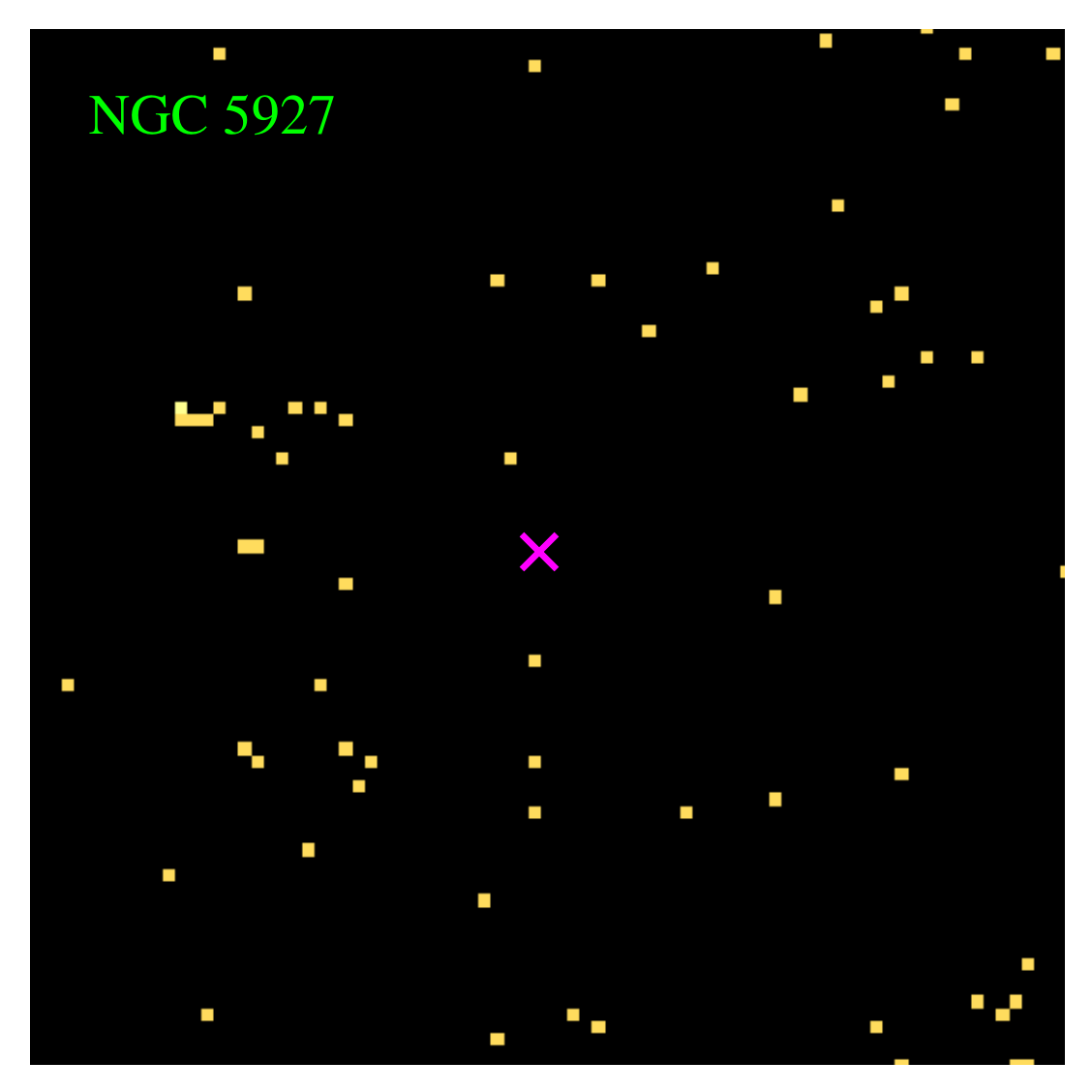}{0.2\textwidth}
        \afig{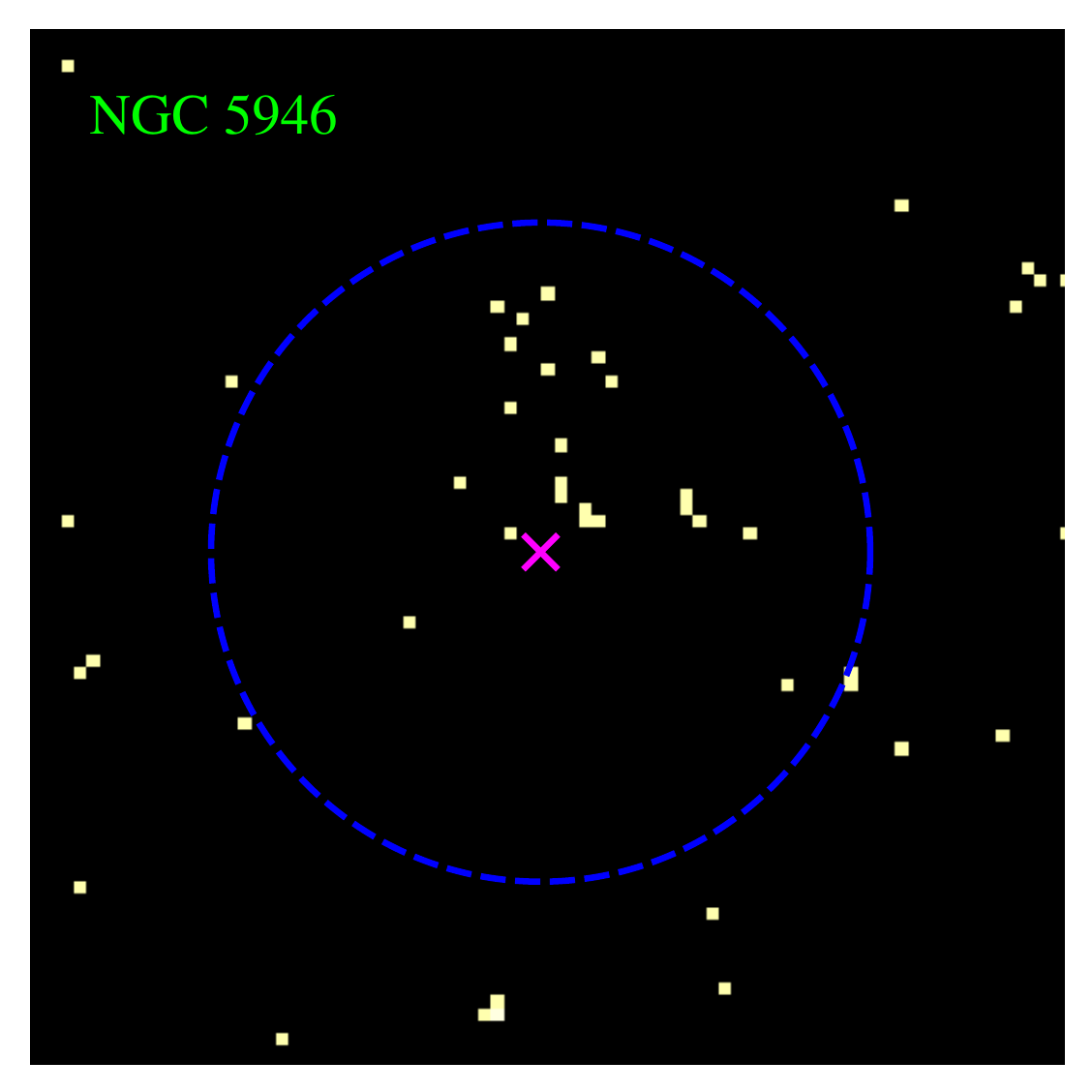}{0.2\textwidth}
        \afig{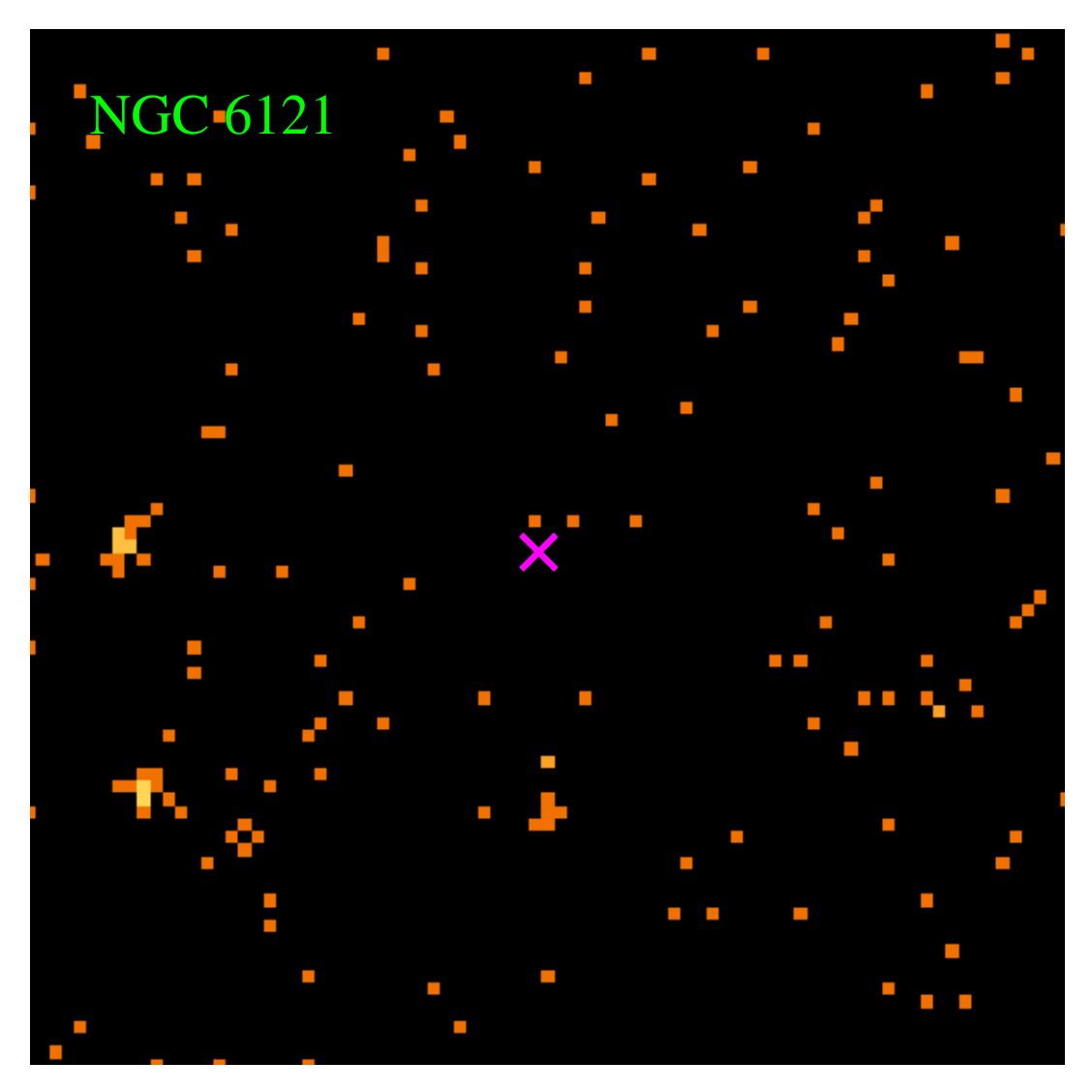}{0.2\textwidth}
        \afig{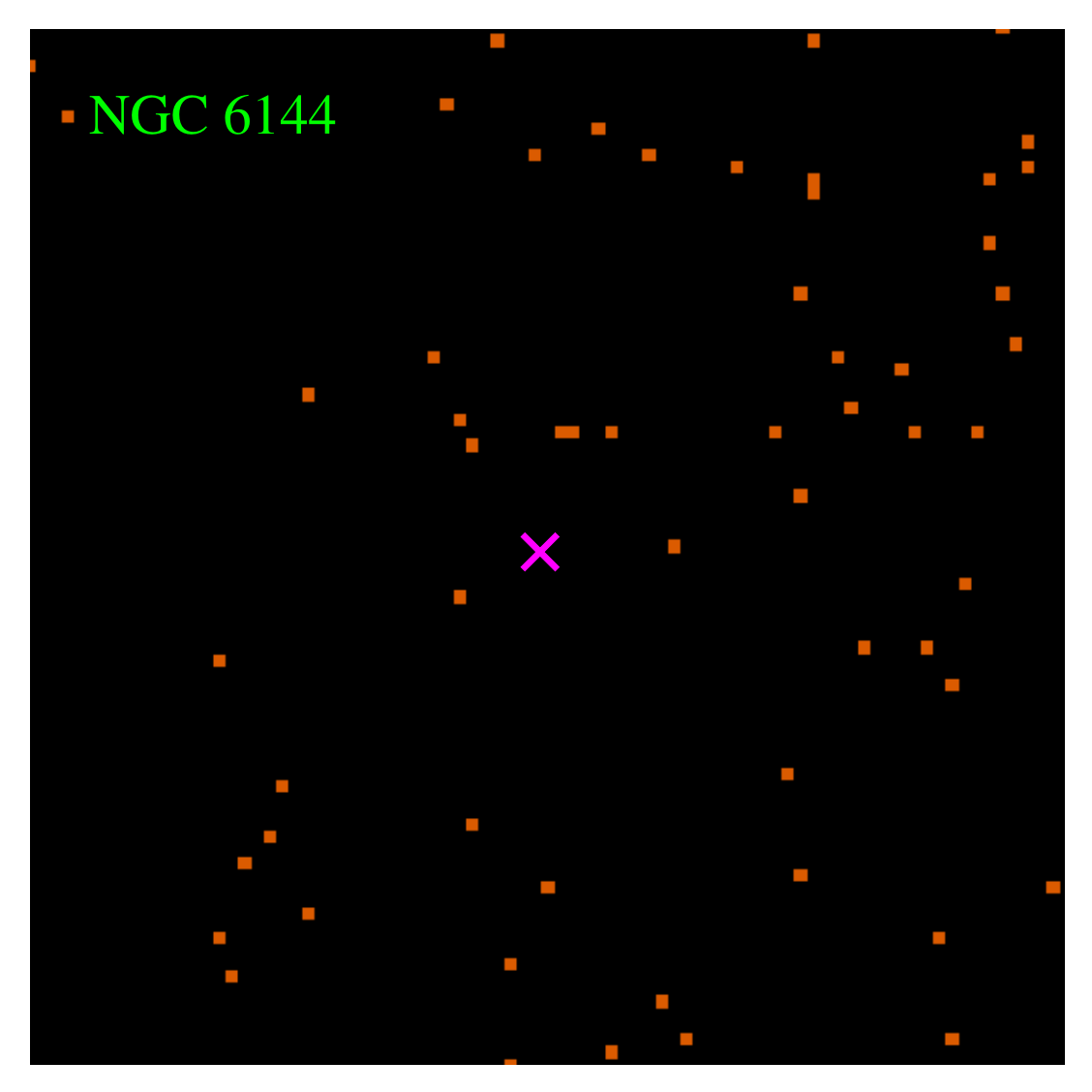}{0.2\textwidth}
    }
\agridline{
        \afig{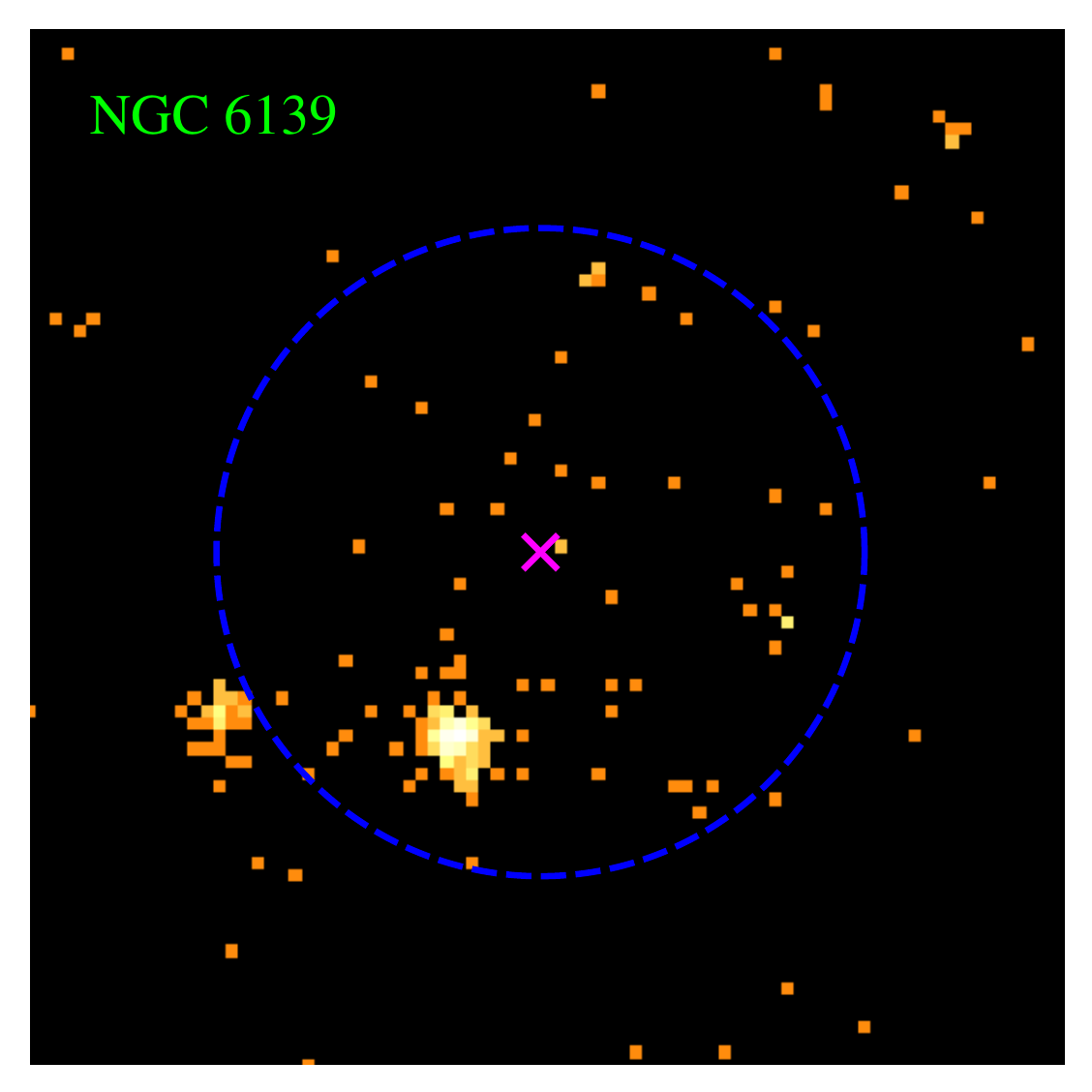}{0.2\textwidth}
        \afig{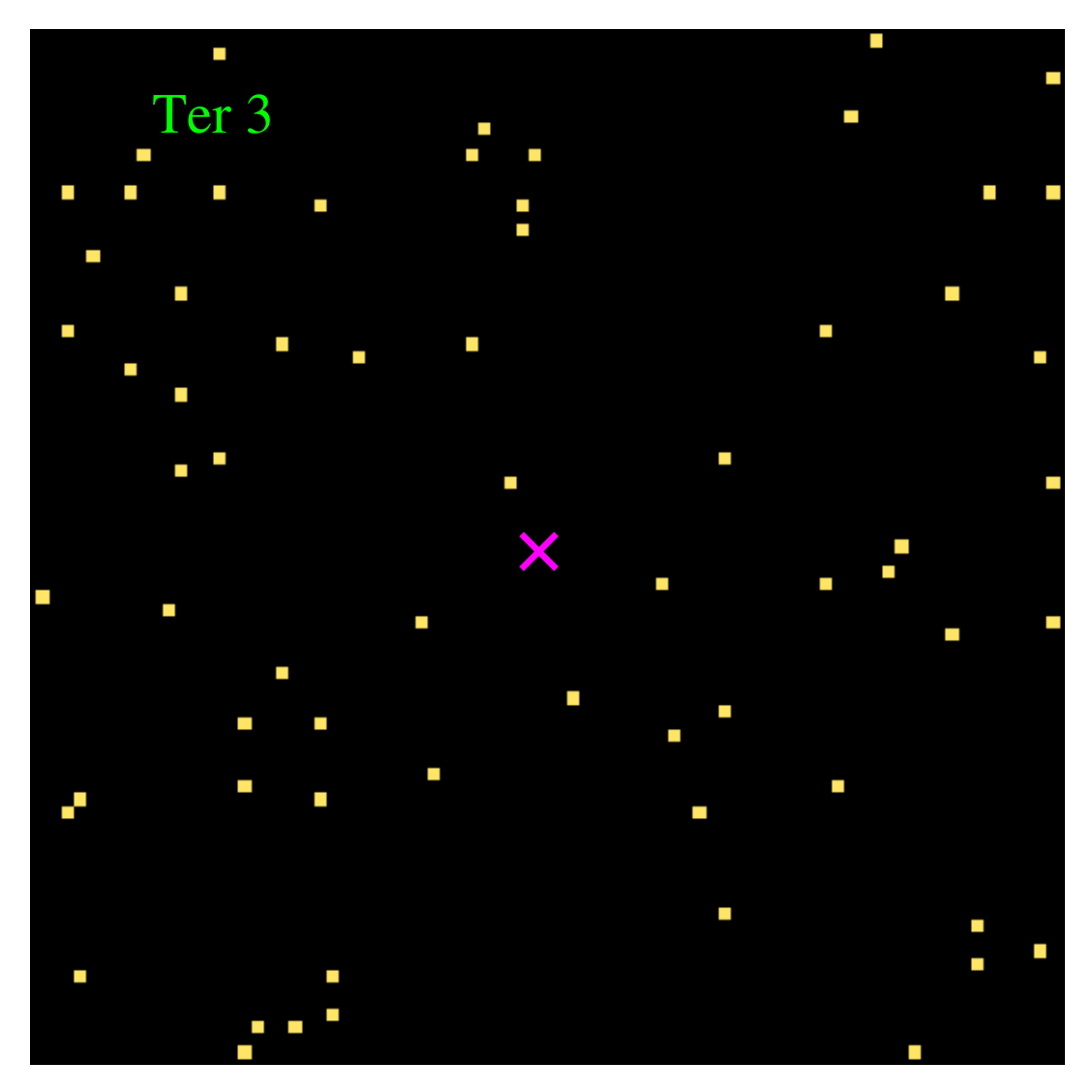}{0.2\textwidth}
        \afig{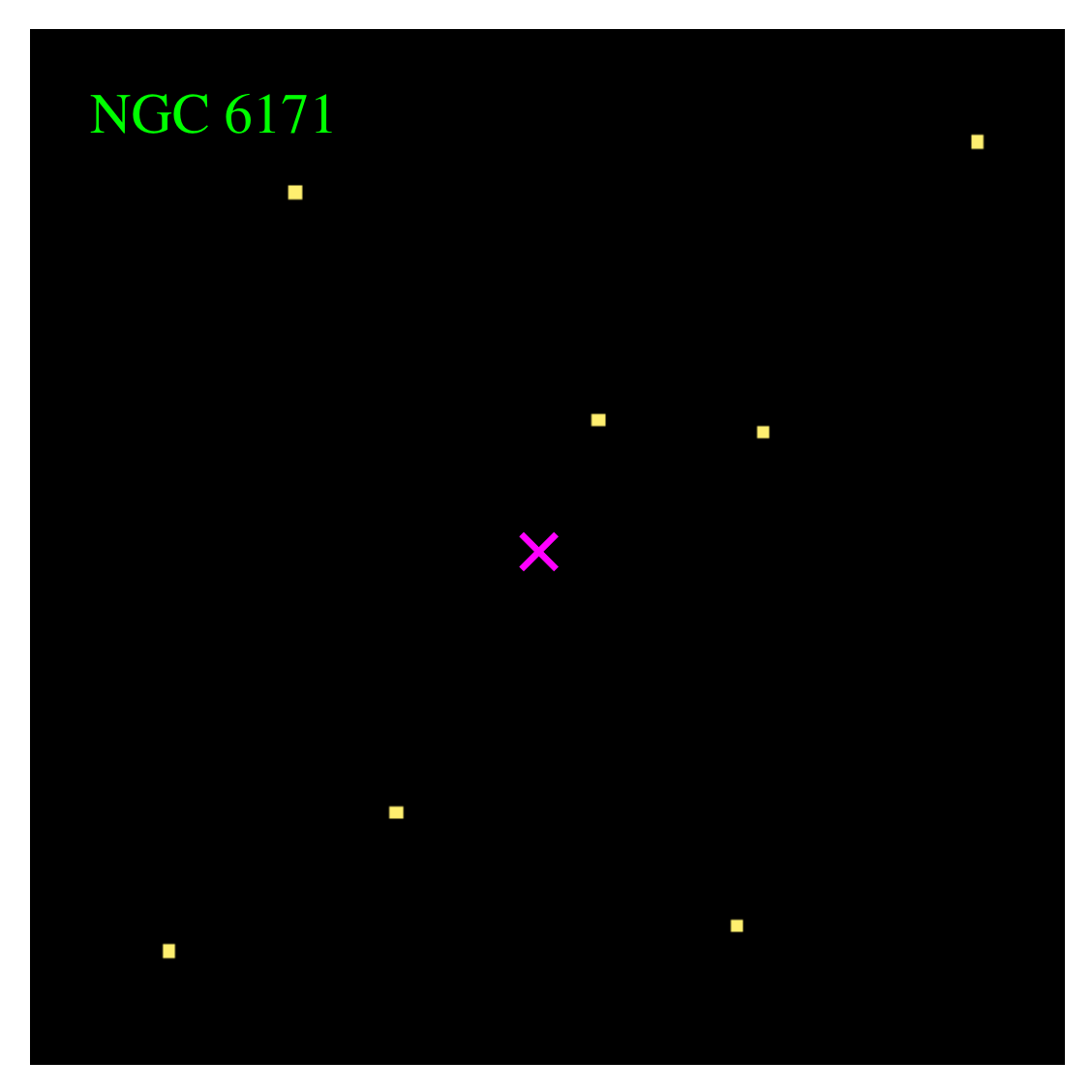}{0.2\textwidth}
        \afig{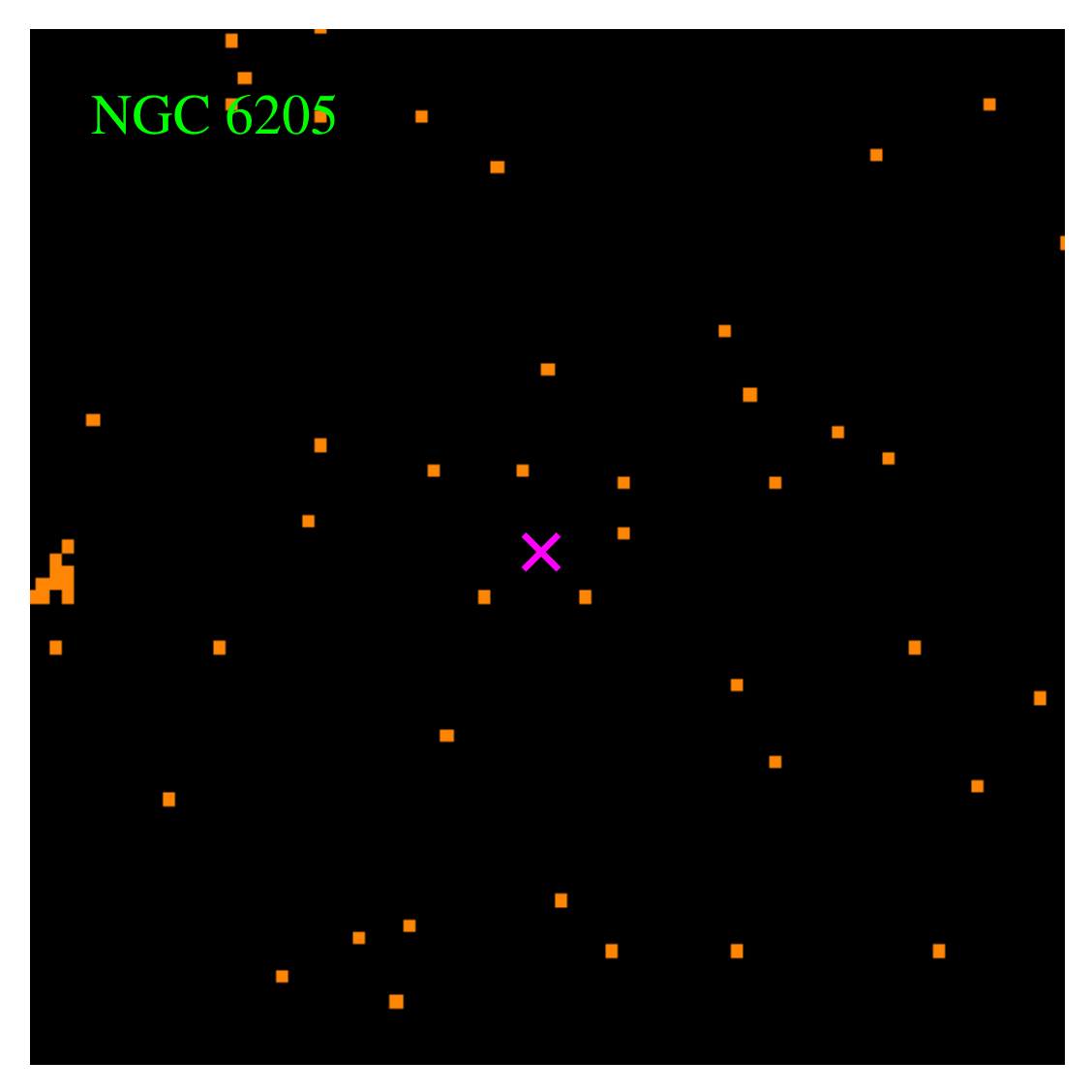}{0.2\textwidth}
        \afig{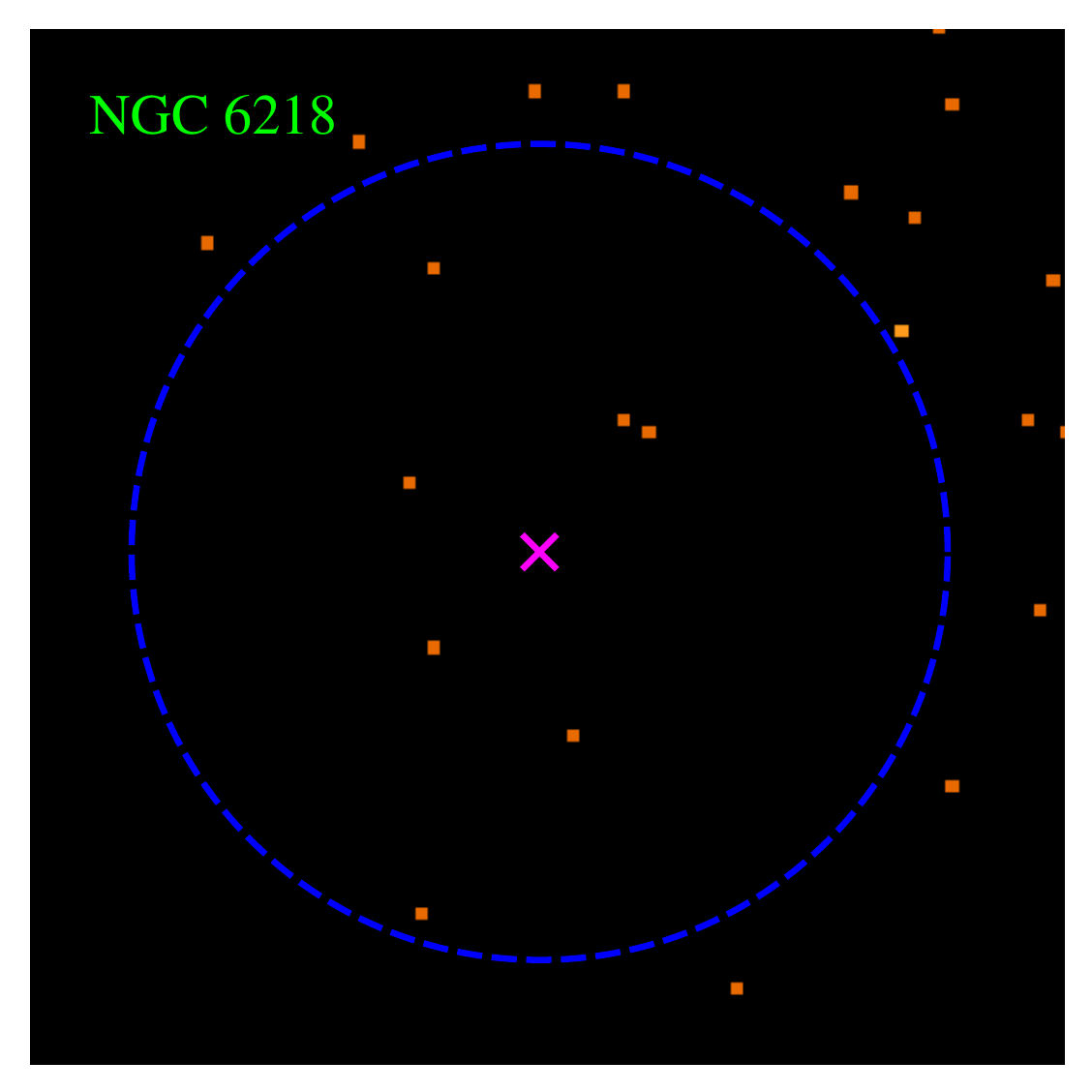}{0.2\textwidth}
    }
\agridline{
        \afig{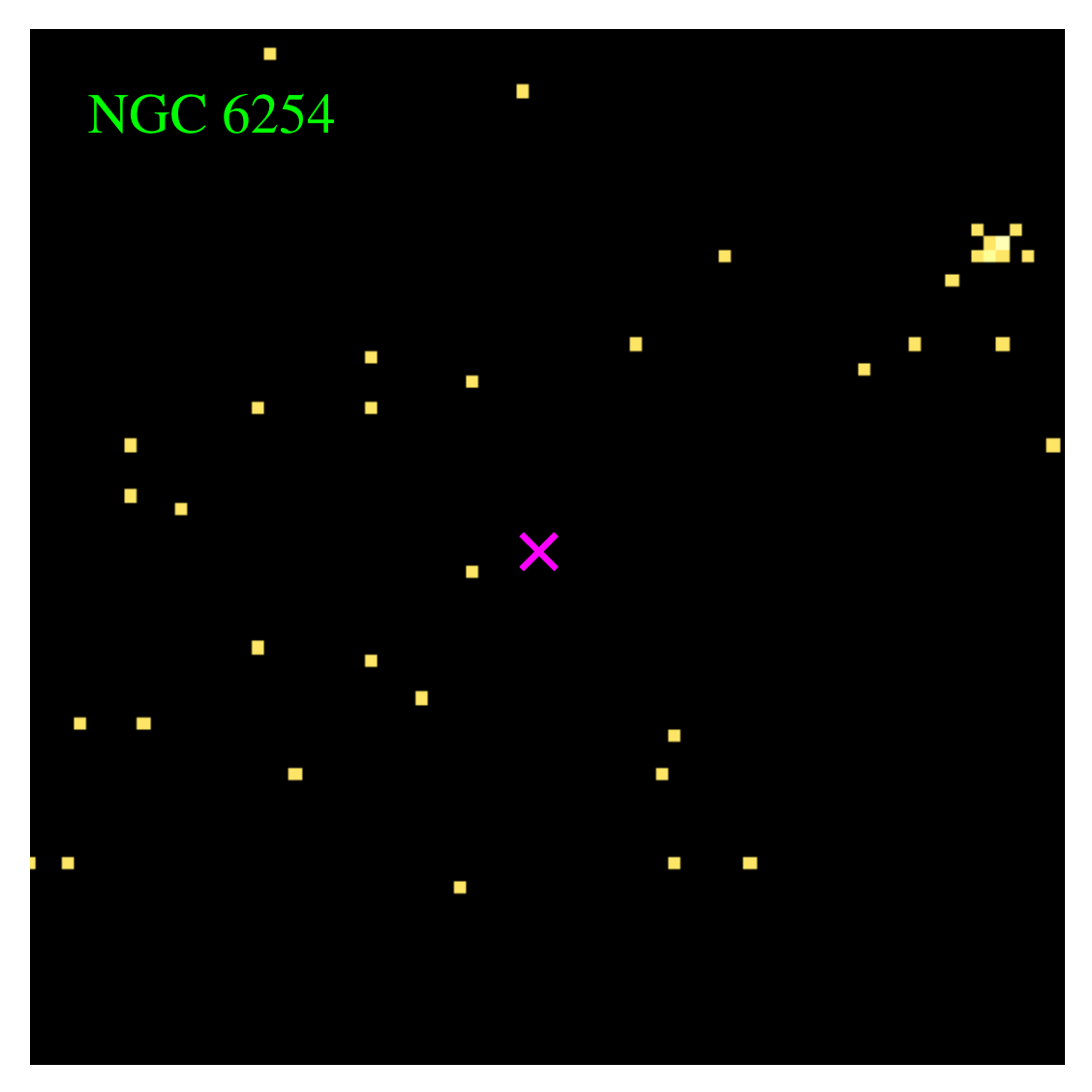}{0.2\textwidth}
        \afig{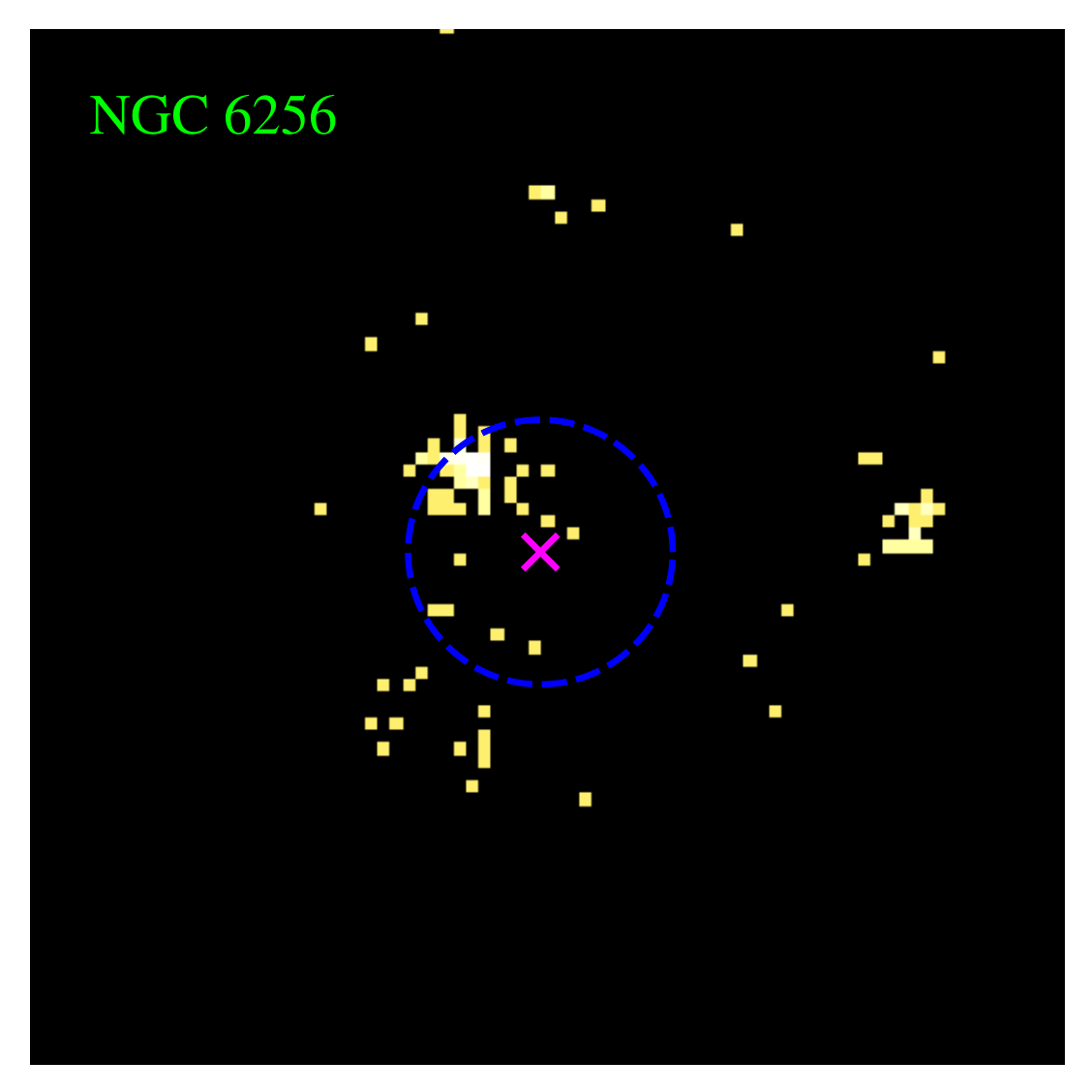}{0.2\textwidth}
        \afig{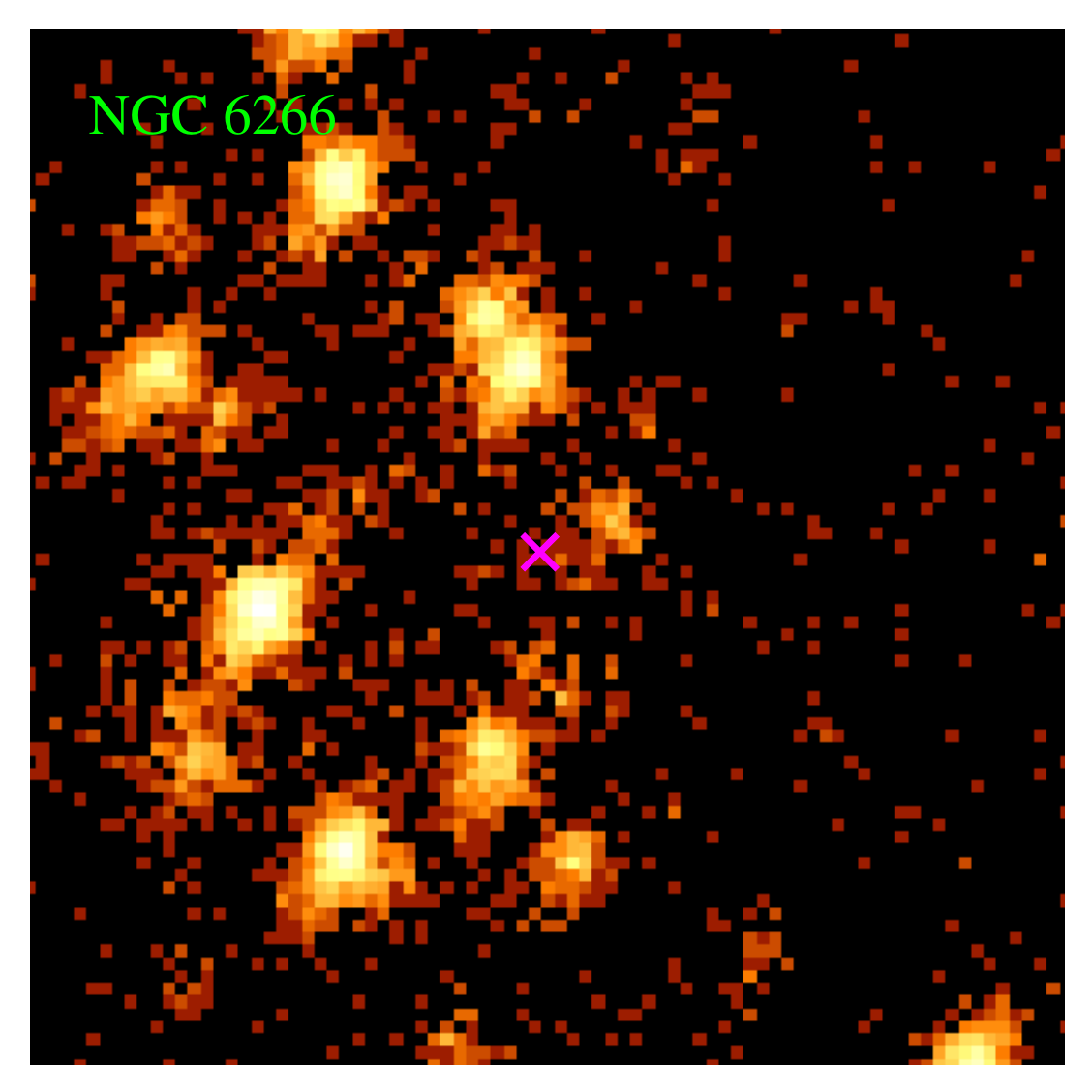}{0.2\textwidth}
        \afig{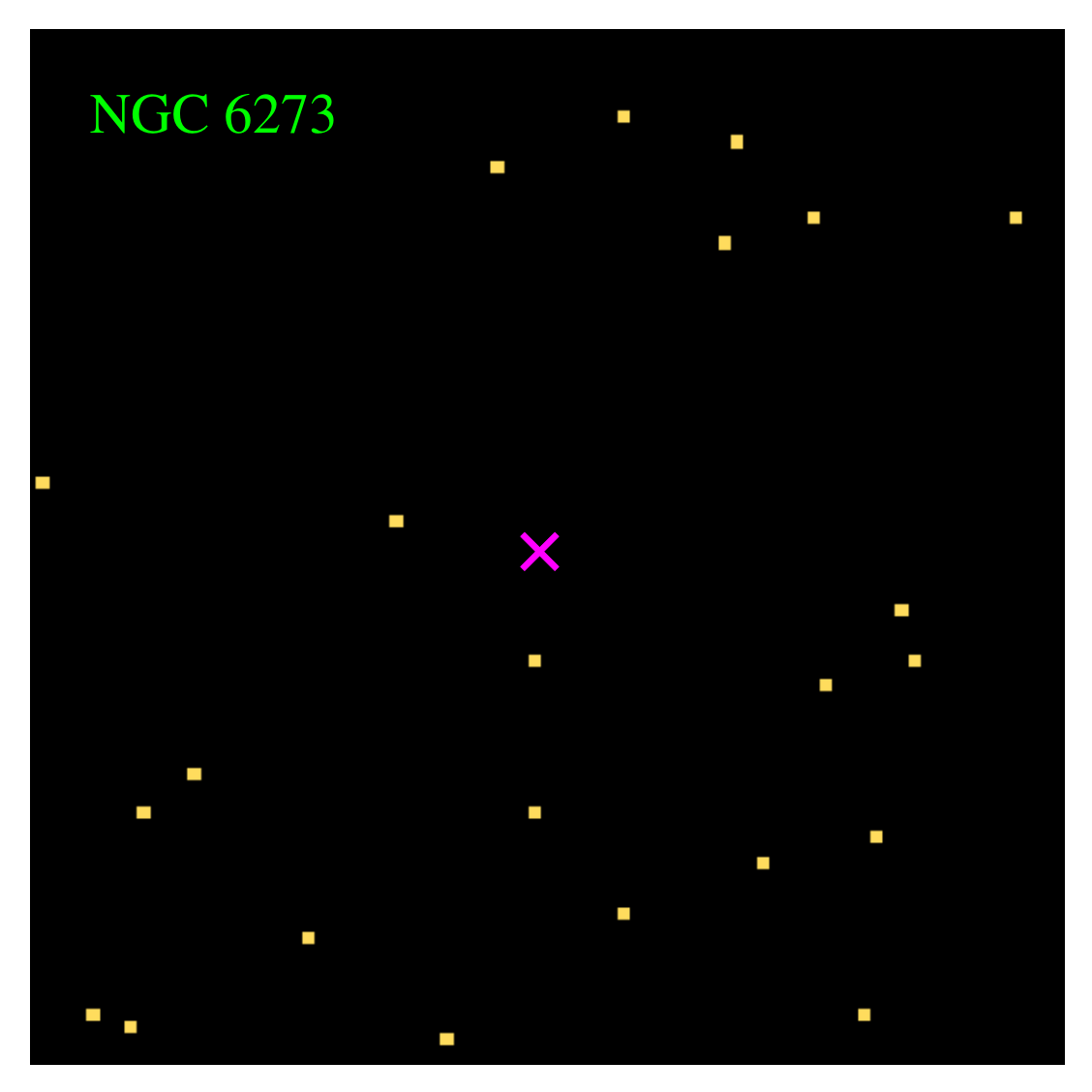}{0.2\textwidth}
        \afig{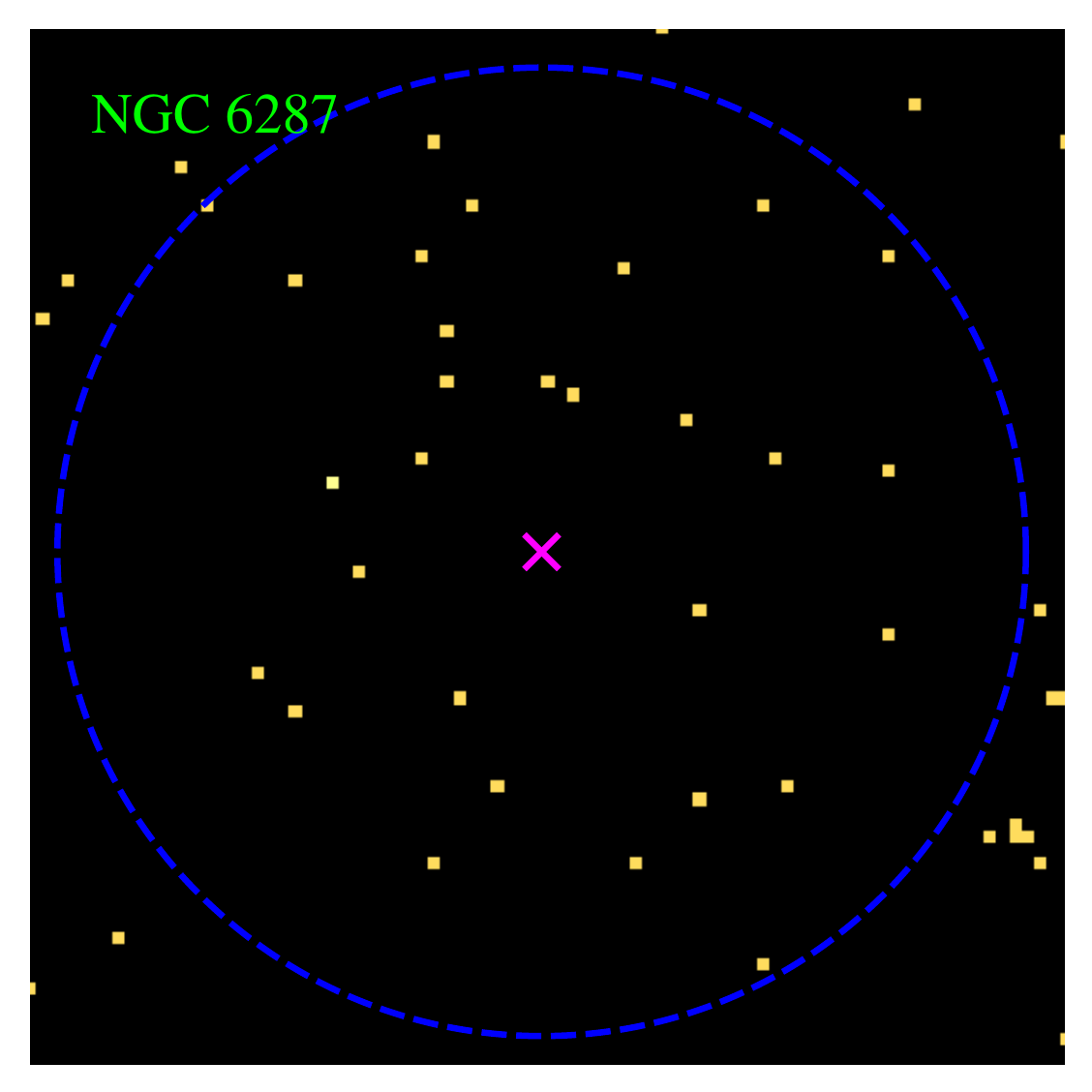}{0.2\textwidth}
    }
 	\caption{{\it Chandra}/ACIS 0.5--8 keV counts images of the 75 GCs without a detected source coincident with the cluster center. 
    The images have a size of $20\arcsec \times 20 \arcsec$ and a binning of 1/2 natal ACIS pixel.
    In each panel, the magenta cross marks the putative cluster center. The blue dashed circle marks the core radius; panels without the blue circle are due to a large core radius that lies beyond the image.\label{fig:nondetection}}
    \end{figure*}
    \begin{figure*}
        \centering

\agridline{
        \afig{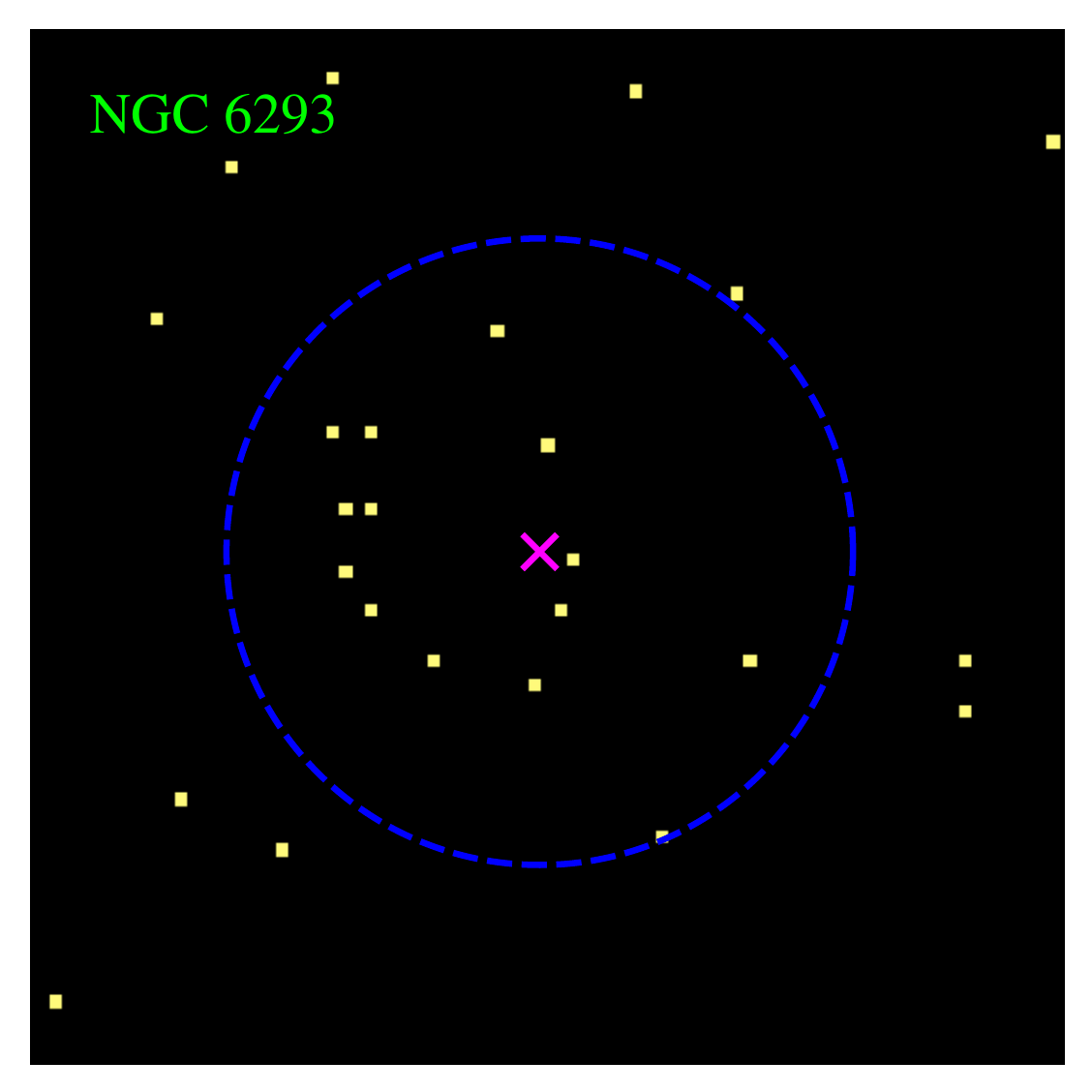}{0.2\textwidth}
        \afig{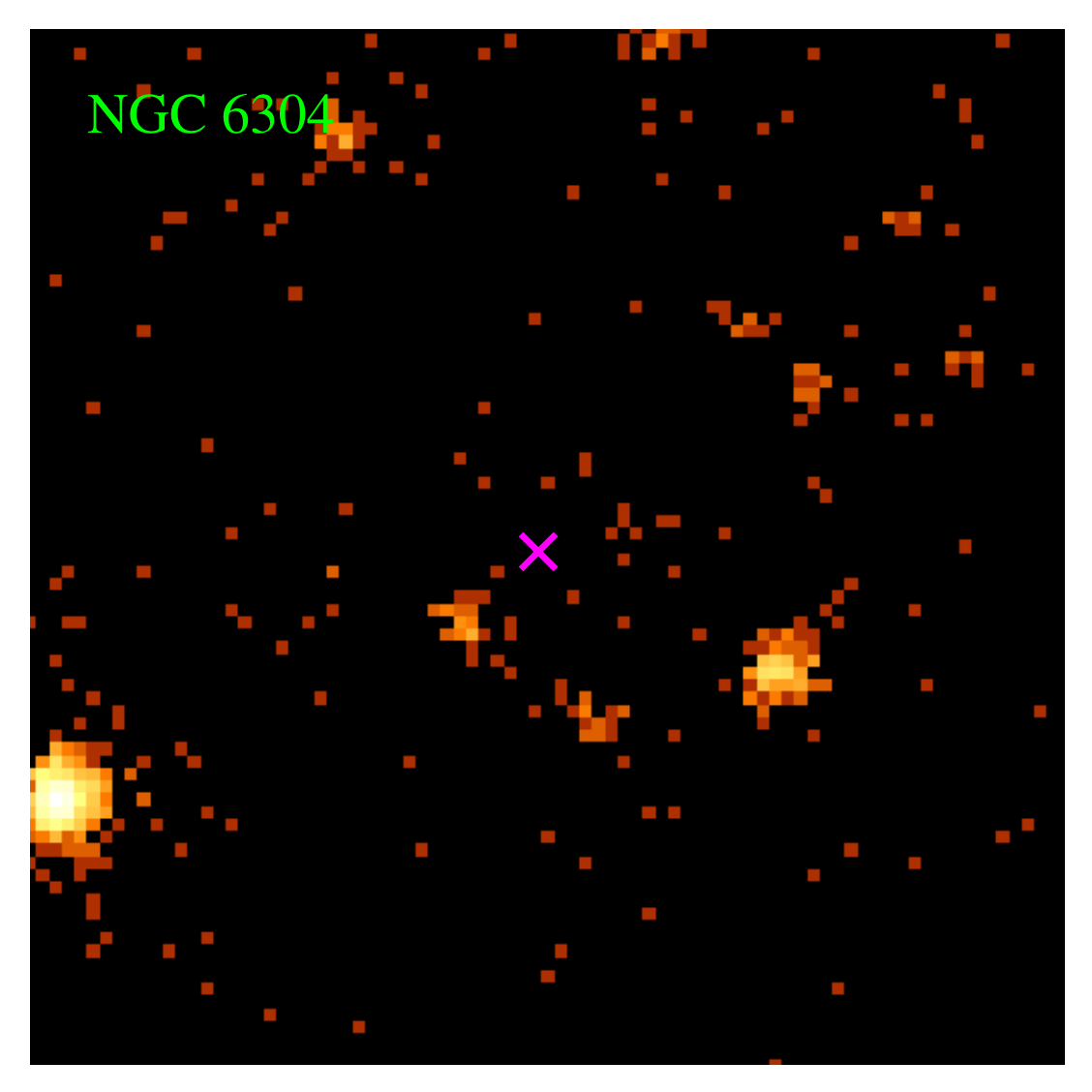}{0.2\textwidth}
        \afig{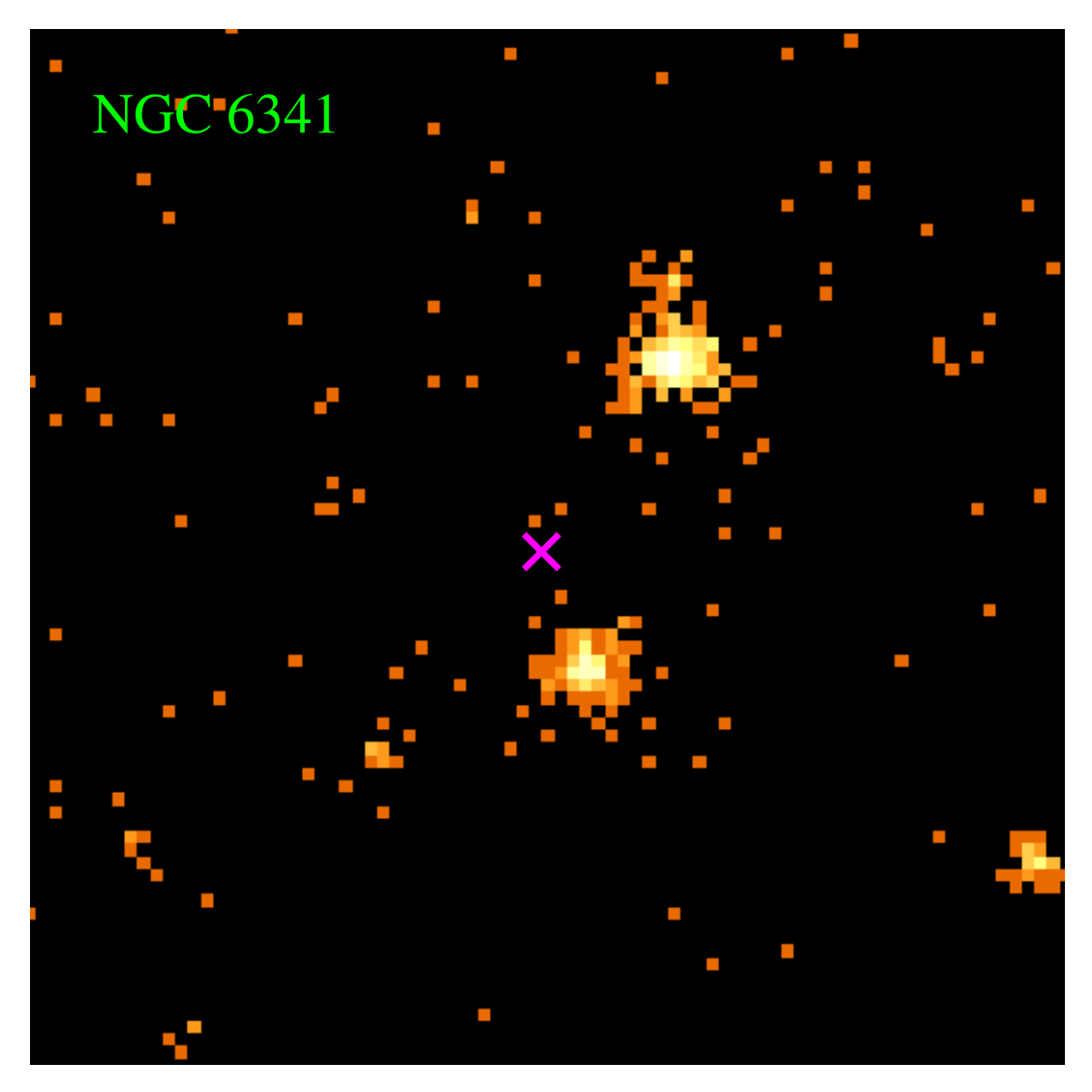}{0.2\textwidth}
        \afig{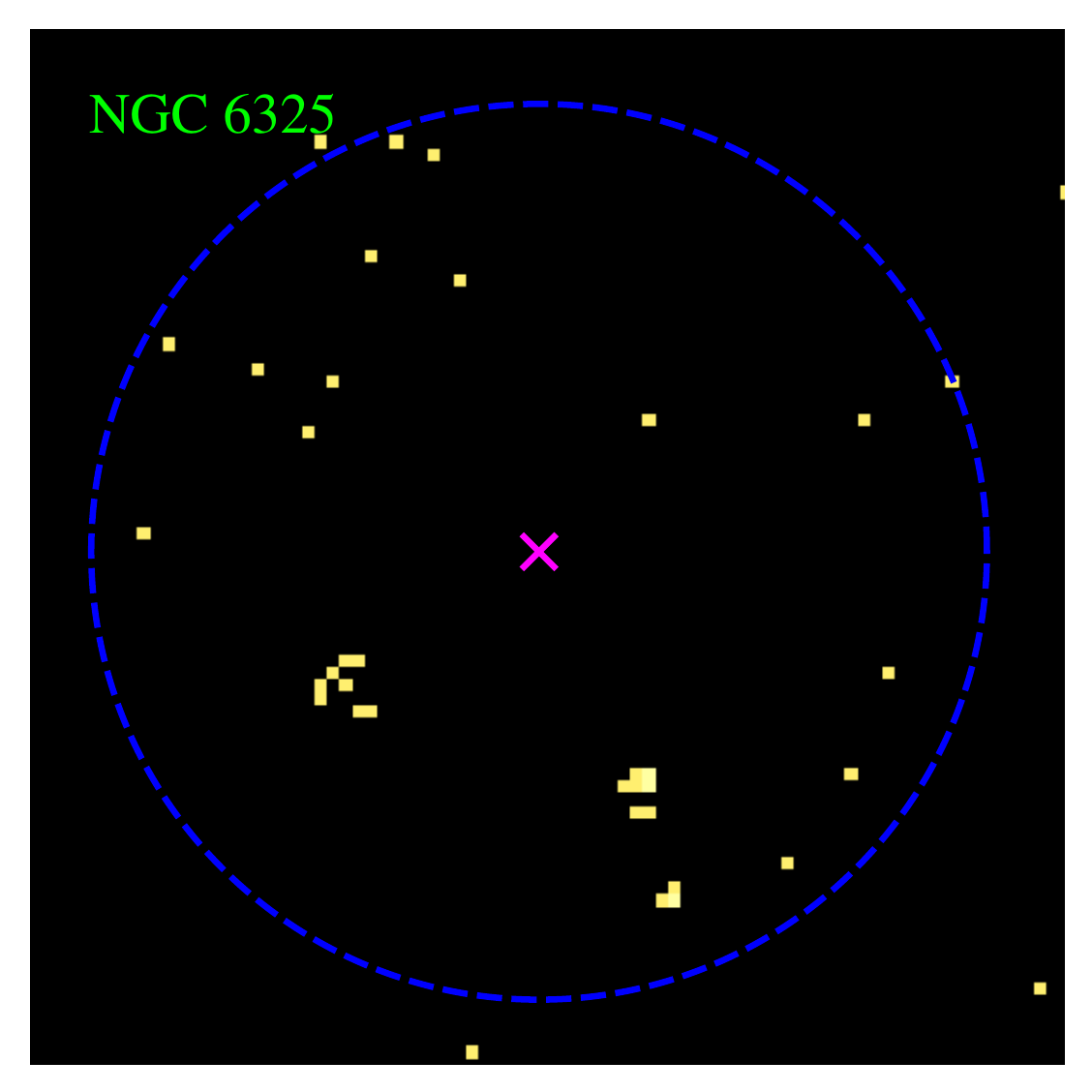}{0.2\textwidth}
        \afig{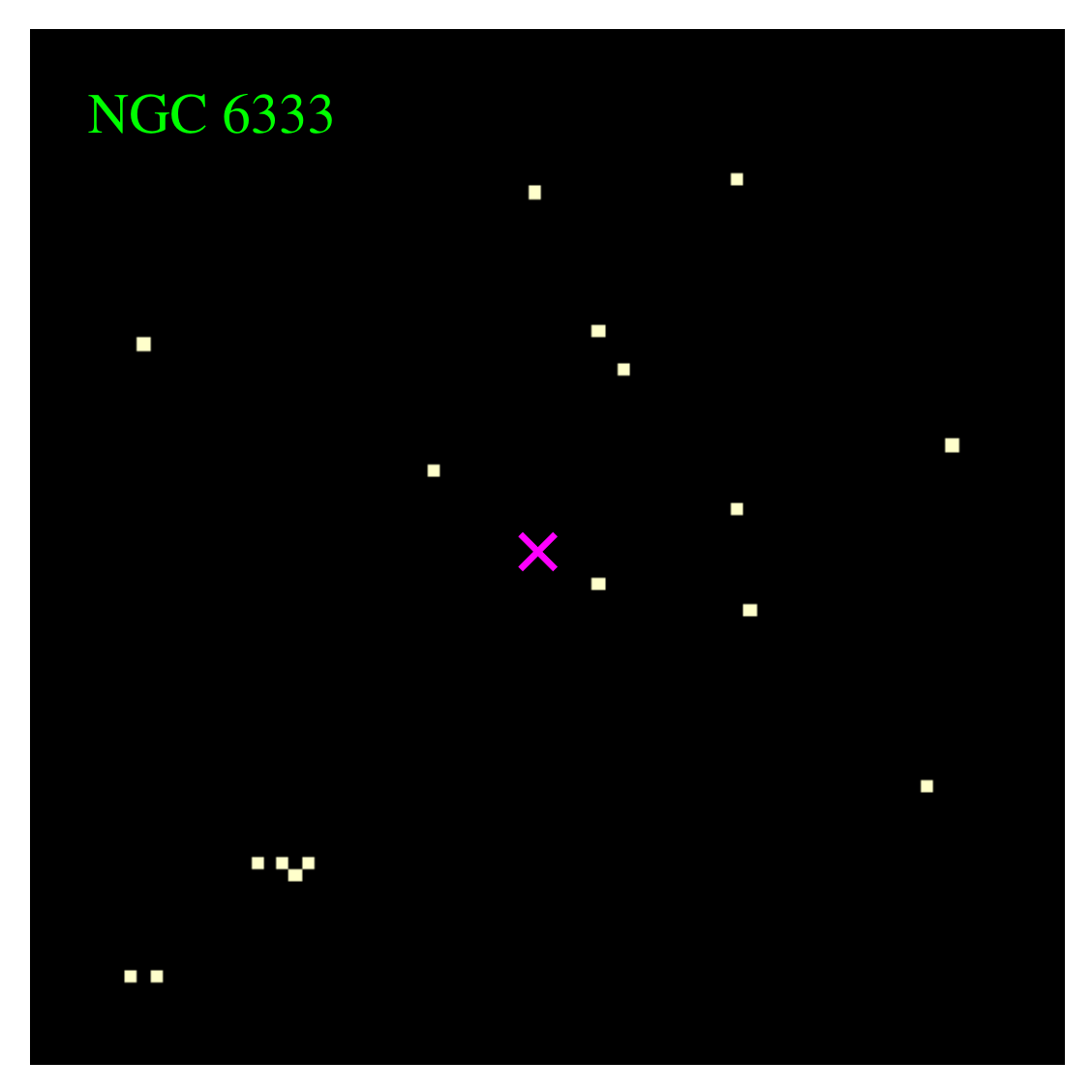}{0.2\textwidth}
    }
\agridline{
        \afig{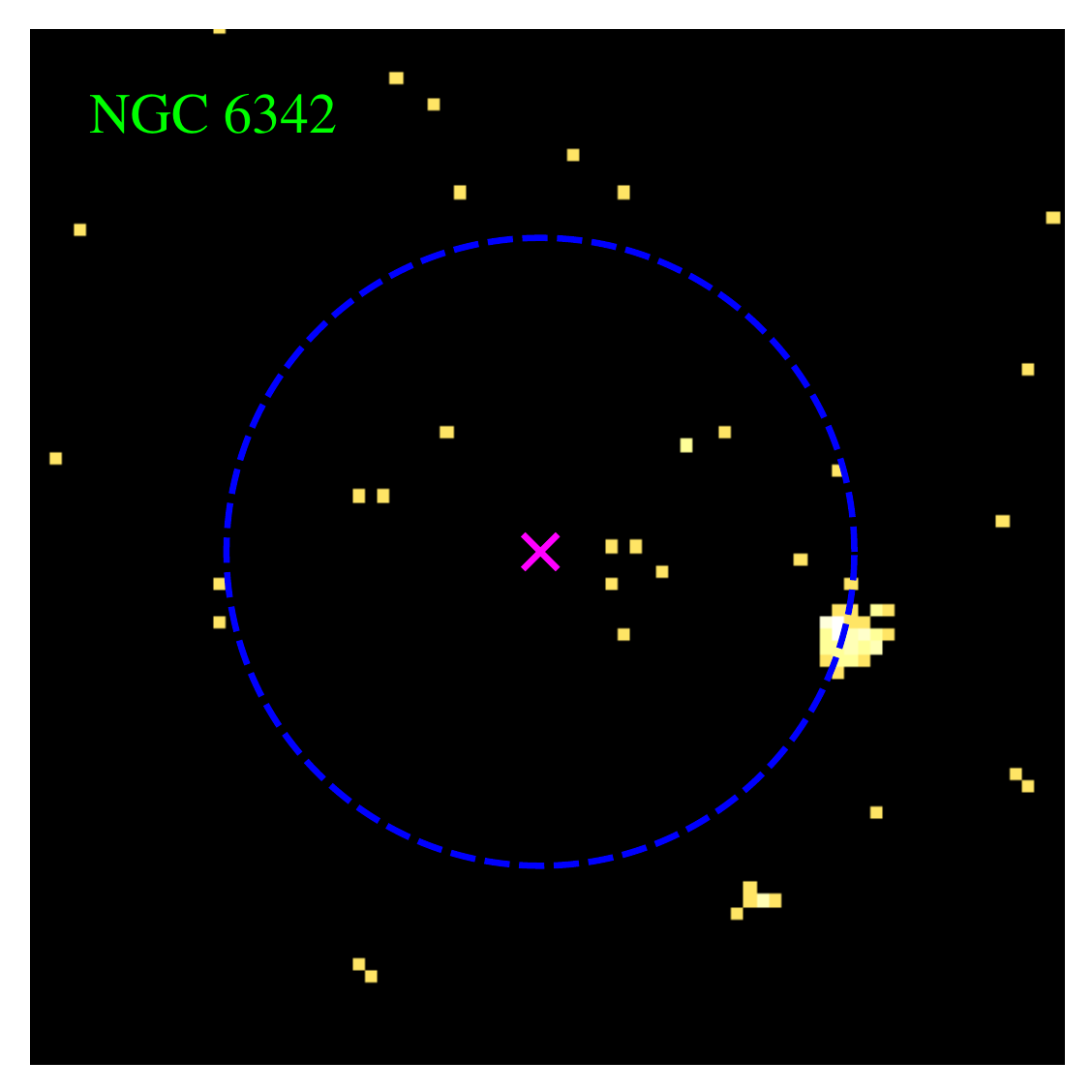}{0.2\textwidth}
        \afig{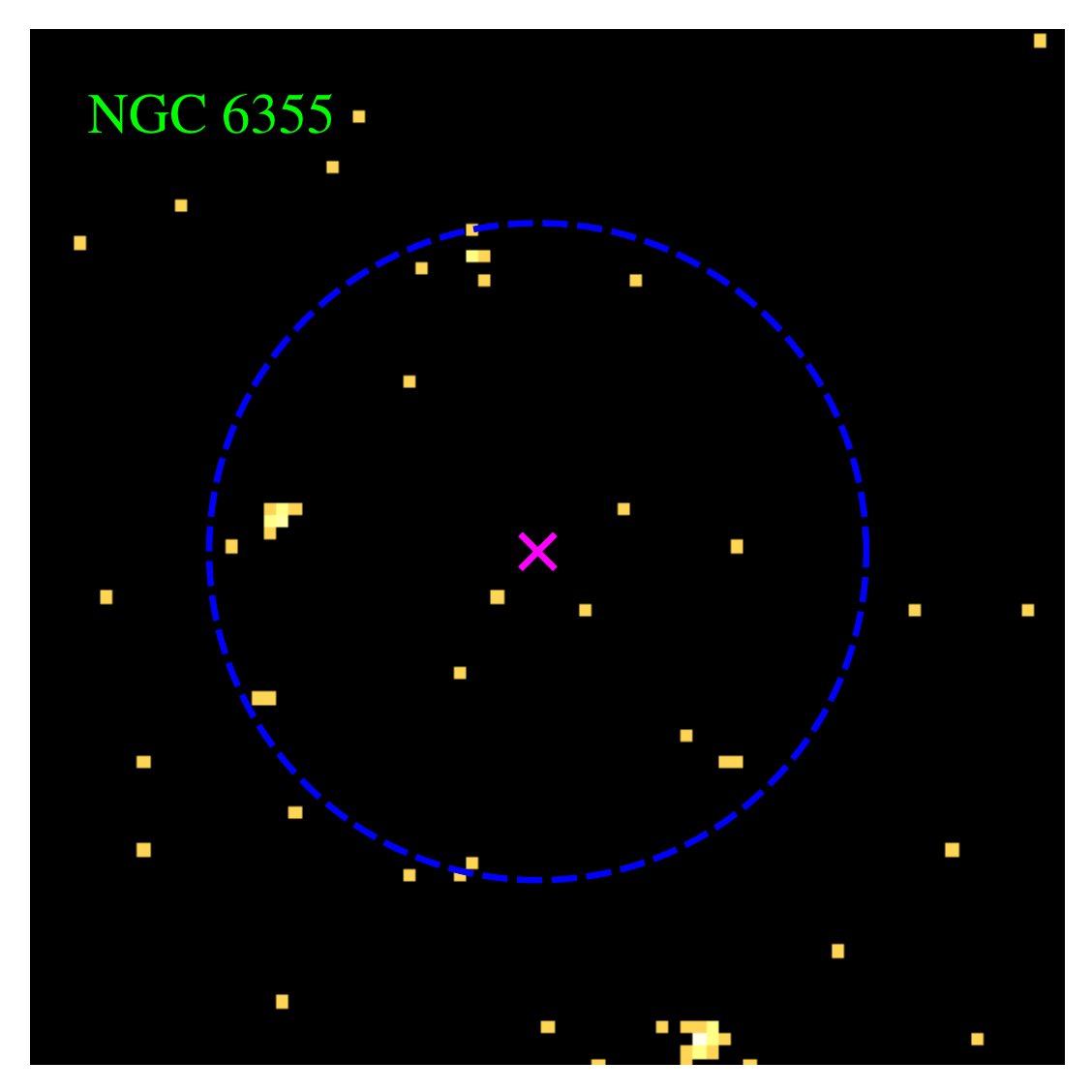}{0.2\textwidth}
        \afig{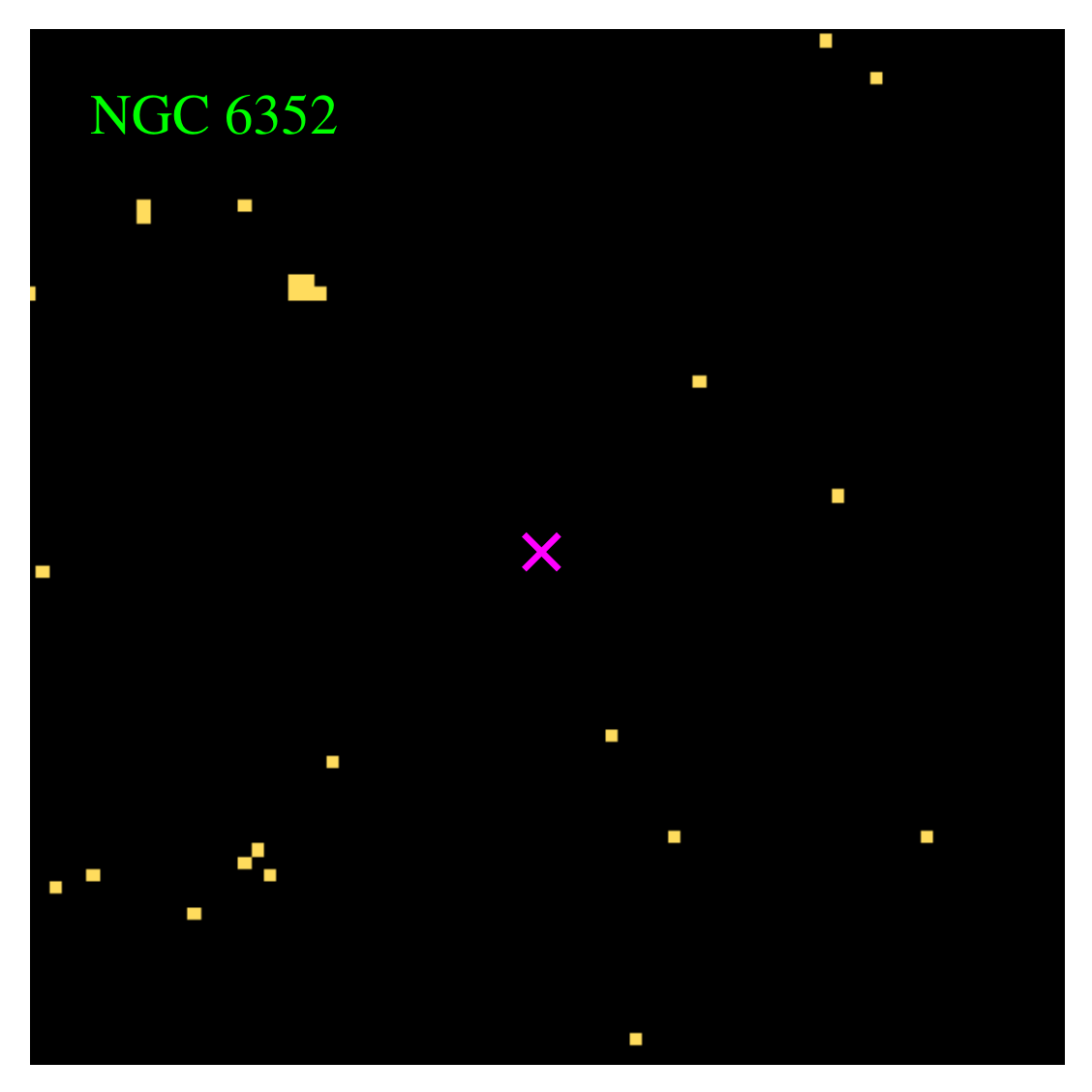}{0.2\textwidth}
        \afig{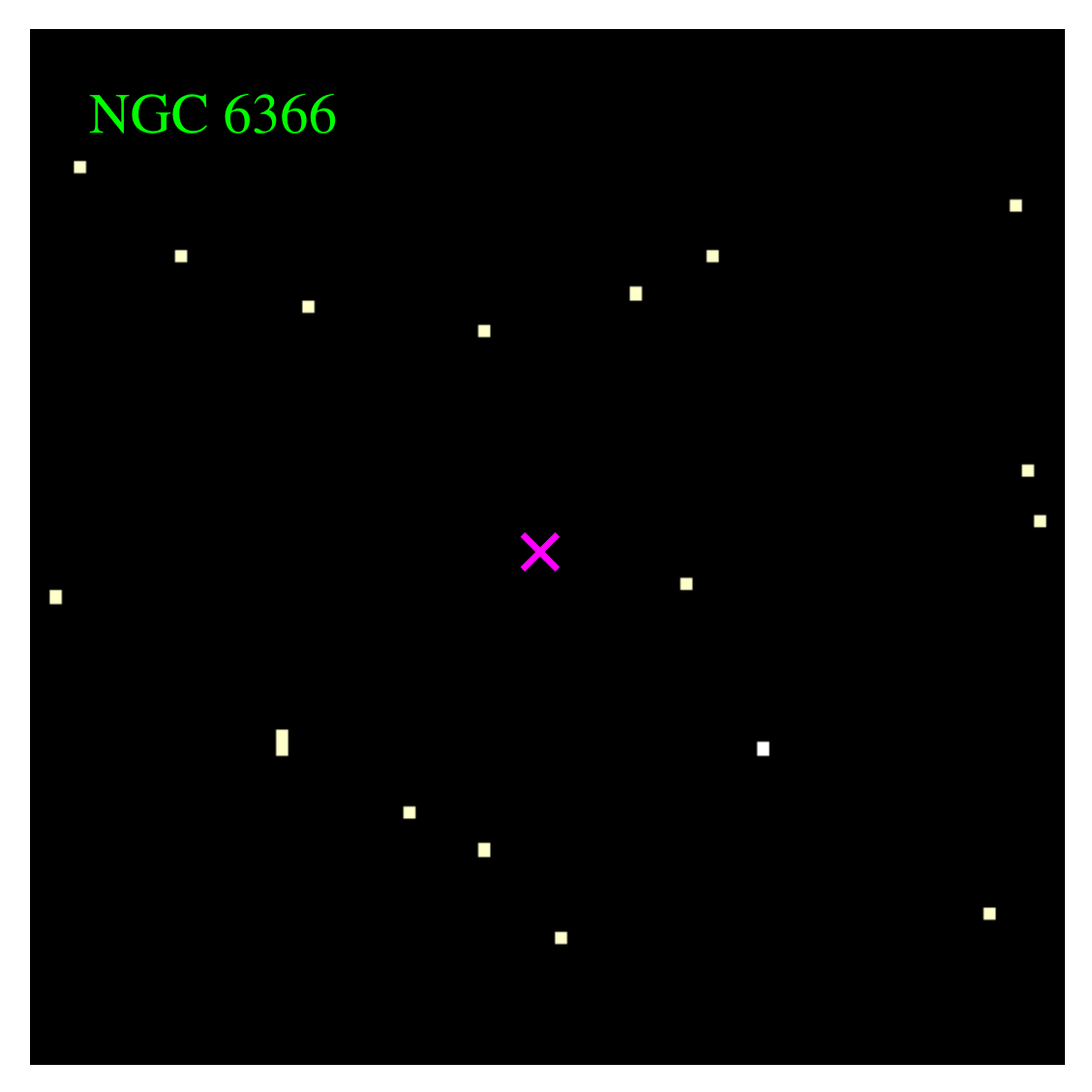}{0.2\textwidth}
        \afig{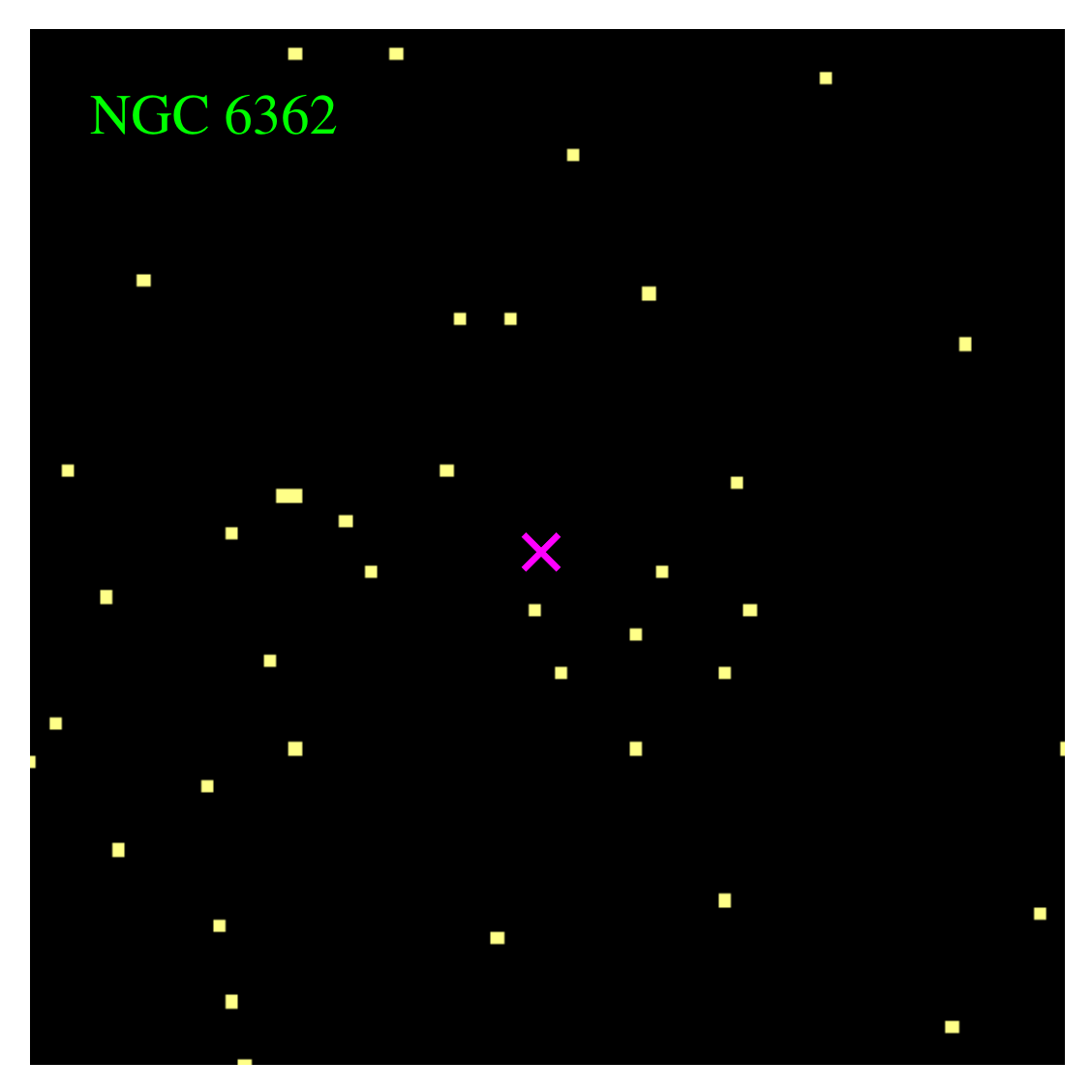}{0.2\textwidth}
    }
\agridline{
        \afig{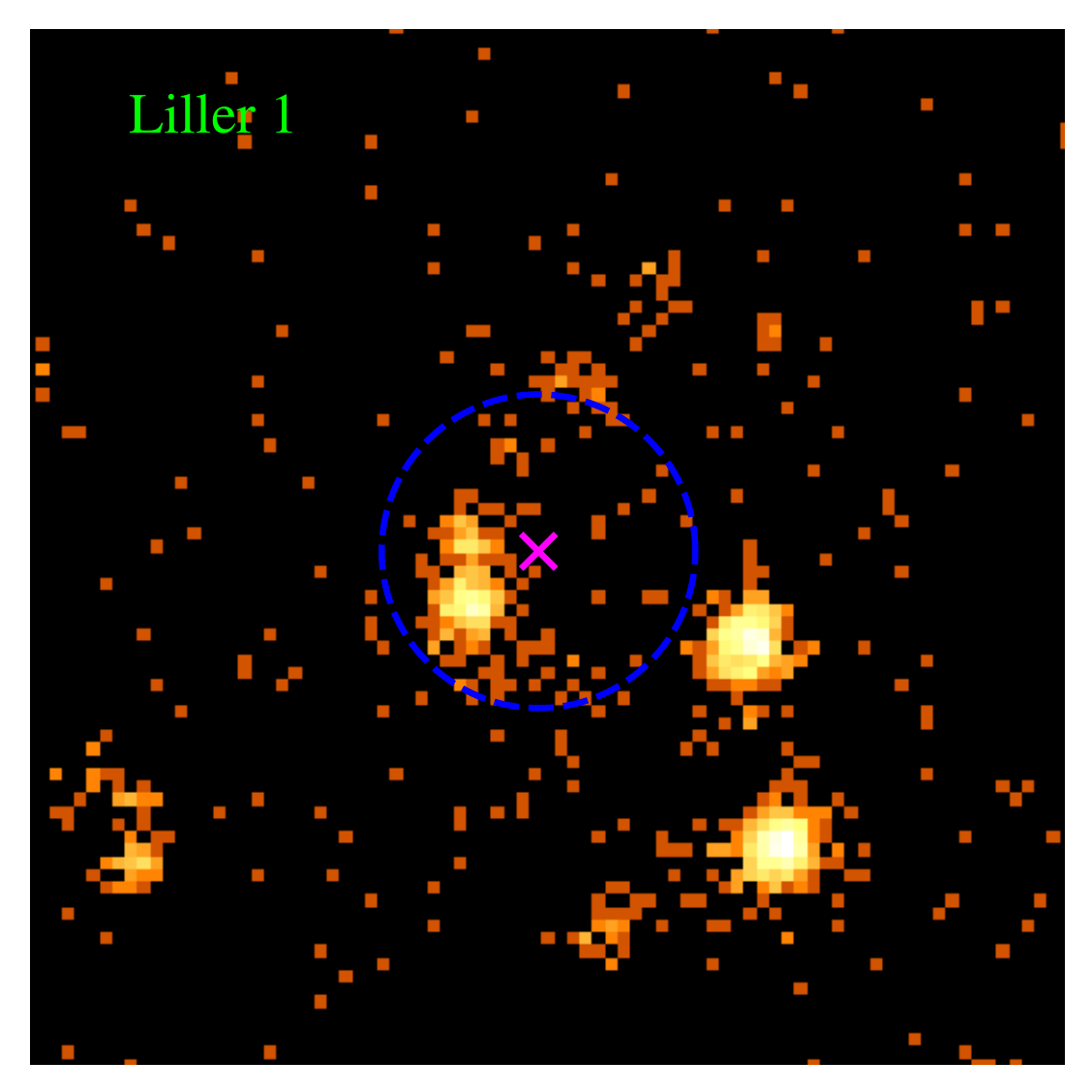}{0.2\textwidth}
        \afig{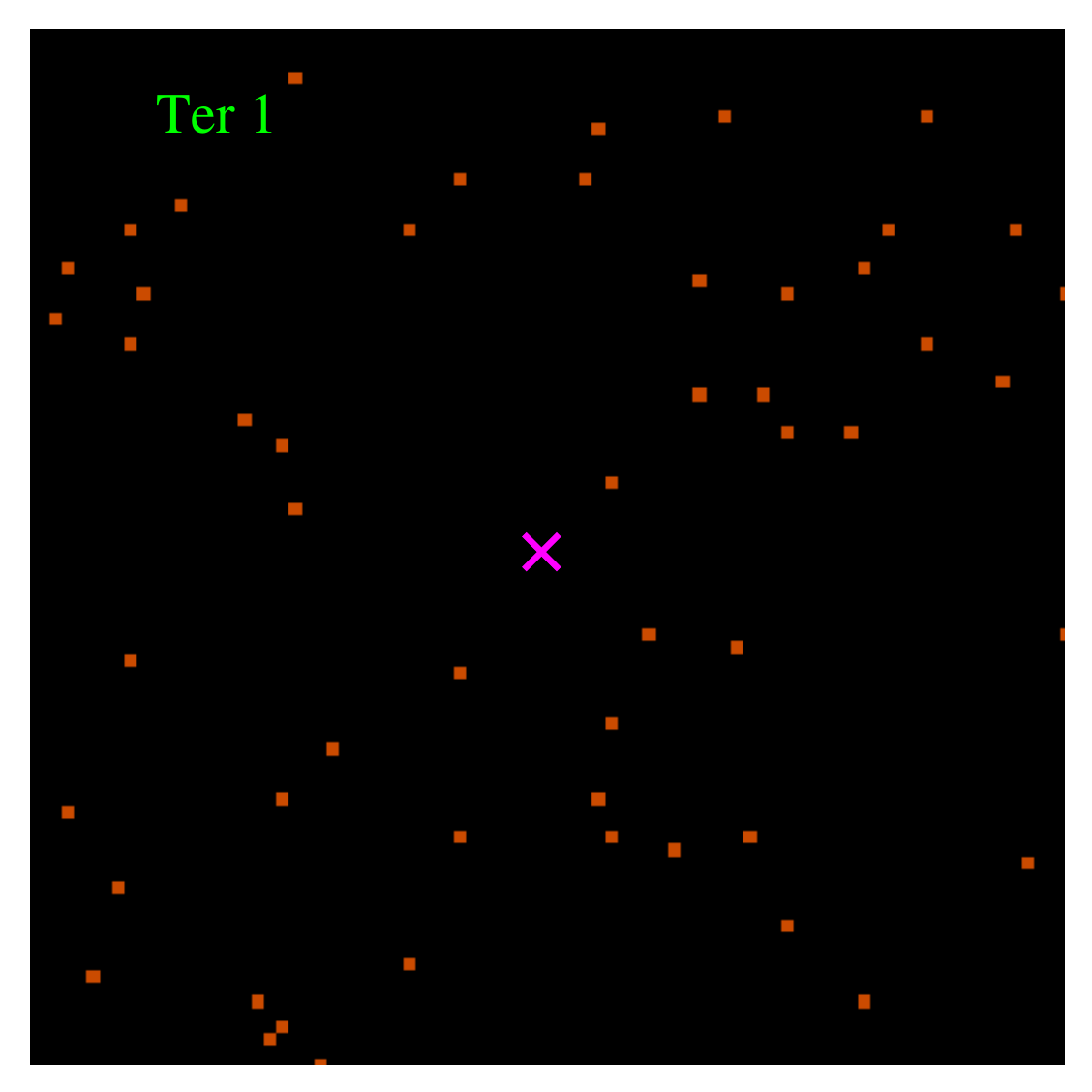}{0.2\textwidth}
        \afig{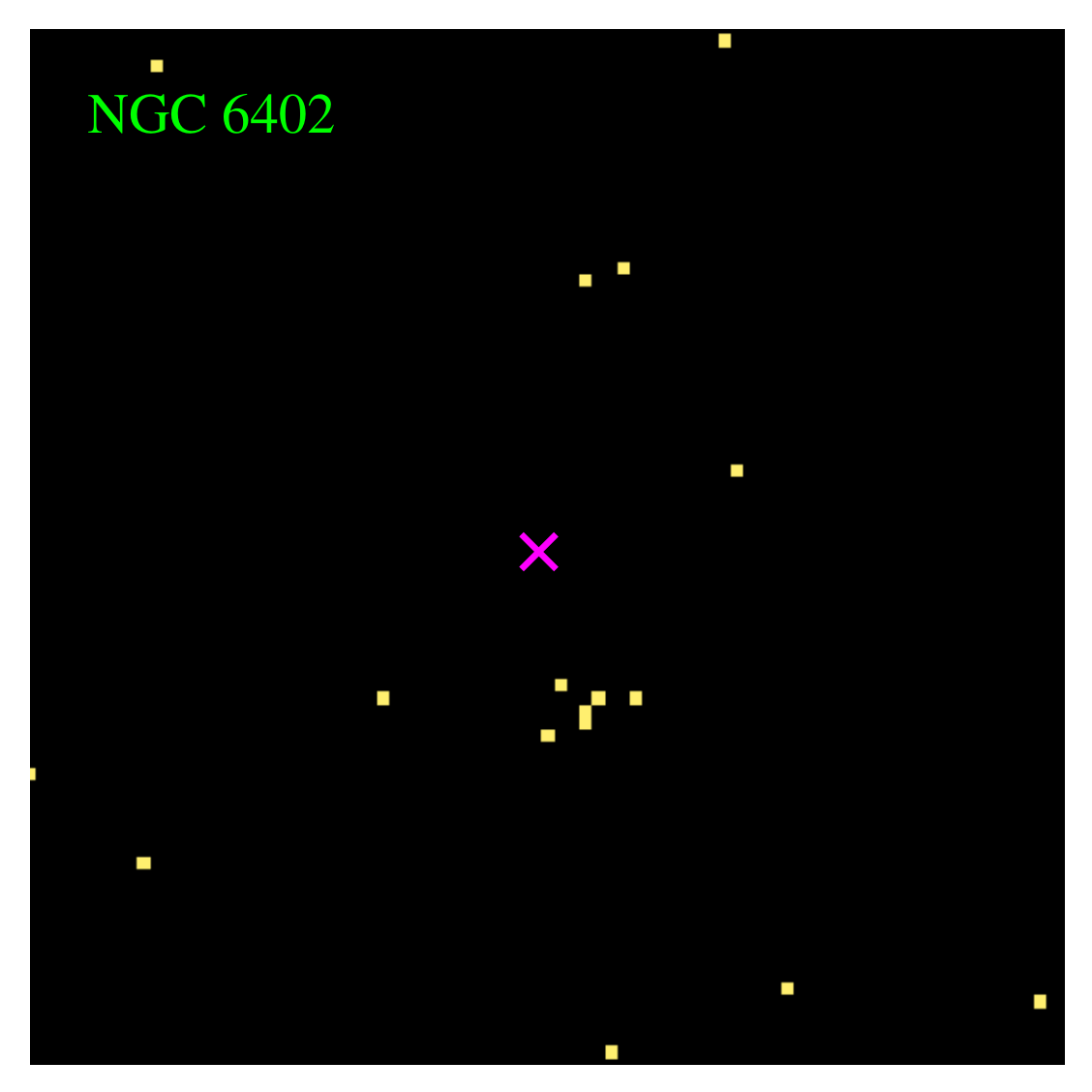}{0.2\textwidth}
        \afig{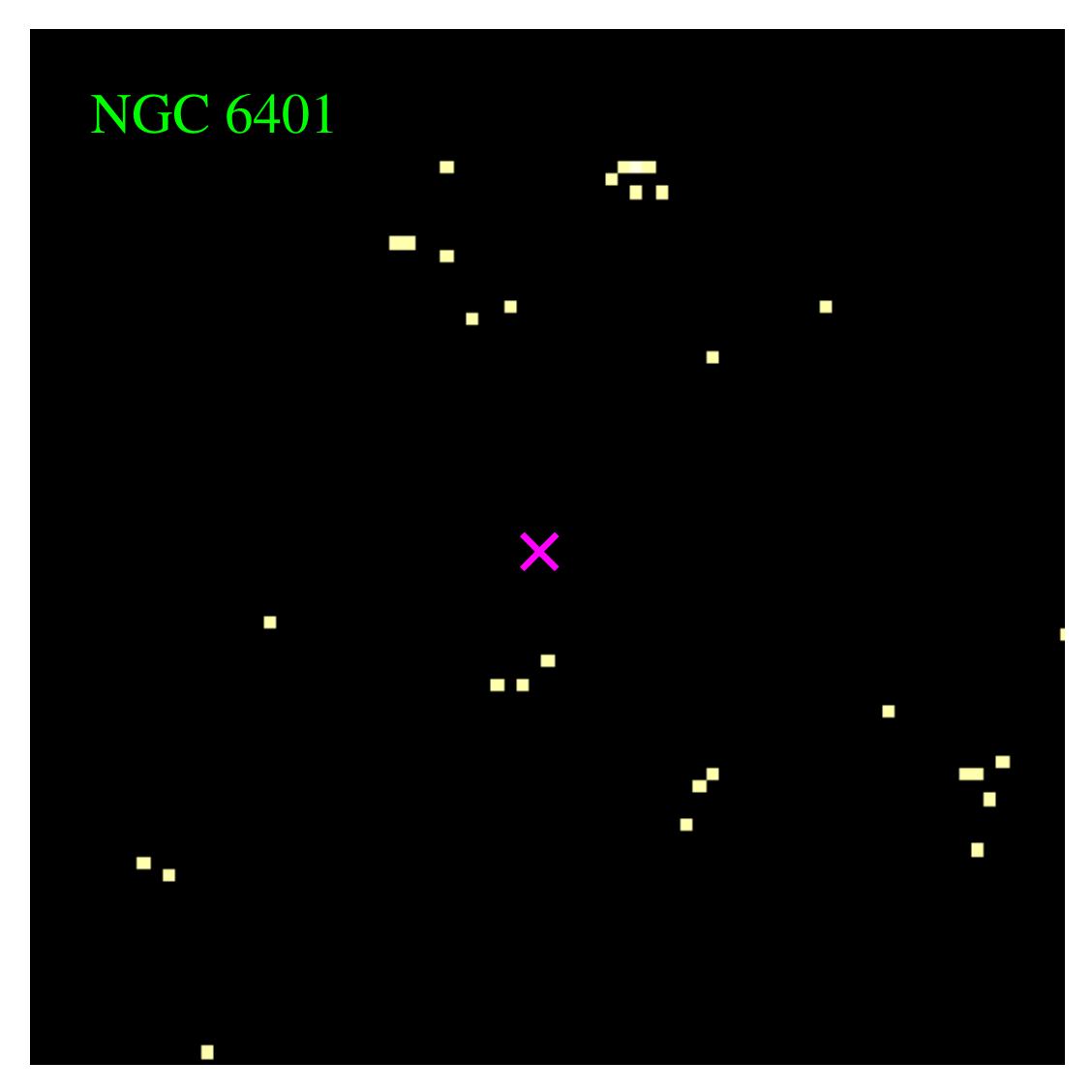}{0.2\textwidth}
        \afig{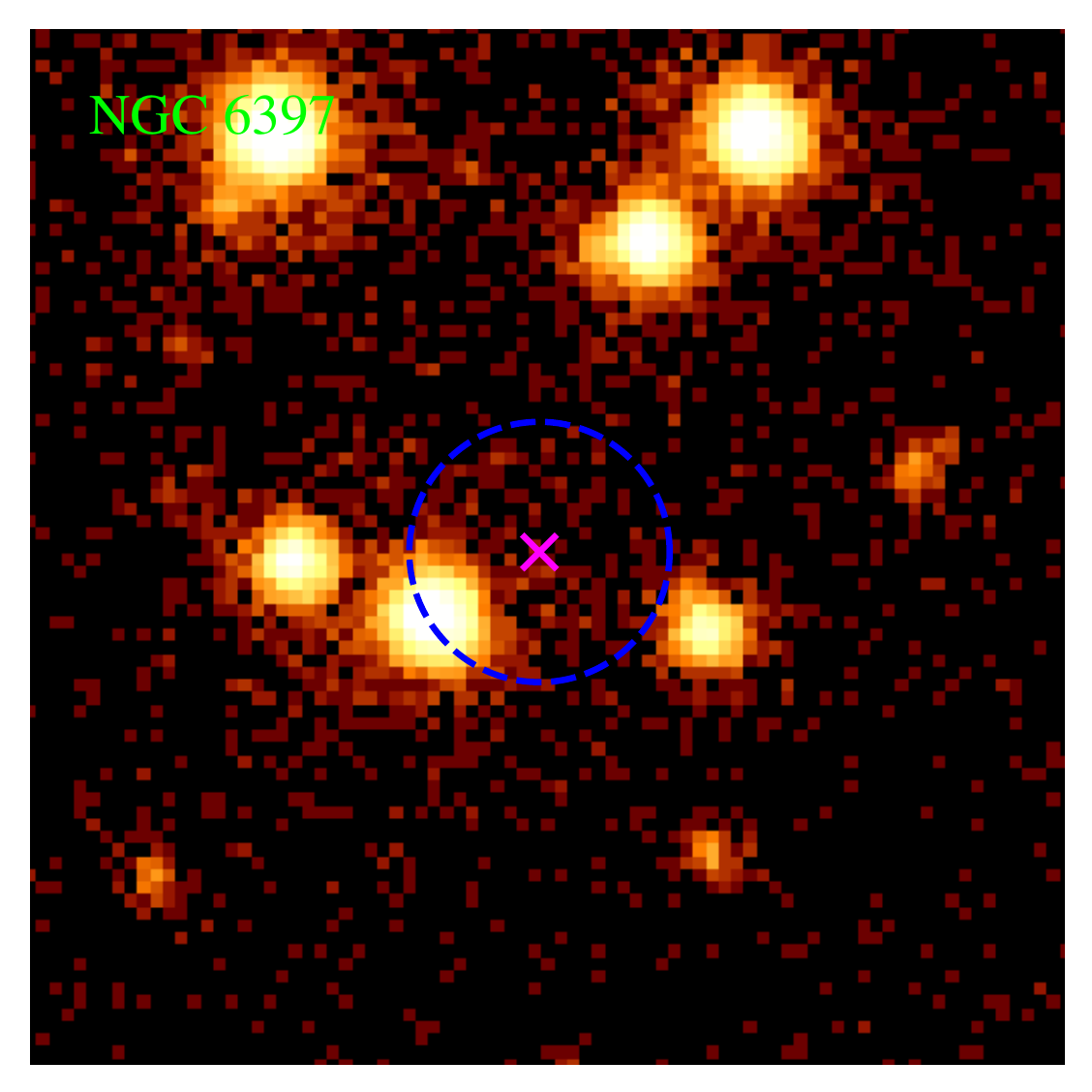}{0.2\textwidth}
    }
\agridline{
        \afig{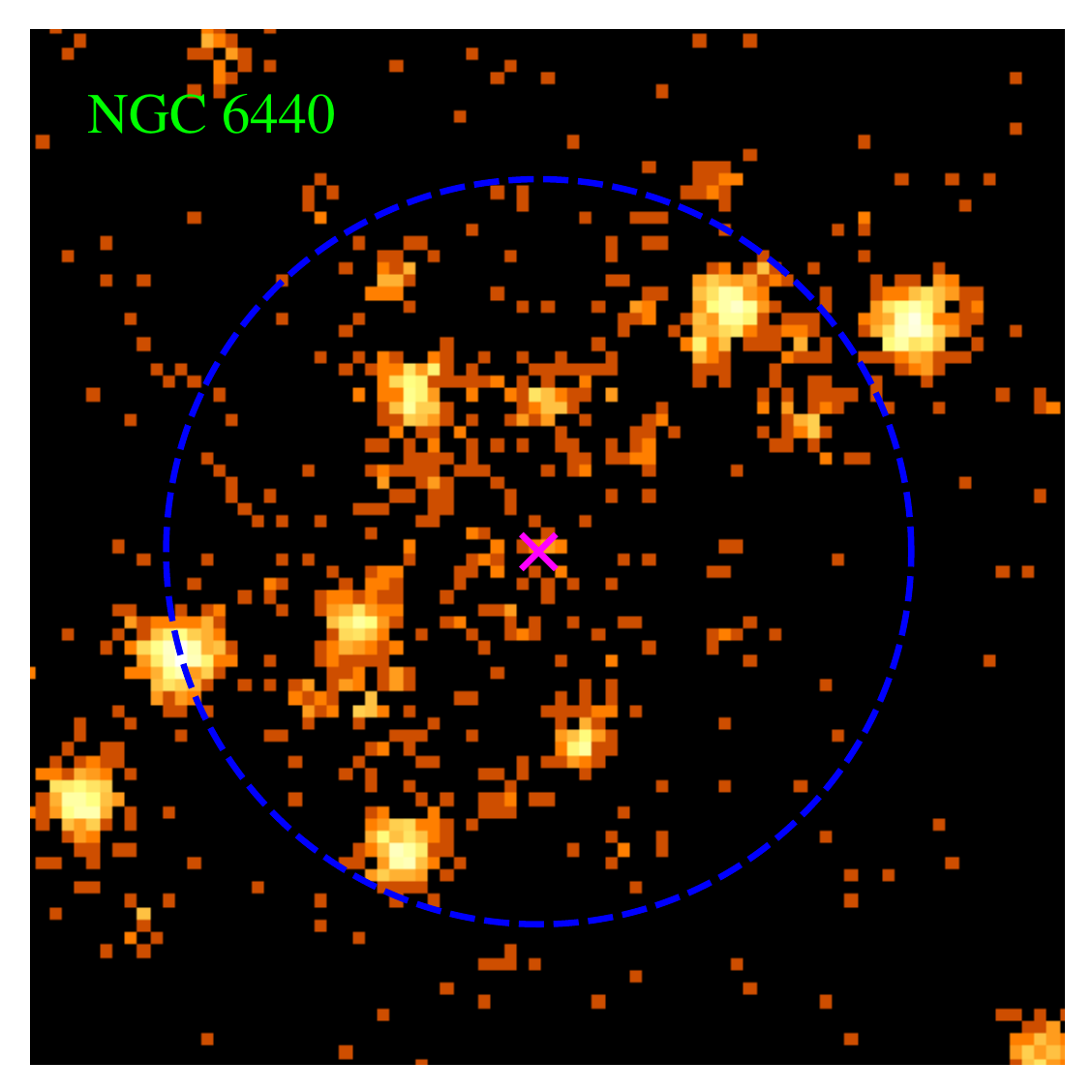}{0.2\textwidth}
        \afig{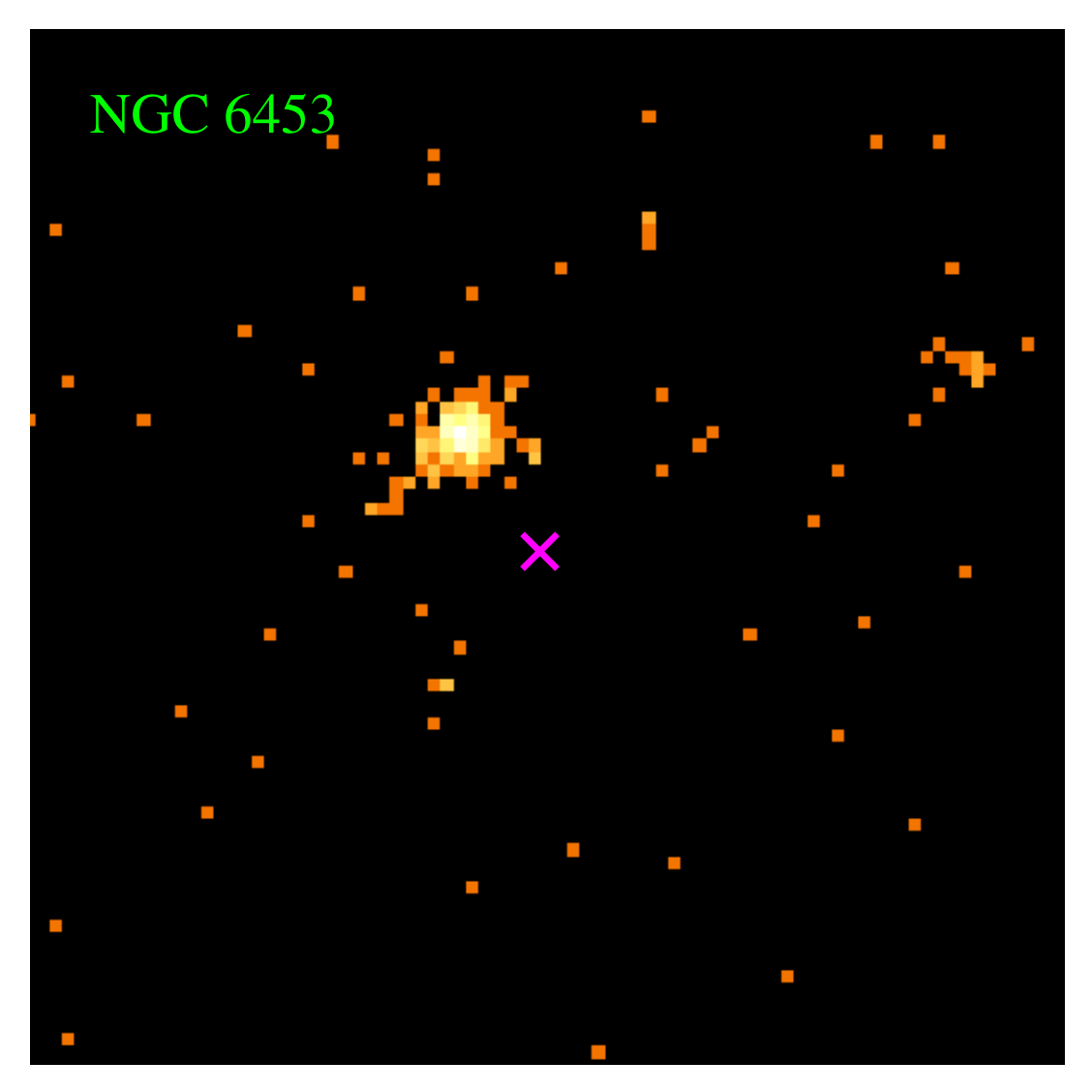}{0.2\textwidth}
        \afig{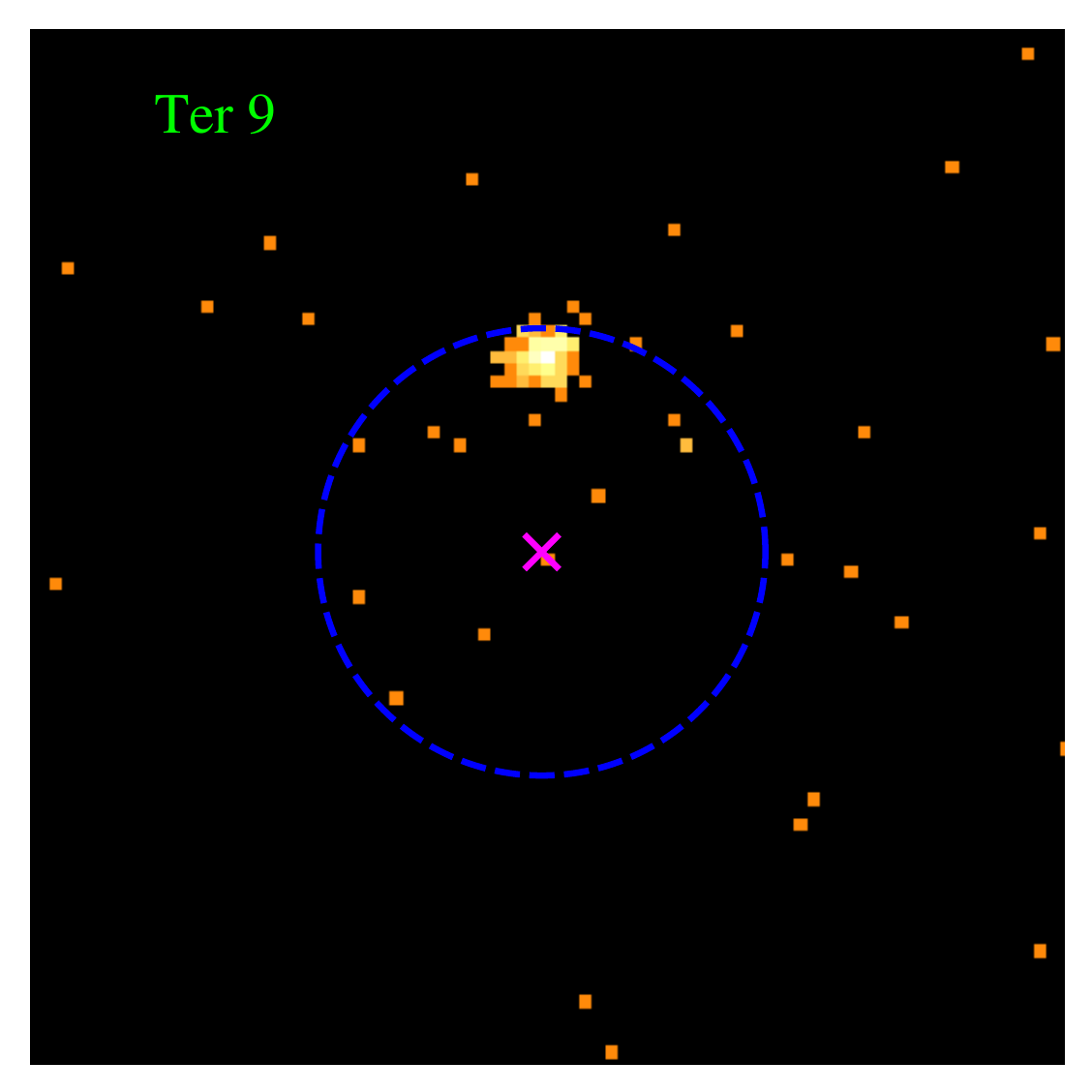}{0.2\textwidth}
        \afig{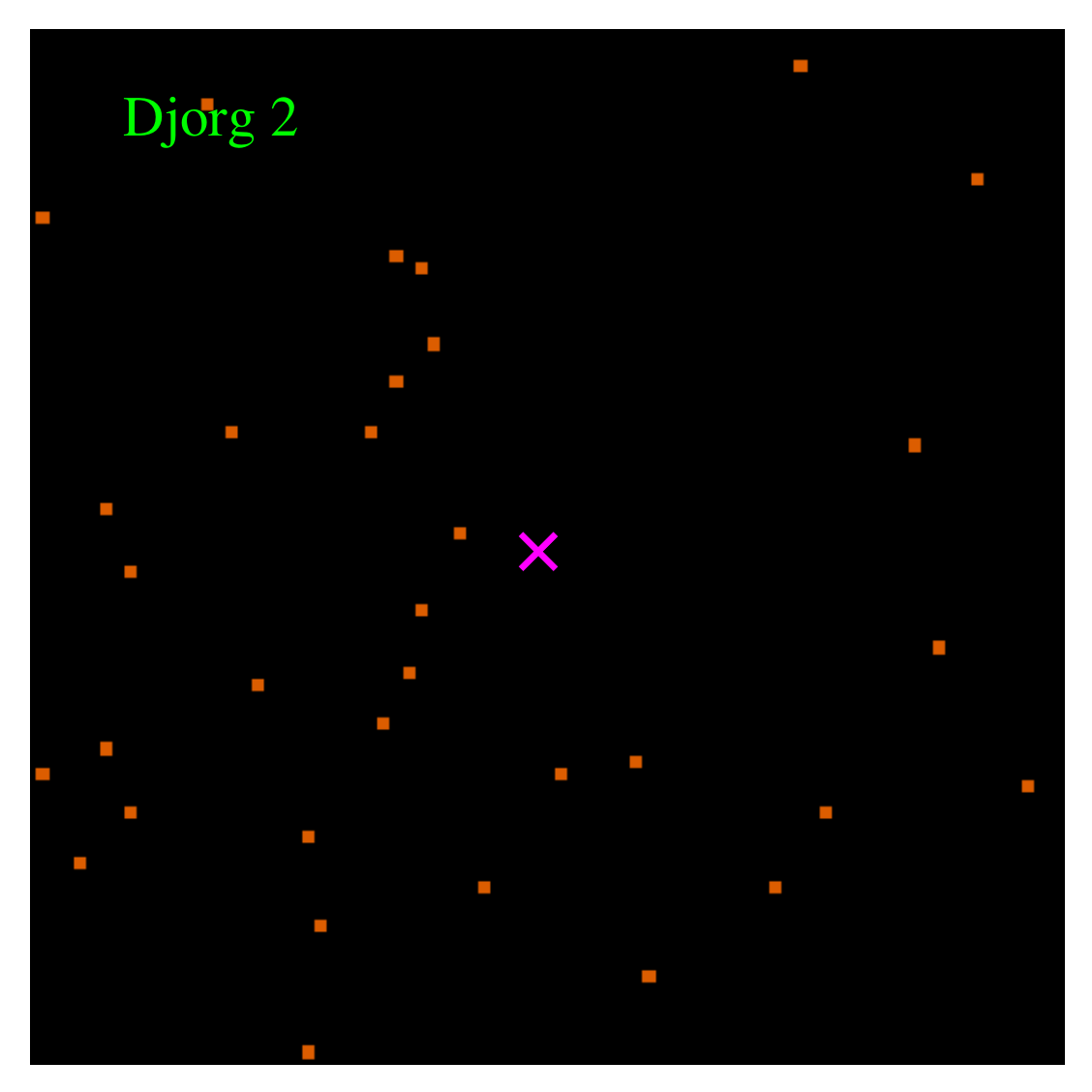}{0.2\textwidth}
        \afig{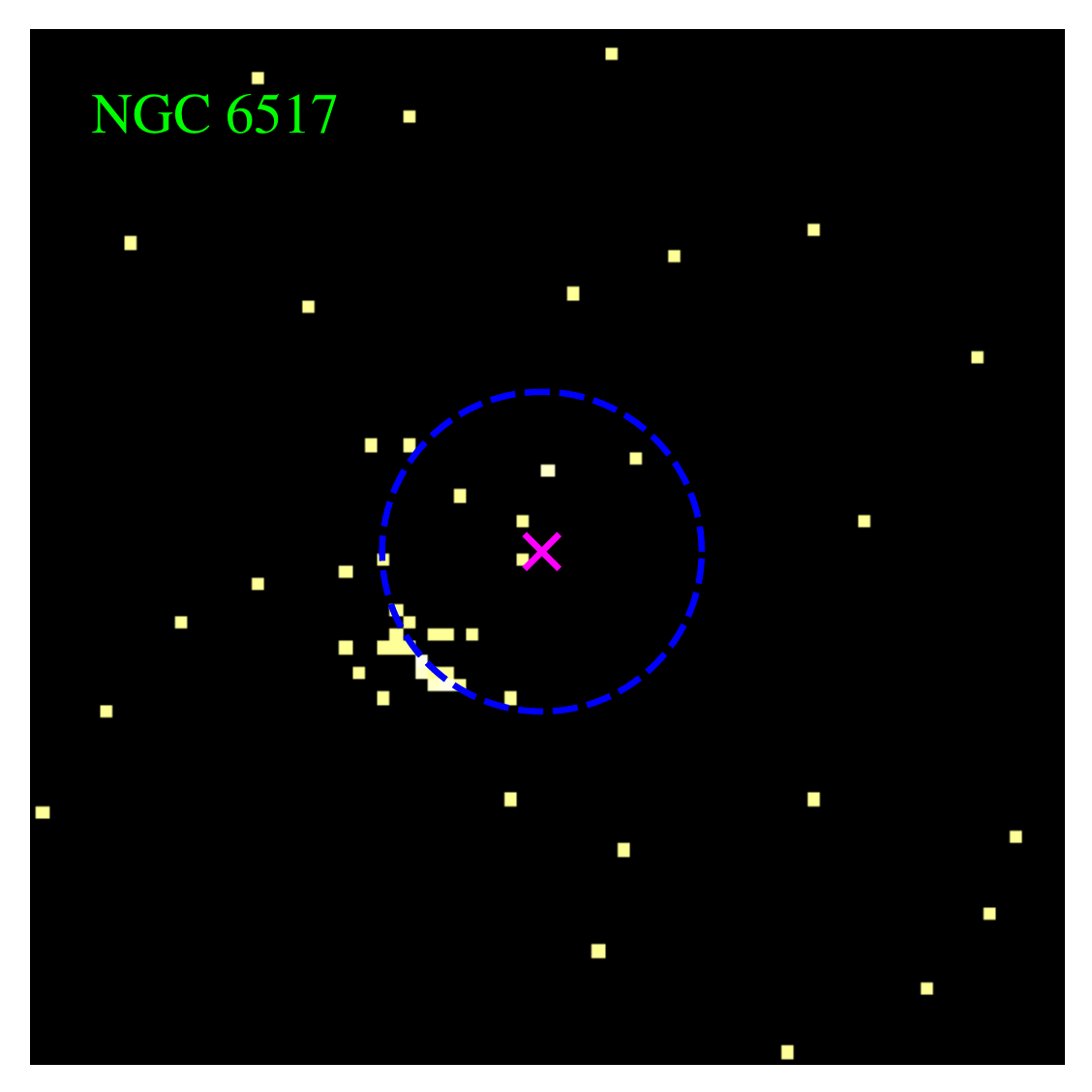}{0.2\textwidth}
    }
    
\agridline{
        \afig{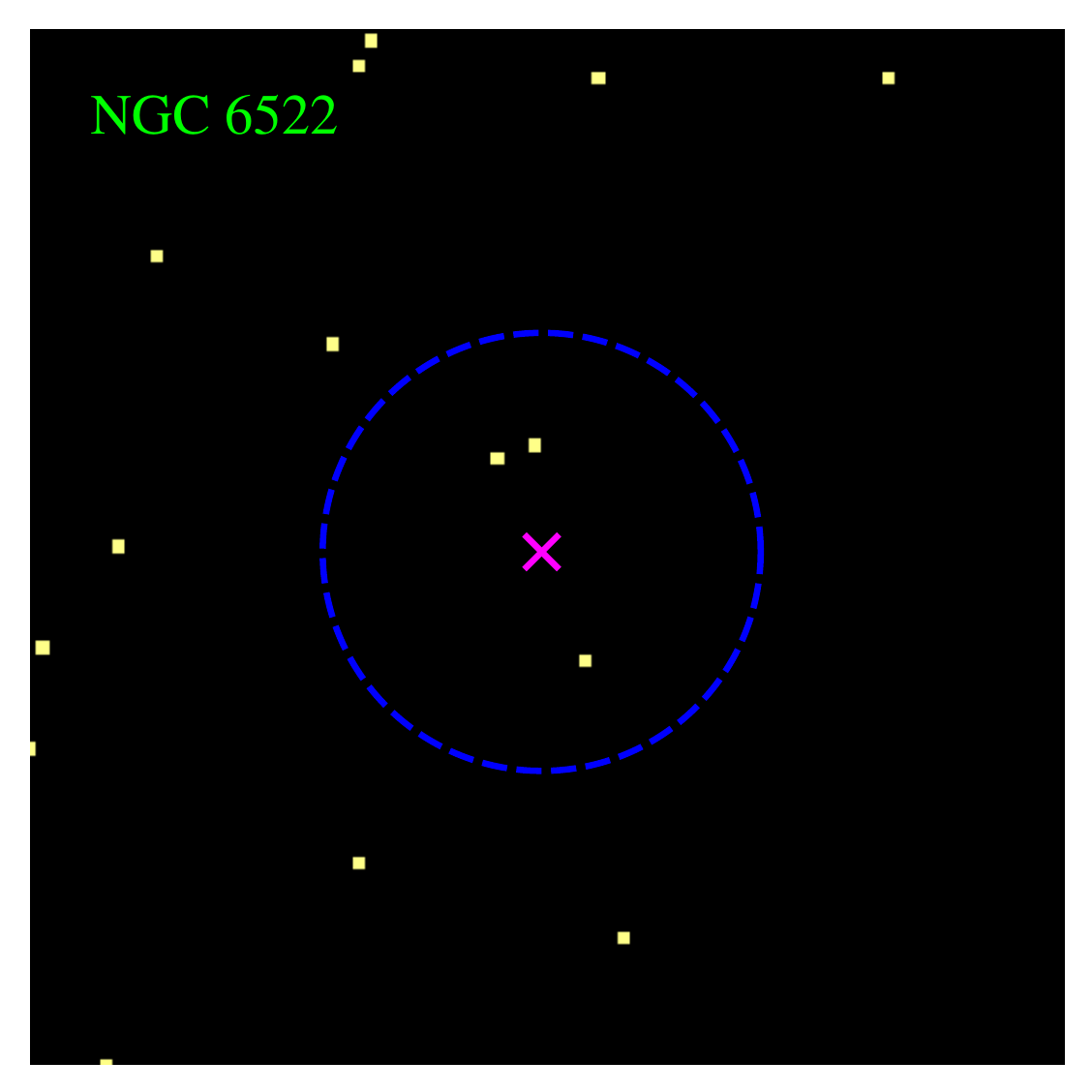}{0.2\textwidth}
        \afig{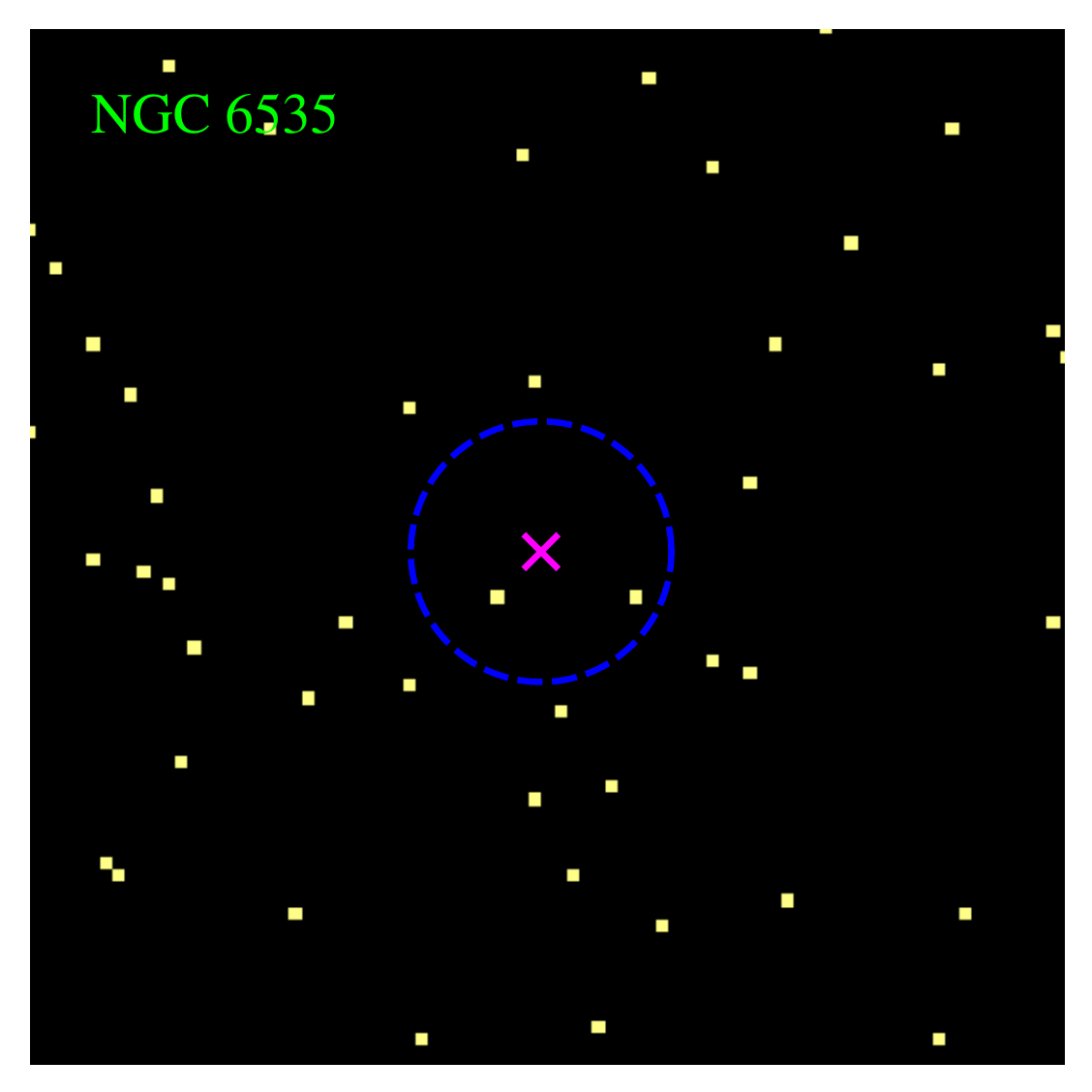}{0.2\textwidth}
        \afig{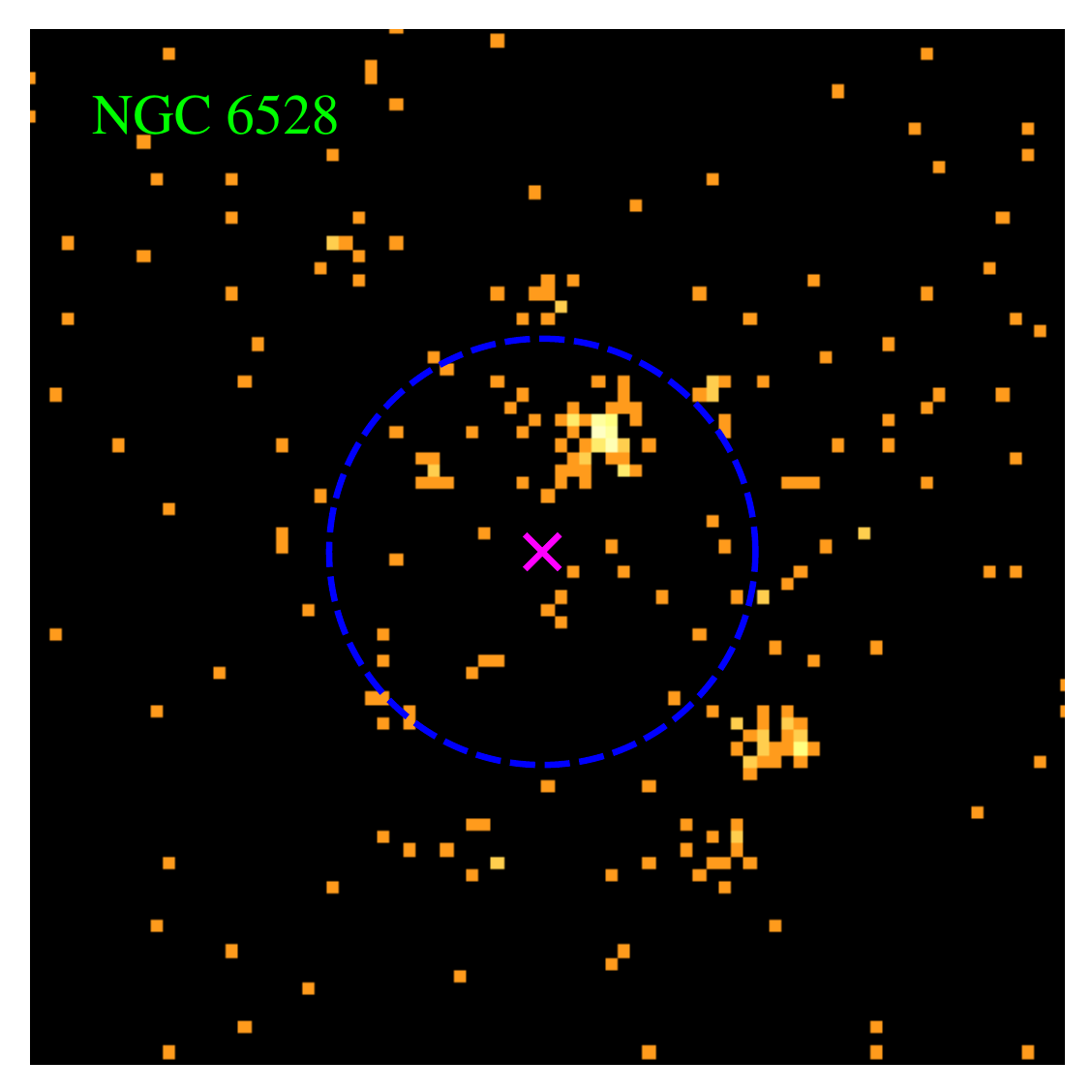}{0.2\textwidth}
        \afig{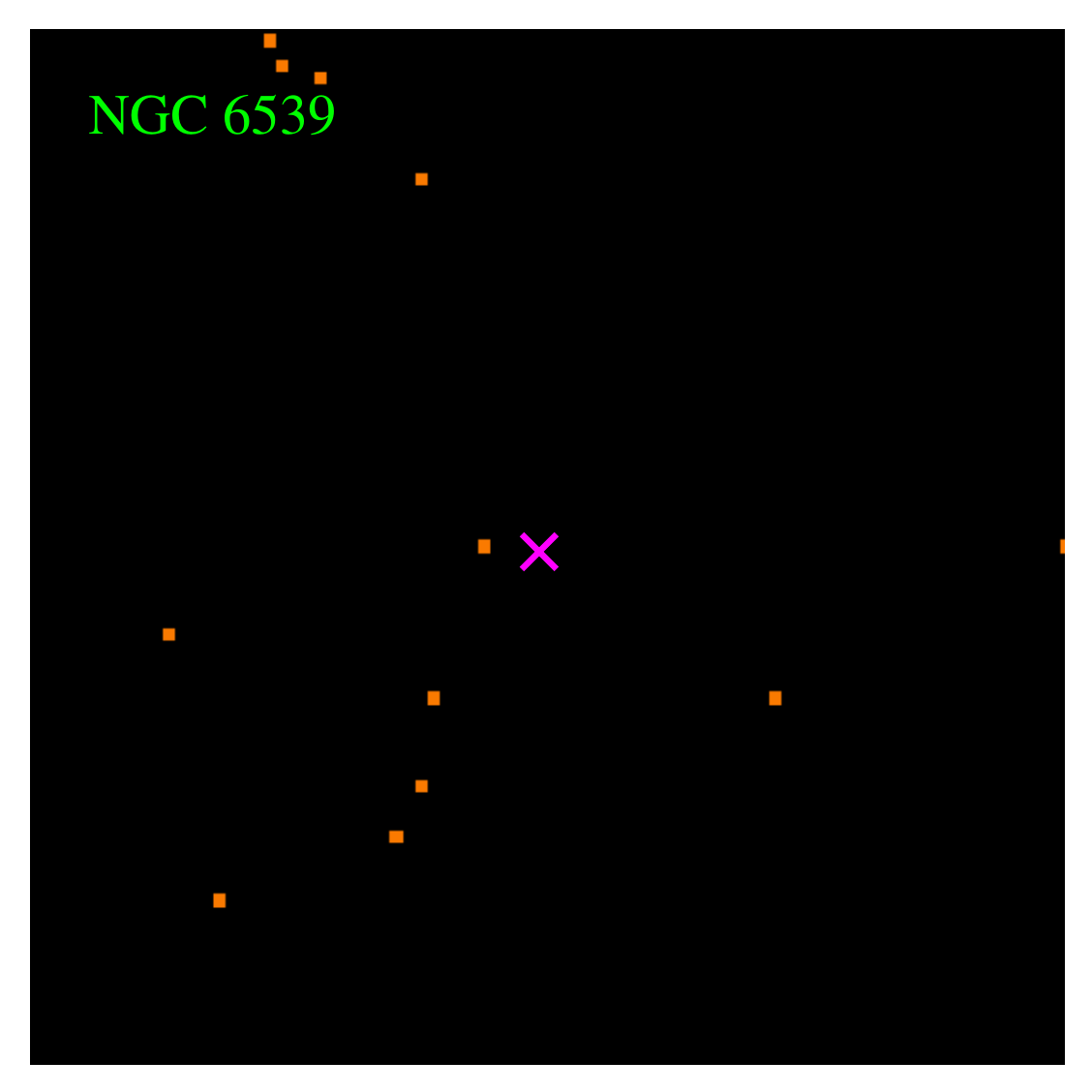}{0.2\textwidth}
        \afig{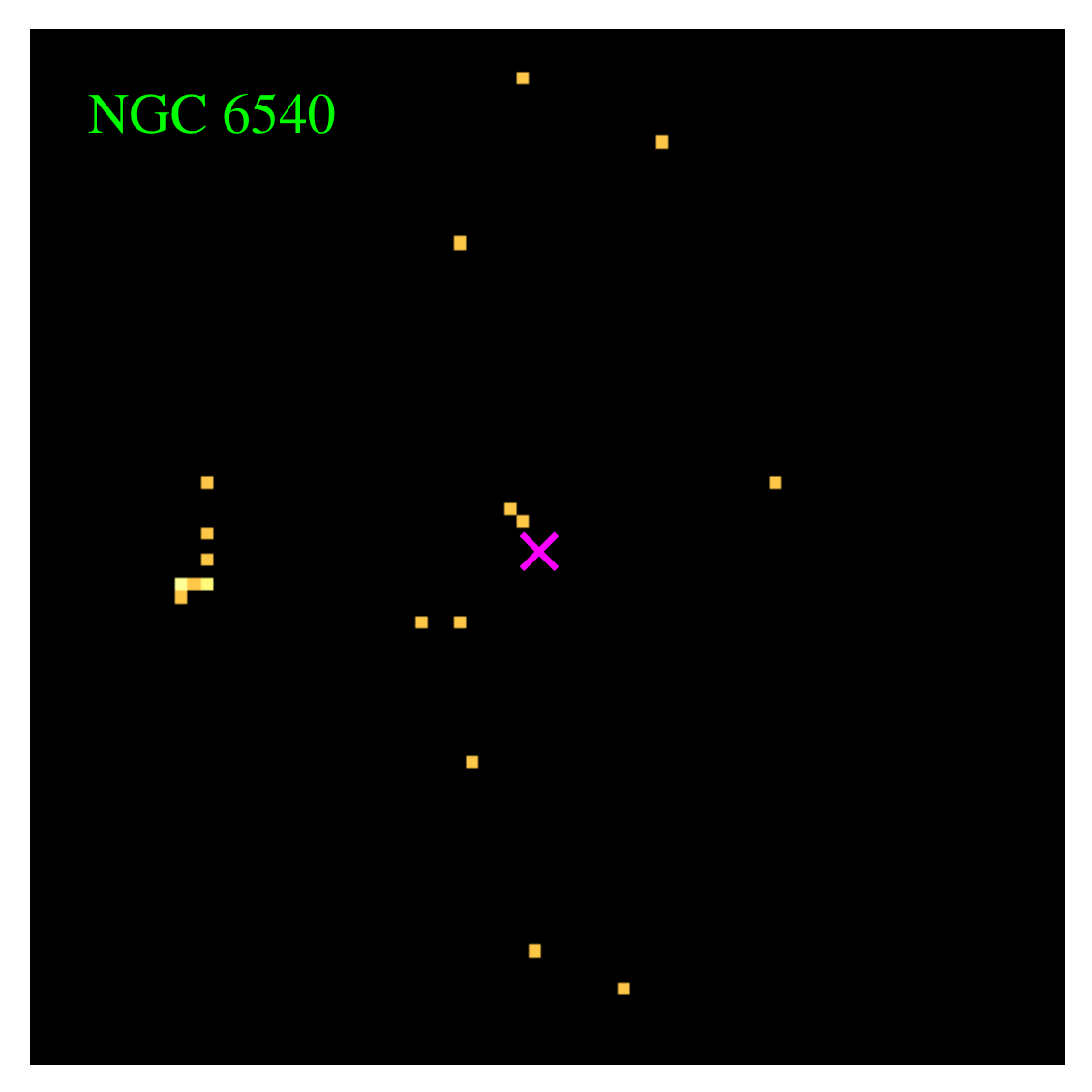}{0.2\textwidth}
    }
\agridline{
        \afig{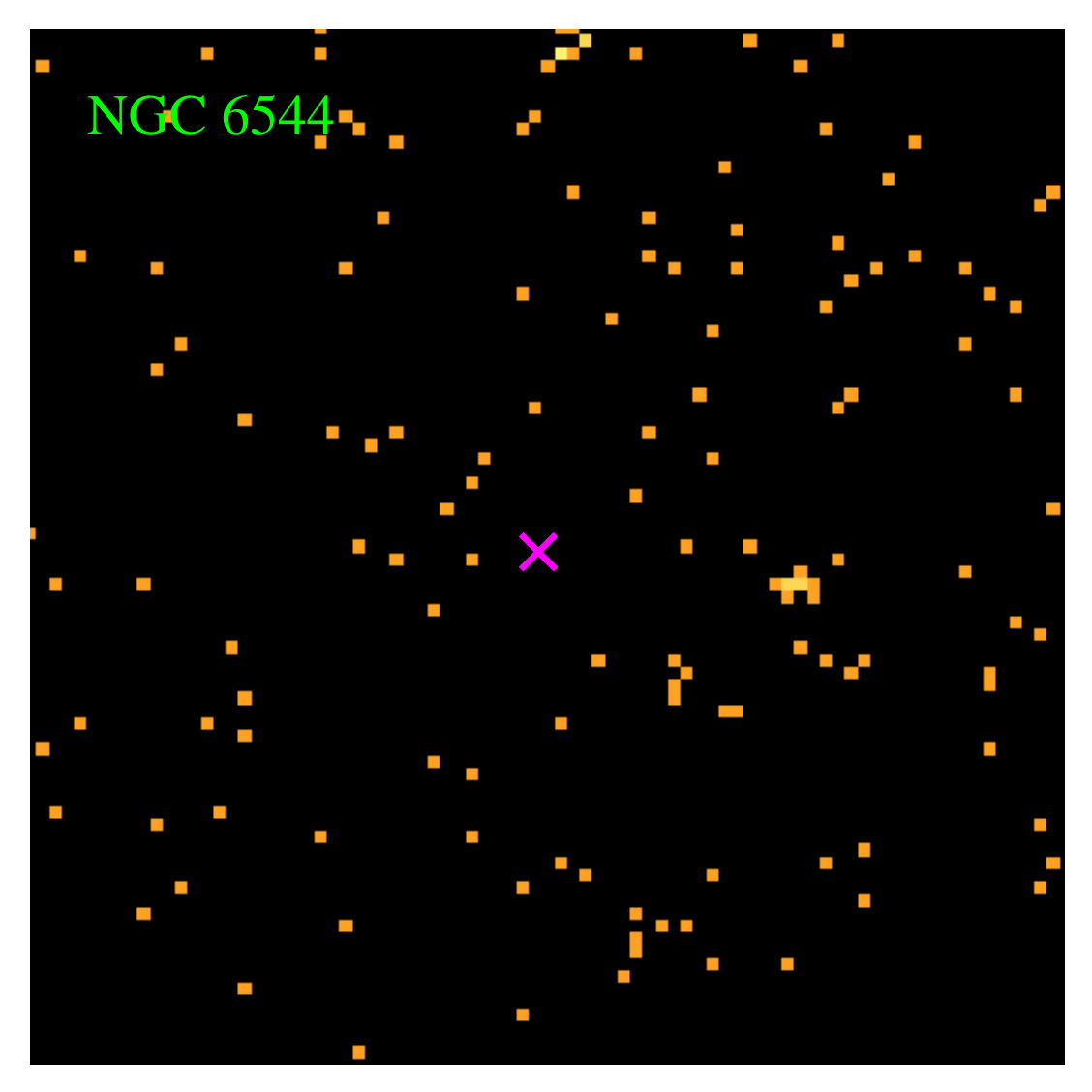}{0.2\textwidth}
        \afig{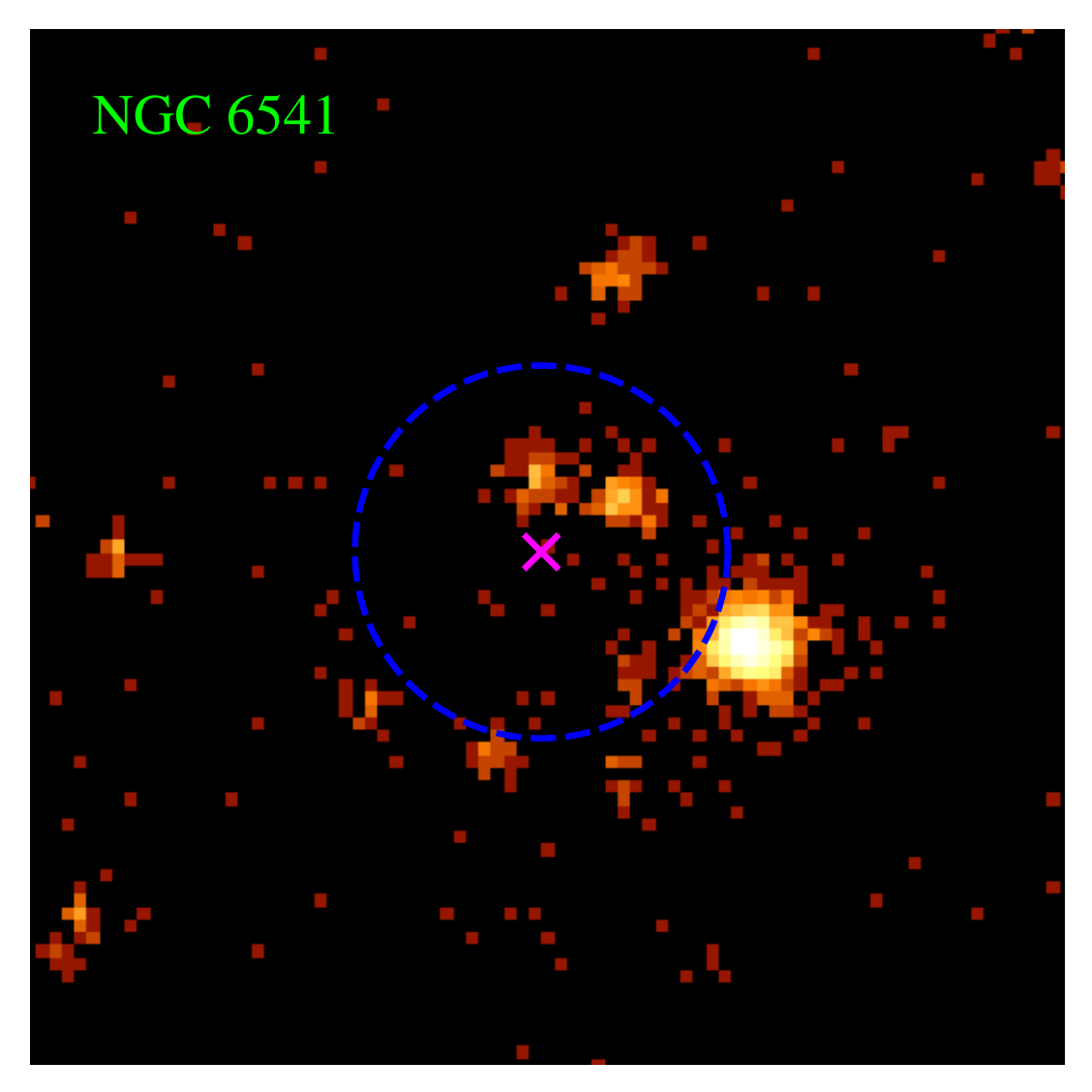}{0.2\textwidth}
        \afig{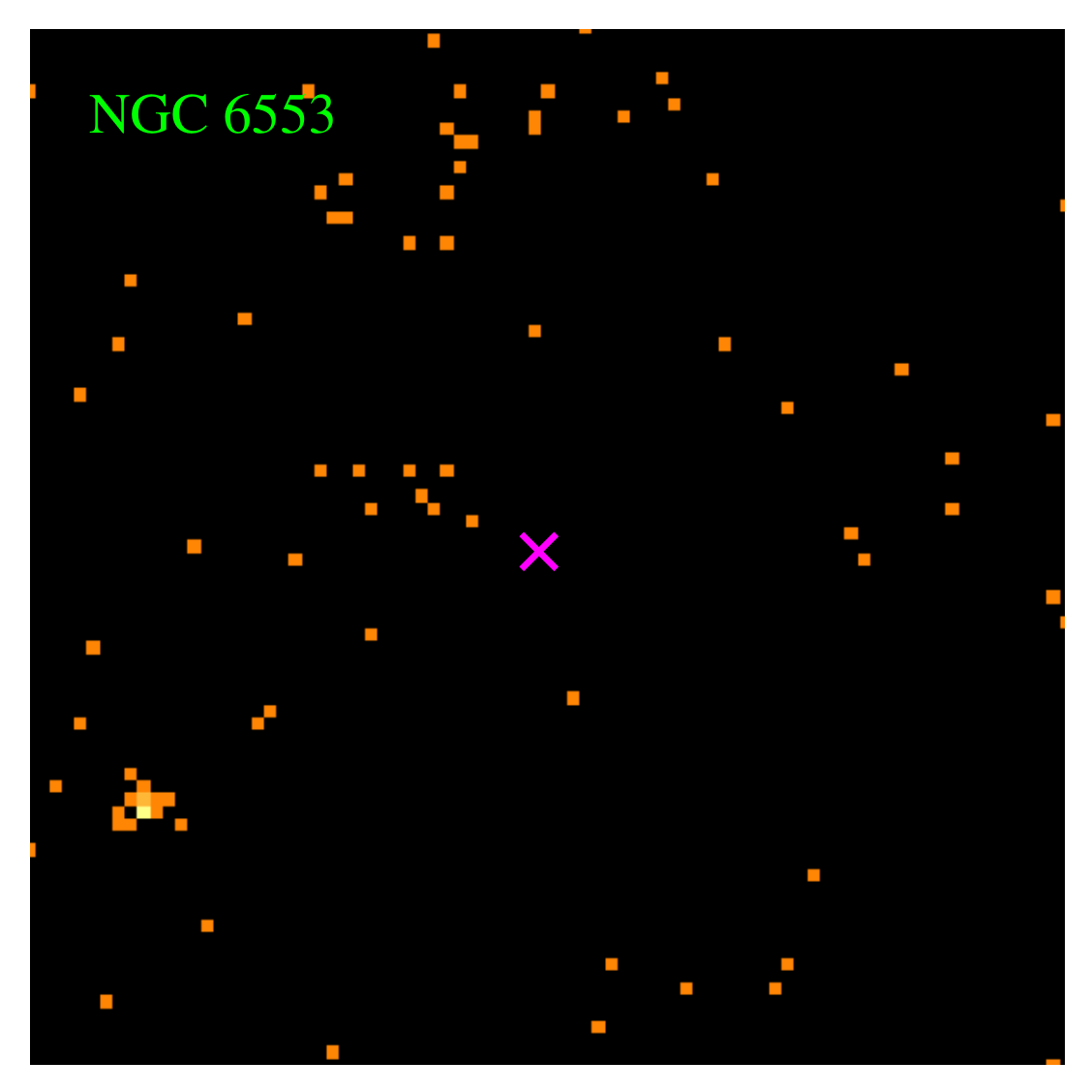}{0.2\textwidth}
        \afig{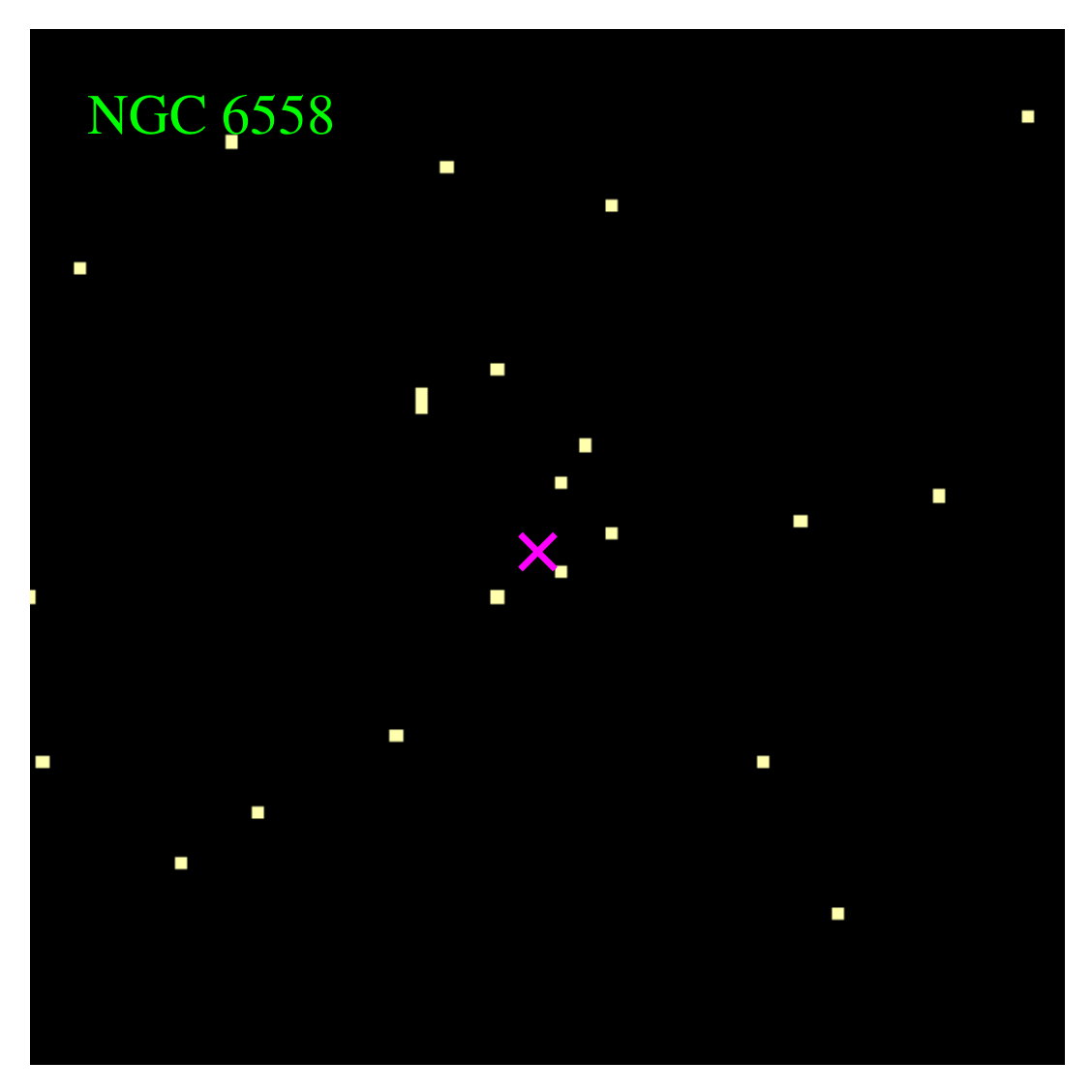}{0.2\textwidth}
        \afig{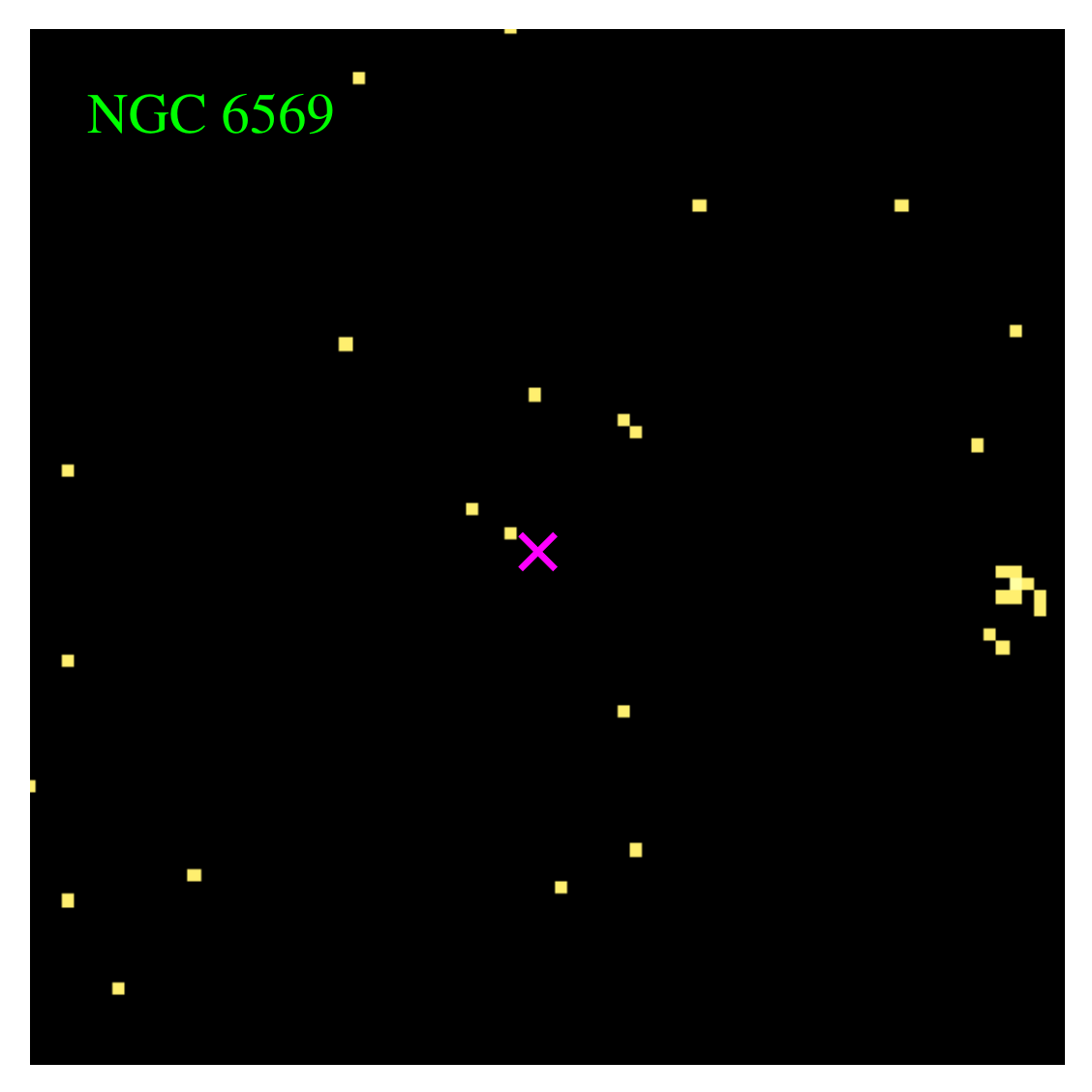}{0.2\textwidth}
    }
    \contcaption{}
    \end{figure*}
    \begin{figure*}
        \centering

\agridline{
        \afig{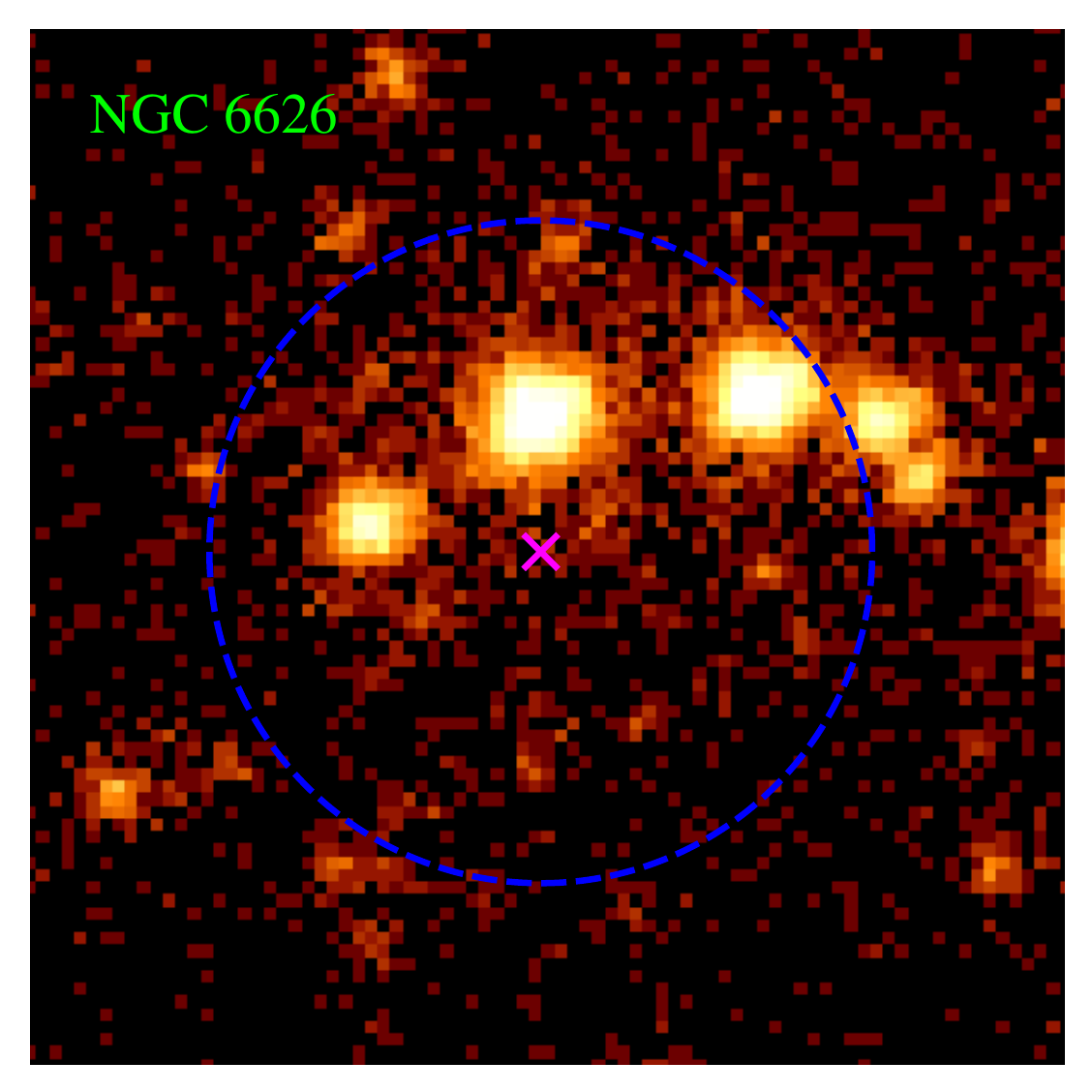}{0.2\textwidth}
        \afig{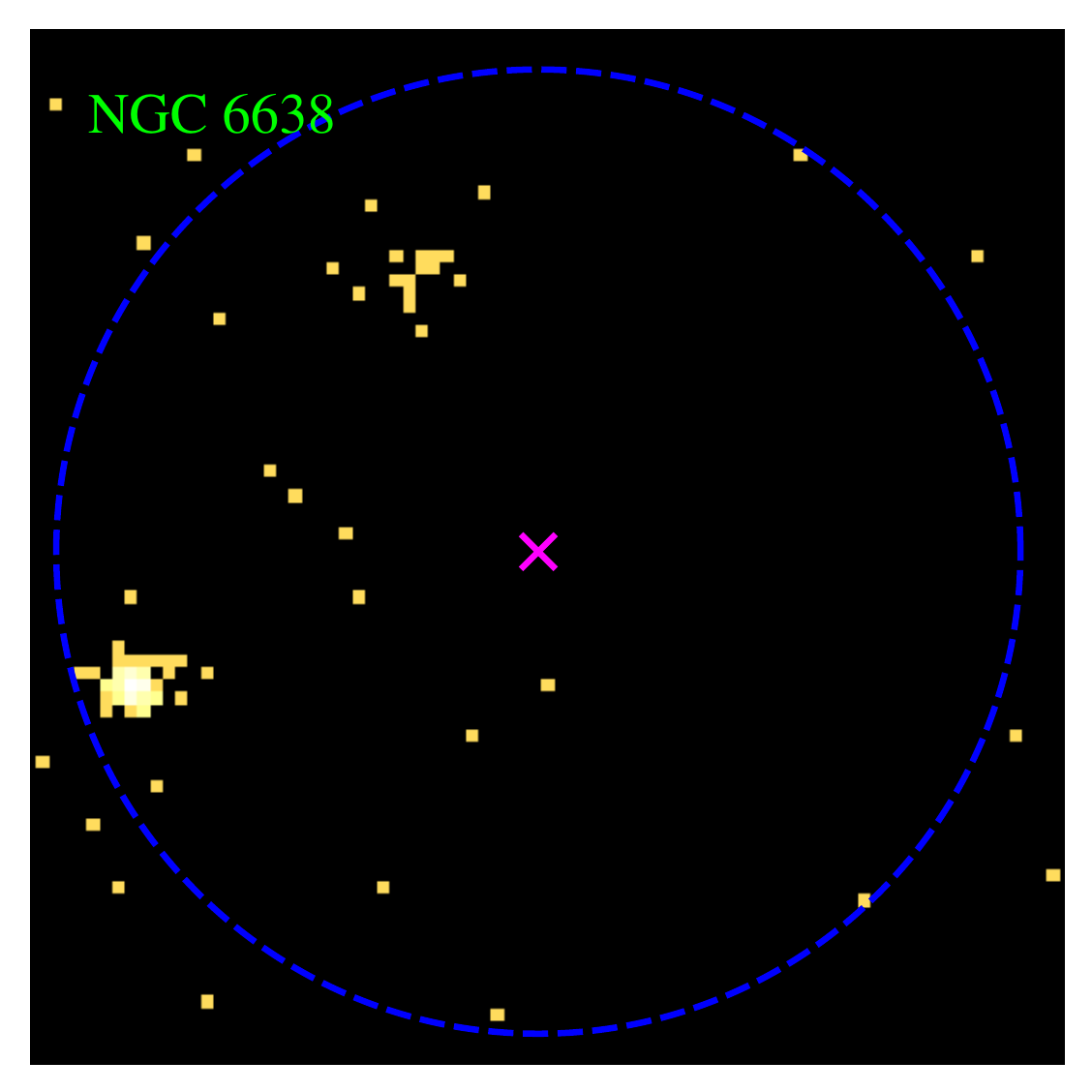}{0.2\textwidth}
        \afig{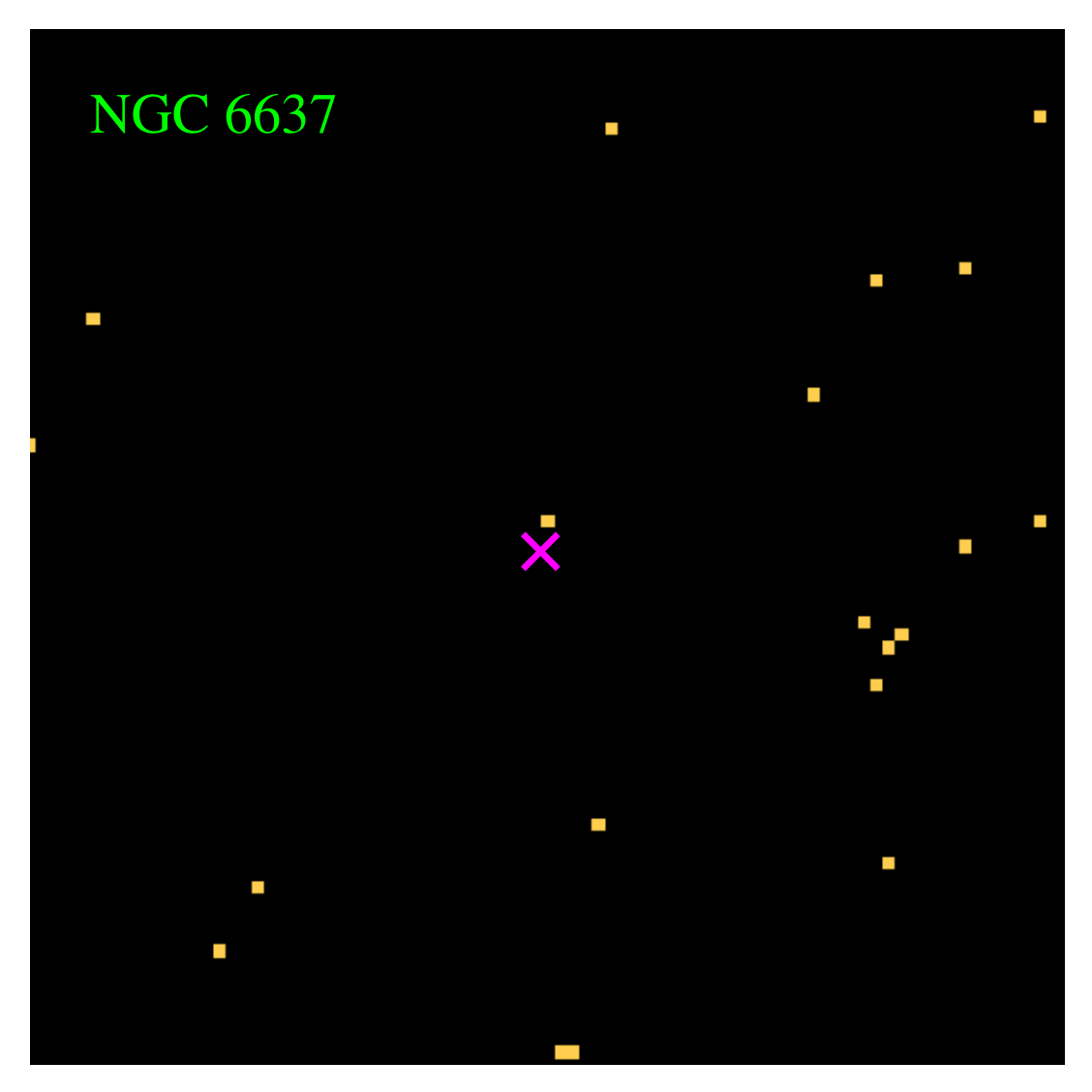}{0.2\textwidth}
        \afig{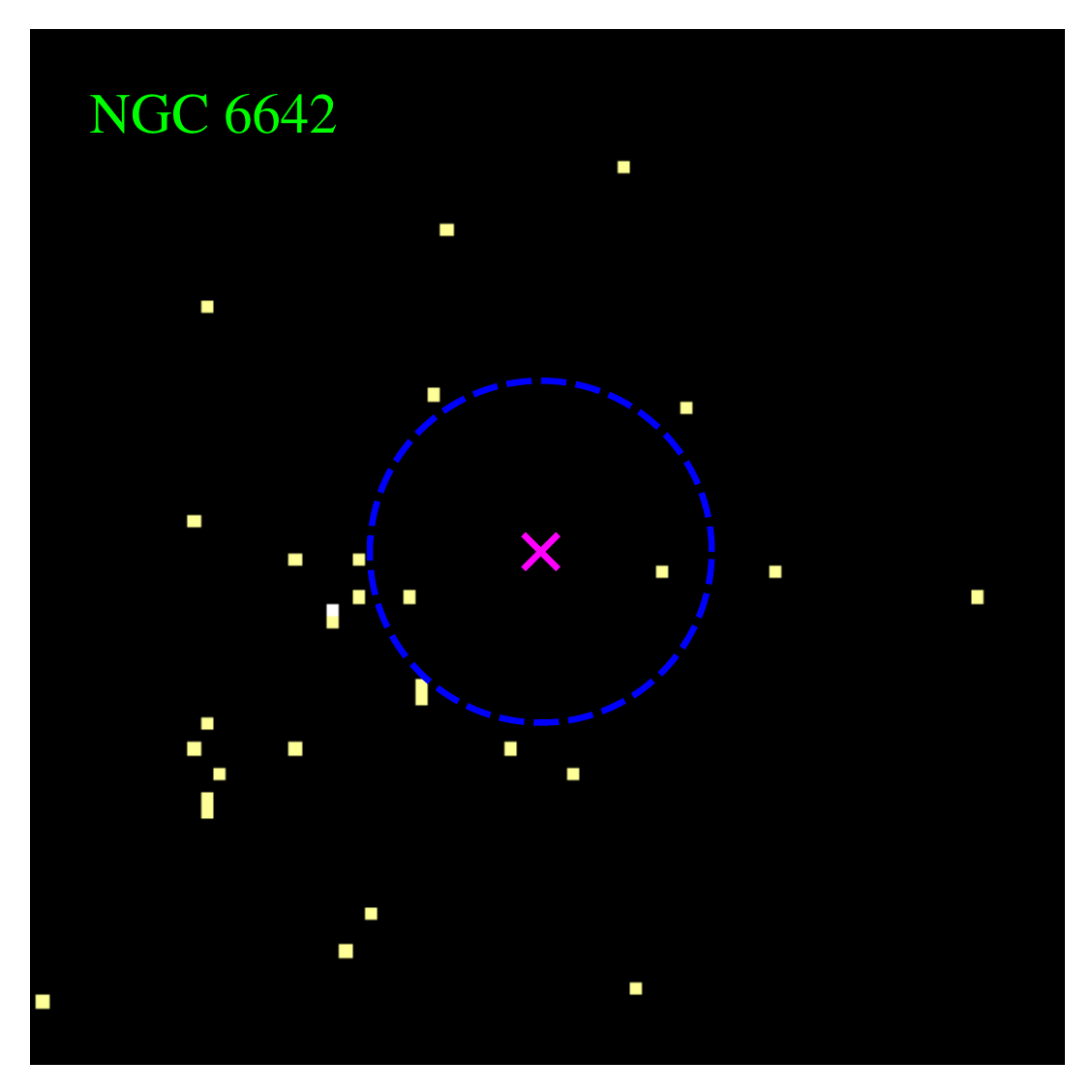}{0.2\textwidth}
        \afig{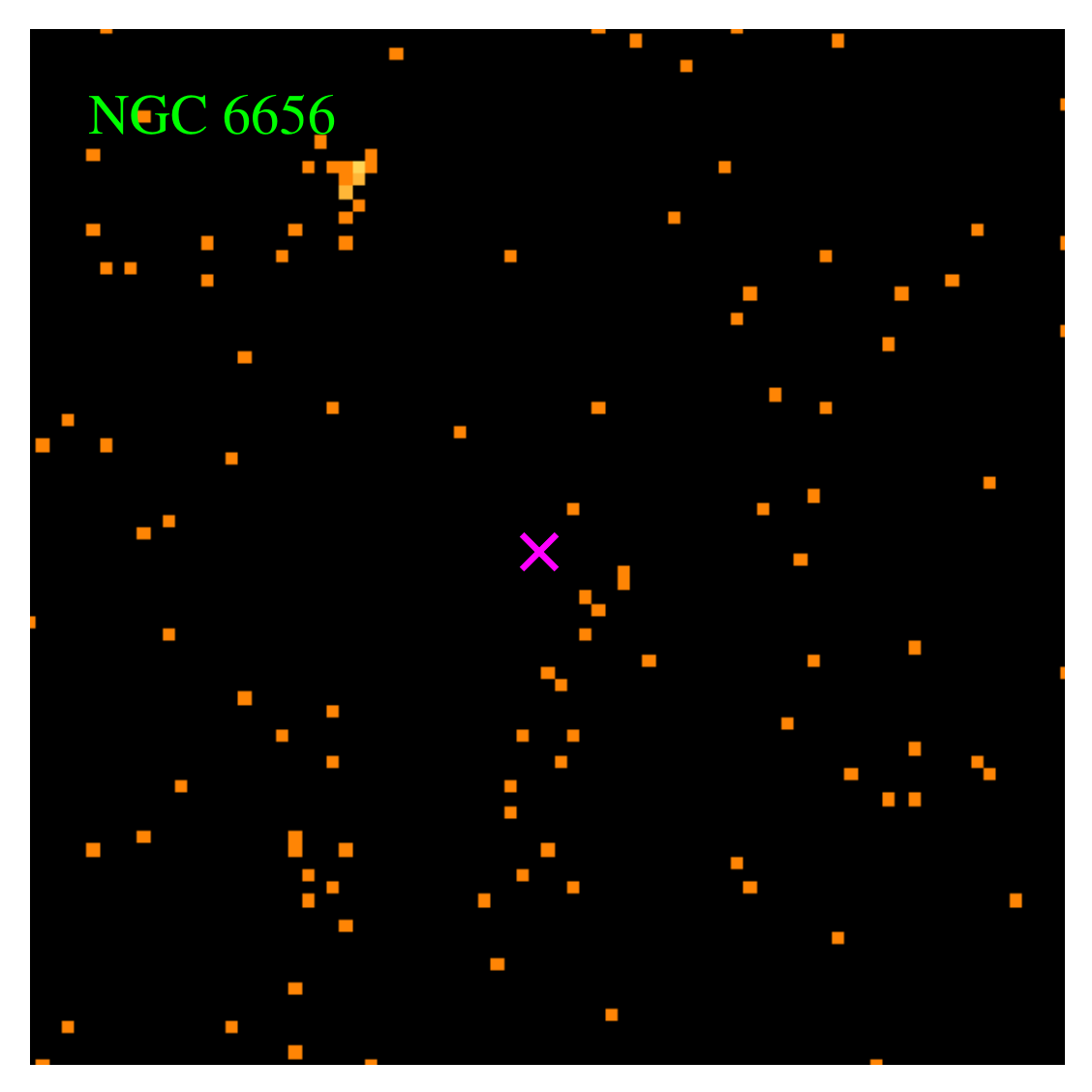}{0.2\textwidth}
    }
\agridline{
        \afig{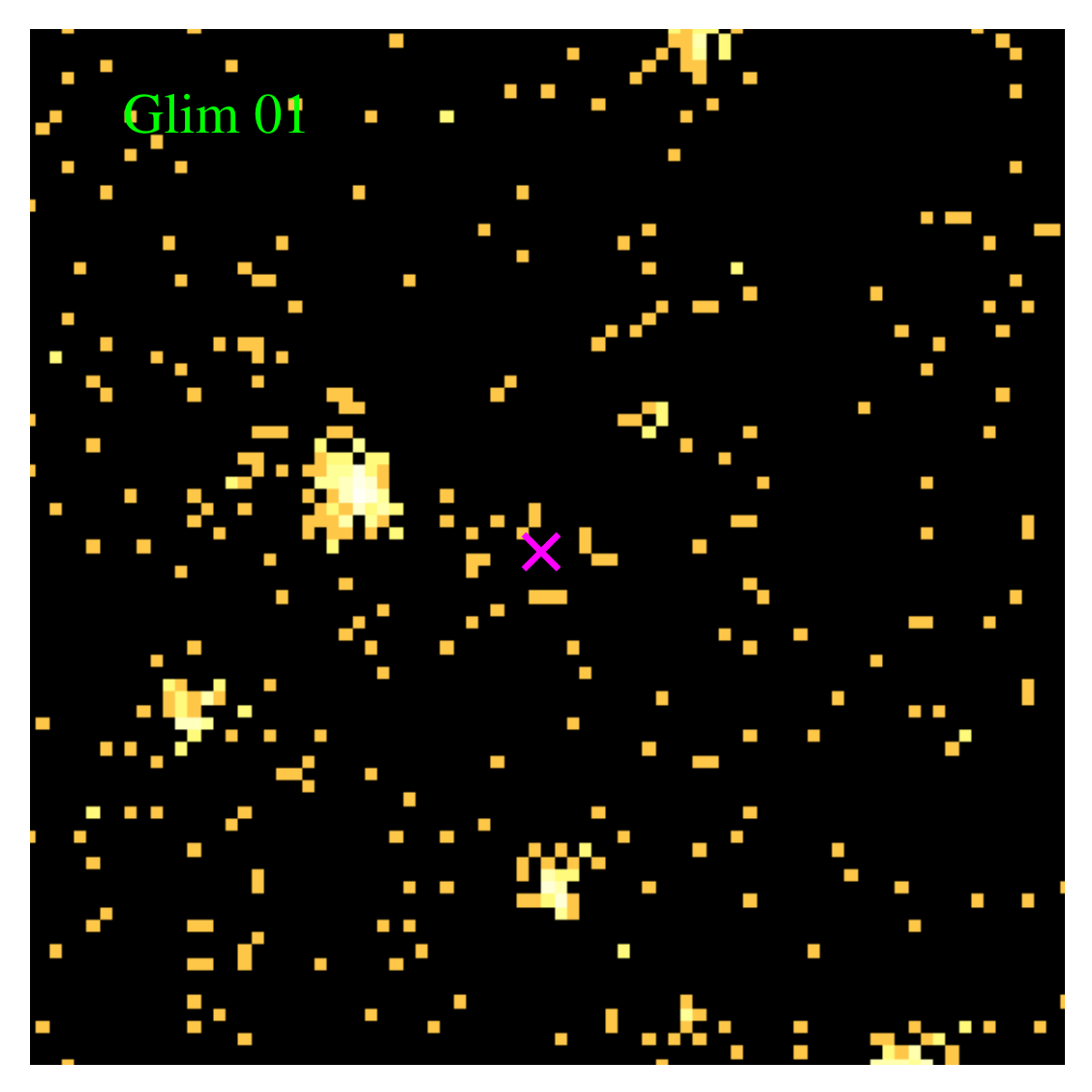}{0.2\textwidth}
        \afig{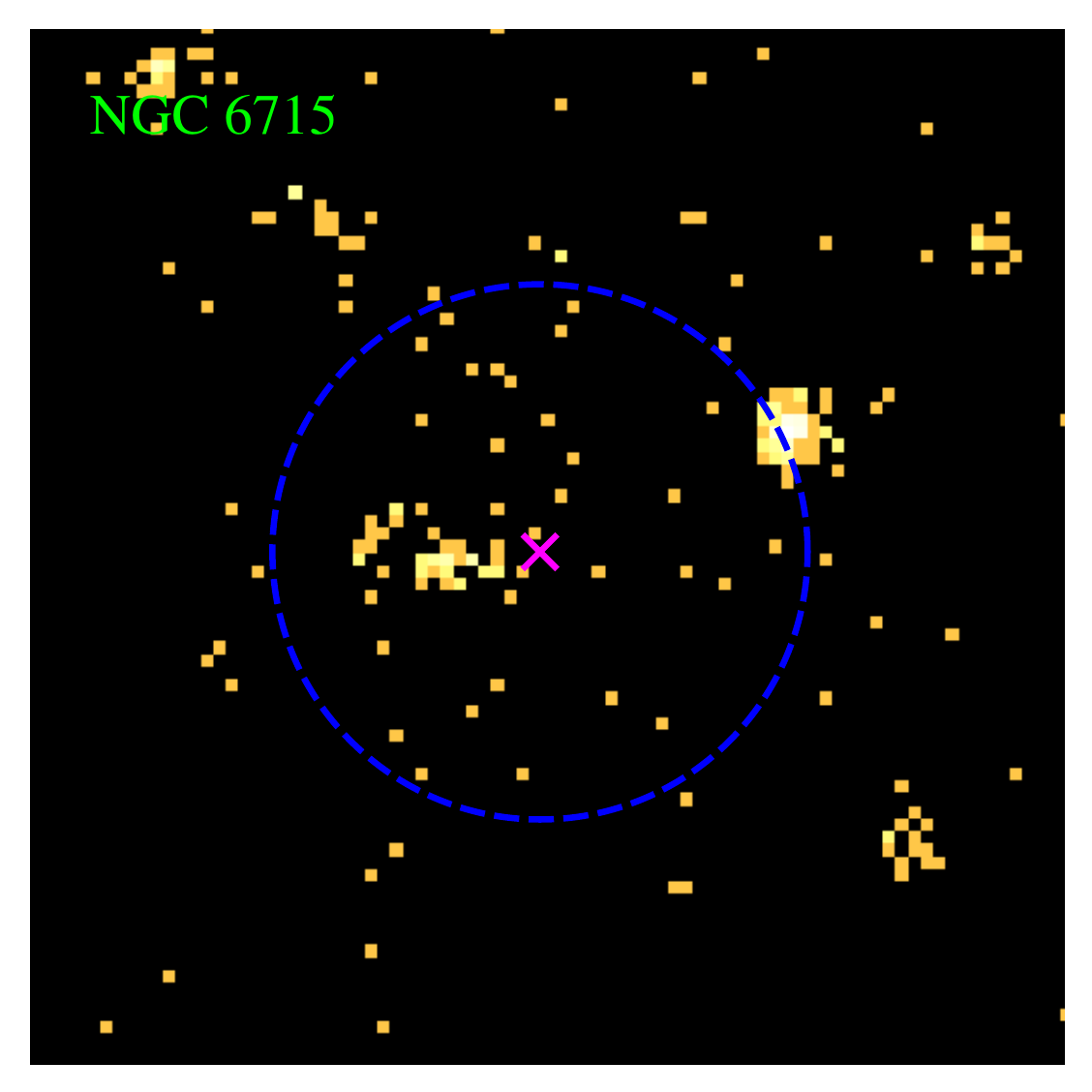}{0.2\textwidth}
        \afig{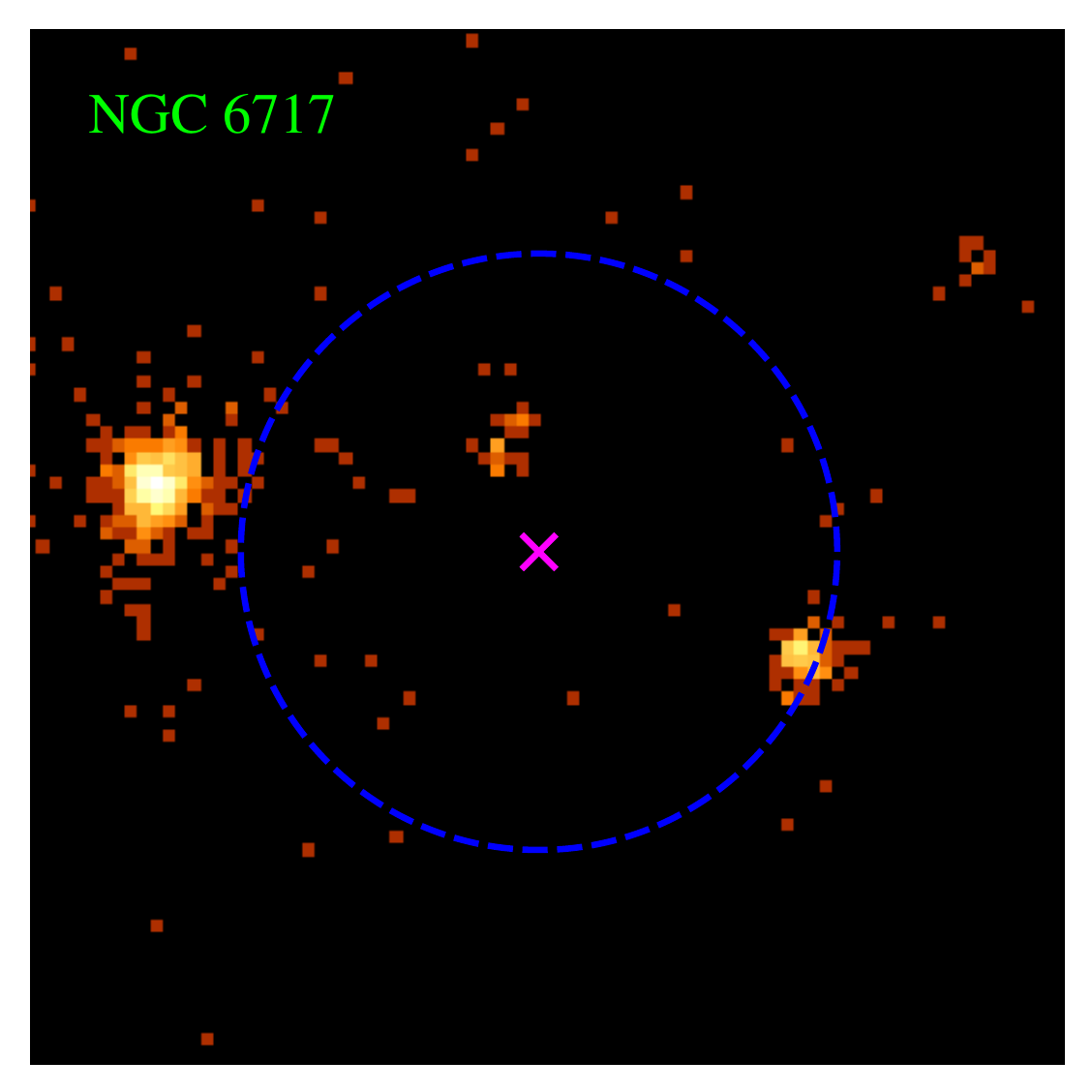}{0.2\textwidth}
        \afig{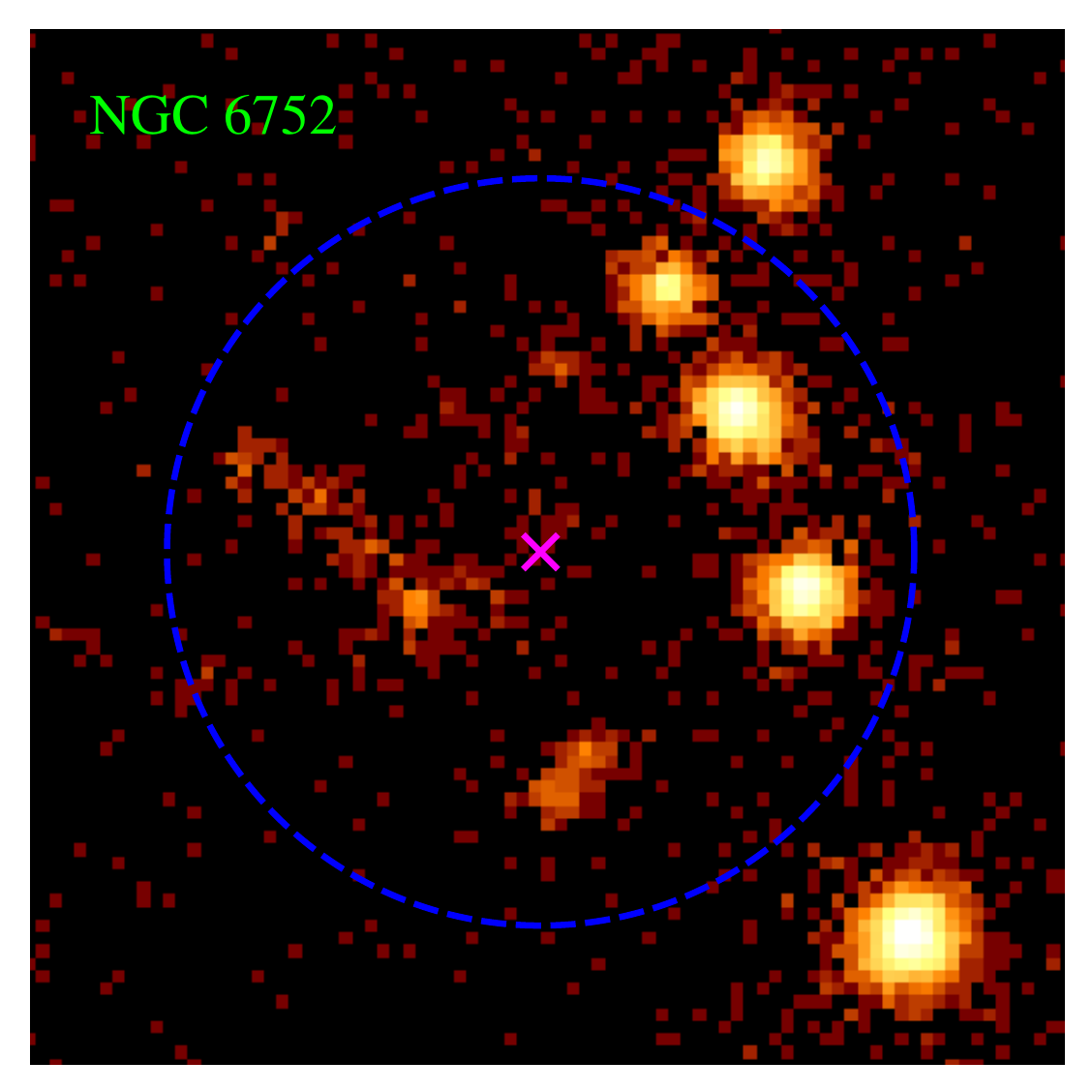}{0.2\textwidth}
        \afig{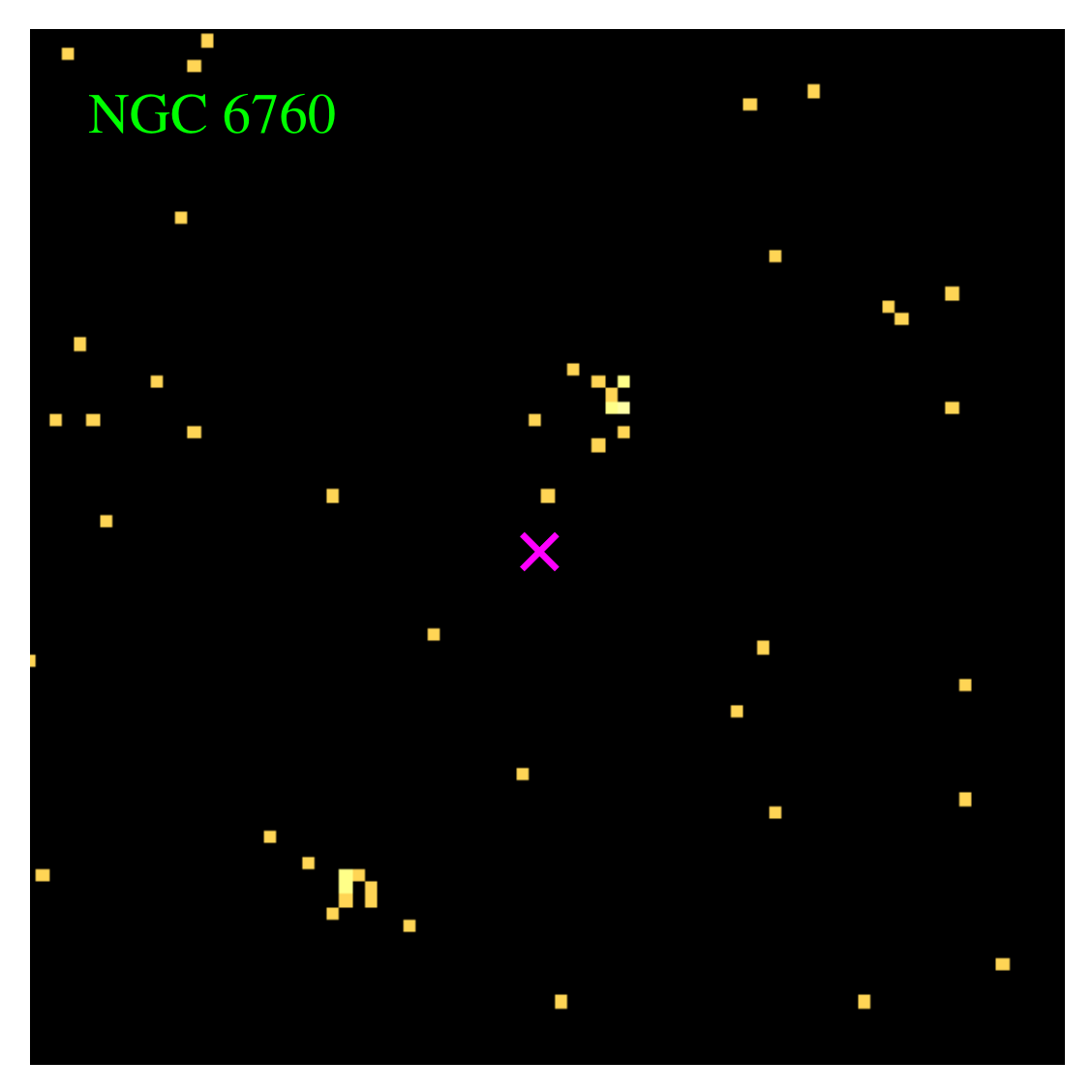}{0.2\textwidth}
    }
\agridline{
        \afig{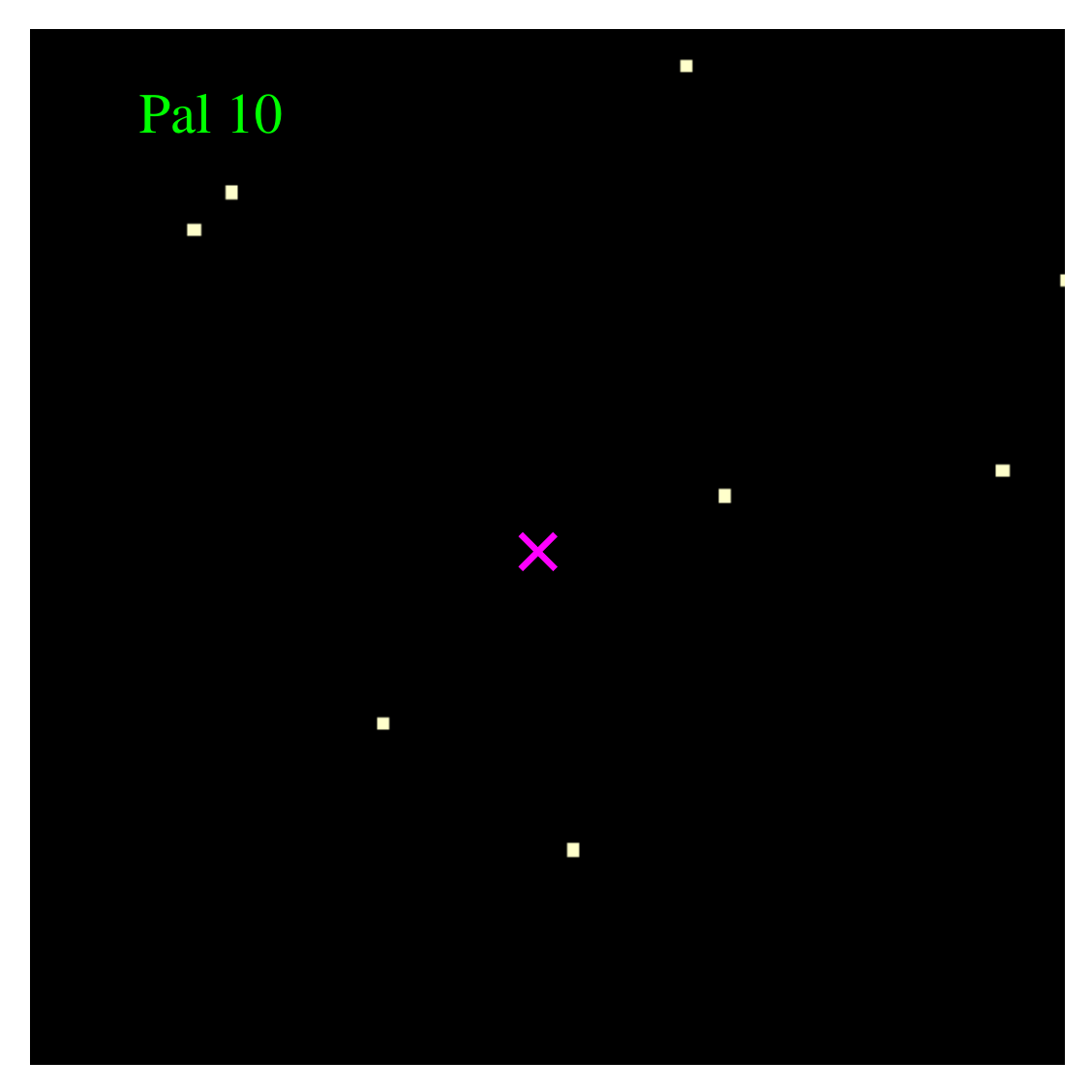}{0.2\textwidth}
        \afig{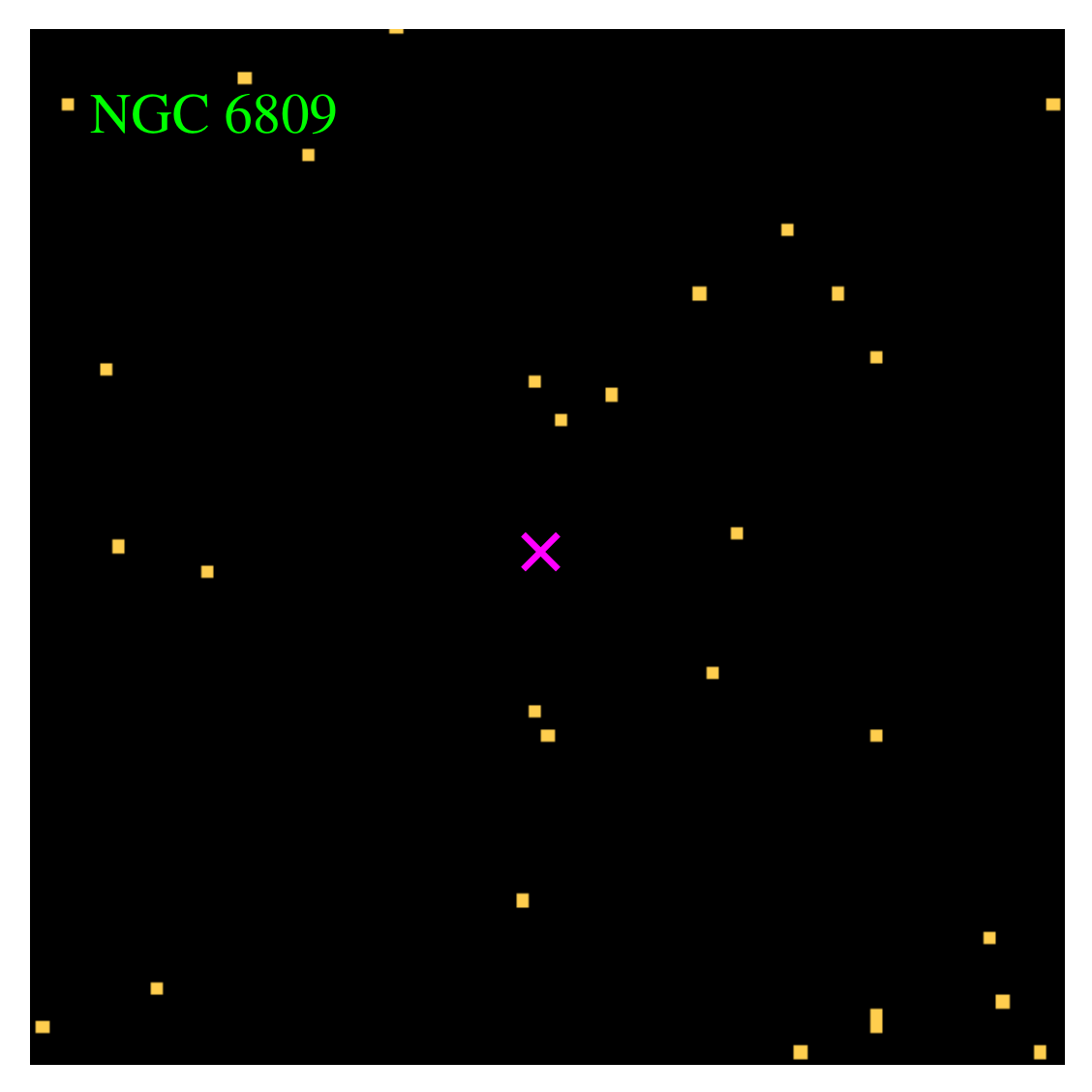}{0.2\textwidth}
        \afig{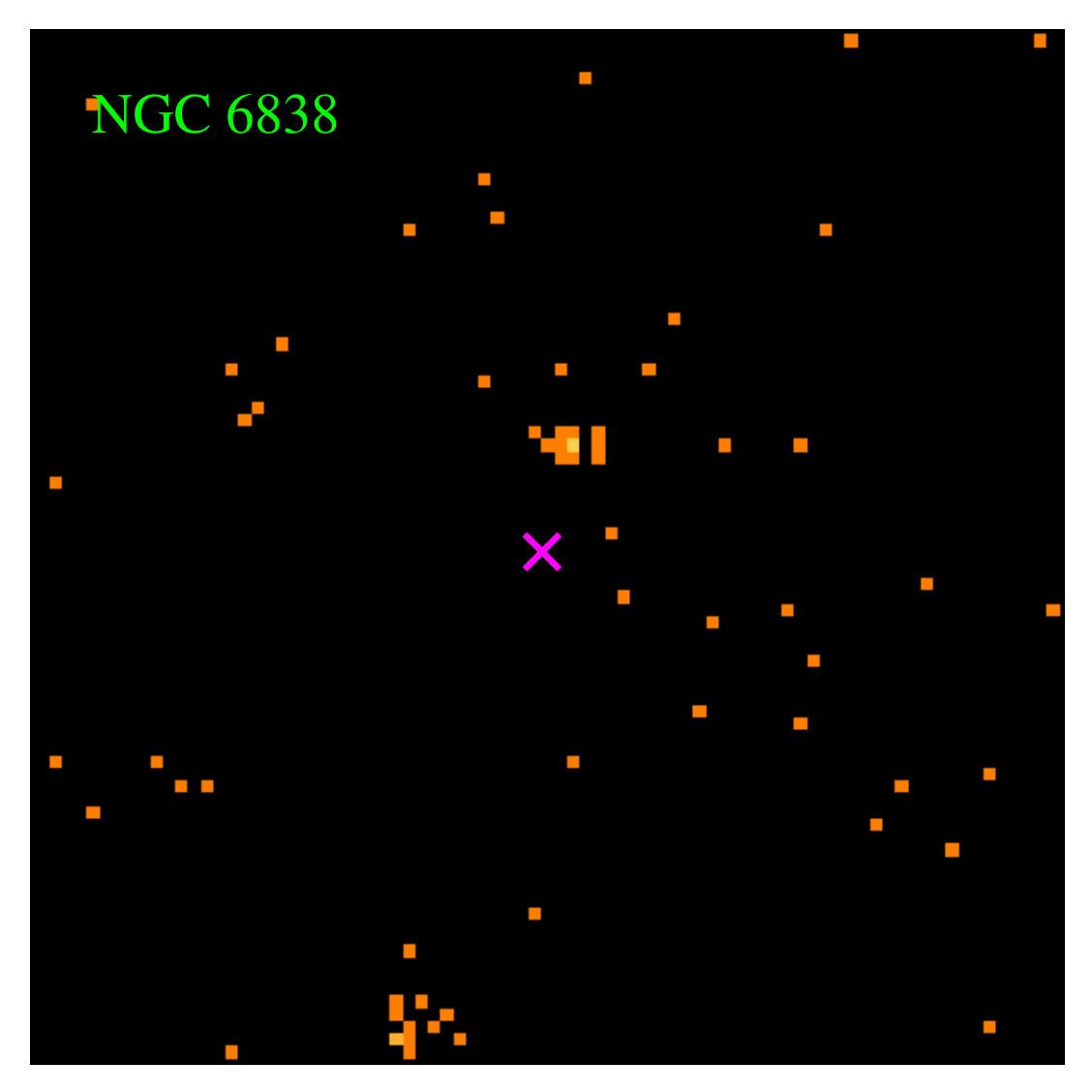}{0.2\textwidth}
        \afig{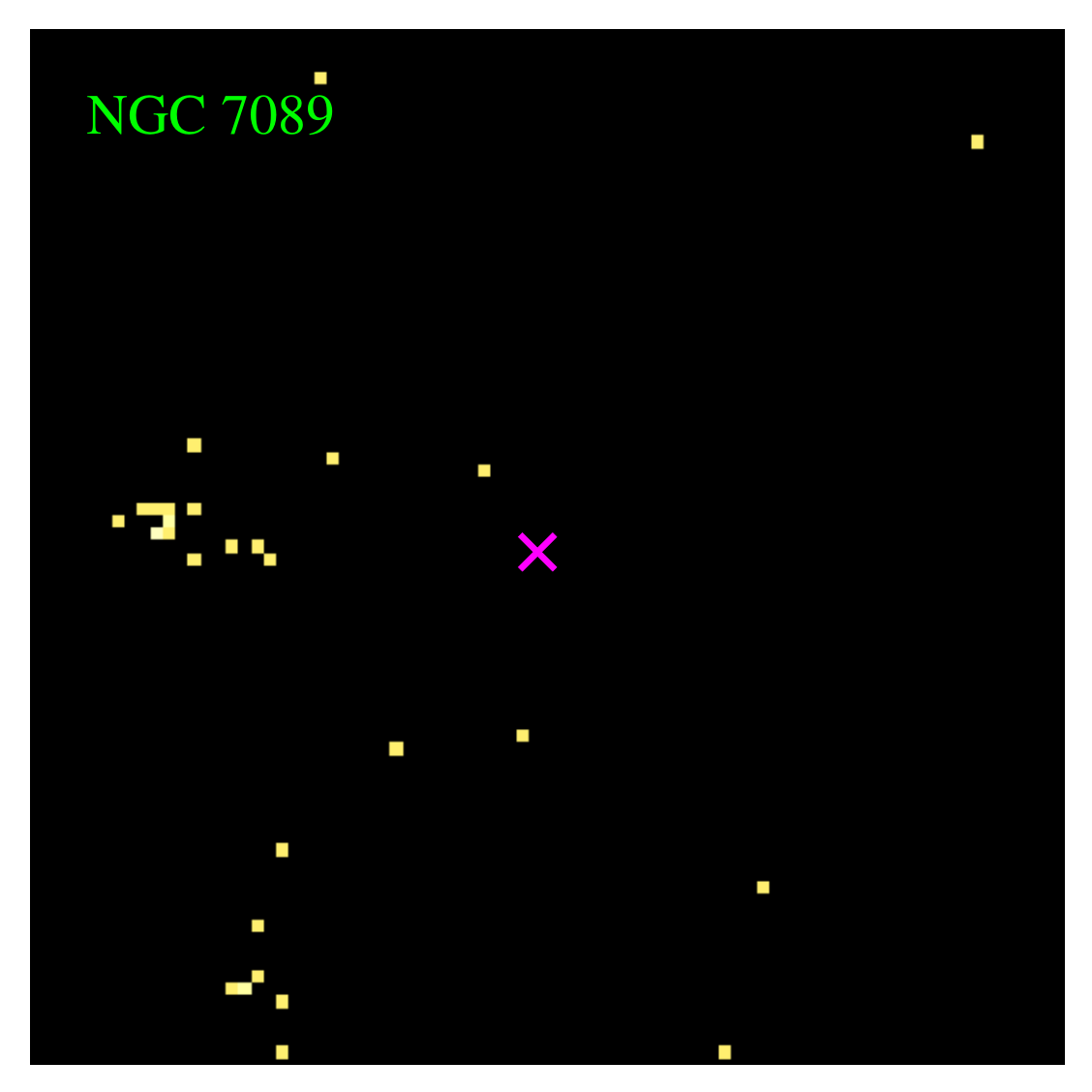}{0.2\textwidth}
        \afig{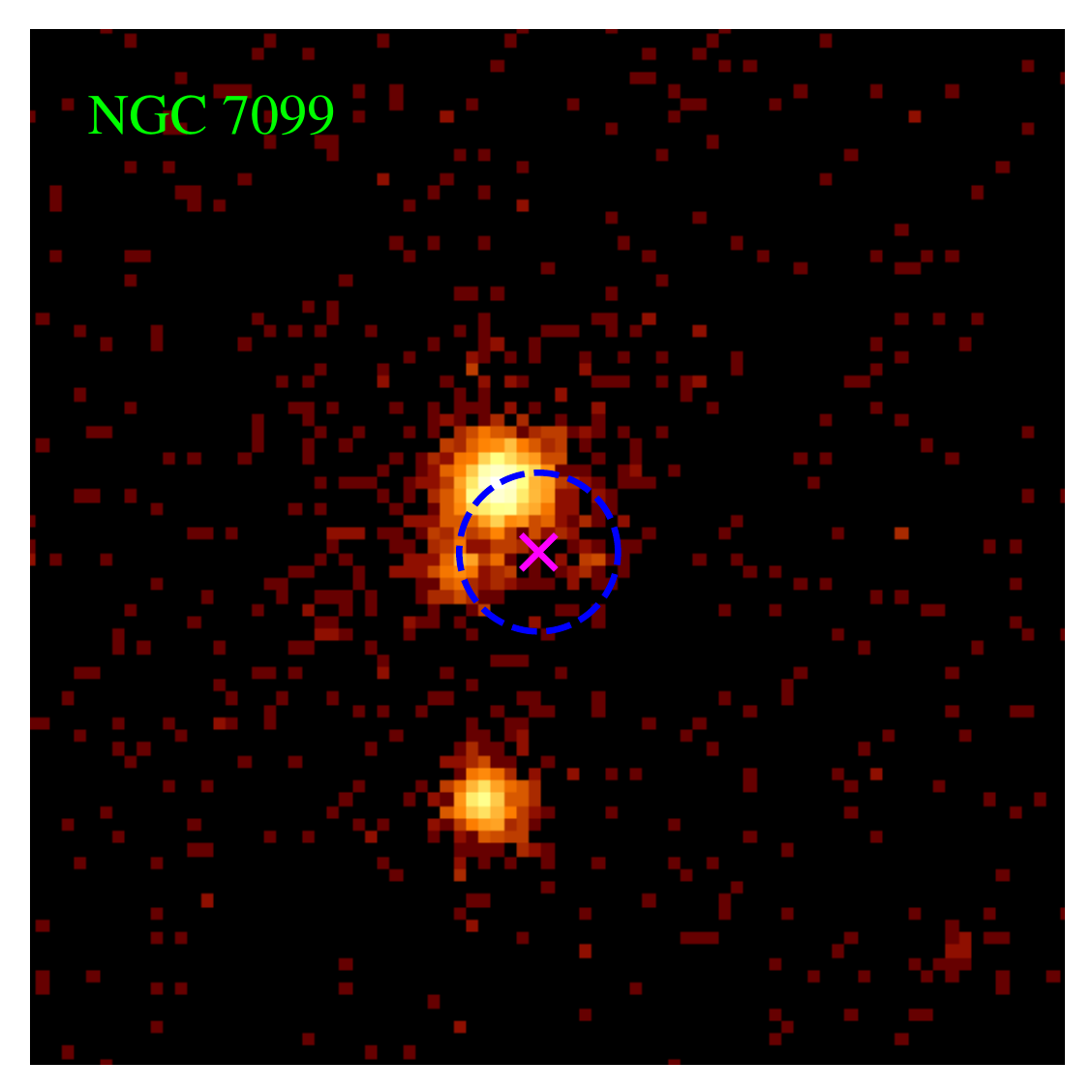}{0.2\textwidth}
    }
	\contcaption{}
    \end{figure*}
    
\bsp	
\label{lastpage}
\end{document}